\RequirePackage[T1]{fontenc}
\documentclass[12pt]{article}

\usepackage[height=8.85in,width=6.45in]{geometry}

\usepackage[utf8]{inputenc}
\usepackage{amsmath}
\usepackage{amssymb}
\usepackage{mathtools}
\usepackage{physics}
\numberwithin{equation}{section}
\usepackage{slashed}
\usepackage{braket}
\usepackage[svgnames]{xcolor}
\usepackage{soul}
\usepackage[nolist,printonlyused]{acronym}
\def\acroswithtooltips{1}
\if\acroswithtooltips1
\usepackage{pdfcomment}
\newcommand{\tooltipacro}[2]{\acro{#1}[\pdftooltip{#1}{#2}]{#2}}
\fi

\hypersetup{colorlinks,citecolor=DarkGreen,linkcolor=FireBrick,linktocpage}

\usepackage{varioref}
\usepackage[nameinlink,capitalise]{cleveref}

\crefname{table}{table}{tables}
\crefname{table}{Table}{Tables}
\crefname{figure}{figure}{figures}
\crefname{figure}{Figure}{Figures}

\usepackage{cite}
\usepackage{graphicx}
\usepackage{tikz}
\usepackage{tikz-cd}
\usepackage{times}
\usepackage{courier}
\usepackage{bm}
\usepackage{subfig}
\usepackage{ascmac}
\usepackage{tikzsymbols}
\usepackage{tensor}

\usepackage{mdframed}

\makeatletter
\AddToHook{cmd/appendix/before}{\def\cref@section@alias{appendix}}
\makeatother

\renewenvironment{figure}[1][]{
  \begin{originalfigure}[#1]
    \begin{mdframed}[linecolor=black!0,backgroundcolor=black!1]
}{
    \end{mdframed}
  \end{originalfigure}
}
\renewenvironment{table}[1][]{
  \begin{originaltable}[#1]
    \begin{mdframed}[linecolor=black!0,backgroundcolor=black!1]
}{
    \end{mdframed}
  \end{originaltable}
}

\newenvironment{eqaed}
    {\begin{equation}
    \begin{aligned}
    }
    { 
    \end{aligned}
    \end{equation}
    \ignorespacesafterend
    }

\definecolor{RAgreenColor}{cmyk}{0.64,0,0.95,0.40}

\begin{document}

\begin{titlepage}

\vskip 3cm

\renewcommand*{\thefootnote}{\fnsymbol{footnote}}

\begin{center}

{\Large \bfseries Scale-separated quasi-dS and the Swampland}

\vskip 1cm
Rafael \'{A}lvarez-Garc\'{i}a$^{1,2,}$\footnote{\url{ralvarezgarcia@fas.harvard.edu}}, Ivano Basile$^{1,3,}$\footnote{\url{ivanobasile@fas.harvard.edu}}, Shi Chen$^{1,}$\footnote{\url{shichen@g.harvard.edu}} and Cumrun Vafa$^{1,}\footnote{\url{vafa@g.harvard.edu}}$
\vskip 1cm

\begin{tabular}{ll}
$^1$ & \emph{Jefferson Physical Laboratory, Harvard University}\\ & \emph{Cambridge, MA 02138, USA} \\
$^2$ & \emph{Department of Physics, Northeastern University}\\
& \emph{Boston, MA 02115, USA} \\
$^3$ & \emph{INFN sezione di Torino}\\
& \emph{via Pietro Giuria 1, 10125 Torino, Italy}

\end{tabular}

\end{center}

\begin{abstract}
   \noindent Ideas from the Swampland program (dS conjecture and TCC) suggest that it is impossible to obtain solutions of string theory/M-theory in the {\it weak-coupling regions} describing our current universe, with $V>0$, $\norm{\nabla V}/V < 2/\sqrt{d-2}$ and $V \ll m^2$ (scale-separated ``quasi-dS") where $m$ is the mass of the light tower. We test this expectation in M-theory compactifications and find that even after including the quantum effect of Casimir energy, for hyperbolic and (rigid) Riemann-flat manifolds, de Sitter critical points under EFT control are excluded: the slope of the potential cannot drop below the asymptotic value predicted by the TCC $\norm{\nabla V}/V\gtrsim 2/\sqrt{d-2}$ except perhaps in an order-one region in the bulk of moduli space, where EFT control is lost. At the parametric level, our results hold at one-loop level in supergravity; using the scalar-field Casimir energy as a proxy, we also obtain explicit numerical bounds to provide a ballpark estimate of how much control one can expect to achieve in such settings. 
\end{abstract}

\end{titlepage}

\setcounter{tocdepth}{2}
\tableofcontents

\renewcommand*{\thefootnote}{\arabic{footnote}}
\setcounter{footnote}{0}


\section{Introduction}\label{sec:introduction}

The quest for a realistic model of our universe within \ac{UV}-complete quantum gravity is paramount to connect fundamental physics to observations, and despite decades of effort it remains among the most important open problems of theoretical physics. String/M-theory provides a wealth of controlled constructions, which led to a great deal of insight into the general principles behind \ac{UV}-complete quantum gravity. Distilling, motivating and connecting these principles is one of the goals of the Swampland program \cite{Vafa:2005ui, Brennan:2017rbf, Palti:2019pca, vanBeest:2021lhn, Grana:2021zvf, Agmon:2022thq}. Among the key lessons we have learned from this research endeavor is that \emph{bona fide} \ac{UV}-completable \acp{EFT} of gravity are few and far between---in a sense, they form a measure-zero subset of the \acp{EFT} that survive purely \ac{IR} constraints. \\

\noindent For the purposes of finding a realistic model of our universe, this is a double-edged sword. On the one hand, it tells us that the pool of options we need to search through is infinitely smaller than what one could have naively imagined. On the other hand, the decades-long difficulties in realizing a (metastable) \ac{dS} state in string theory with the tools at our disposal \cite{Kachru:2003aw, Burgess:2003ic, Balasubramanian:2005zx,Westphal:2006tn, Danielsson:2009ff, Cicoli:2012fh, Moritz:2017xto, Cordova:2018dbb, Kallosh:2018nrk, Bena:2018fqc, Gautason:2018gln, Blaback:2019zig, Hamada:2019ack, Cordova:2019cvf, Carta:2019rhx, Gautason:2019jwq, Montero:2020rpl, Brahma:2020tak, McAllister:2024lnt, Chakravarty:2024pec, Brahma:2024wtt, Dasgupta:2025ypg} point to the idea that, if a realistic configuration is indeed part of the landscape, it might be located outside of the limited corners that we understand physically and control mathematically\footnote{An alternative to a gravitational \ac{EFT} with a quasi-\ac{dS} configuration is dark bubble cosmology \cite{Banerjee:2018qey, Banerjee:2019fzz, Banerjee:2020wix, Banerjee:2021yrb, Danielsson:2021tyb, Danielsson:2022lsl, Danielsson:2022fhd, Basile:2023tvh, Banerjee:2023uto, Danielsson:2023alz, Danielsson:2025ffk, Danielsson:2026wqp, Danielsson:2026qch}, which however seems to have some issues \cite{Basile:2025lek, Basile:2025lwx} at least with standard boundary conditions \cite{Danielsson:2025ffk, Danielsson:2026wqp}.} \cite{Obied:2018sgi, Bedroya:2019snp, Bedroya:2022tbh, Lust:2022lfc, Bena:2024are}. \\

\noindent Given this state of affairs, it was natural to expand the search to more general viable cosmologies \cite{Agrawal:2019dlm,Andriot:2023wvg,Shiu:2023fhb,Shiu:2023nph,Andriot:2024jsh,Casas:2024xqy,Andriot:2026lac}. Strikingly, this attitude was recently given stronger support by observations \cite{DESI:2025zgx}. In this context, Swampland ideas suggest a more general coupling between dark matter and dark energy \cite{Agrawal:2019dlm} which is realized in the context of the dark dimension scenario \cite{Montero:2022prj} and leads to a predictive model which anticipated the observations of DESI and fits very well with the current data \cite{Bedroya:2025fwh}. \\

\noindent These tantalizing developments notwithstanding, the basic question remains open: \emph{can we find examples of positive vacuum energy which are scale-separated and allow accelerated expansion---in string theory?} More precisely, a description of our universe calls for a region of field space where, in Planck units,
\begin{eqaed}\label{eq:quasi-dS_conditions}
    V > 0\,, \qquad \norm{\frac{\nabla V}{V} }<\frac{2}{\sqrt{d-2}}\,, \qquad V \ll m^2\,,
\end{eqaed}
with $m$ the mass scale of a light tower. We refer to such a configuration as scale-separated ``quasi-\ac{dS}''. Ideas from the Swampland program, namely the \ac{dS} conjectures \cite{Obied:2018sgi} and the \ac{TCC} \cite{Bedroya:2019snp}, suggest that this is impossible in the weakly coupled regions of the landscape, where the potential is too steep: asymptotically, the \ac{TCC} predicts $\norm{\nabla V}/V \gtrsim 2/\sqrt{d-2}$ in $d$ dimensions. One approach has been to argue for the existence of unstable \ac{dS} points using symmetries (including duality symmetries) \cite{Chen:2025rkb, Baykara:2026jzs}. Even though these provide the first concrete examples of (unstable) \ac{dS} in string theory, they lead to $V\sim M_\text{pl}^d$ and so are not scale-separated, and we have to look for additional ingredients to realize scale-separated dS. \\

\noindent Finding solutions to \cref{eq:quasi-dS_conditions} if we flip the sign of $V<0$ is not difficult if we relax the condition of being scale-separated and allow $|V|\sim m_\text{KK}^2$. In particular classical solutions of supergravity give many such solutions as is familiar in the context of supersymmetric \ac{AdS} vacua and holography. If we relax the requirement of supersymmetry, it has been shown that one can even get scale-separated unstable \ac{AdS} \cite{Luca:2022inb, Aparici:2025kjj}. So one might think that it should not be difficult to do the same for $V>0$.  However this is not the case and there are a number of no-go results \cite{Kutasov:2015eba, Maldacena:2000mw, Basile:2020mpt}. The difficulty seems to be finding good sources of negative energy in string theory which is a unitary theory.  These questions motivated attempts to begin a systematic investigation of simple supergravity compactifications supplemented by the quantum effect of \emph{Casimir energy} \cite{Luca:2022inb, Bento:2025wuu, DallAgata:2025jii, ValeixoBento:2025qih, Aparici:2025kjj, Aparici:2026wdm, Aparici:2026qme} which could lead to potentially unbounded negative energy contributions. The advantages of this approach are clear: conceptually, the setup involves well-understood ingredients of \ac{EFT}, without the complications introduced by branes, orientifolds, the associated warping and singularities, non-perturbative effects and so on. Technically, the Casimir energy is in principle a concrete and calculable quantity; moreover, this paradigm is applicable to a much wider range of compactification geometries (with or without fluxes) than what is allowed when one seeks to preserve low-energy supersymmetry. Unfortunately, none of these solutions have so far led to a point where the \ac{EFT} can be trusted, nor to cases where $V$ is scale-separated. Nevertheless, one can carry out a more complete search, to see whether for example arbitrarily large topological complexity (like large-genus Riemann surfaces) changes the situation. In this paper we perform such a search for M-theory compactifications on negative-curvature manifolds, as well as on \acp{RFM}, including the effect of Casimir energy. This brings us to the main question motivating this work:

\begin{center}
    \emph{Can a scale-separated \ac{dS} solution supported by \\ Casimir energy be argued to exist for any manifold?}
\end{center}

\noindent To investigate this question, we first develop a general intuition for how the Casimir energy behaves as a function of the compactification geometry. We identify three key ways that the Casimir energy can enhance \ac{EFT} control for the resulting \ac{dS} solutions of the two-derivative quantum effective action. The first two ways are the most natural, namely increasing the curvature or decreasing the length of short closed geodesics. However, these two mechanisms highlight a clear risk: enhancing the Casimir energy to improve control in this way also directly impacts the control parameters of the \ac{EFT} in the opposite fashion! This structural tension is at the heart of the quantitative results that we will present. The third possibility to improve \ac{EFT} control in these settings is to find families of manifolds (and critical points of the effective potential) whose global topology enhances the Casimir energy without affecting the curvatures and geodesic lengths directly. At first glance, there seems to be no issue: high-degree covers or large-genus Riemann surfaces are obvious candidates. In particular, one can imagine sufficiently complicated topology having many small cycles, each of which is much bigger than the Planck length. If we have enough growth for such cycles in topology, the contribution to the Casimir energy could be large enough to avoid losing \ac{EFT} control due to Planckian cycles. \\  

\noindent However, this turns out not to work. We study in detail the case of Riemann surfaces, higher-dimensional hyperbolic manifolds, and finally \acp{RFM}. In all these cases there are upper bounds on the control parameters, meaning that global topology does not help, at least within these classes of manifolds. In particular, for hyperbolic and rigid Riemann-flat manifolds, controlled \ac{dS} critical points are excluded even after including the Casimir energy. More specifically, for our quantitative bounds we work with the Casimir energy produced by a scalar field, with the expectation that its complete supergravity counterpart behaves similarly; in \cref{app:bounding_sugra_casimir} we provide a parametric extension to the actual supergravity spectrum. In this setting, we establish numerical upper bounds on the average curvature radius and the length of the shortest cycles in eleven-dimensional Planck units for M-theory compactifications on Riemann surfaces, higher-dimensional hyperbolic manifolds and rigid \acp{RFM} (by which we mean those for which a critical point is forced by enhanced symmetry). For general \ac{RFM} critical points, we also obtain a no-go statement, but in a parametric form. In other words, we were not able to find an explicit useful numerical bound, although it does exist. Finally, we extend the analysis to the slope of the effective potential away from critical points. The conclusions are similar: in the controlled regime, the slope cannot fall below the asymptotic value $2/\sqrt{d-2}$ predicted by the \ac{TCC} \cite{Bedroya:2019snp}. It can do so only within an order-one region in the bulk of moduli space, where we obtain upper bounds on the \ac{EFT} control parameters, so that control is lost. \\

\noindent Our results confirm, at least in this setting, the Swampland expectation based on the \ac{dS} conjecture and the \ac{TCC} that the conditions in \cref{eq:quasi-dS_conditions} cannot be met in a controlled regime. They compel us to expand the search for scale-separated quasi-\ac{dS} solutions---no matter how unstable---to more exotic corners of the landscape. The general mechanism we highlight entails a basic tension between the behavior of the Casimir energy and the features of a controlled \ac{dS} solution. It is perhaps conceptually intriguing that genuinely stringy, or otherwise \ac{UV} ingredients may be required to complete the quest for a realistic model; nonetheless, there are considerable difficulties in undertaking this endeavor. Whatever the right path will turn out to be, it will likely hold deep and exciting lessons on quantum gravity. \\

\noindent \textbf{Structure of the paper.} For the reader's convenience, our presentation is thematically divided in two parts:

\begin{itemize}
    \item \textbf{Overview:} In \cref{sec:setup} we describe the setup and derive the properties of \ac{dS} solutions supported by Casimir energy. In \cref{sec:casimir_general} we present a geometric and physical picture of how the Casimir potential behaves for general manifolds. With some simplifying assumptions, we prove some general bounds which highlight the enhancement loopholes discussed above. We also briefly outline how to extend these results to the Casimir energy stemming from the supergravity spectrum.

    \item \textbf{Numerical and parametric bounds:} The systematic investigation of the aforementioned loopholes is divided into three concrete settings: in \cref{sec:casimir_riemann_surfaces} we discuss Riemann surfaces, starting from the rigid case and then moving on to the general case. In \cref{sec:hyperbolic_manifolds} we discuss higher-dimensional hyperbolic manifolds, which are rigid but feature a richer geometry. After some brief considerations on Ricci-flat manifolds, in \cref{sec:flat_manifolds} we study \acp{RFM} in detail. Once again, we begin with the rigid case, and then generalize the no-go result (albeit this time only parametrically) to the general case. In \cref{sec:potential_slopes} we forsake critical points and analyze the whole field space, seeking controlled regions where the potential slope lies below the \ac{TCC} value. Even in this context, for Riemann surfaces and \acp{RFM}, we find an obstruction to a controlled \ac{EFT} description.
\end{itemize}

\noindent In \cref{sec:conclusions} we end with some concluding remarks about what the present work suggests about pathways to construct scale-separated quasi-\ac{dS} vacua. We relegate some technical details to the appendices. In \cref{app:packing} we explain the packing methods employed to bound the sums over geodesics in the Casimir energy. In \cref{app:laplacian_bounds_geodesic_variations} we derive some results on moduli-space gradients for Riemann surfaces, used to bound the gradient of the Casimir potential. In \cref{app:spectral_analysis_supergraviton_fluctuations}, we provide a spectral analysis for supergravity one-loop determinants, in order to extend the scalar-field case discussed in the main text to parametric bounds for supergravity; we do so in \cref{app:bounding_sugra_casimir}.

\section{de Sitter solutions in supergravity}\label{sec:setup}

Before introducing the sort of \ac{dS} solutions we have discussed above, it is instructive to draw a comparison with the more familiar \ac{AdS} solutions supported by fluxes. Consider a $D$-dimensional \ac{EFT} with an $n$-form field strength $F_n$. The relevant part of the classical effective action is given by
\begin{eqaed}\label{eq:AdS_flux_eff_action}
    S_\text{classical} = \int d^Dx \sqrt{-g_D} \left(M_D^{D-2} R_D - M_D^{D-2n} \norm{F_n}^2 \right) ,
\end{eqaed}
where $M_D$ denotes the $D$-dimensional Planck scale. The standard approach is to seek a solution of the form $\text{AdS}_d \times M_n$, where $M_n$ is a compact Einstein manifold of positive Ricci curvature $R_{ij} = \frac{n-1}{r^2} \, g_{ij}$. The curvature radius $r$ and the \ac{AdS} radius $L$ are dynamically determined by the flux number $\int_M F_n = N$, and the resulting solution is controlled for large $N$. Indeed, one finds the Freund-Rubin solution
\begin{eqaed}\label{eq:AdS_Freund-Rubin_solution}
    r = \frac{n-1}{d-1} \, L\,, \qquad M_D L \propto N^\frac{1}{n-1}
\end{eqaed}
up to a numerical prefactor. \\

\noindent We can get a more intuitive understanding of this solution by looking at the effective potential $V$. Rescaling to the $d$-dimensional Einstein frame and working in units of the $d$-dimensional Planck scale $M_d$, we have
\begin{eqaed}\label{eq:AdS_Freund-Rubin_eff_potential}
    \frac{V}{M_d^d} \propto (M_d r)^{-\frac{dn}{d-2}} \left( \frac{N^2}{(M_d r)^n} - n(n-1) \, (M_d r)^{n-2} \right) ,
\end{eqaed}
highlighting the competition between curvature and flux contributions to the energy. In particular, the flux term gives a positive contribution and dominates at small radius, while the curvature term gives a negative contribution and dominates at large radius. This produces the familiar \ac{AdS} minimum for the radius. This simple structure is rather suggestive: What if we could flip the sign of these terms? More precisely, how can we find a similar two-term structure with opposite signs? The curvature term can be modified simply by choosing an internal manifold with negative curvature, as we shall explain below. In order to have a negative source of energy that dominates at small radius, the natural candidate is the Casimir energy. This suggests that the combination of negative curvature and Casimir energy (produced by breaking supersymmetry via compactification) can lead to a \ac{dS} maximum in the radial direction. This idea was indeed explored in \cite{DeLuca:2021pej}; in the following we introduce a simpler version of this construction, by averaging over the internal manifold. For the sake of concreteness and to avoid additional scalar fields to stabilize, we work in eleven-dimensional M-theory. \\

\noindent Concretely, we consider compactifications of M-theory on uniform-curvature (Einstein) hyperbolic manifolds supported by Casimir energy. 
A \emph{negative Einstein manifold} is a Riemannian manifold whose Ricci tensor obeys\
$$R_{ij}=-\frac{n-1}{r^{2}}g_{ij}\,,$$
 where \(n\) is the dimension and \(r>0\) sets the curvature scale. This fixes the averaged curvature in every direction, but does not require every sectional curvature to be negative. A \emph{hyperbolic manifold} satisfies the stronger condition \
 $$R_{ijkl}=-\frac{1}{r^{2}}(g_{ik}g_{jl}-g_{il}g_{jk})\,,$$
 meaning that every sectional curvature equals \(-1/r^{2}\). Both have scalar curvature 
 $$R=g^{ij}R_{ij}=-\frac{n(n-1)}{r^{2}}\,,$$
 but only the hyperbolic condition fixes the full curvature tensor. \\

\noindent In supergravity the Casimir energy of a non-supersymmetric compactification on flat manifolds is \ac{UV}-insensitive due to higher-dimensional supersymmetry. On curved manifolds, the local breaking of supersymmetry can produce \ac{UV}-divergences, which we renormalize via zeta-function regularization. We disregard internal inhomogeneities: although in general they will appear by backreaction due to the Casimir energy-momentum tensor, in the controlled regime one expects a leading-order homogeneous solution with Einstein metric. However, in order to exclude this scenario, we can also consider more general metrics and discuss the purely lower-dimensional, dimensionally reduced equations in terms of the averaged internal curvature. If solutions to those cannot be under control, neither will the full higher-dimensional solutions, if any. \\

\noindent To see what the control conditions are, consider a general compactification of $D = d + n$ dimensions on a negative-curvature manifold $M_n$. (As anticipated, in \cref{sec:casimir_riemann_surfaces,sec:hyperbolic_manifolds} we will take such $M_n$ to be hyperbolic.) Letting its curvature radius be $r$, we parametrize

\begin{eqaed}\label{eq:parameters}
    & R = - \, \frac{n(n-1)}{r^2}\,, \\
    & \rho_\text{Cas} = - \, \frac{A}{r^D}\,, \\
    & m_\text{KK} = \frac{B}{r}\,, \\
    & \text{Vol}(M_n) = v \, r^n\,.
\end{eqaed}
The dimensionless parameters $v,A,B$ depend on the geometry of the internal manifold: $v$ is the volume in units of the curvature radius $r$; $A$ is the averaged Casimir energy density in units of $r$; finally, $B$ is the \ac{KK} mass gap in units of $r$, which is related to the spectral gap of the Laplacian. Once again, in the following we denote with $M_{D,d}$ the Planck mass in $D,d$ dimensions respectively, which enter the Einstein-Hilbert term. We look for solutions of the form $\text{dS}_d \times M_n$, where the \ac{dS} factor has curvature radius $L$. Evaluated on this ansatz, the effective action is\footnote{With respect to the usual normalization, we rescaled $M_D^{D-2}/2$ to $M_D^{D-2}$ for convenience. All control parameters should thus be rescaled by a factor of $2^{\frac{1}{9}} \approx 1.08$ to be compared with the standard conventions. This will not matter for our arguments.}
\begin{eqaed}\label{eq:eff_action}
    S_\text{eff} & = \int d^Dx \sqrt{-g_D} \left(M_D^{D-2}R_D - \rho_\text{Cas} \right) \\
    & \propto v L^d r^n \left(M_D^{D-2} \, \frac{d(d-1)}{L^2} - M_D^{D-2} \, \frac{n(n-1)}{r^2} + \frac{A}{r^D} \right) .
\end{eqaed}
Extremizing over the \ac{dS} radius $L$ and $r$, one finds a \ac{dS} maximum with scalings
\begin{eqaed}\label{eq:solution}
    M_D r \propto M_D L \propto A^\frac{1}{D-2}\,, \qquad m_\text{KK} L \propto B\,,
\end{eqaed}
from which it follows that, in units of $M_d^{d-2} = M_D^{D-2} v \, r^n$,
\begin{eqaed}\label{eq:alpha_relation}
    \frac{m_\text{KK}}{M_d} \propto \left(\frac{V_*}{M_d^d}\right)^{\!\alpha}\,, \quad \alpha = \frac{1}{2}\left(1 - (d-2) \, \frac{\log B}{\log v A}\right)\,,
\end{eqaed}
where $V_*$ is the critical value of the potential and $\alpha$ is defined asymptotically for a family of solutions. Therefore, scale separation is controlled by $B$, \ac{EFT} corrections are controlled by $A$, and the precise scaling exponent $\alpha$ is controlled by $v$.  Another consideration is that the Planck scale generally receives a finite renormalization beyond tree level; if a solution is to be controlled, once we rescale everything to Planck units this effect ought to be subleading.   We will see that for M-theory compactification on negatively curved Einstein manifolds there is no parametric scale separation, as $A\rightarrow \infty$ and $\alpha=\frac{1}{2}$. \\

\noindent Before moving on, we observe that even when a full homogeneous\footnote{We consider unwarped geometries in this paper. It would be interesting to study how the effects of warping could change the story qualitatively \cite{DeLuca:2021pej}.} $D$-dimensional solution does not exist, in order to show that a more general solution would be uncontrolled it is sufficient to perform a dimensional reduction over $M_n$. In this more general case, the Casimir parameter is to be interpreted as the average over $M_n$, and similarly the radius $r$ entering the solution in \cref{eq:solution} is replaced by the average curvature radius
\begin{eqaed}\label{eq:average_radius}
    - \frac{n(n-1)}{r^2} = \frac{1}{\text{Vol}(M_n)} \int_{M_n} R \, \text{vol}_{M_n}\,,
\end{eqaed}
where, here and in the following, $\text{vol}_{M_n}$ denotes the volume form on $M_n$ induced by the metric.

\subsection{Casimir energy without scale separation}\label{sec:casimir_no_scale}

In the above analysis, taking $A$ to be constant \emph{a priori} requires scale separation, because the Casimir energy density on $\mathbb{R}^{d-1,1} \times M_n$ is only an approximation of that on $dS_d \times M_n$ when $L \gg r$. We vary the geometry while maintaining the physical couplings and normalization conditions fixed. This prescription is implemented using zeta regularization with a fixed reference mass $\mu$; see, e.g., \cite{Vassilevich:2003xt}\footnote{A geometry-dependent renormalization scale may be used to evaluate the same action as long as the compensating dependence of the couplings and normalization is retained. For the scalar prescription used below, replacing $\lambda/\mu^2$ by $r^2\lambda$ with the rest of the action unchanged adds a $\log(\mu r)$ term and changes its common-scale derivative by $1$. Such a subtraction cannot be an ordinary local counterterm: on a connected finite cover, local integrals grow with the covering degree, whereas the scalar still has one constant zero-mode. A cancellation involving other fields or zero-mode measures may occur, but must first be derived from the complete path integral, which we do not do here.}. \\

\begin{figure}[!ht]
    \centering
    \includegraphics[width=0.75\linewidth]{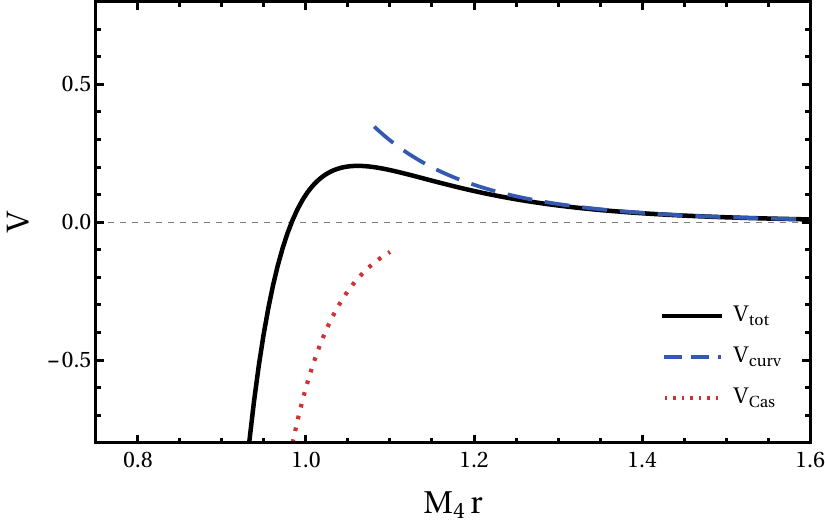}

    \caption{A depiction of the Einstein-frame effective potential for the volume radius $r$ (fixing the other moduli, if any) for a Casimir compactification on a negatively curved (Einstein) manifold. For concreteness, we chose $d=4, n=7$ compactifications from eleven to four dimensions. The plot also shows how each term dominates for small or large $r$. A \ac{dS} critical point for the radius does exist in the two-derivative approximation, but is not under control nor scale-separated. The coefficients were chosen for visual clarity, but the general shape does not change when using the correct coefficients.}
    \label{fig:pat_plat}
\end{figure}

\noindent Define $x\equiv r/L$ and consider a common rescaling $(L,r)\mapsto(aL,ar)$, which leaves the ratio $x$ fixed. On the Euclidean continuation of the geometry $S^d\times M_n$, every non-zero scalar eigenvalue then transforms as $\lambda\mapsto a^{-2}\lambda$. Hence, the zeta function of the non-zero spectrum obeys $\zeta_{aL,ar}(s)=a^{2s}\zeta_{L,r}(s)$. In odd total dimension, as is relevant for our setup, the scalar heat trace has no constant term in its small proper time expansion. Subtracting the single constant zero-mode on the connected product introduces the constant term $-1$, such that $\zeta_\Delta(0)=-1$. For the determinant with the zero eigenvalue omitted, which we will denote by a prime, the rescaling shifts $\log\det'\Delta$ by $2\log a$. With the sign convention of \cref{eq:eff_action}, the scalar contribution is $-\frac{1}{2}\log\det'(\Delta/\mu^2)$ and therefore contains $-\log(\mu r)$, where $\mu$ is a fixed reference mass. More generally, allow for a coefficient $b$ multiplying this term. As long as the number of zero-modes remain fixed, so is $b$, which for a real scalar is $b=1$. With $\omega_d=\operatorname{Vol}(S^d_1)$ and $A(x)$ defined by the spectrum at unit internal radius, the effective action reads
\begin{eqaed}\label{eq:scaling_action}
	S_\text{eff}=\omega_d v x^{-d}\Big[(M_Dr)^{D-2}\big(d(d-1)x^2-n(n-1)\big)+A(x)\Big]-b\log(\mu r)\,.
\end{eqaed}
Extremizing with respect to $r$ at fixed $x$ gives
\begin{eqaed}\label{eq:x_extremum}
	d(d-1)x^2-n(n-1)=\frac{b x^d}{(D-2)\omega_d v(M_Dr)^{D-2}}\,.
\end{eqaed}
If the $b$ contribution were absent, \cref{eq:x_extremum} would have the simple algebraic solution
\begin{eqaed}\label{eq:x_nice_extremum}
x_*\equiv\sqrt{\frac{n(n-1)}{d(d-1)}}\,.
\end{eqaed}
However, for the scalar prescription, $b=1$ implies $x>x_*$ and hence $L/r<1/x_*$. We conclude that $L/r$ is bounded by a number depending only on $d$ and $n$, which excludes parametric curvature-scale separation. The stationary equation for $x$ at fixed $r$, together with \cref{eq:x_extremum}, fixes the radii:
\begin{eqaed}\label{eq:L_solution}
	(M_D L)^{D-2}&=x^{2-D}(M_D r)^{D-2} \\
	&=\frac{x^{-D}}{2(d-1)}\left[A(x)-\frac{x}{d}A'(x)+\frac{b x^d}{(D-2)\omega_d v}\right].
\end{eqaed}
The bounds below estimate both contributions in the expression above over the allowed stationary ratios. When flux-dependent terms in the supergravity kinetic operators are neglected, no sign assumption on $b$ is needed: \cref{app:bounding_sugra_casimir} bounds it uniformly and proves $x\to x_*$ whenever all required physical length scales become large.
So the absence of scale separation expressed by \cref{eq:x_extremum} modifies the control parameter depending on the additional derivative term. Since the Casimir energy depends on $L$, the role of $V$ is now played by the Hubble curvature $1/L^2$ in $d$-dimensional Planck units. This result can be understood visually as in \cref{fig:pat_plat}. \\

\noindent If the solution is parametrically controlled, so in particular $M_DL\gg1$, it makes sense to write a scaling relation of the form \cref{eq:alpha_relation}, and in this case we would have $\alpha = 1/2$ as befits a non-scale-separated solution. Inverting the logic, if it had been the case that $\alpha < 1/2$, we could have simply removed the dependence on $x$ in the argument of $A$, since scale separation would have allowed us to approximate it by the Casimir energy density on $\mathbb{R}^{d-1,1} \times M_n$. This would have removed the additional $A'$ term in \cref{eq:L_solution}. A posteriori, we see from \cref{eq:x_extremum} that this does not work, and scale separation cannot be achieved. However, the additional $A'$ term in \cref{eq:L_solution} will prove to be important when deriving bounds on \ac{EFT} control parameters in \cref{sec:casimir_riemann_surfaces,sec:hyperbolic_manifolds,sec:RFMs}. \\

\noindent \textbf{A comment on why fluxes do not help.} Adding $n$-form flux for $n<d$ preserves the existence of a \ac{dS} maximum, and one could \emph{a priori} tune it against the Casimir parameter $A$ to achieve scale separation. However, increasing the flux \emph{decreases} the critical radius, and thus this possibility requires that $A$ be very large. As we shall see, this is impossible; in particular, in \cref{sec:hyperbolic_manifolds} we cover the case $n=4$ in the flat-space approximation, which is the relevant case for a scale-separated \ac{dS} maximum supported by flux in M-theory. A more sophisticated approach is to thread several four-cycles with independent fluxes, but the resulting effective potential introduces additional moduli dependence and is not universal. Still, dominant balance rules out parametric control: if both curvature and Casimir energy are important, the solution (if any) will stabilize so that they are of the same order. Therefore, control still requires a large Casimir contribution, which we show to be impossible irrespective of fluxes. A possible loophole is thus the case in which the internal manifold is flat; in \cref{sec:RFMs} we investigate this possibility for Riemann-flat manifolds in detail, and we also derive a no-go result in this case.

\section{Generalities on Casimir potentials}\label{sec:casimir_general}

In the preceding section we discussed the role of the Casimir potential in concocting \ac{EFT} solutions with broken supersymmetry. For the purposes of this paper, we mostly focus on the Casimir potential produced by a scalar field, since the bounds we will obtain are not evaded by the partial cancellation between bosonic and fermionic polarizations of supergravity fields. On the other hand, each contribution could be enhanced relative to the scalar case due to curvature couplings, so we cannot use this as a rigorous replacement but merely as a numerical proxy. This provides a heuristic expectation on what numbers one is expected to obtain for the actual supergravity spectrum; in \cref{app:bounding_sugra_casimir} we will extend it to a parametric bound. In the settings described by \cref{eq:eff_action}, where the ingredients are the internal curvature and the Casimir energy, the parametric dependence of \cref{eq:solution} on the dimensionless coefficient $A$ shows that control requires $A \gg 1$. This turns out to also be the case with fluxes and flat manifolds \cite{Bento:2025wuu,Aparici:2025kjj,ValeixoBento:2025qih,Aparici:2026wdm,Aparici:2026qme}, as we shall discuss in detail in \cref{sec:flat_manifolds}. \\

\noindent As we have anticipated in \cref{eq:average_radius}, in general the internal space is not homogeneous enough to treat the full eleven-dimensional energy density as constant. However, a necessary condition for the existence of a controlled solution in higher dimensions is the existence of a controlled solution in the dimensionally reduced \ac{EFT}. This leads us to consider $A$ as (minus) the averaged energy density over the internal space $M_n$, so that $\mathcal{C}_\text{dS} = \text{Vol}(M_n)A$ is given by the (regularized) Schwinger integral of the (trace of the) heat kernel of the scalar Laplacian,
\begin{eqaed}\label{eq:general_casimir_expression}
    \mathcal{C}_\text{dS} = \frac{1}{2} \int_0^\infty \frac{dt}{t} \, \frac{K_{S^d \times M_n}(t)}{\text{Vol}(S^d)}\,,
\end{eqaed}
where we assumed a solution of the form of \cref{eq:solution}. In the scale-separated limit, the \ac{dS} contribution (which is Wick-rotated to $S^d$ in \cref{eq:general_casimir_expression}) can be approximated by that of flat space. As shown in \cref{eq:solution}, in the presence of curvature this is impossible, and one should really account for the \ac{dS} curvature in the heat kernel. Nonetheless, as we will show in \cref{sec:casimir_riemann_surfaces,sec:hyperbolic_manifolds,sec:flat_manifolds}, this contribution does not affect the qualitative conclusions. An intuitive explanation for this is that replacing flat space with the sphere produces exponentially suppressed massive modes, together with an effectively lower-dimensional zero-mode. The former do not qualitatively affect the control issues stemming from the Casimir potential, while the latter behaves similarly to its higher-dimensional counterpart with some quantitative scaling differences. Mathematically, this intuition will be realized as numerical bounds relating the heat kernel on $S^d$ and its flat-space counterpart, together with a separate treatment of the zero-mode contribution. In light of these considerations, to keep this section more readable and focused on the general conceptual picture, we shall temporarily approximate \cref{eq:general_casimir_expression} with its flat-space counterpart
\begin{eqaed}\label{eq:general_casimir_expression_flat}
    \mathcal{C} = \frac{1}{2(4\pi)^\frac{d}{2}} \int_0^\infty \frac{dt}{t^{1+\frac{d}{2}}} \, K_{M_n}(t)\,.
\end{eqaed}
The Schwinger integral in \cref{eq:general_casimir_expression_flat} ought to be regularized, reflecting the \ac{UV}-sensitivity of vacuum energy. When the higher-dimensional theory is supersymmetric, the full Casimir energy is \ac{UV}-insensitive, and the integral converges. The upshot is that the scalar-field counterpart of the full Casimir energy of supergravity is accompanied by a subtraction of the first $\lfloor D/2 \rfloor+1$ local terms of the small-$t$ heat kernel expansion \cite{Vassilevich:2003xt}. In the flat-space approximation, the same thing can be achieved by zeta-function regularization, which leads to the spectral zeta function
\begin{eqaed}\label{eq:zeta_reg_casimir_flat_general}
    \mathcal{C} = \frac{\Gamma(-\frac{d}{2})}{2(4\pi)^\frac{d}{2}} \, \zeta_{M_n}\!\left(-\frac{d}{2}\right) .
\end{eqaed}
The purpose of this section is to develop a physical intuition for these expressions, and a general picture of what mechanisms can cause $\mathcal{C}$ to grow large or prevent it from getting large. To this end, we will discuss two ways of expressing \cref{eq:general_casimir_expression_flat} that highlight different aspects of the relevant physics.

\subsection{Path-integral expression}\label{sec:path-integral_expression}

To begin with, it is insightful to recast the heat kernel of \cref{eq:general_casimir_expression_flat} as a worldline Euclidean path integral on the circle,
\begin{eqaed}\label{eq:path_integral_heat_kernel}
    K_{M_n}(t) = \int_{x(0)=x(1)}\mathcal{D}x \, e^{-\frac{1}{4t}\int_0^1 \, g(\dot x, \dot x) \, d\tau} 
\end{eqaed}
where we extracted the Schwinger proper-time parameter $t$ by rescaling the worldline coordinate $\tau$. Here $g$ is the metric on $M_n$, and the worldline action in \cref{eq:path_integral_heat_kernel} is that of a one-dimensional non-linear sigma model. This theory is interacting unless $M_n$ is flat. \cref{eq:path_integral_heat_kernel} is the Lagrangian counterpart of the canonical partition function, which can be used to glean some insights in certain cases \cite{Harlow:2021trr, Kang:2022orq, Benjamin:2023qsc}. \\

\noindent One can express \cref{eq:path_integral_heat_kernel} in terms of the universal cover $\widetilde{M_n} \to M_n$ using the method of images. This requires the full heat kernel $K_{\widetilde{M_n}}(y,y';t)$, whose trace gives the function entering \cref{eq:general_casimir_expression_flat}. One has\footnote{Strictly speaking, \cref{eq:sum_over_images} should contain an integral over a fundamental domain for $\widetilde{M_n} \to M_n$. We will retain the convenient abuse of notation, occasionally pointing it out for clarity.}
\begin{eqaed}\label{eq:sum_over_images}
    K_{M_n}(t) = \int_{M_n} \text{vol}_{M_n}(y) \sum_{g \in \pi_1(M_n)} K_{\widetilde{M_n}}(t;y, g \cdot y)\,. 
\end{eqaed}

\begin{figure}[ht]
    \centering
    \includegraphics{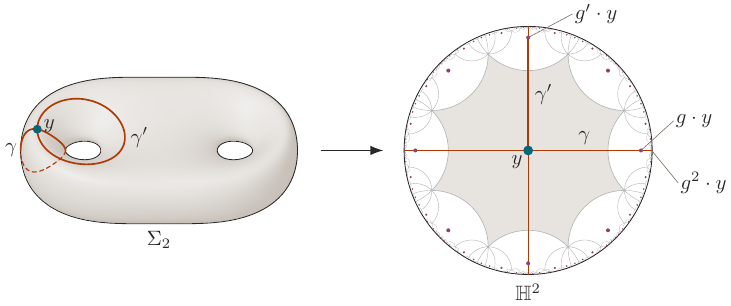}
    \caption{Closed geodesics on a hyperbolic Riemann surface lift to geodesic segments joining a point to its images in the universal cover.}
    \label{fig:universal-cover}
\end{figure}

\noindent This expression can be approximated in the semiclassical limit $t \ll 1$, where classical solutions $x_0(\tau)$ correspond to geodesics $\gamma$. This is depicted in \cref{fig:universal-cover} for hyperbolic Riemann surfaces. For simplicity, we discuss the case where geodesics are isolated; the purpose of this section is to provide some physical intuition for the general mechanism that prevents \ac{EFT} control in \ac{dS} solutions with Casimir energy. At any rate, this is the relevant case for the concrete hyperbolic examples we will study in \cref{sec:casimir_riemann_surfaces,sec:hyperbolic_manifolds}. (We also separately treat the case of Riemann-flat manifolds in \cref{sec:RFMs}.) The leading action is simply the distance $\text{dist}_{\widetilde{M_n}}(y, g_\gamma \cdot y )^2/(4t)$ produced by the group element associated with the geodesic $\gamma$. The one-loop fluctuations assemble into the Van Vleck-Morette determinant $\Delta_\text{VM}$ \cite{zannias},
\begin{eqaed}\label{eq:van-vleck_det}
    K_{\widetilde{M}_n}(t;y,g_\gamma \cdot y) \overset{t \ll 1}{\sim} \frac{\Delta_\text{VM}^\frac{1}{2}(y, g_\gamma \cdot y)}{(4\pi t)^\frac{n}{2}} \, e^{-\frac{\text{dist}_{\widetilde{M_n}}(y, g_\gamma \cdot y )^2}{4t}} \left(1 + \dots \right) .
\end{eqaed}
The Van Vleck-Morette determinant arises from the Gel'fand-Yaglom formula for the Jacobian factor in the path integral measure, due to the variation of the action (given by the Synge world function) and the endpoint boundary conditions on the worldline. Upon integration over $M_n$, as in \cref{eq:sum_over_images}, the classical exponentials localize the integral to the shortest separation, i.e., the length $\ell_\gamma$ of the geodesic, with additional Gaussian fluctuations associated with Jacobi fields. The combined formula in \cref{eq:sum_over_images} can then be recast as a sum over closed geodesics \cite{MURATOREGINANNESCHI2003299,ludewig2018heat}
\begin{eqaed}\label{eq:sum_over_images_semiclassical}
    K_{M_n}(t) \overset{t \ll 1}{\sim} K^\text{id}_{M_n}(t) + \sum_{\gamma} \frac{\ell_{\gamma_0}}{\sqrt{4\pi t}} \, \frac{e^{-\frac{\ell_\gamma^2}{4t}}}{\sqrt{\abs{\det(I_{2(n-1)} - M_\gamma)}}} \left( 1 + \dots \right) ,
\end{eqaed}
where $\gamma_0$ is the \emph{primitive} geodesic within the set of iterates $\gamma=\gamma_0^k$, and the monodromy\footnote{The $2(n-1) \times 2(n-1)$ matrix $M_\gamma^{\perp}$ exponentiates the linearized geodesic flow in the phase space transverse to the translational zero-mode. Equivalently, the one-loop fluctuations are controlled by a quadratic form whose eigenvalue problem can be converted as a first-order problem in phase space \cite{MURATOREGINANNESCHI2003299}.} $M_\gamma^{\perp}$ acts on the transverse Jacobi deformations of the closed geodesic $\gamma$  \cite{MURATOREGINANNESCHI2003299}. The integration over the longitudinal mode---the translation along $\gamma$---is performed separately, yielding the prefactor $\propto \ell_{\gamma_0}/\sqrt{t}$. The sum in \cref{eq:sum_over_images_semiclassical} can be further split into primitive geodesics and windings, which will be useful later on. The identity contribution $K^\text{id}_{M_n}(t)$ from constant maps is local, and affords an asymptotic expansion of the form \cite{Vassilevich:2003xt}
\begin{eqaed}\label{eq:identity_term_local_heat-kernel_expansion}
    K^\text{id}_{M_n}(t) \overset{t \ll 1}{\sim} \frac{1}{(4 \pi t)^\frac{n}{2}} \sum_{k \geq 0} a_k \, t^k\,,
\end{eqaed}
where the Seeley-deWitt coefficients $a_k$ are local functionals of the metric; in particular, we have \mbox{$a_0 = \text{Vol}(M_n)$}. This contribution naively leads to \ac{UV}-divergences in the Schwinger integral for the Casimir energy, which are simply a symptom of \ac{UV}-sensitivity. Since we explicitly deal with the contribution from a single real scalar field, in practice we treat \ac{UV}-divergences by zeta-function regularization, which effectively subtracts the terms with $k \leq \lfloor \frac{D}{2}\rfloor = 5$ from \cref{eq:identity_term_local_heat-kernel_expansion} in the \ac{UV} region of the Schwinger integral. In the examples we study, it turns out that the identity contribution is several orders of magnitude smaller than the bounds we obtain for the remainder. \\

\noindent The expression in \cref{eq:sum_over_images_semiclassical} is corrected by curvature terms in each ``instanton'' sector, corresponding to the fact that the non-linear sigma model is interacting when the metric on $M_n$ is curved. Still, representing the Casimir energy in terms of a sum over geodesics is conceptually instructive, and we shall make extensive use of it in \cref{sec:casimir_riemann_surfaces,sec:hyperbolic_manifolds,sec:RFMs}. For the purposes of this section, the main point we emphasize is that, in order to obtain a large Casimir energy, the geometry of $M_n$ must realize either of the following:

\begin{itemize}
    \item Some sectional curvatures are much larger than the average Ricci curvature;

    \item There are closed geodesics that are significantly shorter than the average curvature radius and the diameter of $M_n$;

    \item The local geometry and short geodesics of $M_n$ are ``tame'', but the global topology is ``wild'' in the sense that there are overwhelmingly more geodesics than the exponential factor in \cref{eq:sum_over_images_semiclassical} can suppress. 
\end{itemize}

\begin{figure}[t]
    \centering
    \includegraphics{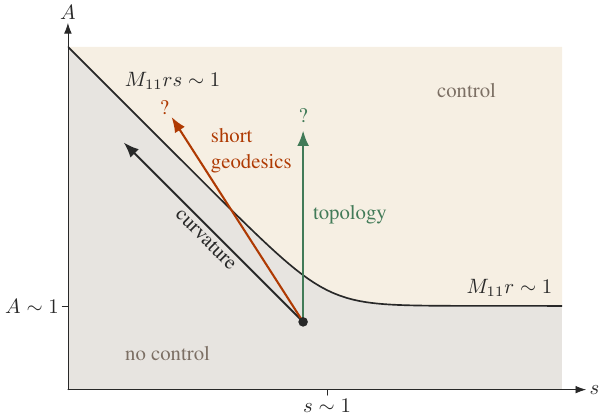}
    \caption{Schematic illustration of the relation between Casimir enhancement and \ac{EFT} control. The question marks indicate the routes to control examined below.}
    \label{fig:control-plane}
\end{figure}

\noindent The situation is summarized visually in \cref{fig:control-plane}. The first two mechanisms seem natural, and can be arranged to occur easily. In fact, even at small curvatures, the second mechanism is transparent---the sum over geodesics is lower-bounded by the contribution from systoles (the shortest closed geodesics). However, these two mechanisms also create a tension with \ac{EFT} control: on the one hand, a controlled solution requires a large Casimir coefficient $\mathcal{C}$, on account of \cref{eq:solution}; on the other hand, if $\mathcal{C} \gg 1$ due to large curvatures or short geodesics, the control of the \ac{EFT} is at risk. The question is how well this trade-off can be balanced. The third mechanism is essentially a fine-tuning of global topology which could skew the trade-off in favor of \ac{EFT} control. It is not intuitively obvious why this cannot work.  The main quantitative no-go results in this work show that, at least for hyperbolic and Riemann-flat manifolds, this mechanism is not realized.

\subsection{Heat-kernel estimates}\label{sec:heat-kernel_estimates}

Before moving on to the specific settings where we can prove no-go statements on the controlled \ac{dS} solutions of \cref{eq:solution}, we can get some more mileage out of the general expression in \cref{eq:general_casimir_expression_flat} using heat-kernel methods. This approach supports the mechanisms that we have outlined above. To proceed, we split the Schwinger integral of \cref{eq:general_casimir_expression_flat} in a \ac{UV} region with $t \leq t_*$ and its \ac{IR} complement, and write the zeta-function renormalized expression
\begin{eqaed}\label{eq:schwinger_split}
    \mathcal{C} & = \frac{1}{2(4\pi)^\frac{d}{2}} \int_0^{t_*} \frac{dt}{t^{1+\frac{d}{2}}} \left(K_{M_n}(t) - \frac{1}{(4\pi t)^\frac{n}{2}} \sum_{k=0}^{\left\lfloor \frac{D}{2} \right\rfloor} a_k \, t^k \right) \\
    & + \frac{1}{2(4\pi)^\frac{d}{2}} \int_{t_*}^\infty \frac{dt}{t^{1+\frac{d}{2}}} \, K_{M_n}(t) + \frac{1}{2(4\pi)^\frac{D}{2}}\sum_{k=0}^{\left\lfloor \frac{D}{2} \right\rfloor} \frac{a_k \, t_*^{k-\frac{D}{2}}}{k-\frac{D}{2}} \\
    & \equiv \mathcal{C}_\text{UV} + \mathcal{C}_\text{IR} + \Delta \mathcal{C}\,.
\end{eqaed}
The last term $\Delta \mathcal{C}$ arises by the analytic continuation of the otherwise \ac{UV}-divergent terms, and it only depends on local curvatures. The goal is then to bound $\mathcal{C}_\text{UV} , \mathcal{C}_\text{IR}$ in terms of curvature and geodesic lengths, as anticipated above. If both factors are (uniformly) bounded, the only way to produce a large Casimir coefficient is to find a family of manifolds whose global topology enhances it. \\

\noindent \textbf{Bounding the \ac{UV} contribution.} The (local diagonal) heat-kernel expansion in \cref{eq:identity_term_local_heat-kernel_expansion} tells us that the difference between the full heat kernel and a finite number of local terms is asymptotic to the successive term in the expansion \cite{Vassilevich:2003xt}. In a compact interval $t \in [0, t_*]$, one is naturally led to look for a bound of this form \cite{berline2004heat}. Indeed, there is a constant $C$ such that \cite{Ludewig2019}
\begin{eqaed}\label{eq:general_heat-kernel_UV_bound}
    \abs{K_{M_n}(t) - \frac{1}{(4\pi t)^\frac{n}{2}} \sum_{k=0}^N a_k \, t^k} \leq \frac{\text{Vol}(M_n)}{(4\pi t)^\frac{n}{2}} \, C \, t^{N+1}\,, \qquad t \leq t_*\,.
\end{eqaed}
The natural expectation is that $C$ be controlled by the overall/average curvature radius $r$, but in principle it could be made larger by highly non-uniform sectional curvatures or degeneration limits. This is how the first mechanism we described can produce a large Casimir coefficient, but it can also jeopardize the \ac{EFT} control. As we have anticipated, the examples that we discuss in detail in \cref{sec:casimir_riemann_surfaces,sec:hyperbolic_manifolds,sec:RFMs} are uniform enough that this does not happen. The upshot of \cref{eq:general_heat-kernel_UV_bound} is that
\begin{eqaed}\label{eq:general_casimir_UV_bound}
    \abs{\mathcal{C}_\text{UV}} \leq \frac{\text{Vol}(M_n)}{2(4\pi)^\frac{D}{2}} \, C \, \int_0^{t_*} t^{\lfloor \frac{D}{2} \rfloor-\frac{D}{2}} \, dt\,,
\end{eqaed}
where the integral gives a factor of $t_*$ for $D$ even and $2\sqrt{t_*}$ for $D$ odd. \\

\noindent \textbf{Bounding the \ac{IR} contribution.} In order to bound the \ac{IR} contribution $\mathcal{C}_\text{IR}$, we employ the Li-Yau upper estimate \cite{LiYau1986,BaumgarthDevyverGuneysu2023}. For Ricci curvature bounded below by $-K \leq 0$, there are constants $C_1, C_2$ such that\footnote{In fact, assuming that the sectional curvature be bounded leads to a stronger bound \cite{cheng-yau}.}
\begin{eqaed}\label{eq:general_heat-kernel_li-yau_bound}
    K_{M_n}(t) \leq C_1 \, e^{C_2 K t} \int_{M_n} \frac{\text{vol}_{M_n}(y)}{\text{Vol}(B_y(\sqrt{t}))}\,,
\end{eqaed}
where $B_y(\sqrt{t})$ denotes the ball of radius $\sqrt{t}$ centered at $y$. Since the heat kernel is decreasing in $t$, we can write
\begin{eqaed}\label{eq:general_casimir_IR_bound}
    \mathcal{C}_\text{IR} & \leq \frac{C_1}{2(4\pi)^\frac{d}{2}} \, e^{C_2 K t_*} \int_{M_n} \frac{\text{vol}_{M_n}(y)}{\text{Vol}(B_y(\sqrt{t_*}))} \int_{t_*}^\infty \frac{dt}{t^{1+\frac{d}{2}}} \\
    & = \frac{C_1}{d(4\pi t_*)^\frac{d}{2}} \, e^{C_2 K t_*} \int_{M_n} \frac{\text{vol}_{M_n}(y)}{\text{Vol}(B_y(\sqrt{t_*}))}\,.
\end{eqaed}
Intuitively, the length scale $\sqrt{t_*}$ controlling the split into \ac{UV} and \ac{IR} contributions should be related to the scale at which some global properties of $M_n$ become visible, namely the systole length $\text{sys}(M_n)$. More precisely, if the sectional curvature of $M_n$ is non-positive, the global injectivity radius is $\text{sys}(M_n)/2$ \cite{Klingenberg1959}, and G\"{u}nther's comparison theorem bounds the volumes from below by their Euclidean counterparts $\omega_n \, t_*^\frac{n}{2}$. As a result, for $\sqrt{t_*} = \text{sys}(M_n)/2$, one obtains
\begin{eqaed}\label{eq:general_casimir_IR_volume_gunther_bound}
    \mathcal{C}_\text{IR} & \leq \frac{C_1}{d(4\pi)^\frac{d}{2}} \frac{\text{Vol}(M_n)}{\omega_n} \, e^{C_2 K t_*} \, t_*^{-\frac{D}{2}}\,.
\end{eqaed}
Merely assuming a bound on the Ricci curvature does not grant these estimates, which is one of the reasons why general Einstein manifolds are a much more complicated class to deal with. Nevertheless, \cref{eq:general_casimir_UV_bound,eq:general_casimir_IR_bound} give an intuitive sense of what must happen in order that $\mathcal{C} \gg 1$. Namely, the most natural enhancement mechanisms---curvature and short geodesics---are precisely those endangering control of the \ac{EFT}. The issue is to understand to what extent, and whether global topology can rescue it.

\subsection{Bounds for covering spaces}\label{sec:covering_spaces}

Given the preceding considerations, the first place to look for an enhancement of the Casimir parameter due to global topology is high-degree coverings. Letting
\begin{eqaed}\label{eq:cover_def}
    M_n^N \overset{\pi}{\longrightarrow} M_n
\end{eqaed}
be a (non-trivial) connected $N$-sheeted cover, we have $\text{Vol}(M_n^N) = N \, \text{Vol}(M_n)$. If the Casimir coefficient $\mathcal{C}_N \sim N \, \mathcal{C}_1$ grows linearly with $N \gg 1$, this approach cannot work, since the extensive factor of $N$ drops out of \cref{eq:solution}. (However, it does affect the lower-dimensional Planck scale, although there is no scale separation due to \cref{eq:x_extremum}.) {For the compact Euclidean product, every connected cover has a single constant scalar zero-mode still, such that the coefficient $b=1$ in \cref{eq:L_solution} is unchanged. Since its contribution to the radius equation is divided by the internal volume, it decreases like $N^{-1}$ at fixed $x$. This additional term supplies, therefore, no enhancement from increasing the covering degree alone.} \\

\noindent We are therefore led to ask: can we arrange a cover such that $\mathcal{C}_N$ grows \emph{superlinearly} in $N$? The answer, unfortunately, is negative. To see this, we apply the analog of \cref{eq:sum_over_images} to the heat kernel of the cover,
\begin{eqaed}\label{eq:cover_heat_kernel}
    K_{M_n}(t;\pi(y),\pi(y)) = \sum_{y' \in \pi^{-1}(\pi(y))} K_{M_n^N}(t;y,y')\,,
\end{eqaed}
so that integrating over $M_n^N$ gives
\begin{eqaed}\label{eq:integrated_cover_heat_kernel}
    N \, K_{M_n}(t) & = K_{M_n^N}(t) + \int_{M_n^N} \text{vol}_{M_n^N}(y) \!\!\!\!\sum_{y' \in \pi^{-1}(\pi(y)) : y' \neq y}\!\!\!\! K_{M_n^N}(t;y,y') \\
    & \geq K_{M_n^N}(t)\,.
\end{eqaed}
As a result, we also have\footnote{The \ac{UV}-sensitive part is immaterial in deriving \cref{eq:casimir_cover_inequality}, because taking the difference $K_{M_n^N} - N \, K_{M_n}$ between the heat kernels cancels them.}
\begin{eqaed}\label{eq:casimir_cover_inequality}
    \mathcal{C}_N < N \, \mathcal{C}_1\,.
\end{eqaed}
In fact, for large $N$ one gets asymptotic linearity along a nested tower of coverings, at least if the identity contribution $A^\text{id}_\infty$ to the (averaged) Casimir density of the universal cover is positive. Indeed, the (averaged) density $A_N = \mathcal{C}_N/\text{Vol}(M_n^N) \geq A^\text{id}_\infty$ is non-increasing along the tower and bounded below by this contribution, and thus converges to some positive limit for $N \to \infty$. \\

\noindent The finite-$N$ deficit from linearity/extensivity can also be understood intuitively from the Hamiltonian expression of the heat kernel, namely the canonical partition function of the particle. In this language, for regular covers \cref{eq:cover_heat_kernel} can be recast as a sum over insertions of deck transformations $g \in \text{Deck}(\pi)$ in the trace, connecting to the results of \cite{Harlow:2021trr,Kang:2022orq,Benjamin:2023qsc}. The semiclassical (small-$t$) approximation to the ``twisted'' contributions is then weighted by $\exp(-\frac{\delta_g^2}{4t})$. The (squared) minimal separation $\delta_g^2$ in the cover plays the role, in the worldline case, of the twisted-sector gap in the settings of \cite{Harlow:2021trr,Kang:2022orq,Benjamin:2023qsc}. This tells us that the \ac{UV} contribution to the extensivity deficit, driven by eigenmodes that do not descend to the base, is suppressed. \\

\noindent This result illustrates the general mechanisms we put forth: the simplest way to enhance the Casimir coefficient $\mathcal{C}$ without touching the local geometry and geodesic distribution does not work.

\subsection{From scalars to supergravitons}\label{sec:from_scalars_to_supergravitons}

The preceding arguments are based on geometric estimates of the Casimir potential for scalar fields, with some simplifying physical assumptions on the internal manifold. On the one hand, this qualitative approximation can be expected to provide at least a rough estimate for the full supergravity case, given the general nature of the mechanisms we have outlined. On the other hand, in practice the complete computation introduces several technical complications: to begin with, one must include the external \ac{dS} spacetime explicitly, since otherwise there is no critical point where the effective potential becomes gauge-independent. Then, one must address gauge-fixing, ghosts, control over (classically) off-shell gauge dependence, and---most importantly for us---connection and curvature couplings in the heat kernel. Schematically, the scalar Laplacian is replaced by Laplace-like operators of the form $D_f = (\nabla+A_f)^\dagger (\nabla+A_f) + \mathcal{E}_f$ acting on the (single-particle) Hilbert space $\mathcal{H}_f$ of each field type $f$. The presence of curvature couplings makes it harder to estimate the \ac{IR} part of the Schwinger integral for the Casimir energy, while gauge-fixing complicates the estimation of the \ac{UV} part due to a more intricate cancellation of \ac{UV}-divergences within the BRST complex. In \cref{app:bounding_sugra_casimir} we discuss these technical issues in more detail; for the time being, we provide an outline of how the general mechanism discussed above can survive in the supergravity case, with some simplifying assumptions for ease of exposition. The same considerations indicate how the detailed settings studied in \cref{sec:casimir_riemann_surfaces,sec:hyperbolic_manifolds,sec:flat_manifolds} generalize (at least parametrically) to this case. \\

\noindent The basic idea is that, after gauge-fixing, the supergraviton BRST complex contains (the horizontal projection of) graviton, gravitino, and three-form fluctuations, along with their ghosts. If the corresponding curvature endomorphisms $\mathcal{E}_f \geq -E$ are bounded below by a non-positive constant $-E \leq 0$, one can estimate \cite{Hess:1977gb, OuhabazOuhabaz+2005, güneysu2012generalizedschrodingersemigroups}
\begin{eqaed}\label{eq:general_heat_kernel_sugra_hsu_estimate}
    \abs{e^{-D_f t}} \lesssim e^{E t} \, K_{S^d \times M_n}
\end{eqaed}
where (here and in the following) $\lesssim$ denotes inequality up to constant positive prefactors. Crucially, \cref{eq:general_heat_kernel_sugra_hsu_estimate} holds \emph{pointwise} in the heat kernel, before taking the trace. As such, this bound can be used to bound the \ac{UV} part of the Casimir energy\footnote{More precisely, one should bound the combination in \cref{eq:L_solution}, due to the presence of the external \ac{dS} spacetime. We do so in more detail in \cref{app:bounding_sugra_casimir}. Also, since the (renormalized) identity part is universal, we only consider the winding terms.} by integrating up to $t = t_*$. This argument works for local fluctuation operators $D_f$, and gauge-fixing can jeopardize the validity of \cref{eq:general_heat_kernel_sugra_hsu_estimate}. Addressing this issue leads to a more involved analysis, which we defer to \cref{app:bounding_sugra_casimir}. {For the \ac{IR} part, the fixed-$r$ combination $\mathcal A=A-\frac{x}{d}\partial_xA$ removes the constant subtraction $n_\text{zero-modes}^f$ from the heat trace. Its common-scale effect remains in the separate $b$ term of \cref{eq:L_solution} and its flux-supported counterpart.} Then, if the spectrum of eigenvalues $\Lambda \geq \Lambda_\text{gap} > 0$ has a positive gap $\Lambda_\text{gap}$, the combinations $e^{\Lambda_\text{gap}t}(K^f_{S^d \times M_n}(t)-n_\text{zero-modes}^f)$ are decreasing. Hence, we can bound
\begin{eqaed}\label{eq:general_heat_kernel_IR_bound_gap}
    \int_{t_*}^\infty \frac{dt}{t} \, \abs{K^f_{S^d \times M_n}(t)-n_\text{zero-modes}^f} \leq \int_{t_*}^\infty \frac{dt}{t} \, e^{-\Lambda_\text{gap}(t-t_*)} \abs{K^f_{S^d \times M_n}(t_*)-n_\text{zero-modes}^f}\,,
\end{eqaed}
so that we can re-use the bound of \cref{eq:general_heat_kernel_sugra_hsu_estimate} sector-by-sector. If the local fluctuation operator is merely non-negative, the \ac{IR} part of the Casimir parameter of \cref{eq:L_solution} can still be bounded by studying the $x$-dependence of the eigenvalues, which we do in \cref{app:bounding_sugra_casimir}. Once again, the detailed analysis is significantly complicated by the gauge-fixing, whence the above local bounds do not apply. Moreover, the presence of negative modes that survive gauge-fixing violates the above simplifying assumption, and must be handled carefully; this is related to the possibility of growing exponentials in \cref{eq:general_heat_kernel_sugra_hsu_estimate}. We address both issues in \cref{app:bounding_sugra_casimir} by treating negative modes via the standard steepest-descent prescription (see, e.g., \cite{Witten:2010cx}) and bounding the potentially offending eigenvalues employing the direct spectral analysis of \cref{app:spectral_analysis_supergraviton_fluctuations}. \\

\noindent This concludes our general overview of the Casimir potential, and our geometric and physical arguments for why \ac{EFT} control is expected to be obstructed. In the next sections, we will restrict to specific classes of manifolds---namely, Riemann surfaces, higher-dimensional hyperbolic manifolds and finally flat manifolds---to derive numerical and parametric upper bounds on how well the trade-off between the Casimir potential and geodesic lengths can enhance \ac{EFT} control. The analysis will encompass each class of manifolds uniformly, showing that global topology does not help.

\section{Casimir energy on rigid Riemann surfaces}\label{sec:casimir_riemann_surfaces}

In this section, we begin our systematic derivation of numerical bounds on \ac{EFT} control parameters by considering M-theory on Riemann surfaces. The moduli space of constant negative curvature on a Riemann surface of genus $g>1$ has $3g-3$ complex moduli, as well as one real parameter corresponding to its curvature radius. In order for the potential to be critical as a function of the $3g-3$ complex moduli,  we pick rigid Riemann surfaces $\Sigma_g$ with $g>1$. Here and in the following, by ``rigid'' we do not necessarily mean that there be no volume-preserving moduli---indeed, in this case there are---rather, we assume an enhanced symmetry that freezes all of them at a critical point. In other words, in these cases the particular point on moduli has enough symmetry that all $3g-3$ complex moduli are charged under it, leading to $d_\text{moduli}V=0$ in these directions by symmetry. For hyperbolic Riemann surfaces, these are Riemann surfaces which are regular branched covers $\Sigma_g\to\mathbb{P}^1$ with three branched points. Indeed, letting $G$ be the automorphism group of $\Sigma_g$, this induces a branched cover $\Sigma_g\mapsto \Sigma_g/G\simeq \Sigma_{g_0}$ with $n$ branch points; then, the $G$-invariant tangent moduli space has dimension
\begin{eqaed}\label{eq:tangent_moduli}
    \dim_\mathbb{C} (T_{\Sigma_g}\mathcal{M}_g)^G = \dim_\mathbb{C} (T_{\Sigma_{g_0}}\mathcal{M}_{g_0,n}) = 3g_0-3+n\,,
\end{eqaed}
which vanishes for $g_0 = 0$ and $n = 3$. Such covers are classified by the ramification indices $(p,q,r)$ and a choice of Fuchsian normal subgroup $K \subset \Delta(p,q,r)$ of the triangle group, whereby $G = \Delta(p,q,r)/K$ \cite{wolfart}. The systole length is bounded below by $2 \sinh^{-1}\sqrt{\cos\frac{3\pi}{7}} \approx 0.911$ \cite{beardon1983geometry, Leininger_2004}, and in fact the smallest systole length for triangle orbifolds is about $0.983$ and is attained by a $(2,3,7)$-surface \cite{Suzzi_Valli_2015}. For our purposes, a (strict) lower bound of $0.9$ will be sufficient, although we will obtain an improved bound for smooth surfaces. \\

\noindent Since rigid Riemann surfaces exist, their complex structure extremizes the Casimir energy, and the analysis of \cref{sec:casimir_no_scale} applies. We will first consider the flat-space approximation, and then include the contribution of the external dS spacetime to the Casimir term. Moreover, we will restrict to the contribution of a single scalar field for simplicity. As we mentioned throughout the analysis, a heuristic expectation is that the complete Casimir energy to be no more than a few orders of magnitude higher than the scalar-field contribution. The actual supergravity computation contains a character factor in the trace formula, whose contribution can be bounded by summing over ranks. However, as we shall discuss in more detail in \cref{app:spectral_analysis_supergraviton_fluctuations}, the additional curvature couplings of tensor and spinor fields complicate matters significantly. In \cref{app:bounding_sugra_casimir} we provide a parametric extension to the supergravity one-loop Casimir potential of the numerical bounds derived in the scalar-field case.

\subsection{Flat-space approximation}\label{sec:flat-space_approximation}

In the flat-space approximation (ignoring the one-loop renormalization of the Planck mass which we will remedy later), the reduced Casimir energy $V_\text{Cas} = \text{Vol}(\Sigma_g) \, \rho_\text{Cas} = - 4\pi(g-1)A/r^9$ is given in terms of the positive scalar Laplacian $\Delta_{\Sigma_g}$ on the unit-radius surface\footnote{This means that the Gaussian curvature $K=R/2=-1$, which is the normalization that is commonly used to write formulae for the heat kernel.} by
\begin{eqaed}\label{eq:casimir_flat_def}
    - \, 4\pi(g-1) \, A & = \frac{1}{2} \int_{\mathbb{R}^9} \frac{d^9p}{(2\pi)^9} \, \Tr \log\left(p^2 + \Delta_{\Sigma_g}\right) \\
    & = - \, \frac{1}{2(2\pi)^9} \int \frac{dt}{t} \int d^9p \, e^{-tp^2} \, K_{\Sigma_g}(t) \\
    & = - \, \frac{1}{2(4\pi)^\frac{9}{2}} \int \frac{dt}{t^\frac{11}{2}} \, K_{\Sigma_g}(t) \\
    & = - \, \frac{\Gamma\left(-\frac{9}{2}\right)}{2(4\pi)^\frac{9}{2}} \, \zeta_{\Sigma_g}\left(-\frac{9}{2}\right) \\
    & = \frac{1}{30240\pi^4} \, \zeta_{\Sigma_g}\left(-\frac{9}{2}\right) ,
\end{eqaed}
where the necessary (zeta-function) regularization has been performed. The prefactor is approximately $3.39 \times 10^{-7}$. \\

\noindent Parametrizing the eigenvalues $\lambda_j = r_j^2+\frac{1}{4}$, the Selberg trace formula \cite{marklof2008selbergstraceformulaintroduction} for an even function $h(r)$ whose Fourier transform is $g(\omega) = \frac{1}{2\pi}\int dr h(r)e^{-i\omega r}$ gives
\begin{eqaed}\label{eq:selberg_formula}
    \sum_j h(r_j) & = \frac{4\pi (g-1)}{2\pi} \int_0^\infty h(\rho) \tanh(\pi \rho) \rho \, d\rho \\
    & + \sum_{k > 0} \sum_{\gamma \in \mathcal{P}} \frac{\ell_\gamma \, g(k \ell_\gamma)}{2\sinh(k \ell_\gamma/2)}\,,
\end{eqaed}
where $\mathcal{P}$ denotes the set of oriented prime geodesics $\gamma$ whose length is denoted $\ell_\gamma$. Inserting $h(r) = (r^2+\frac{1}{4})^{-s}$, we need to analytically continue the identity term (the first line) before setting $s=-\frac{9}{2}$. The integral converges absolutely for $\Re s > 1$. Since the derivative of the hyperbolic tangent vanishes exponentially at infinity, it is useful to integrate by parts to obtain it. Since $\rho \, h \, (\rho) d\rho = \frac{1}{2(1-s)} \, d(\rho^2+\frac{1}{4})^{1-s}$, the boundary terms vanish and we arrive at
\begin{eqaed}\label{eq:zeta_riemann_s}
    \zeta_{\Sigma_g}\left(-\frac{9}{2}\right) & = - \, \frac{4\pi (g-1)}{22} \int_0^\infty \left(\rho^2 + \frac{1}{4}\right)^{\! \frac{11}{2}}\sech^2(\pi \rho) \, d\rho \\
    & - \, \frac{945}{64\pi} \sum_{k>0} \sum_{\gamma \in \mathcal{P}} \frac{K_5(k \ell_\gamma/2)}{k^5 \ell_\gamma^4 \sinh(k \ell_\gamma/2)}\,,
\end{eqaed}
where we used the Fourier transform $g(\omega) = \frac{\omega^{s-\frac{1}{2}}}{\sqrt{\pi} \, \Gamma(s)}K_{\frac{1}{2}-s}(\omega/2)$ for $\omega > 0$. \\

\noindent The identity term numerically evaluates to $-4\pi(g-1) \times (3.26 \times 10^{-3})$. To bound the second term, let $\ell_0 \geq 0.9$ be a lower bound for the length of all prime geodesics, and $N(L)$ be the number of prime geodesics of length below $L$. Write the second line as $-\frac{945}{64\pi} \sum_{\gamma \in \mathcal{P}} F(\ell_\gamma)$. Then, for $x \geq 0.9$ we have $K_5(x/2) \leq 33000 \, e^{-x/2}$. Hence, we can write
\begin{eqaed}\label{eq:F_bound}
    F(x) & \equiv \sum_{k>0} \frac{K_5(kx/2)}{k^5x^4 \sinh(kx/2)} \\
    & \leq \frac{33000}{x^4} \sum_{k>0} \frac{e^{-kx/2}}{k^5 \sinh(kx/2)} \\
    & \leq \frac{33000 \, e^{-x/2}}{x^4\sinh(x/2)} \sum_{k>0} \frac{1}{k^5} \\
    & = \frac{33000\zeta(5)}{x^4} \, \frac{e^{-x/2}}{\sinh(x/2)}\,.
\end{eqaed}
Using the fact that $N(L) \leq 4\pi(g-1) \, k \, e^L$, with $k \leq 72$ (see \cref{app:riemann_surfaces}), we can integrate by parts in the following steps,
\begin{eqaed}\label{eq:F_kernel}
    \sum_{\gamma \in \mathcal{P}} F(\ell_\gamma) & = \int_{\ell_0}^\infty F(\ell) \, dN(\ell) \\
    & = \int_{\ell_0}^\infty N(\ell) (-F'(\ell)) d\ell \\
    & \leq 4\pi(g-1) \, k \int_{\ell_0}^\infty e^\ell (-F'(\ell)) d\ell \\
    & = 4\pi(g-1) \, k \left(e^{\ell_0} \, F(\ell_0) + \int_{\ell_0}^\infty e^\ell \, F(\ell) \, d\ell \right) \\
    & \leq 4\pi(g-1) \times 33000 \zeta(5) \, k \left(5.14 + \int_{\ell_0}^\infty \frac{e^{\ell/2} \, d\ell}{\ell^4 \sinh(\ell/2)}\right) \\
    & \leq 4\pi(g-1) \times 221053 \, k\,.
\end{eqaed}

\begin{figure}[t]
    \centering
    \subfloat[The weight $F(\ell)$, with $\ell<\ell_0=0.9$ excluded by the systole bound.\label{fig:geodesic-weight}]{\includegraphics{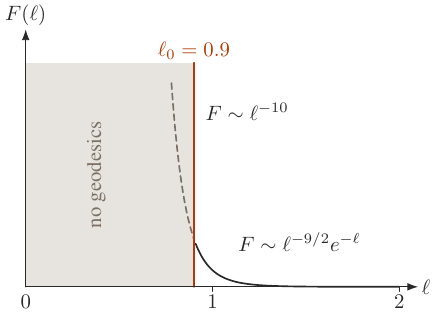}}
    \qquad
    \subfloat[Disjoint balls around images separated by at least $\ell_0$, illustrating the packing estimate for $N(L)$.\label{fig:packing}]{\includegraphics{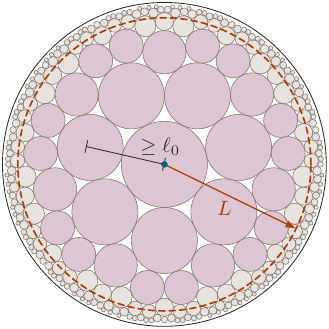}}
    \caption{Geodesic weights showing the systole cutoff and hyperbolic packing in the Casimir bound for rigid Riemann surfaces.}
    \label{fig:casimir-bound}
\end{figure}

\noindent This packing technique is depicted in \cref{fig:casimir-bound}. Putting everything together, from \cref{eq:zeta_riemann_s,eq:constant_C} we obtain
\begin{eqaed}\label{eq:zeta_bound}
    \abs{\zeta_{\Sigma_g}\left(-\frac{9}{2}\right)} \leq 4\pi(g-1) \times (1.04 \times 10^6) \, k\,,
\end{eqaed}
which on account of \cref{eq:casimir_flat_def} leads to
\begin{eqaed}\label{eq:A_bound}
    A \leq 0.36 \, k \leq 26\,.
\end{eqaed}
Hence, due to the power $1/9$ with which $A$ enters the \ac{EFT} control parameters, even an enhancement of several orders of magnitude in the full supergravity computation would not alter our conclusions. This derivation of the scalar-field bound of \cref{eq:A_bound} was chosen for simplicity of exposition; using the techniques from \cref{sec:hyperbolic_manifolds} it can be improved by four orders of magnitude.

\subsection{Including de Sitter curvature}\label{sec:casimir_curved}

As explained in \cref{sec:setup}, the correct Casimir energy for our problem must include the external \ac{dS} curvature, since there is no scale separation. {We start by computing the relevant Casimir quantities at the convenient reference ratio $x_*=1/6$. The scalar stationary ratio is, however, slightly larger, on account of \cref{eq:x_extremum}. After obtaining the reference estimates, we extend the bounds to the allowed interval and find no significant deviation from the reference results.} \\

\noindent To this end, since we also need its derivative, due to \cref{eq:L_solution}, we set $r=1$ and $L=1/x$ in the following. Then, subtracting the zero-mode contribution, \cref{eq:casimir_flat_def} generalizes to
\begin{eqaed}\label{eq:casimir_curved_def}
    - \, 4\pi(g-1) \, A = - \, \frac{1}{2 \, \text{Vol}(S^9)} \int_{0}^{\infty} \frac{dt}{t} \left( K_{S^9}(t) \, K_{\Sigma_g}(t) - 1 \right) .
\end{eqaed}
The volume of the nine-sphere is given by $\text{Vol}(S^9) = \frac{\pi^5}{12} \, x^{-9}$, while its heat kernel is given by \cite{Kluth:2019vkg}
\begin{eqaed}\label{eq:heat_kernel_sphere}
    K_{S^9}(t) = \sum_{l \geq 0} \frac{(l+7)!(2l+8)}{8!l!} \, e^{-t \, x^2 \, \, l(l+8)}\,.
\end{eqaed}
Notice that the Casimir coefficient $A$ takes the form $A(x) = \frac{1}{\text{Vol}(S^9)} \, \widetilde{A}(x)$, so that the curved-space counterpart of $A$ from the flat-space approximation, due to \cref{eq:L_solution}, reads
\begin{eqaed}\label{eq:A_combination_curved}
    A(x_*) - \frac{x_*}{d} \, A'(x_*) = -\, \frac{x_*}{d \, \text{Vol}(S^9)} \, \widetilde{A}'(x_*) \equiv \mathcal{A}\,.
\end{eqaed}
{The constant zero-mode subtraction drops out of this derivative at fixed $r$. The logarithmic common-scale contribution is nonetheless retained in \cref{eq:x_extremum,eq:L_solution}.} \\

\noindent The winding terms in \cref{eq:casimir_curved_def}, arising from the second line of \cref{eq:selberg_formula}, are automatically \ac{UV}-finite; the identity term is \ac{UV}-divergent due to the first six local curvature contributions. However, these terms mostly cancel out due to higher-dimensional supersymmetry; the only possible exception is the $R^4$-term (and the supersymmetric completion thereof), and applying zeta-function regularization effectively subtracts this divergence (up to scheme dependence). We can thus bound the subtracted expression directly. Before differentiating with respect to $x$, the (formally divergent) expression is
\begin{eqaed}\label{eq:subtracted_identity}
    \frac{1}{2 \, \text{Vol}(S^9)} \int_0^\infty \frac{dt}{t} \left(K_{S^9}(t) - \, \frac{\text{Vol}(S^9)}{(4\pi t)^\frac{9}{2}}\sum_{k=0}^5 a_k \, (t x^2)^k \right) K_{\Sigma_g}(t)\,,
\end{eqaed}
to which we add back the analytically continued \ac{UV}-divergent terms as before. Here $a_k$ are the heat-kernel coefficients of the unit $S^9$, which are determined by the generating function (see appendix C of \cite{Kluth:2019vkg}\footnote{The relevant expressions in \cite{Kluth:2019vkg}, namely Eq. (98) and Eq. (C9), contain a typo: in the second line, $I_{d,2n}(R \, t)$ ought to be replaced by $I_{d,2n+2}(R \, t)$. This reproduces their explicit results correctly.})
\begin{eqaed}\label{eq:heat_kernel_coeffs_litim}
    \sum_{k \geq 0} a_k \, t^k = \left(1 - 4 t + \frac{28}{5} \, t^2 - \, \frac{96}{35} \, t^3\right) e^{16 t}\,.
\end{eqaed}
This contribution arises from zeta-function regularization, and cancels the \ac{UV}-divergent part in the Schwinger integral of \cref{eq:subtracted_identity}. \\

\noindent The winding term can instead be integrated termwise, obtaining
\begin{eqaed}\label{eq:winding_curved}
    & \frac{1}{2 \, \text{Vol}(S^9)} \sum_{l \geq 0} \frac{(l+7)! (2l+8)}{8!l!}\sum_{k>0} \sum_{\gamma \in \mathcal{P}} \frac{\ell_\gamma}{2 \sinh(k \ell_\gamma/2)} \int_0^\infty \frac{dt}{t} \, \frac{e^{-t \left(\frac{1}{4} + x^2 \, l(l+8)\right) - \frac{(k \ell_\gamma)^2}{4t}}}{\sqrt{4\pi t}} \\
    & = \frac{1}{2 \, \text{Vol}(S^9)} \sum_{l \geq 0} \frac{(l+7)! (2l+8)}{8!l!}\sum_{k>0} \sum_{\gamma \in \mathcal{P}} \frac{\ell_\gamma}{2 \sinh(k \ell_\gamma/2)} \, \frac{e^{-k \ell_\gamma \sqrt{\frac{1}{4} + x^2 \, l(l+8)}}}{k \ell_\gamma}\,.
\end{eqaed}
We now bound the corresponding combinations in \cref{eq:subtracted_identity,eq:winding_curved}. \\

\noindent \textbf{The identity term.} We have
\begin{eqaed}\label{eq:identity_A_curved}
    \mathcal{A}_\text{identity} & = \frac{1}{2\pi} \int_0^\infty H\!\left(\rho^2+\frac{1}{4}\right) \tanh(\pi \rho) \, \rho \, d\rho + \Delta \mathcal{A}\,,
\end{eqaed}
where $\Delta \mathcal{A}$ is the analytically continued part, which is numerically bounded by $\approx 10^{-9}$ in absolute value, and
\begin{eqaed}\label{eq:H_function}
    H(\lambda) \equiv \frac{1}{18} \int_0^\infty \frac{dt}{t} \left(\frac{1}{(4\pi t)^{\frac{9}{2}}} \sum_{k=0}^5 (2k-9) \, a_k \, (t x_*^2)^k - \, \frac{x_* \partial_x K_{S^9}(t)}{\text{Vol}(S^9)}\bigg|_{x=x_*} \right) e^{-t \lambda}\,.
\end{eqaed}
For the integral over $0<t\leq1$, it suffices to consider $z=x_*^2t\leq\frac{1}{36}$.
\\

\noindent Using the Poisson identity
\begin{equation}
	\sum_{m\in\mathbb Z}e^{-zm^2}
	=\sqrt{\frac{\pi}{z}}\sum_{q\in\mathbb Z}e^{-\frac{\pi^2q^2}{z}}\,,
\end{equation}
define
\begin{equation}
	P(z)=1-4z+\frac{28}{5}z^2-\frac{96}{35}z^3\,,
\end{equation}
which is the polynomial in \cref{eq:heat_kernel_coeffs_litim}.
The unit-radius heat-kernel density can be expressed by
\begin{equation}
	k(z)\equiv\left.\frac{K_{S^9}(z)}{\text{Vol}(S^9)}\right|_{x=1}
	=\frac{e^{16z}P(z)}{(4\pi z)^{\frac{9}{2}}}+E(z)\,,
\end{equation}
where the remainder term $E(z)$ is given by
\begin{equation}
	E(z)=\frac{e^{16z}}{1680\pi^{\frac{9}{2}}}
	\sum_{q=1}^{\infty}\big(\partial_z^4+14\partial_z^3+49\partial_z^2+36\partial_z\big)F_q(z)\,.
\end{equation}
with $F_q(z)\equiv z^{-\frac{1}{2}}e^{-\frac{\pi^2q^2}{z}}$ by the Poisson summation formula.\\

\noindent Using $16z<1$, one can check that
\begin{equation}
	\begin{aligned}
		\frac{\abs{2zE'(z)}}{z^{\frac{3}{2}}}
		&\leq\frac{10\pi^{\frac{11}{2}}}{21}e^{16z}z^{-11}
		\sum_{q=1}^{\infty}q^{10}e^{-\frac{\pi^2q^2}{z}} \\
		&\leq\frac{52\pi^{\frac{11}{2}}}{35}z^{-11}e^{-\frac{\pi^2}{z}}
		<10^{-70}\,.
	\end{aligned}
\end{equation}
Define now $f(z)=e^{16z}[(32z-9)P(z)+2zP'(z)]$ and denote its degree-five Taylor polynomial at zero by $T_5f(z)$. The six local subtraction terms in \cref{eq:H_function}, evaluated at $x=1$, are precisely $(4\pi z)^{-\frac{9}{2}}T_5f(z)$. Differentiating on $0\leq z\leq\frac{1}{36}$ yields
\begin{equation}
	0<f^{(6)}(z)\leq6!\,\frac{557056}{105}e^{\frac{4}{9}}\,.
\end{equation}
Taylor's theorem bounds the full subtracted derivative by
\begin{equation}
	\abs{\frac{T_5f(z)}{(4\pi z)^{\frac{9}{2}}}-2zk'(z)}
	\leq\left(\frac{557056e^{\frac{4}{9}}}{105(4\pi)^{\frac{9}{2}}}+10^{-70}\right)z^{\frac{3}{2}}
	<0.1\,z^{\frac{3}{2}}\,.
\end{equation}
Restoring $x=x_*$ multiplies this estimate by $x_*^9$, giving the bound $0.1\,x_*^{12}t^{\frac{3}{2}}$ for the parenthesis term in \cref{eq:H_function}. Furthermore, to bound the region $t \geq 1$, we use the fact that the difference between the full heat kernel term and the local terms with $k \leq 4$ is negative. By dimensional analysis, reinstating $x=x_*$ effectively amounts to rescaling $t \mapsto x_*^2 t$ and multiplying everything by $x_*^9$. Moreover, we can simplify the heat-kernel derivative according to
\begin{eqaed}\label{eq:xdK_bound}
    - \, \frac{x_* \partial_x K_{S^9}(t)}{\text{Vol}(S^9)}\bigg|_{x=x_*} & = \frac{4 \, t \, x_*^2}{8! \text{Vol}(S^9)} \sum_{l > 0} \frac{(l+8)!(l+4)}{(l-1)!} \, e^{-t \, x_*^2 l(l+8)} \\
    & \geq \frac{4 \, t \, x_*^2}{8! \text{Vol}(S^9)} \sum_{l=1}^{15} \frac{(l+8)!(l+4)}{(l-1)!} \, e^{-t \, x_*^2 l(l+8)}\,,
\end{eqaed}
where 15 terms are retained simply for numerical convenience. Since, by singling out the $k=5$ term, the remaining local terms are actually negative and larger than the above expression, we can obtain an absolute upper bound by truncating the sum. With these ingredients, we split the Schwinger integral according to
\begin{eqaed}\label{eq:H_bound}
    \abs{H(\lambda)} & \leq \frac{x_*^{12}}{180} \int_0^1 dt \, \sqrt{t} \, e^{-t\lambda} + \int_1^\infty \frac{dt}{t}  \frac{a_5}{18(4\pi t)^{\frac{9}{2}}} (t x_*^2)^5 e^{-t\lambda} \\
    & + \int_1^\infty \frac{dt}{t} \left(\sum_{k=0}^4 \frac{(9-2k)a_k}{18(4\pi t)^{\frac{9}{2}}} (t x_*^2)^k - \frac{t \, x_*^{11}}{\pi^5} \sum_{l=1}^{15} \frac{(l+8)!(l+4)}{15120(l-1)!} \, e^{-t \, x_*^2 l(l+8)} \right) e^{-t \lambda}\,.
\end{eqaed}
Inserting this into \cref{eq:identity_A_curved}, the contribution from the first line is bounded by $1.02 \times 10^{-12}$, whereas the contribution from the second line is bounded by $1.5 \times 10^{-7}$. As a result, $\abs{\mathcal{A}_\text{identity}}$ is bounded by, e.g., $10^{-6}$. \\

\noindent \textbf{The winding term.} We have
\begin{eqaed}\label{eq:winding_A_curved}
    4\pi (g-1) \, \mathcal{A}_\text{winding} & = \sum_{l \geq 0} \frac{(l+7)! (2l+8)}{36 \, \text{Vol}(S^9) \times 8!l!}\sum_{k>0} \sum_{\gamma \in \mathcal{P}} \frac{\ell_\gamma \, x_*^2 \, l(l+8)}{\sinh(k \ell_\gamma/2)} \, \frac{e^{-k \ell_\gamma \sqrt{\frac{1}{4} + x_*^2 \, l(l+8)}}}{\sqrt{\frac{1}{4} + x_*^2 \, l(l+8)}} \\
    & = \frac{2x_*^{11}}{3\pi^5} \sum_{l > 0} \frac{(l+8)!(l+4)}{8!(l-1)!}\sum_{k>0} \sum_{\gamma \in \mathcal{P}} \frac{\ell_\gamma}{\sinh(k \ell_\gamma/2)} \, \frac{e^{-k \ell_\gamma \sqrt{\frac{1}{4} + x_*^2 \, l(l+8)}}}{\sqrt{\frac{1}{4} + x_*^2 \, l(l+8)}} \\
    & \equiv \frac{x_*^{11}}{60480\pi^5} \sum_{\gamma \in \mathcal{P}} \mathcal{F}(\ell_\gamma)\,.
\end{eqaed}
As in \cref{eq:F_bound}, we write
\begin{eqaed}\label{eq:curved_F_bound}
    \mathcal{F}(z) & \equiv \sum_{l,k > 0} \frac{(l+8)!(l+4)}{(l-1)!} \, \frac{z}{\sinh(k z/2)} \, \frac{e^{-k z \sqrt{\frac{1}{4} + x_*^2 \, l(l+8)}}}{\sqrt{\frac{1}{4} + x_*^2 \, l(l+8)}} \\
    & \leq \frac{1}{x_*} \, \frac{z}{\sinh(z/2)} \sum_{l > 0} \frac{(l+8)^{9}}{e^{z \sqrt{\frac{1}{4}+x_*^2 \, l(l+8)}} - 1}\,.
\end{eqaed}
Using the fact that $4x_*^2 \, l(l+8) = \frac{l(l+8)}{9} \geq 1$, and that in the range $l\geq 20$ we have $\sqrt{1+\frac{l(l+8)}{9}}-1>\frac{l}{3}+a$ for $0 < a \leq \sqrt{569}/3-23/3 \approx 0.284$, we bound
\begin{eqaed}\label{eq:exp_bound}
    \frac{1}{e^{\frac{z}{2} \sqrt{1 + 4x_*^2l(l+8)}} - 1} & \leq \frac{1}{e^{\frac{z}{2} \sqrt{1 + \frac{l(l+8)}{9}}} - e^{\frac{z}{2} \left( \sqrt{1 + \frac{l(l+8)}{9}}-1\right)}} \\
    & = \frac{e^{-\frac{z}{2} \left( \sqrt{1 + \frac{l(l+8)}{9}}-1\right)}}{e^{\frac{z}{2}} - 1}\,,
\end{eqaed}
leading to
\begin{eqaed}\label{eq:curved_F_curved_further_bound}
    \mathcal{F}(z) & \leq \frac{6z \, e^{- \frac{z}{2}}}{\sinh(\frac{z}{2})(1-e^{-\frac{z}{2}})} \sum_{l > 0} (l+8)^9 \, e^{-\frac{z}{2} \left(\sqrt{1+\frac{l(l+8)}{9}}-1\right)} \\
    & \leq \frac{6z \, e^{- \frac{z}{2}}}{\sinh(\frac{z}{2})(1 - e^{-\frac{z}{2}})} \left(\sum_{l=1}^{20}(l+8)^9 \, e^{-\frac{z}{2} \left(\sqrt{1+\frac{l(l+8)}{9}}-1\right)} + \sum_{l>20} (l+8)^9 \, e^{-a\frac{z}{2}} \, e^{-\frac{z}{6}l} \right) .
\end{eqaed}
Employing the method in \cref{eq:F_kernel} once again with the maximum value of $a$, we obtain
\begin{eqaed}\label{eq:sum_curved_F_bound}
    \sum_{\gamma \in \mathcal{P}} \mathcal{F}(\ell_\gamma) \leq 4\pi (g-1) \times (2 \times 10^{16}) \, C\,,
\end{eqaed}
from which \cref{eq:winding_A_curved} is bounded according to
\begin{eqaed}\label{eq:winding_A_curved_bound}
    \mathcal{A}_\text{winding} \leq 3 \, C \leq 216\,,
\end{eqaed}
using again \cref{eq:constant_C}. Despite some degree of enhancement, the qualitative conclusion remains unaltered, and the overall bound $\mathcal{A} \leq 216$ holds analogously to \cref{eq:A_bound}. In the more complicated settings studied in later sections, we will use a simpler approach to bounding the \ac{dS} contribution in terms of the flat-space approximation directly, although applied to this case the resulting bound would have been weaker by around a factor of $20$. \\

\noindent To conclude the argument, we still need to bound the Casimir contribution at the stationary ratio $x$. For $M_{11}L\geq1$ and $g\geq2$, \cref{eq:x_extremum} gives
\begin{equation}
	x^2=\frac{1}{36}+\frac{1}{216\pi^6(g-1)(M_{11}L)^9}\,,
	\qquad \frac{1}{6}<x<\overline{x}\equiv\frac{83341}{500000}\,.
\end{equation}
Since $\overline{x}^2<\frac{1}{35}$, the Poisson and Taylor estimates above provide a bound for the parenthesis in \cref{eq:H_function} by $0.1x^{12}t^{\frac{3}{2}}$, valid in the interval $0<t\leq1$. Consider now the same quantity $\mathcal A$ for general values of $x$. We split its identity contribution at $t=1$ and bound the hyperbolic heat density by $\frac{e^{-\frac{t}{4}}}{4\pi t}$. For $t\geq1$, we bound the local subtraction and spectral terms separately. The same Poisson formula gives $\frac{K_{S^9}(1)}{\text{Vol}(S^9)}<\frac{2.6}{(4\pi)^{\frac{9}{2}}}$. Together with $\int_1^\infty\frac{e^{-at}}{t}\,dt\leq\frac{e^{-a}}{a}$ for $a>0$, this yields
\begin{equation}
\begin{aligned}
	\abs{\mathcal A_\text{identity}(x)}
	&\leq\frac{0.1\overline{x}^{12}}{36\pi}
	+\frac{1}{72\pi(4\pi)^{\frac{9}{2}}}\sum_{k=0}^5
	\frac{\abs{2k-9}a_k\overline{x}^{2k}}{\frac{11}{2}-k} \\
	&\quad+\frac{2.6e^{-\frac{1}{4}}}{36\pi(4\pi)^{\frac{9}{2}}}
	+2.551\times10^{-8}<4\times10^{-7}\,.
\end{aligned}
\end{equation}
The last term bounds the finite contribution restored by analytic continuation. After integrating by parts, using $\sqrt{\rho^2+\frac{1}{4}}\leq\rho+\frac{1}{2}$ and $\operatorname{sech}^2(\pi\rho)\leq4e^{-2\pi\rho}$ reduces its estimate to integrals of polynomial terms times $e^{-2\pi\rho}$. \\

\noindent After removing the prefactor $x^{11}$ in \cref{eq:winding_A_curved}, each positive summand decreases with $x$. Bounding this prefactor at $\overline{x}$ and the remaining sum at $x_*=\frac{1}{6}$, \cref{eq:sum_curved_F_bound} gives
\begin{equation}
\begin{aligned}
	\mathcal A_\text{winding}(x)
	&\leq\frac{\overline{x}^{11}(2\times10^{16})}{60480\pi^5}\,C
	<2.99C\leq215.28\,, \\
	\mathcal A(x)&<215.28+4\times10^{-7}<216\,, \\
	(M_{11}r)^9&=\frac{\mathcal A(x)}{16x^2}
	+\frac{x^7}{48\pi^6(g-1)}<484.381<486\,.
\end{aligned}
\end{equation}
In the last line we used \cref{eq:L_solution}, and $\frac{L}{r}<6$ also gives a bound for the external radius. Points with $M_{11}L<1$ already have uncontrolled external curvature. \\

\noindent All in all, we have proven that the Casimir energy is bounded even upon inclusion of the \ac{dS} contribution due to the lack of scale separation. In order to enhance the Casimir contribution one needs to move away from critical points in moduli space, leading to likely steep runaways and/or additional complications, and most prominently a loss of control in the physical systole length. This will play an important role in the following section.

\subsection{Beyond rigid Riemann surfaces}\label{sec:beyond_rigid_riemann_surfaces}

We now extend the above analysis beyond rigid Riemann surfaces. The basic mechanism we have identified is that a uniform bound on the (dimensionless) systole length $s = \text{sys}(\Sigma_g)$ is enough to bound the Casimir energy, at least in the approximation in which the scalar-field Casimir energy replaces the full supergravity combination---in \cref{app:bounding_sugra_casimir} we extend the resulting numerical bounds for this case to the supergravity spectrum. In this section, we show that this is not exclusive to rigid Riemann surfaces; all critical point of the Casimir potential are bounded by this mechanism. Concretely, we find that the Casimir potential has no critical point when $s \leq 1$.  Once again, we emphasize that this bound on the \ac{EFT} control parameters was derived by approximating the Casimir energy of supergravity by that of scalar fields. In the rigid case, the partial cancellation between bosonic and fermionic degrees of freedom can only worsen the situation, and therefore the bounds should remain valid---we expect that $A$ cannot be enhanced more than a few orders of magnitude. However, in the discussion below on moduli derivatives of $A$, these cancellations could in principle go in the opposite direction and provide more non-rigid critical points than the ones we will be able to control. Thus, strictly speaking, the following considerations do not apply directly to supergravity \emph{per se}. In \cref{app:bounding_sugra_gradient} we address this issue, albeit only parametrically. \\

\begin{figure}[ht]
    \centering
    \includegraphics{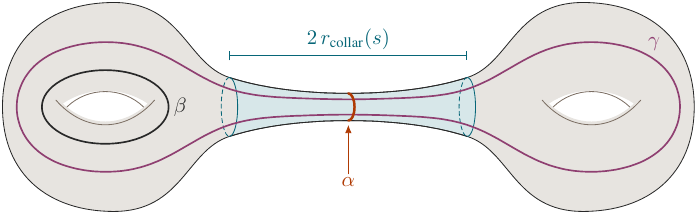}
    \caption{Collar around a systole $\alpha$, with a crossing geodesic $\gamma$ and a disjoint geodesic $\beta$.}
    \label{fig:collar}
\end{figure}

\noindent \textbf{Improved bound on rigid Riemann surfaces.} To begin with, as a warm-up for the ensuing proof strategy, we can improve the previous bound by showing that rigid Riemann surfaces have a stronger lower bound on their systole length, given by $\text{sys}(\Sigma_g) \geq 2 \sinh^{-1}(1) \equiv s_0$. Intuitively, the idea is that short enough systoles do not intersect due to the ``collar lemma" \cite{Keen1974COLLARSOR, Buser1978} (explained shortly), and therefore can participate independently in a pair-of-pants decomposition, which has moduli. In detail, the collar lemma implies that systoles are surrounded by pairwise disjoint tubes of radius
\begin{eqaed}\label{eq:collar_lemma}
    r_\text{collar}(s_0) & = \sinh^{-1}\!\left(\frac{1}{\sinh{s_0/2}}\right) \\
    & = \frac{s_0}{2}\,.
\end{eqaed}
This is depicted in \cref{fig:collar}. As a result, if $\text{sys}(\Sigma_g) < s_0$, any two systoles $\alpha \neq \beta$, whose length $\ell_\alpha = \ell_\beta = \text{sys}(\Sigma_g)$ are disjoint. Letting $G = \text{Aut}(\Sigma_g)$, the automorphism orbit of $\alpha$ is a set $\mathcal{O}_\alpha$ of curves in $\Sigma_g$. Then, the function
\begin{eqaed}\label{eq:sum_of_lengths}
    S : \mathcal{T}_g & \to \mathbb{R}_{\geq 0} \\
    u & \mapsto \sum_{\gamma \in \mathcal{O}_\alpha} \ell_\gamma(u)
\end{eqaed}
on Teichm\"{u}ller space is $G$-invariant at the fixed point $u=u_*$ corresponding to our chosen rigid Riemann surface. Since the $G$-invariant tangent space at this point is trivial, $dS$ must vanish there. However, evaluating this one-form on the vector field $\nabla \ell_\alpha$, the results of \cite{Riera2005,wolpert2008behaviorgeodesiclengthfunctionsteichmueller,Wolpert2012Geodesic} for disjoint geodesics and their length gradients imply
\begin{eqaed}\label{eq:riera}
    dS(\nabla \ell_\alpha) & = \sum_{\gamma \in \mathcal{O}_\alpha} \langle \nabla \ell_\gamma, \nabla \ell_\alpha \rangle \\
    &= \norm{\nabla \ell_\alpha}^2 + \sum_{\gamma \in \mathcal{O}_\alpha \backslash \{ \alpha \} } \langle \nabla \ell_\gamma, \nabla \ell_\alpha \rangle \\
    & > \frac{2}{\pi} \, \ell_\alpha\,,
\end{eqaed}
which is a contradiction. Therefore, rigid Riemann surfaces have $\text{sys}(\Sigma_g) \geq s_0$. This improved bound on $\text{sys}(\Sigma_g)$ does not cover orbifolds, but strengthens \cref{eq:A_bound,eq:winding_A_curved_bound}. \\

\noindent \textbf{Casimir extrema have bounded systole length.} Now we apply the same strategy to the effective potential itself. Since complex-structure moduli extremize the effective potential if and only if they extremize the winding term in the Casimir energy\footnote{The identity term does not depend on complex-structure moduli.}, we can consider the gradient of the trace formula with respect to the shape moduli, which has the schematic form
\begin{eqaed}\label{eq:trace_formula_all_geodesics}
    A = \sum_{\gamma \in \mathcal{P}} \sum_{k \geq 1} \frac{1}{k} \, F(k\ell_\gamma)
\end{eqaed}
with $F$ decreasing. As a function on moduli space,
\begin{eqaed}\label{eq:casimir_differential}
    dA & = - \sum_{\gamma \in \mathcal{P}} \sum_{k \geq 1} (-F'(k\ell_\gamma)) \, d\ell_\gamma \\
    & \equiv - \sum_{\gamma \in \mathcal{P}} w(\ell_\gamma) \, d\ell_\gamma\,,
\end{eqaed}
and $dA=0$ at a critical point. We will now show that if $s=\text{sys}(\Sigma_g)$ is bounded above as, say, $s \leq 1$, then $dA$ does not vanish. \\

\noindent Letting $\alpha$ be a systole and denoting $\ell_\alpha = \text{sys}(\Sigma_g)=s \leq 1$, we compute
\begin{eqaed}\label{eq:dA_contraction}
    dA(\nabla \ell_\alpha) & = - \sum_{\gamma \in \mathcal{P}} w(\ell_\gamma) \, \langle \nabla \ell_\gamma , \nabla \ell_\alpha \rangle \\
    & \leq - \frac{2s}{\pi} \, w(s) + \frac{4}{\pi} \sum_{\gamma : \gamma \cap \alpha \neq \varnothing} w(\ell_\gamma) \, I(\alpha, \gamma)\,,
\end{eqaed}
where $I(\alpha,\gamma)$ denotes the number of crossings $c \in \gamma \cap \alpha$ between distinct\footnote{Alternatively, we can say that the geodesics intersect transversely.} intersecting geodesics $\alpha$ and $\gamma$. To derive \cref{eq:dA_contraction} we used two facts \cite{Riera2005, wolpert2008behaviorgeodesiclengthfunctionsteichmueller,Wolpert2012Geodesic}: first, disjoint geodesics contribute positively to the inner product of length gradients. Second, for distinct intersecting geodesics, the inner product of the length gradients is bounded below by the sum over crossings
\begin{eqaed}\label{eq:crossings_angles}
    \sum_{c \in \gamma \cap \alpha} \left(u_c \, \log \frac{1+u_c}{1-u_c} - 2\right) \geq -2 \, I(\alpha,\gamma)
\end{eqaed}
with $u_c$ the cosine of the crossing angle. Due to the collar lemma, a geodesic $\gamma$ crossing a systole $\alpha$ a number $I(\alpha,\gamma)$ of times must traverse the full width collar of radius $r_\text{tube}(s)$ that many times, and thus $\ell_\gamma \geq 2I(\alpha,\gamma) r_\text{tube}(1)$. This also implies that $\ell_\gamma \geq 2r_\text{tube}(1) \approx 2.8$. Then, we can use the packing estimate for crossing geodesics of \cref{eq:crossing_packing_bound,eq:crossing_packing_bound_18} to obtain
\begin{eqaed}\label{eq:dA_bound}
    \frac{\pi}{2} \, dA(\nabla \ell_\alpha) & \leq - s \, w(s) + \frac{1}{r_\text{tube}(s)} \sum_{\gamma : \gamma \cap \alpha \neq \varnothing} \ell_\gamma \, w(\ell_\gamma) \\
    & \leq - s \, w(s) + \frac{1}{r_\text{tube}(s)} \int_{2r_\text{tube}(s)}^\infty \ell \, w(\ell) \, dN_\text{cross}(\ell) \\
    & \leq - s \, w(s) + \frac{18}{s^2 r_\text{tube}(1)} \int_{2r_\text{tube}(1)}^\infty e^\ell \partial_\ell(-\ell w(\ell)) \, d\ell \\
    & = -s \, w(s) + \frac{36e^{2r_\text{tube}(1)}}{s^2} \, w(2r_\text{tube}(1)) + \frac{18}{s^2 r_\text{tube}(1)} \int_{2r_\text{tube}(1)}^\infty \ell \, e^\ell \, w(\ell) \, d\ell\,,
\end{eqaed}
where we used that $-\ell F'(k\ell)$ is decreasing in $\ell$. Finally, we bound $w(\ell)$ on both sides to simplify each term of \cref{eq:dA_bound}. Since $2r_\text{tube}(1) \approx 2.81$, splitting the estimates for arguments $\ell \leq 1$ and $\ell > 2$ is sensible. For the specific choice\footnote{Since we seek to prove that $dA \neq 0$, we are free to conveniently rescale $F$ to simplify the resulting expressions. The overall neglected prefactor is $(4096\pi^5v)^{-1}$.} $F(\ell) = \frac{2K_5(\ell/2)}{\ell^4 \sinh(\ell/2)}$, we have
\begin{eqaed}\label{eq:F_split_1}
    -F'(\ell) & = \frac{\ell \, K_6(\ell/2)+\left(\ell \coth(\ell/2)-2\right) K_5(\ell/2)}{\ell^5 \sinh(\ell/2)} \\
    & \leq \frac{K_6(\ell/2)+K_5(\ell/2)}{\ell^4 \sinh(\ell/2)}\,,
\end{eqaed}
and therefore, up to an irrelevant prefactor,
\begin{eqaed}\label{eq:w_sum_upper_bound}
    w(\ell) & \leq \sum_{k \geq 1} \frac{(k\ell)^5 \left(K_6(k\ell/2)+K_5(k\ell/2)\right)}{(k\ell)^9 \sinh(k\ell/2)} \\
    & \leq \frac{\ell^5\left(K_6(\ell/2)+K_5(\ell/2)\right)}{\ell^9 \sinh(\ell/2)} \sum_{k \geq 1} \frac{1}{k^9} \\
    & = \zeta(9) \, \frac{\left(K_6(\ell/2)+K_5(\ell/2)\right)}{\ell^4 \sinh(\ell/2)}\,,
\end{eqaed}
where we used that $z^5K_{5,6}(z/2)$ is decreasing. Furthermore, since $-F'(\ell) \sim 491520 \, \ell^{-11}$ for small $\ell$, one can check that $-\ell^{11}F'(\ell)$ is decreasing and thus $-F'(\ell) \geq 469578 \, \ell^{-11}$ for $\ell \leq 1$. As a result, retaining the first term in the winding sum, $w(\ell) \geq 469578 \ell^{-11}$ in this range. \\

\noindent All in all, if $s \leq 1$, up to an irrelevant positive prefactor that we have neglected we have
\begin{eqaed}\label{eq:dA_bound_2}
    \text{const.} \times dA(\nabla \ell_\alpha) & \leq - \, \frac{469578}{s^{10}} + \frac{3642}{s^2} \leq -465936 \,,
\end{eqaed}
and thus $dA \neq 0$. As anticipated, extrema of the effective potential in complex-structure moduli space must have $s > 1$, and thus the bounds we derived (in fact, even stronger bounds, as we have seen above) apply even beyond rigid Riemann surfaces. As we have anticipated, strictly speaking a full derivation of this result in supergravity requires controlling the (partial) Bose-Fermi cancellation in the gradient, which could undermine the above argument. However, the weaker divergence surviving the cancellation can in principle still obstruct parametric control. Indeed, in \cref{app:bounding_sugra_gradient} we address this point for the supergravity spectrum at a parametric level. \\

\noindent \textbf{Including \ac{dS} curvature.} To conclude the proof, we must extend it to external \ac{dS} spacetime. To this end, in the expression for $F(\ell)$ we replace the function $g(\ell)$ by its counterpart appearing in \cref{eq:casimir_curved_def,eq:winding_curved}. Up to irrelevant prefactors,
\begin{eqaed}\label{eq:F_dS_riemann}
    F(\ell) = \int_0^\infty \frac{dt}{t^\frac{3}{2}} \, \frac{K_{S^9}(t)}{\text{Vol}(S^9)} \, \frac{\ell \, e^{-\frac{t}{4}-\frac{\ell^2}{4t}}}{\sinh(\ell/2)}\,,
\end{eqaed}
so that
\begin{eqaed}\label{eq:Fprime_dS_riemann}
    -F'(\ell) & = \int_0^\infty \frac{dt}{t^\frac{5}{2}} \, \frac{K_{S^9}(t)}{\text{Vol}(S^9)} \, \frac{\ell^2+(\ell \coth(\ell/2)-2)t}{2\sinh(\ell/2)} \, e^{-\frac{t}{4}-\frac{\ell^2}{4t}} \equiv \int_0^\infty \frac{dt}{t^\frac{5}{2}} \, \frac{K_{S^9}(t)}{\text{Vol}(S^9)} \, G(\ell,t)\,.
\end{eqaed}
The combination $\ell \coth(\ell/2)-2$ multiplying $t$ is non-negative, and thus $-F'(\ell) \geq 0$ as expected. In order to apply the preceding results, we can split the Schwinger integral at $t=1$ and employ upper and lower bounds for $K_{S^9}(t)$ in terms of its flat-space counterpart. To begin with, we have $K_{S^9}(t) \geq 1$ due to the zero-mode. Then, since the sphere has positive curvature, the Li-Yau inequality \cite{LiYau1986} implies that $K_{S^9}(t)/\text{Vol}(S^9) \geq (4\pi t)^{-9/2}$. Moreover, we have the upper bounds
\begin{eqaed}\label{eq:K_S9_upper_bounds}
    K_{S^9}(t) & \leq 2.6 \, \frac{\text{Vol}(S^9)}{(4\pi t)^{\frac{9}{2}}}\,, \qquad t \leq 1\,, \\
    K_{S^9}(t) & \leq 1 + 2.6 \, t^\frac{9}{2} \, e^{-\lambda_1 (t-1)} \, \frac{\text{Vol}(S^9)}{(4\pi t)^{\frac{9}{2}}}\,, \qquad t > 1\,.
\end{eqaed}
The former bound in \cref{eq:K_S9_upper_bounds} follows from the global upper bound \cite{baudoin2009perelmansentropydoublingproperty, baudoin2011notelowerboundsestimates}
\begin{eqaed}\label{eq:K_Sd_global_bound}
    K_{S^d_1}(t) \leq \left(1 - e^{- \frac{2(d-1)t}{3}}\right)^{-\frac{d}{2}}
\end{eqaed}
for the unit $d$-sphere, optimizing the ratio to the flat-space heat kernel over the range $t \leq 1$. The latter bound in \cref{eq:K_S9_upper_bounds} follows from recasting \cref{eq:heat_kernel_sphere} into the spectral sum
\begin{eqaed}\label{eq:K_S9_large-t_bound}
    K_{S^9}(t) & = 1 + e^{-\lambda_1 t}\sum_{l > 0} D_l \, e^{-(\lambda_l-\lambda_1)t} \\
    & \leq 1 + e^{-\lambda_1 t} \sum_{l > 0} D_l \, e^{-(\lambda_l-\lambda_1)} \\
    & \leq 1 + e^{-\lambda_1 (t-1)} \, K_{S^9}(1) \\
    & \leq 1 + 2.6 \, t^\frac{9}{2} \, e^{-\lambda_1 (t-1)} \, \frac{\text{Vol}(S^9)}{(4\pi t)^{\frac{9}{2}}}\,,
\end{eqaed}
Here $\lambda_1=9x^2\geq1/4$. The same $2.6$ heat-kernel bound holds throughout $x\leq\overline{x}$: its ratio to the flat density increases with $x^2t$ and at $x^2t=\overline{x}^{2}$ it is still below $2.6$. Except for the constant zero-mode term in \cref{eq:K_S9_upper_bounds}, these bounds can be used to modify \cref{eq:dA_bound,eq:dA_bound_2}. Since the integrand in \cref{eq:Fprime_dS_riemann} is positive, the counterpart of the first, negative, term in \cref{eq:dA_bound_2} is bounded below by the flat-space approximation. To bound the positive terms, we isolate the constant zero-mode contribution in \cref{eq:K_S9_upper_bounds} by splitting $K_{S^9}(t) = 1 + (K_{S^9}(t)-1)$ and apply \cref{eq:K_S9_upper_bounds}. Up to an irrelevant overall prefactor that we shall ignore in the following, this results in
\begin{eqaed}\label{eq:new_w}
    w(\ell) & = (4\pi)^\frac{9}{2}\int_0^\infty \frac{dt}{t^\frac{5}{2}} \, \frac{K_{S^9}(t)}{\text{Vol}(S^9)} \sum_{k \geq 1} G(k\ell,t) \\
    & \leq 2.6 \int_0^1 \frac{dt}{t^\frac{5}{2}} \, \frac{1}{t^{\frac{9}{2}}} \sum_{k \geq 1} G(k\ell,t) + (4\pi)^\frac{9}{2}\int_1^\infty \frac{dt}{t^\frac{5}{2}} \, \frac{K_{S^9}(t)}{\text{Vol}(S^9)} \sum_{k \geq 1} G(k\ell,t) \\
    & \leq 2.6 \, w_\text{flat}(\ell) + (4\pi)^\frac{9}{2}\int_1^\infty \frac{dt}{t^\frac{5}{2}} \, \frac{K_{S^9}(t)-1}{\text{Vol}(S^9)} \, \sum_{k \geq 1} G(k\ell,t) + w_\text{z.m.}(\ell)\,.
\end{eqaed}
In order to simplify this expression, for $t \geq 1$ we bound
\begin{eqaed}\label{eq:G_sum_bound}
    \sum_{k \geq 1} G(k\ell,t) & = \sum_{\geq 1} \frac{k^2\ell^2+(k\ell \coth(k\ell/2)-2)t}{2\sinh(k\ell/2)} \, e^{-\frac{t}{4}-\frac{k^2\ell^2}{4t}} \\
    & \leq \frac{1}{2\sinh(\ell/2)} \sum_{k \geq 1} \left(k^2\ell^2+k\ell t\right) e^{-\frac{t}{4}-\frac{k^2\ell^2}{4t}} \\
    & \leq \frac{\ell(\ell+t)}{2\sinh(\ell/2)} \sum_{k \geq 1} k^2 \, e^{-\frac{t}{4}-\frac{k\ell^2}{4t}} \\
    & = \frac{\ell(\ell+t) \, \cosh(\frac{\ell^2}{8t}) \, e^{-\frac{t}{4}}}{8 \sinh(\frac{\ell}{2})\sinh^3(\frac{\ell^2}{8t})} \equiv \mathcal{G}(\ell,t)\,.
\end{eqaed}
Thus, using \cref{eq:K_S9_upper_bounds} and separating the constant zero-mode term $w_\text{z.m.}(\ell)$, \cref{eq:new_w} becomes
\begin{eqaed}\label{eq:new_w_bound}
    w(\ell) & \leq 2.6 \, w_\text{flat}(\ell) + w_\text{z.m.}(\ell) + 2.6 \int_1^\infty \frac{dt}{t^\frac{5}{2}} \, e^{-\frac{t-1}{4}} \, \mathcal{G}(\ell,t)\,.
\end{eqaed}
Rescaling $A \mapsto \widetilde{A}$ to remove the irrelevant positive prefactor that we ignored in the above steps, \cref{eq:new_w_bound} can be evaluated numerically to yield
\begin{eqaed}\label{eq:dA_curved_bound}
    d\widetilde{A}(\nabla \ell_\alpha) & \leq -0.999s \, w_\text{flat}(s) + \frac{36e^{2r_\text{tube}(1)}}{s^2} \, w(2r_\text{tube}(1)) \\
    & + \frac{18}{s^2 r_\text{tube}(1)} \int_{2r_\text{tube}(1)}^\infty \ell \, e^\ell \, (w(\ell)-w_\text{z.m.}(\ell)) \, d\ell +  \delta\widetilde{A}_\text{z.m.} \\
    & \leq - \, \frac{469108}{s^{10}} + \frac{28639}{s^2} + \delta\widetilde{A}_\text{z.m.}\,,
\end{eqaed}
where the coefficient $0.999$ follows from the bound $w(s)-w_\text{z.m.}(s)\geq0.999\,w_\text{flat}(s)$ for $0<s\leq1$, established below. The following zero-mode formulae are evaluated at the reference ratio $x_*=\frac{1}{6}$. At general $x$, their right-hand sides acquire the factor $(6x)^9$. The estimate in \cref{eq:G_sum_bound} is insufficient to treat the integrated zero-mode contribution, since the resulting integral diverges. In fact, even the packing estimate used in \cref{eq:dA_bound} is too coarse in this sense. Due to the choice of overall rescaling in \cref{eq:dA_curved_bound} from the choice of $F$, the zero-mode term with the correct relative prefactor is given by
\begin{eqaed}\label{eq:w_zm_correct}
    w_\text{z.m.}(\ell) = \frac{4}{6561} \sum_{k \geq 1} \frac{1}{\sinh^2(k\ell/2)}\,.
\end{eqaed}
Actually, the terms with $k \geq 2$ are unproblematic, and can be treated as above with the packing estimate. Their contribution to \cref{eq:dA_curved_bound} is approximately $1.3 \times 10^{-2} \, s^{-2}$, so we could safely remove these terms from \cref{eq:dA_curved_bound}. However, it turns out to be useful to exploit the geometric structure of the full contribution. Tracking the relative normalizations of the terms in \cref{eq:dA_contraction,eq:dA_bound}, the additional term in \cref{eq:dA_curved_bound} is
\begin{eqaed}\label{eq:dA_zm}
    \delta \widetilde{A}_\text{z.m.} & = - \, \frac{\pi}{2} \sum_{\gamma \in \mathcal{P}} w_\text{z.m.}(\ell_\gamma) \, \langle \nabla \ell_\gamma , \nabla \ell_\alpha \rangle \\
    & = - \, \frac{2\pi}{6561} \sum_{\gamma \in \mathcal{P}} \sum_{k \geq 1} \frac{\langle \nabla \ell_\gamma , \nabla \ell_\alpha \rangle}{\sinh^2(k\ell_\gamma/2)}\,. 
\end{eqaed}
Since the contribution to $w_\text{z.m.}(\ell)$ stems from differentiating the winding sum of $F_\text{z.m.}(\ell) \equiv \frac{16}{6561} (e^\ell - 1)^{-1}$, the expression $\delta \widetilde{A}_\text{z.m.}$ is (up to the overall prefactor) the directional derivative along the vector $X_\alpha \equiv \nabla \ell_\alpha$ of the regularized sum\footnote{The specific regularization employed in \cref{eq:zero-mode_zeta} is immaterial, since the singular part is moduli-independent.}
\begin{eqaed}\label{eq:zero-mode_zeta}
    - \sum_{\gamma \in \mathcal{P}} \sum_{k \geq 1} \frac{1}{k} \, \frac{e^{k\ell_\gamma(1-z)}}{e^{k\ell_\gamma} - 1} & = \log \prod_{\gamma \in \mathcal{P}} \prod_{m \geq 0} \left(1 - e^{-(m+z)\ell_\gamma} \right) 
    & = \log Z_{\Sigma_g}(z)
\end{eqaed}
at $z=1$, in terms of the Selberg zeta function\footnote{The result in \cref{eq:zero-mode_zeta} can be derived by the standard trick of expanding the logarithm and exchanging the order of summations.} $Z_{\Sigma_g}(z)$. Specifically, we have
\begin{eqaed}\label{eq:dA_zeta}
    \delta \widetilde{A}_\text{z.m.} & = - \, \frac{8\pi}{6561} \, X_\alpha(\log Z_{\Sigma_g}(z)) \Big|_{z\to 1^+} \\
    & = - \, \frac{8\pi}{6561} \, X_\alpha(\log Z'_{\Sigma_g}(1)) \\
    & = - \, \frac{8\pi}{6561} \, X_\alpha(\log \text{det}' \Delta_{\Sigma_g})\,,
\end{eqaed}
in terms of the (zeta-regularized) scalar Laplacian determinant \cite{DHoker1986}. Using the bound in \cref{eq:final_selberg-zeta_bound}, we obtain
\begin{eqaed}\label{eq:dA_zeta_final_bound}
    s \, \abs{\delta \widetilde{A}_\text{z.m.}} & \leq \frac{8s^2}{6561} \, \coth^2\!\left(\frac{s}{4}\right) \leq 0.11\,.
\end{eqaed}
{Let us now show that $dA\neq0$ for $s\leq1$ throughout $\frac{1}{6}<x<\overline{x}$. Denote by $\widetilde A$ the Casimir coefficient in the Bessel normalization fixed before \cref{eq:F_split_1}, multiplied by $\frac{\pi}{2}$ as in \cref{eq:dA_bound}. Removing the spherical zero-mode from the global heat-kernel comparison yields $w_{\text{n.z.}}=w-w_{\text{z.m.}}\geq w_{\text{flat}}-w_{\text{z.m.}}$. Given that
\begin{equation}
	w_{\text{z.m.}}(s)\leq\frac{32(6\overline{x})^9}{6561s^2}\,,\qquad
	w_{\text{flat}}(s)\geq\frac{469578}{s^{11}}\,,
\end{equation}
we obtain the bound $w_{\text{n.z.}}(s)\geq0.999w_{\text{flat}}(s)$ for $0<s\leq1$. For crossing geodesics, the remaining large-time contribution can be estimated by
\begin{equation}
\begin{aligned}
	T(\ell)&=2.6\int_1^\infty t^{-\frac{5}{2}}e^{-\frac{t-1}{4}}\mathcal G(\ell,t)\,dt \\
	&\leq\frac{2.6e^{\frac{1}{4}}\sqrt\pi}{\sinh(\frac{\ell}{2})}
	\sum_{k\geq1}\left(\frac{2\sqrt k}{\ell}+\sqrt2 k+k^{\frac{3}{2}}\right)e^{-\ell\sqrt{\frac{k}{2}}}\,.
\end{aligned}
\end{equation}
The above expression is obtained by expanding the hyperbolic functions in \cref{eq:G_sum_bound} and extending the positive Gaussian integrals to $t=0$. This bounds the weight at each fixed length $\ell$. To estimate the sum over all geodesics crossing the systole, set $a=2r_{\text{tube}}(1)>\frac{14}{5}$ and employ the packing bound of \cref{eq:dA_bound}. This leads to the inequality
\begin{equation}
	\frac{1}{r_{\text{tube}}(s)}
	\sum_{\substack{\gamma\ne\alpha\\\gamma\cap\alpha\ne\varnothing}}
	\ell_\gamma T(\ell_\gamma)
	\leq\frac{1}{s^2}
	\left[36e^aT(a)+\frac{36}{a}\int_a^\infty\ell e^\ell T(\ell)\,d\ell\right].
\end{equation}
To obtain an estimate for the term in brackets, we substitute the positive series bound for $T$ into both summands and we integrate term by term. Grouping the resulting sum over $k$ into ranges $m^2\leq k<(m+1)^2$, the first three groups contribute less than $170954$, $20010$ and $5871$, respectively. The fourth and all later groups are bounded by a geometric series with first term $1662$ and ratio $0.3$. Hence the coefficient can be bounded by
\begin{equation}
	170954+20010+5871+\frac{1662}{1-0.3}<200000\,.
\end{equation}
Since $(6x)^9\leq(6\overline{x})^9$, the zero-mode estimate of \cref{eq:dA_zeta_final_bound} gives $\abs{\delta\widetilde A_{\text{z.m.}}}<\frac{0.11}{s}$. Combining this with the flat crossing bound and the large-time estimate above, we obtain
\begin{equation}
\begin{aligned}
	d\widetilde A(\nabla\ell_\alpha)
	&\leq-\frac{0.999\cdot469578}{s^{10}}+
	\frac{2.6\cdot3642+200000}{s^2}+\frac{0.11}{s}\\
	&<-\frac{259000}{s^{10}}<0\qquad(0<s\leq1)\,.
\end{aligned}
\end{equation}
We conclude that scalar critical points with $M_{11}L\geq1$ must have $s>1$.
}

\section{Higher-dimensional hyperbolic manifolds}\label{sec:hyperbolic_manifolds}

In the preceding section, we established a uniform numerical bound on \ac{EFT} control parameters for critical points of M-theory compactified on Riemann surfaces. As anticipated in \cref{sec:casimir_general}, the proof strategy highlights a general mechanism: achieving a large physical systole $M_{11} r s \gg 1$ requires a small dimensionless systole length $s \ll 1$, but this cannot happen at a critical point; in other words, critical Riemann surfaces cannot be too degenerate. On the other hand, higher-dimensional hyperbolic manifolds are Mostow-rigid, so one could try finding a family of such manifold whose dimensionless systole length tends to zero and avoid the large-gradient problem discussed in \cref{sec:beyond_rigid_riemann_surfaces} for Riemann surfaces. In this section we thus investigate this possibility. \\

\noindent Since higher-dimensional hyperbolic manifolds are both rigid and allow arbitrarily small systole lengths, a uniform bound of the value of the Casimir energy requires controlling the distribution of geodesics; in particular, \emph{a priori} one could conceive a proliferation of short geodesics along a suitable sequence of manifolds. This presents us with a new challenge, because a direct application of the packing method employed in \cref{sec:casimir_riemann_surfaces} turns out to be insufficient---the volume of hyperbolic balls in higher dimensions leads to much weaker bounds. To solve this problem, we make use of more refined features of hyperbolic geometry: such manifolds decompose into ``thick'' and ``thin'' parts, which are characterized by their injectivity radius\footnote{Roughly speaking, the injectivity radius at a point is the maximal radius for which the exponential map is a diffeomorphism. Intuitively, in our case it tells us the scale at which the geometry around that point looks like hyperbolic space.}, and the ``thin part'' comprises pairwise disjoint tubes, called \emph{Margulis tubes}. The key property is that these tubes are collars around short closed geodesics, and there exist universal bounds on the size of these tubes. \\

\noindent The upshot is that the asymptotic behavior of the physical systole in higher dimensions is less forgiving, and sending $s \to 0$ results in $M_{11} r s \leq O(1)$. This tells us that the geometry of higher-dimensional hyperbolic manifolds prevents controlled \ac{dS} solutions in a slightly different way than Riemann surfaces, namely by realizing a tension between physical and geometric (dimensionless) systole lengths. More generally, we will derive numerical upper bounds on \ac{EFT} control parameters that are uniform in topology, meaning that controlled \ac{dS} solutions of this type cannot be found by complicating the global topology of the manifold independently of the systole length. \\

\noindent Concretely, we shall proceed by analogy with the preceding section. We begin in \cref{sec:flat-space_nd} with the flat-space approximation, which is again physically unjustified but useful as a mathematical warm-up. After having laid out the groundwork, we conclude the proof for scalar-field Casimir potentials in \cref{sec:flat-space_nd}, and include the external \ac{dS} contribution in \cref{sec:dS_curvature_higher_dim}. \\

\noindent We begin by defining hyperbolic manifolds as quotients $M_n = \mathbb{H}^n/\Gamma$ of hyperbolic space in dimension $3 \leq n \leq 7$. (In fact, the case $n=3$ still covers general negative-curvature Einstein manifolds: the sectional curvatures are determined by the Ricci tensor, and thus any such manifold is hyperbolic.) We will use the fact that the eigenvalues of the scalar Laplacian take the form $\lambda_j = (n-1)^2/4 + r_j^2$. With the same notation as in \cref{eq:selberg_formula}, the sectional curvature is still $K=-1$ and the trace formula for even $h$ now reads \cite{Bytsenko:1994bc, Williams1998Meromorphic, bourque2019kissingnumbersclosedhyperbolic}
\begin{eqaed}\label{eq:nd_trace_formula}
    \sum_j h(r_j) = \frac{2v}{(4\pi)^{\frac{n}{2}}\Gamma(n/2)} \int_0^\infty h(\rho) \, p_n(\rho) \, d\rho + \sum_{k \geq 1} \sum_{\gamma \in \mathcal{P}} \frac{\ell_\gamma \, g(k\ell_\gamma)}{D(\gamma^k)}\,,
\end{eqaed}
where $D(\gamma^k) = \abs{\det(e^{k \ell_\gamma/2}I_{n-1} - e^{-k\ell_\gamma/2}A_\gamma^k)}$ is controlled by a rotation matrix $A_\gamma \in SO(n-1)$ and the kernel
\begin{eqaed}\label{eq:pn_kernel}
    p_n(\rho) = \begin{cases}
    \prod_{m=0}^{\frac{n-3}{2}}(\rho^2+m^2)\,, \qquad n \text{ odd}\,, \\
    \rho \tanh (\pi \rho) \prod_{m=1}^{\frac{n-2}{2}}(\rho^2+\left(m-\frac{1}{2}\right)^2)\,, \qquad n \text{ even}\,.
\end{cases}
\end{eqaed}

\subsection{Flat-space approximation}\label{sec:flat-space_nd}

We now bound the two contributions in \cref{eq:nd_trace_formula} individually. Since the overall result is negative, we choose $h$ without the overall minus sign for convenience, and denote the integrated Casimir coefficient by $C$. Then the relevant control parameter is $A = C/v$. We begin with the flat-space $\mathbb{R}^{d-1,1} \times M_n$ approximation. \\

\noindent \textbf{The identity contribution in the flat-space approximation.} In order to bound the identity contribution for all $3 \leq n \leq 7$, it is inconvenient to integrate by parts as in \cref{eq:zeta_riemann_s}. Rather, we keep the zeta-function argument $s$ in the convergence domain, and we write $\tanh(\pi r) = 1 - \frac{2}{e^{2\pi r}+1}$ for even $n$. Then, the first term can be integrated in closed form and analytically continued to $s = -d/2 = (n-11)/2$, while the second term can be integrated numerically setting $s = -d/2$ directly. The final results are given in \cref{tab:identity_higher_dim}.
\begin{table}[ht!]
    \centering
    \begin{tabular}{c|c}
        $n$ & $C_\text{identity}/v$ \\
        \hline
        2 & $1.11 \times 10^{-9}$ \\
        3 & $4.91 \times 10^{-9}$ \\
        4 & $2.90 \times 10^{-8}$ \\
        5 & $8.38 \times 10^{-7}$ \\
        6 & $1.14 \times 10^{-5}$ \\
        7 & $9.02 \times 10^{-5}$ \\
    \end{tabular}
    \caption{The identity contribution to the Casimir coefficient in units of the curvature radius for hyperbolic manifolds of dimensions $2 \leq n \leq 7$.}
    \label{tab:identity_higher_dim}
\end{table}
Therefore, the maximum value divided by $v$ is attained at $n=7$ and is bounded by $10^{-4}$. \\

\noindent \textbf{The winding contribution in the flat-space approximation.} We can proceed as in \cref{sec:casimir_riemann_surfaces}, but we must be more careful when treating the contribution from short geodesics. This is because the systole length is now unbounded below despite rigidity. Neglecting this fact would lead to an incorrect asymptotic scaling for vanishing systole length in the bound. Accordingly, we split the sum into a ``short'' and a ``long'' contribution, depending on whether $\ell_\gamma \leq \ell_*$ for some fixed $\ell_*$,
\begin{eqaed}\label{eq:short_long}
    C_\text{short} \equiv \sum_{k \geq 1} \sum_{\gamma \in \mathcal{P} : \ell_\gamma \leq \ell_*} \frac{\ell_\gamma \, g(k\ell_\gamma)}{D(\gamma^k)}\,, \\
    C_\text{long} \equiv \sum_{k \geq 1} \sum_{\gamma \in \mathcal{P} : \ell_\gamma   > \ell_*} \frac{\ell_\gamma \, g(k\ell_\gamma)}{D(\gamma^k)}\,.
\end{eqaed}
We choose $\ell_*=1$ throughout, where $s\equiv\text{sys}(M_n)$. The ``long'' contribution can be treated as in \cref{sec:casimir_riemann_surfaces}. Since $A_\gamma$ is a rotation, its eigenvalues have norm one, and thus we can bound $D(\gamma^k) \geq (2 \sinh(k \ell_\gamma/2))^{n-1}$. The relevant choices of $g(\ell)$ are the heat-kernel Fourier transform $e^{-\frac{(n-1)^2}{4}t-\frac{\ell^2}{4t}}/\sqrt{4\pi t}$ integrated against the heat kernel (density) of the external spacetime. In the flat-space $\mathbb{R}^{d-1,1} \times M_n$ approximation,
\begin{eqaed}\label{eq:g_function_flat}
    g(\ell) = \frac{1}{2} \int_0^\infty \frac{dt}{t} \, \frac{1}{(4\pi t)^\frac{d}{2}} \, \frac{e^{-\frac{(n-1)^2}{4} t -\frac{\ell^2}{4t}}}{\sqrt{4\pi t}} = \frac{(n-1)^\frac{d+1}{2}}{(4\pi \ell)^{\frac{d+1}{2}}} \, K_{\frac{d+1}{2}} \! \left(\frac{n-1}{2} \, \ell \right)
\end{eqaed}
is such that the sum over $k$
\begin{eqaed}\label{eq:bound_sum_k_long}
    F(\ell) \equiv \sum_{k \geq 1} \frac{\ell \, g(k \ell)}{(2 \sinh(k \ell/2))^{n-1}}
\end{eqaed}
decays exponentially in $\ell$ as $e^{-(n-1)\ell}$, at the same rate at which the number of prime geodesics grows (see \cref{eq:vol_bound}). In addition, there is a power-law decay $\ell^{-d/2} \leq \ell^{-2}$ which makes the integrals converge absolutely. Hence, using the results in \cref{app:higher_dim}, we can integrate by parts as in \cref{eq:F_kernel}, obtaining
\begin{eqaed}\label{eq:long_bound}
    C_\text{long} & \leq \int_{\ell_*}^\infty F(\ell) \, d(N(\ell)-N(\ell_*)) \\
    & = \int_{\ell_*}^\infty (N(\ell)-N(\ell_*)) (-F'(\ell)) d\ell \\
    & \leq v \int_{\ell_*}^\infty k_n^\text{packing} \, e^{(n-1)\ell} (-F'(\ell)) d\ell \\
    & = k_n^\text{packing} \, v \left(e^{(n-1)\ell_*} F(\ell_*) + (n-1)\int_{\ell_*}^\infty e^{(n-1)\ell} F(\ell) \, d\ell \right) .
\end{eqaed}
To proceed, we use that $K_{\frac{d+1}{2}}(\frac{n-1}{2} 10 \ell) \leq 5 \, e^{-\frac{n-1}{2} \, 10 \ell}/\sqrt{\frac{n-1}{2} 10 \ell}$ for $\ell \geq 1$ and $3 \leq n \leq 7$ (which translates into $4 \leq d \leq 8$). We bound
\begin{eqaed}\label{eq:F_bound_higher_dim}
    F(\ell) & \leq \frac{(n-1)^\frac{d+1}{2}}{(4\pi \ell)^{\frac{d+1}{2}}} \sum_{k \geq 1} \frac{\ell \, K_{\frac{d+1}{2}}\! \left(\frac{n-1}{2} \, k\ell \right)}{(2 \sinh(k\ell/2))^{n-1}k^\frac{d+1}{2}} \\
    & \leq \frac{(n-1)^\frac{d+1}{2}}{(4\pi \ell)^{\frac{d+1}{2}}} \left(\sum_{k = 1}^{10} \frac{\ell \, K_{\frac{d+1}{2}}\! \left(\frac{n-1}{2} \, k\ell \right)}{(2 \sinh(k\ell/2))^{n-1}k^\frac{d+1}{2}} + \sum_{k \geq 11} \frac{5\ell \, e^{-\frac{n-1}{2}k \ell}}{(2 \sinh(5\ell))^{n-1} \sqrt{\frac{n-1}{2}10 \ell}} \right) ,
\end{eqaed}
where the pre-truncated geometric series can be evaluated in closed form. Plugging \cref{eq:F_bound_higher_dim} in \cref{eq:long_bound}, we finally arrive at
\begin{eqaed}\label{eq:long_bound_final}
    C_\text{long} \leq k_n^\text{packing} \, a_n \, v\,,
\end{eqaed}
where numerically (rounding up) $a_3 \approx 0.274, a_4 \approx 0.676, a_5 \approx 1.674, a_6 \approx 4.100, a_7 \leq 9.82$ using the results in \cref{app:packing}. As discussed in \cref{app:higher_dim,app:packing_short}, since $N(\ell)=0$ for $\ell < s$ when $s \geq \ell_*=1$ it is convenient to take the packing cutoff $\ell_0 = \max(\mu_n, \min(s,1))$. Then we will maximize the bounds with respect to $s$. \\

\noindent We now turn to bounding $C_\text{short}$. Due to the presence of short geodesics, the method used above does not work. Therefore, we leverage Margulis' lemma \cite{ballmann1985manifolds}---there exists a constant $\mu_n$ such that prime geodesics of length below $\mu_n$ are the cores of disjoint tubes. For $n=3$, one can take $\mu_3 = 0.104$ \cite{Meyerhoff_1987}. For $n=4,5$, one can take $\mu_4=\mu_5=\frac{\sqrt{3}}{9\pi} \approx 0.0612$, and for any $n \geq 2$ one can use the bound \cite{Kellerhals2004}
\begin{eqaed}\label{eq:kellerhals_bound}
    \mu_n \geq \frac{2^{\nu+1}}{3^{\nu+1}\pi^\nu} \, \frac{\Gamma(\frac{\nu+2}{2})^2}{\Gamma(\nu+2)}\,, \qquad \nu \equiv \Big\lfloor \frac{n-1}{2} \Big\rfloor\,.
\end{eqaed}
For $n=4,5$ this bound is worse than the above, while for $n=6,7$ we get $\mu_6 \geq \frac{4}{81\pi^2} \approx 5 \times 10^{-3}$ and $\mu_7 \geq \frac{1}{216\pi^2} \approx 4.69 \times 10^{-4}$. \\

\noindent With these ingredients, we proceed as follows. Consider a prime geodesic $\gamma \in \mathcal{P}$ with $\ell_\gamma < \mu_n$. Since the loxodromic isometry corresponding to $\gamma$ does not move its points more than $\ell_\gamma$, the whole geodesic belongs to a connected component of the ``thin part'' of $M_n$. Such connected components are disjoint \emph{Margulis tubes} $T_\gamma$, whose volumes therefore satisfy\footnote{The factor of two in \cref{eq:margulis_tube_bound} arises because we are counting oriented prime geodesics.}
\begin{eqaed}\label{eq:margulis_tube_bound}
    \sum_{\gamma \in \mathcal{P} : \ell_\gamma < \mu_n} \!\!\!\! \text{Vol}(T_\gamma) \leq 2v\,.
\end{eqaed}
We now split $C_\text{short}$ into the sum $C_\text{tube}$ over geodesics with length below $\alpha \mu_n$ and its complement $C_\text{compl}$, where $0 < \alpha < 1$ is a parameter that we optimize over to strengthen the bounds. For each prime geodesic $\gamma \in \mathcal{P}$ with $\ell_\gamma < \mu_n$, choose a Fermi coordinate patch $(s,y=\sinh(r)u)$ where $s$ is the arclength coordinate, $r$ the transverse radial coordinate and $u$ a transverse unit vector. Then the volume form is locally $ds \wedge d^{n-1}y$, and the primitive loxodromic isometry $g_\gamma \in \Gamma$ corresponding to $\gamma$ acts on the covering space by translating $s$ by $\ell_\gamma$ and rotating $y$ by $A_\gamma \in SO(n-1)$. Using the embedding hyperboloid model for $\mathbb{H}^n \subset \mathbb{R}^{n,1}$, the distance $\text{dist}_{\mathbb{H}^n}(p,q)$ between two points on $\mathbb{H}^n$ parametrized by ambient-space vectors $X_p, X_q \in \mathbb{R}^{n,1}$ satisfies
\begin{eqaed}\label{eq:hyperboloid_distance}
    \cosh \text{dist}_{\mathbb{H}^n}(p,q) = \abs{\langle X_p, X_q \rangle}\,.
\end{eqaed}
Hence, for $p \in T_\gamma$ we can choose $X_p = (\cosh(r) \cosh(s), \cosh(r) \sinh(s), y)$, and the distance $\text{dist}(p,g_\gamma^k \cdot p) < \mu_n$ satisfies
\begin{eqaed}\label{eq:hyperboloid_distance_gk}
    \cosh \text{dist}_{\mathbb{H}^n}(p,g_\gamma^k \cdot p) & = \abs{\langle X_p, X_{g_\gamma^k \cdot p} \rangle} \\
    & = \cosh^2(r) \cosh(k \ell_\gamma) - \sinh^2(r) (u^T \! A_\gamma^k \, u) \\
    & = \cosh(k \ell_\gamma) + \sinh^2(r) \left( \cosh(k \ell_\gamma) - u^T \! A_\gamma^k \, u \right) \\
    & = \cosh(k \ell_\gamma) + y^T \left(\cosh(k\ell_\gamma)I_{n-1} - \, \frac{1}{2}\left(A_\gamma^k + A_\gamma^{-k}\right)\right) y \\
    & = \cosh(k \ell_\gamma) + \frac{1}{2} \, y^T M_{\gamma^k}^T M_{\gamma^k} y\,, \qquad M_{\gamma^k} \equiv e^{\frac{k\ell_\gamma}{2}} I_{n-1} - \, e^{-\frac{k\ell_\gamma}{2}} A_\gamma^k\,.
\end{eqaed}
Notice that $D(\gamma^k) = \abs{\det(M_{\gamma^k})}$. Therefore, rearranging terms, we find that the domain $T_\gamma^{(k)} \subseteq T_\gamma$ where $\text{dist}_{\mathbb{H}^n}(p,g_\gamma^k \cdot p) < \mu_n$ is the (solid interior of an) $(n-1)$-ellipsoid defined by the inequality
\begin{eqaed}\label{eq:ellipsoid}
    \frac{1}{2} \, y^T M_{\gamma^k}^T M_{\gamma^k} y < \cosh(\mu_n) - \cosh(k\ell_\gamma)
\end{eqaed}
times the axis $\gamma \subset T_\gamma$. Therefore,
\begin{eqaed}\label{eq:volume_ellipsoid}
    \text{Vol}(T_\gamma^{(k)}) = \frac{(2\pi)^\frac{n-1}{2}}{\Gamma(\frac{n+1}{2})} \, \frac{\ell_\gamma}{D(\gamma^k)} \, (\cosh(\mu_n) - \cosh(k\ell_\gamma))^{\frac{n-1}{2}}\,,
\end{eqaed}
and thus
\begin{eqaed}\label{eq:ellipsoid_tube}
    C_\text{tube} & = \frac{\Gamma(\frac{n+1}{2})}{(2\pi)^{\frac{n-1}{2}}}\sum_{k \geq 1} \sum_{\gamma \in \mathcal{P} : k\ell_\gamma \leq \alpha\mu_n} \frac{\text{Vol}(T_\gamma^{(k)}) \, g(k \ell_\gamma)}{(\cosh(\mu_n) - \cosh(k\ell_\gamma))^{\frac{n-1}{2}}} \\
    & \leq \frac{\Gamma(\frac{n+1}{2})}{(2\pi)^{\frac{n-1}{2}}} \sum_{\gamma \in \mathcal{P}} \sum_{1 \leq k \leq \frac{\alpha\mu_n}{\ell_\gamma}} \frac{\text{Vol}(T_\gamma) \, g(k \ell_\gamma)}{(\cosh(\mu_n) - \cosh(k\ell_\gamma))^{\frac{n-1}{2}}} \\
    & \leq \frac{\Gamma(\frac{n+1}{2})/(2\pi)^{\frac{n-1}{2}}}{\left(\cosh(\mu_n) - \cosh(\alpha\mu_n)\right)^{\frac{n-1}{2}}} \sum_{\gamma \in \mathcal{P}} \text{Vol}(T_\gamma) \!\!\!\! \sum_{1 \leq k \leq \frac{\alpha\mu_n}{\ell_\gamma}} \!\!\!\! g(k \ell_\gamma) \\
    & \equiv \frac{\Gamma(\frac{n+1}{2})/(2\pi)^{\frac{n-1}{2}}}{c(\mu_n)} \sum_{\gamma \in \mathcal{P}} \text{Vol}(T_\gamma) \!\!\!\! \sum_{1 \leq k \leq \frac{\alpha\mu_n}{\ell_\gamma}} \!\!\!\! g(k \ell_\gamma)\,.
\end{eqaed}
If the systole $s = \text{sys}(M_n)$ is larger than $\alpha \mu_n$, the sum is empty and $C_\text{tube} = 0$. Otherwise, and accounting for this using the Heaviside step function $\theta_s \equiv \theta(\alpha \mu_n - s)$, since \cref{eq:g_function_flat} is decreasing we can use \cref{eq:margulis_tube_bound} to further bound
\begin{eqaed}\label{eq:tube_bound_s}
    C_\text{tube} & \leq \frac{2^{\frac{n+1}{2}} (n-1)^\frac{d+1}{2} v \, \theta_s}{c(\mu_n)} \, \frac{\zeta(\frac{d+1}{2})\Gamma(\frac{n+1}{2})}{(4\pi)^\frac{11}{2} s^{\frac{d+1}{2}}} \, K_{\frac{d+1}{2}} \! \left(\frac{n-1}{2} \, s\right) .
\end{eqaed}
Finally, for $0<s\leq1$, we bound the complement $C_\text{compl}$ by replacing $A_\gamma$ with the identity in $D(\gamma^k)$, according to
\begin{eqaed}\label{eq:compl_bound}
    C_\text{compl} & = \sum_{k \geq 1} \sum_{\gamma \in \mathcal{P} : \frac{\alpha\mu_n}{k} < \ell_\gamma \leq \ell_*} \frac{\ell_\gamma \, g(k\ell_\gamma)}{D(\gamma^k)} \\
    & \leq \sum_{\gamma \in \mathcal{P}: \ell_\gamma \leq \ell_*} \sum_{k > \frac{\alpha\mu_n}{\ell_\gamma}} \frac{\ell_\gamma \, g(k\ell_\gamma)}{(2 \sinh(k\ell_\gamma/2))^{n-1}} \\
    & = \frac{(n-1)^{\frac{d+1}{2}}}{(4\pi)^{\frac{d+1}{2}}} \, \sum_{\gamma \in \mathcal{P} : \ell_\gamma \leq \ell_*} \sum_{k > \frac{\alpha\mu_n}{\ell_\gamma}} \frac{\ell_\gamma \, K_{\frac{d+1}{2}}(\frac{n-1}{2} \, k\ell_\gamma)}{(2 \sinh(k\ell_\gamma/2))^{n-1}(k\ell_\gamma)^{\frac{d+1}{2}}} \\
    & = \frac{(n-1)^{\frac{d+1}{2}}}{2^{n-1}(4\pi)^{\frac{d+1}{2}}} \, \int_{[s,1]} \sum_{k > \frac{\alpha\mu_n}{\ell}} \frac{\ell \, K_{\frac{d+1}{2}}(\frac{n-1}{2} \, k\ell)}{\sinh^{n-1}(k\ell/2)(k\ell)^{\frac{d+1}{2}}} \, dN(\ell)\,.
\end{eqaed}
Letting $n_\ell = \lfloor \alpha\mu_n/\ell \rfloor + 1$ be the first value of $k$ contributing to the sum in the integrand, we shift the summation index according to $k = n_\ell + m$ and use that $n_\ell \ell \geq \max(\alpha\mu_n,s) \equiv \eta_n$. This provides an upper bound on the integrand as follows: for each term in the sum,
\begin{eqaed}\label{eq:k_bound}
    \frac{\ell \, K_{\frac{d+1}{2}}(\frac{n-1}{2} \, k\ell)}{\sinh^{n-1}(k\ell/2)(k\ell)^{\frac{d+1}{2}}} & = \left(\frac{k\ell}{\sinh(k\ell/2)}\right)^{\!n-1} \left((k\ell)^\frac{d+1}{2}K_{\frac{d+1}{2}}\left(\frac{n-1}{2} \, k\ell \right)\right) k^{-11}\ell^{-10}
\end{eqaed}
where each bracketed combination is a decreasing function of $k\ell \geq \eta_n$. Bounding them with this argument, and shifting the summation index as above, we can further bound $n_\ell \ell \geq \max(\eta_n, \ell)$ and the sum over $m$ evaluates to $\ell^{-10}\zeta_{\frac{\max(\eta_n,\ell)}{\ell}}(11)$, where the Hurwitz zeta function $\zeta_a(z) \equiv \sum_{m \geq 0}(a+m)^{-z}$. All in all, up to the $\eta_n$-dependent prefactor, this leaves the integral
\begin{eqaed}\label{eq:int_by_parts_steps}
    \int_{[s,1]} \ell^{-10}\zeta_{\frac{\max(\eta_n,\ell)}{\ell}}(11) \, dN(\ell) & = N(1^+) \, \zeta(11) + \int_s^1 N(\ell) (-\partial_\ell(\ell^{-10}\zeta_{\frac{\max(\eta_n,\ell)}{\ell}}(11))) \, d\ell \\
    & \leq N(1^+) \, \zeta(11) + \int_{\eta_n}^1 N(\ell) (-\partial_\ell(\ell^{-10}\zeta_{\frac{\max(\eta_n,\ell)}{\ell}}(11))) \, d\ell \\
    & = N(1^+) \zeta(11) + 10 \zeta(11) \int_{\eta_n}^1 \frac{N(\ell)}{\ell^{11}} \, d\ell \equiv \mathcal{I}_n(s)\,,
\end{eqaed}
where we used that the function being differentiated is increasing for $\ell \leq \eta_n$ and that $\eta_n \geq s$ by definition. This allows us to avoid a potentially dangerous small-$s$ divergence in the bounds. \\

\noindent Including the prefactors, we thus arrive at
\begin{eqaed}\label{eq:compl_bound_final}
    C_\text{compl} & \leq \frac{(n-1)^{\frac{d+1}{2}}}{2^{n-1}(4\pi)^{\frac{d+1}{2}}} \, \frac{\eta_n^{n-1+\frac{d+1}{2}} K_{\frac{d+1}{2}}(\frac{n-1}{2}\eta_n)}{\sinh^{n-1}(\eta_n/2)} \, \mathcal{I}_n(s)\,.
\end{eqaed}
All that remains is to bound $N(\ell)/v$ for $\ell \leq 1$. Making use of the improved estimates in \cref{app:packing_short}, we bound \cref{eq:int_by_parts_steps,eq:compl_bound_final} by inserting the estimates for $N(\ell)$ and $N(1^+)$ provided by \cref{eq:N_bound_improved}. The bound for $N(1^+)$ follows by taking $L \to 1^{+}$. For $s>1$, all primitive geodesics belong to the long sector, meaning that the tube and complementary contributions vanish and $\mathcal I_n(s)=0$.  \\

\noindent \textbf{Putting things together.} Adding the upper bounds in \cref{eq:long_bound_final,eq:tube_bound_s,eq:compl_bound_final}, we obtain $A = A_\text{identity} + A_\text{winding}$, with $A_\text{identity} \leq 10^{-4}$ and
\begin{eqaed}\label{eq:full_winding_bound}
    A_\text{winding} & \equiv \frac{C_\text{long} + C_\text{tube} + C_\text{compl}}{v} \\
    & \leq k_n^\text{packing} \, a_n + \frac{2^{\frac{n+1}{2}}(n-1)^\frac{d+1}{2}\zeta(\frac{d+1}{2})\Gamma(\frac{n+1}{2})\theta_s}{(4\pi)^\frac{11}{2} s^{\frac{d+1}{2}}c(\mu_n)} \, K_{\frac{d+1}{2}} \! \left(\frac{n-1}{2} \, s\right) \\
    & + \frac{(n-1)^{\frac{d+1}{2}}}{2^{n-1}(4\pi)^{\frac{d+1}{2}}} \, \frac{\eta_n^{n-1+\frac{d+1}{2}} K_{\frac{d+1}{2}}(\frac{n-1}{2}\eta_n)}{\sinh^{n-1}(\eta_n/2)} \, \frac{\mathcal{I}_n(s)}{v} \equiv A_n(s)\,.
\end{eqaed}
Since $A_\text{identity}$ is so small, we can absorb it into the (upper bounds for) $a_n$ as summarized in \cref{tab:identity_higher_dim} and discussed below \cref{eq:long_bound_final}. \\

\noindent According to the general solution obtained in \cref{sec:setup}, the relevant control parameters are the physical curvature radius $r \propto A^{\frac{1}{9}}$ and the physical systole length in eleven-dimensional Planck units. Since they both ought to be large to achieve control, we consider their minimum\footnote{More precisely, the solution to \cref{eq:solution,eq:L_solution} contains an extra prefactor of $(\frac{d}{2n(n-1)})^\frac{1}{9}$. However, in the worst-case scenario $n=7$ it merely changes the answer by a factor of about $0.713$.}.
\begin{eqaed}\label{eq:harmonic_control}
    H_n(s) & \equiv \text{min}\!\left(A_n(s)^\frac{1}{9},s A_n(s)^\frac{1}{9}\right) = \min(1,s) \, A_n(s)^\frac{1}{9}\,.
\end{eqaed}
As a function of $s$ for $3 \leq n \leq 7$, $H_n(s)$ has a maximum, which is attained around $s \approx \alpha\mu_n$. One naively expects roughly an extra factor of $(256)^\frac{1}{9} \approx 1.85$, together with the numerical prefactor in \cref{eq:solution,eq:L_solution}. At any rate, the resulting radius is rather insensitive to such prefactors because of the $1/9$ exponent. Thus we define the control parameter $H_n^\text{phys} \equiv (\frac{256 d}{2n(n-1)})^{\frac{1}{9}} H_n^\text{max}$. We emphasize that, although the bounds on $H_n^\text{phys}$ as defined are valid, using this parameter as a proxy for the control parameters obtained with the full supergraviton Casimir energy is an estimate based on heuristic expectations. This remark also applies to the whole analysis based on scalar-field Casimir energies, which we extend in \cref{app:bounding_sugra_casimir} to parametric bounds for the full supergravity spectrum. \\

\noindent The final analytic results can be evaluated numerically, rounding up beyond the error, by optimizing over the parameters we introduced, and maximizing over $s$ to obtain a bound that is valid uniformly across hyperbolic manifolds. The resulting values are reported in \cref{tab:control_parameter_higher_dim_2}. The maxima with respect to $s$ lie around $s \approx \alpha \mu_n$ with the optimal value $\alpha \approx 0.99$, slightly varying with $n$. Even just choosing these values for $\alpha$ outright, without numerical optimization, leads to values that are quite close to those in \cref{tab:control_parameter_higher_dim_2}.
\begin{table}[ht!]
    \centering
    \begin{tabular}{c|c|c}
        $n$ & $H_n^\text{max}$ & $H_n^\text{phys}$ \\
        \hline
        3 & $2.8$ & $4.8$ \\
        4 & $5.3$ & $8.5$ \\
        5 & $8.4$ & $12.6$ \\
        6 & $89.6$ & $125.9$ \\
        7 & $1571$ & $2074$ \\
    \end{tabular}
    \caption{A numerical approximation to the upper bound for the control parameter for $3 \leq n \leq 7$. The rightmost column includes the heuristic prefactor of $(256)^\frac{1}{9}$, together with the additional numerical prefactor in \cref{eq:solution,eq:L_solution}. This provides a rough expectation for the actual bounds when the scalar-field Casimir energy is replaced by its complete supergravity counterpart (which we analyze explicitly, albeit parametrically, in \cref{app:bounding_sugra_casimir}). The digits are reported by rounding up their numerical approximation.}
    \label{tab:control_parameter_higher_dim_2}
\end{table}

\subsection{Including de Sitter curvature in higher dimensions}\label{sec:dS_curvature_higher_dim}

We now replace the flat-space heat kernel in \cref{eq:g_function_flat} with the spherical one. The winding inequality holds for any radius ratio $x>0$. We evaluate the identity contribution at the reference value $x_*$ first and then extend the bound to the scalar stationary ratio given by \cref{eq:x_extremum}. The case of Riemann surfaces, studied in \cref{sec:casimir_curved}, was sufficiently explicit to allow a more surgical approach to an upper bound. In order to encompass all compact hyperbolic manifolds $M_n$ of dimensions $3 \leq n \leq 7$, as in the preceding section, we shall directly derive an upper bound for the spherical heat kernels (or, more precisely, the combination appearing in \cref{eq:A_combination_curved}) in terms of flat-space quantities. This will allow us to recycle the bounds in \cref{tab:control_parameter_higher_dim_2}. \\

\noindent \textbf{Bounding the spherical heat kernel.} We begin with the general expression \cite{Kluth:2019vkg}
\begin{eqaed}\label{eq:d-sphere_heat_kernel}
    K_{S^d_{x}}(t) = \sum_{l \geq 0} D_l^{(d)} \, e^{- \lambda_l^{(d)} t x^2}
\end{eqaed}
for a $d$-sphere of radius $1/x$, which generalizes \cref{eq:heat_kernel_sphere}. The scalar Laplacian eigenvalues and their degeneracies are given by
\begin{eqaed}\label{eq:eigenvalues_degeneracies}
    \lambda_l^{(d)} = l(l+d-1)\,, \qquad D_l^{(d)} = \frac{(2l+d-1)\Gamma(l+d-1)}{\Gamma(d) \Gamma(l+1)}\,.
\end{eqaed}
The relevant combination contributing to \cref{eq:A_combination_curved} is
\begin{eqaed}\label{eq:sphere_combination}
    \mathcal{K}_d(t) \equiv \frac{K_{S^d_{x_*}}(t)}{\text{Vol}(S^d_{x_*})} - \frac{x_*}{d} \, \partial_x \frac{K_{S^d_x}(t)}{\text{Vol}(S^d_x)}\Bigg|_{x=x_*}\,.
\end{eqaed}
Inserting \cref{eq:d-sphere_heat_kernel,eq:eigenvalues_degeneracies} into \cref{eq:sphere_combination}, and defining $\text{Vol}(S^d_x) \equiv x^{-d} \omega_d$, with $\omega_d = \frac{2\pi^{\frac{d+1}{2}}}{\Gamma(\frac{d+1}{2})}$, one finds
\begin{eqaed}\label{eq:K_simplification}
    \mathcal{K}_d(t) & = \frac{2tx_*^{d+2}}{d\omega_d} \sum_{l > 0} \lambda_l^{(d)} \, D_l^{(d)} \, e^{-\lambda_l^{(d)} t x_*^2} \\
    & = \frac{2tx_*^{d+2}}{d\omega_d} \sum_{l > 0} d(d+1) \, D_{l-1}^{(d+2)} \, e^{-\lambda_l^{(d)} t x_*^2} \\
    & = \frac{4\pi tx_*^{d+2}}{\omega_{d+2}} \sum_{l > 0} D_{l-1}^{(d+2)} \, e^{-\lambda_{l-1}^{(d+2)} t x_*^2} \, e^{-dtx_*^2} \\
    & = \frac{4\pi tx_*^{d+2}}{\omega_{d+2}} \, e^{-dtx_*^2} \sum_{l \geq 0} D_{l}^{(d+2)} \, e^{-\lambda_{l}^{(d+2)} t x_*^2} \\
    & = \frac{4\pi tx_*^{d+2}}{\omega_{d+2}} \, e^{-dtx_*^2} \, K_{S^{d+2}_1}(tx_*^2)\,.
\end{eqaed}
Therefore, in terms of the dimensionless variable $z=tx_*^2$ in \cref{eq:K_simplification}, the ratio between $\mathcal{K}_d(t)$ and the flat-space heat kernel $(4\pi t)^{-d/2}$ reads 
\begin{eqaed}\label{eq:heat_kernel_ratio}
    \frac{\mathcal{K}_d(t)}{(4\pi t)^{-\frac{d}{2}}} = \frac{(4\pi z)^\frac{d+2}{2}}{\omega_{d+2}} \, e^{-dz} \, K_{S^{d+2}_1}(z)\,, \qquad z \equiv t x_*^2\,.
\end{eqaed}
We now write the spectral sum for the unit-sphere heat kernel $K_{S^m_1}(z)$ as a spectral integral
\begin{eqaed}\label{eq:spectral_integral}
    K_{S^m_1}(z) & = \int_0^\infty e^{-\lambda z} \, d\nu(\lambda) = z \int_0^\infty \nu(\lambda) \, e^{-\lambda z} \, d\lambda\,,
\end{eqaed}
motivated by the Weyl asymptotics
\begin{eqaed}\label{eq:weyl_law}
    \nu(\lambda) & \equiv \sum_{l \, : \, l(l+m-1) \leq \lambda} D_l^{(m)} \\
    & \leq \frac{(m+2L_\lambda)\Gamma(m+L_\lambda)}{\Gamma(m+1)\Gamma(L_\lambda+1)} \\
    & \overset{\lambda \to +\infty}{\sim} \frac{\omega_m \, \lambda^{\frac{m}{2}}}{(4\pi)^{\frac{m}{2}}\Gamma(\frac{m}{2}+1)}
\end{eqaed}
where the angular number $L_\lambda$ is defined by $\lambda = L_\lambda (L_\lambda + m - 1)$. \\

\noindent We now derive a bound inspired by \cref{eq:weyl_law}. Since we want to bound the combination in \cref{eq:heat_kernel_ratio}, it is convenient to seek an inequality of the form
\begin{eqaed}\label{eq:lambda_inequality}
    \nu(\lambda) \leq \text{const.} \times \frac{\omega_m \, (\lambda+m-2)^{\frac{m}{2}}}{(4\pi)^{\frac{m}{2}}\Gamma(\frac{m}{2}+1)}\,,
\end{eqaed}
so that the integration variable in \cref{eq:spectral_integral} can be shifted accordingly. To this end, for $6\leq m \leq 10$ one can observe graphically or check analytically that the ratio
\begin{eqaed}\label{eq:nu_ratio}
    \frac{(4\pi)^{\frac{m}{2}}\Gamma(\frac{m}{2}+1)}{\omega_m \, (l(l+m-1)+m-2)^{\frac{m}{2}}} \, \frac{(m+2l)\Gamma(m+l)}{\Gamma(m+1)\Gamma(l+1)}
\end{eqaed}
is decreasing, and bounded by its value at $l=0$. As a result, letting $\nu_m$ be this value, we have
\begin{eqaed}\label{eq:nu_bound_final}
    \nu(\lambda) \leq \nu_m \, \frac{\omega_m \, (\lambda+m-2)^{\frac{m}{2}}}{(4\pi)^{\frac{m}{2}}\Gamma(\frac{m}{2}+1)}\,,
\end{eqaed}
where the numerical values of $\nu_m$ for $6 \leq m \leq 10$ (rounded up) are reported in \cref{tab:nu_bound}.
\begin{table}[ht!]
    \centering
    \begin{tabular}{c|c}
        $m=d+2=13-n$ & $\nu_m$ \\
        \hline
        6 & $5.63$ \\
        7 & $9.02$ \\
        8 & $15.56$ \\
        9 & $28.57$ \\
        10 & $55.38$ \\
        \end{tabular}
    \caption{The constant $\nu_m$ realizing the upper bound in \cref{eq:nu_bound_final} for $6 \leq m \leq 10$, corresponding to compact hyperbolic manifolds of dimension $3 \leq n \leq 7$.}
    \label{tab:nu_bound}
\end{table}
Plugging this result into \cref{eq:heat_kernel_ratio,eq:spectral_integral} we obtain
\begin{eqaed}\label{eq:heat_kernel_ratio_bound}
    \frac{\mathcal{K}_d(t)}{(4\pi t)^{-\frac{d}{2}}} & \leq \nu_{d+2} \, \frac{z^{\frac{d}{2}+2}}{\Gamma(\frac{d}{2}+2)} \int_0^\infty (\lambda+d)^\frac{d+2}{2} \, e^{-(\lambda+d)z} \, d\lambda \\
    & \leq \nu_{d+2} \, \frac{z^{\frac{d}{2}+2}}{\Gamma(\frac{d}{2}+2)} \int_0^\infty \lambda^\frac{d+2}{2} \, e^{- \lambda z} \, d\lambda = \nu_{d+2}\,.
\end{eqaed}

\noindent \textbf{The winding contribution.} As a result of the above analysis, the winding contribution $\mathcal{A}_\text{winding}$ to the relevant Casimir coefficient in \cref{eq:A_combination_curved} is simply bounded as
\begin{eqaed}\label{eq:A_winding_curved_higher-dim}
    \mathcal{A}_\text{winding} \leq \nu_{d+2} \, A_\text{winding}\,.
\end{eqaed}

\noindent \textbf{The identity contribution.} Finally, we turn to the identity contribution, which requires subtracting the \ac{UV}-divergent terms as spelled out in \cref{eq:subtracted_identity}. We will proceed analogously. As before, the first six heat-kernel coefficients are subtracted by renormalization; thanks to higher-dimensional supersymmetry, in the full supergravity counterpart to the scalar computation the subtraction only affects the supersymmetric $R^4$ terms. For $d \leq 6$, these coefficients are given in Table 1 of \cite{Kluth:2019vkg}, while for $d=7,8$ we can use the recursion relation derived in \cref{eq:K_simplification}. Actually, since we need to bound the combination $\mathcal{K}_d(t)$ rather than the heat kernel itself, we can leverage \cref{eq:K_simplification,eq:heat_kernel_ratio} to extract the relevant \ac{UV}-divergent local terms from the heat kernel coefficients of the $(d+2)$-sphere. We thus need to use the recursion relation up to $d+2=10$. Inverting \cref{eq:K_simplification}, we find the recursion
\begin{eqaed}\label{eq:recursion_relation_heat_kernel}
    \frac{K_{S^d_1}(z)}{\omega_d} \equiv \frac{1}{(4\pi z)^\frac{d}{2}} \, \kappa_d(z)\,, \qquad \kappa_{d+2}(z) = e^{dz} \left(\kappa_d(z) - \frac{2z}{d} \, \kappa_d'(z) \right) ,
\end{eqaed}
from which we compute the coefficients $b_k$ in the local expansion
\begin{eqaed}\label{eq:local_expansion_higher-dim}
    \mathcal{K}_d(t) \overset{t \to 0^+}{\sim} \frac{1}{(4\pi t)^\frac{d}{2}} \sum_{k=0}^5 b_k \, (tx_*^2)^k \equiv \frac{1}{(4\pi t)^\frac{d}{2}} \, P_d(tx_*^2)\,.
\end{eqaed}
These are collected in \cref{tab:local_higher_dim}.
\begin{table}[ht!]
    \centering
    \begin{tabular}{c|c}
        $d$ & $P_d(z) = \sum_{k=0}^5 b_k \, z^k$ \\
        \hline
        4 & $1+z-\frac{37}{63}z^3-\frac{149}{315}z^4-\frac{179}{1155}z^5$ \\
        5 & $1+2z+\frac{16}{15}z^2-\frac{16}{15}z^3-\frac{32}{15}z^4-\frac{64}{45}z^5$ \\
        6 & $1+\frac{10}{3}z+4z^2-\frac{833}{135}z^4-\frac{274}{33}z^5$ \\
        7 & $1+5z+\frac{101}{10}z^2+\frac{501}{70}z^3-\frac{2943}{280}z^4-\frac{8991}{280}z^5$ \\
        8 & $1+7z+21z^2+30z^3-\frac{8501}{99}z^5$ \\
        \end{tabular}
    \caption{The \ac{UV}-divergent local heat-kernel coefficients for the combination $\mathcal{K}_d(t)$ in \cref{eq:K_simplification} for $4 \leq d \leq 8$.}
    \label{tab:local_higher_dim}
\end{table}
Armed with these coefficients, we can treat the Schwinger integral for the function $h$ that enters \cref{eq:nd_trace_formula} along the lines of \cref{eq:H_function}. We have
\begin{eqaed}\label{eq:h_integral}
    h(r) = \frac{1}{2} \int_0^\infty \frac{dt}{t} \left( \mathcal{K}_d(t) - \, \frac{1}{(4\pi t)^\frac{d}{2}} \, P_d(tx_*^2) \right) e^{-t \left(\frac{(n-1)^2}{4}+r^2\right)} + \Delta h(r)\,,
\end{eqaed}
where $\Delta h(r)$ is the analytically continued \ac{UV} part, and we split the integral at Schwinger parameter $t=1$. The contribution from the range $t \geq 1$ can be treated by bounding $\mathcal{K}_d(t)$ according to \cref{eq:heat_kernel_ratio_bound}, whereas the polynomial term can be integrated and bounded separately; the latter numerically integrates to values bounded by $10^{-7}$ for all $4 \leq d \leq 8$ for its contribution to the identity term $\mathcal{A}_\text{identity}$. Using \cref{eq:heat_kernel_ratio_bound,tab:nu_bound}, the former term with $\mathcal{K}_d(t)$ numerically yields values bounded by $2 \times 10^{-6}$ for its contribution to the identity term. \\

\noindent For $0<t<1$, write $R_d(t)=\mathcal{K}_d(t)-(4\pi t)^{-\frac{d}{2}}P_d(tx_*^2)$. Let us separately derive the estimate in two subintervals. Consider $0<t\leq\frac{1}{64}$, set $z=tx_*^2$ and define the shifted quantity $q=l+\frac{d+1}{2}$. The spectrum entering \cref{eq:K_simplification} becomes
\begin{equation}
	K_{S^{d+2}_1}(z)=\frac{2e^{\frac{(d+1)^2z}{4}}}{(d+1)!}
	\sum_{q\in\mathcal{L}_d}p_d(q)e^{-zq^2}\,,
	\qquad p_d(q)=q\prod_{j=1}^{d}\left(q+j-\frac{d+1}{2}\right)\,,
\end{equation}
where $\mathcal{L}_d$ consists of positive integers for odd $d$ and positive half-integers for even $d$. The original sum begins at $q=\frac{d+1}{2}$, since $l\geq0$. Extending it to the smaller positive points of $\mathcal{L}_d$ leaves the sum unchanged, since $p_d(q)=0$ at each additional point. Euler-Maclaurin summation reproduces $P_d$, with an error controlled by integrals of derivatives of $p_d(q)e^{-zq^2}$. These derivatives are polynomials times the same Gaussian, so their absolute integrals are bounded termwise using $\int_0^\infty q^j e^{-zq^2}\,dq=\frac{\Gamma(\frac{j+1}{2})}{2z^{\frac{j+1}{2}}}$. The remaining curvature factor $e^{\frac{(d-1)^2z}{4}}$ is expanded with each Taylor tail bounded by its first omitted term times the exponential. These estimates control $t^{\frac{d}{2}-6}R_d(t)$ uniformly down to $t=0$. For $\frac{1}{64}\leq t\leq1$, we retain $1\leq l<L$, with $L=500$, in the first line of \cref{eq:K_simplification}. We bound the remainder of this truncated expression by Taylor polynomials of degree $14$ over many covering intervals, with Cauchy's estimate providing the Taylor errors throughout each interval. The omitted positive spectral tail satisfies
\begin{equation}
	\sum_{l\geq L}\lambda_l^{(d)}D_l^{(d)}e^{-z\lambda_l^{(d)}}
	\leq\frac{2(L+d)^{d+1}e^{-zL^2}}{(d-1)!\left[1-e^{\frac{d+1}{L}-z(2L+1)}\right]}\,.
\end{equation}
This follows from $\lambda_l^{(d)}\geq l^2$ and $\lambda_l^{(d)}D_l^{(d)}\leq\frac{2(l+d)^{d+1}}{(d-1)!}$ by summing a geometric majorant. Including the prefactor in \cref{eq:K_simplification}, these bounds give, throughout $0<t\leq1$,
\begin{eqaed}\label{eq:C_d_constants}
	|R_d(t)|& \leq C_d\,t^{\frac{12-d}{2}}\,,\\
    (C_4,C_5,C_6,C_7,C_8)&=\left(4.1,0.042,2.5\times10^{-4},3\times10^{-6},1.1\times10^{-8}\right)\,.
\end{eqaed}
The analytically continued terms are also small, and overall bounded by $5 \times 10^{-5}$ in absolute value.
{We now extend the estimates from the reference value $x_*$ to the scalar critical point. Let us define the internal control and overall control parameters
\begin{equation}
	h=M_{11}r\min(1,s)\,,\qquad H_n^\text{hyp}=\min(h,M_{11}L)\,.
\end{equation}
Here $H_n^\text{hyp}$ is the smallest of the internal curvature radius, physical systole and external curvature radius, in eleven-dimensional Planck units. An embedded hyperbolic ball of radius $\frac{\min(1,s)}{2}$ satisfies
\begin{equation}
	v\geq c_n\min(1,s)^n\,,\qquad
	c_n=2^{-n}\operatorname{Vol}(B^n_1)\,,
\end{equation}
where $B^n_1$ is the Euclidean unit ball. The stationary equations then imply
\begin{equation}
\begin{aligned}
	x^2&=\frac{x_*^2}{1-\epsilon}\,,\\
	\epsilon&=\frac{1}{9d(d-1)\omega_d v(M_{11}r)^n(M_{11}L)^{d-2}}
	\leq\frac{E_n}{(H_n^\text{hyp})^9}\,,\\
	E_n&=\frac{1}{9d(d-1)\omega_d c_n}<0.01\,.
\end{aligned}
\end{equation}
The radius equation can be written as
\begin{equation}
	(M_{11}r)^9=\frac{d}{2n(n-1)}F_d(\epsilon)\mathcal A(x)\,,\qquad
	F_d(\epsilon)=\frac{1-\epsilon}{1-\frac{d}{2}\epsilon}\,.
\end{equation}
First, consider $H_n^\text{hyp}\geq2$, so that $\epsilon \leq E_n/512$. It holds that
\begin{equation}
\begin{aligned}
	\alpha\equiv\frac{x^2}{x_*^2}&\in[1,\alpha_n]\,,\qquad
	\alpha_n=\frac{1}{1-\frac{E_n}{512}}\,,\\
	R_{d,x}(t)&=\alpha^{\frac{d}{2}}R_d(\alpha t)\,.
\end{aligned}
\end{equation}
Extending the preceding Taylor and spectral-tail estimates to the additional interval $1\leq u\leq\alpha_n$ yields
\begin{equation}
	\alpha^6\sup_{0<u\leq\alpha}\abs{u^{\frac{d}{2}-6}R_d(u)}<C_d
	\qquad(1\leq\alpha\leq\alpha_n)\,.
\end{equation}
The factor $\alpha^6$ results from rescaling $t^{\frac{d}{2}-6}R_{d,x}(t)$. This means that all five constants $C_d$ also bound the remainder at the stationary ratio. \\

\noindent To bound the integrated identity contribution, we employ
\begin{equation}
\begin{aligned}
	K_{\mathbb{H}^n}(t;y,y)&\leq\frac{e^{-\frac{(n-1)^2t}{4}}}{(4\pi t)^{\frac{n}{2}}}Q_n(t)\,,\\
	(Q_3,Q_4,Q_5,Q_6,Q_7)&=
	\left(1,\,1+\frac{t}{4},\,1+\frac{2t}{3},\,
	1+\frac{5t}{4}+\frac{9t^2}{32},\,1+2t+\frac{16t^2}{15}\right)\,.
\end{aligned}
\end{equation}
For odd $n$ the inequality is saturated. For even $n$, the bound follows by replacing $\tanh(\pi r)$ by $1$ in \cref{eq:nd_trace_formula}. By integrating the remainder and the positive large-time bounds and including the finite terms from analytic continuation, one obtains the bound $\abs{\mathcal A_\text{identity}(x)}<10^{-4}$ for all $3\leq n\leq7$. The finite terms are bounded using Gamma-function expressions and elementary integrals of polynomials times decaying exponentials. \\

\noindent Here $\nu_{d+2}^\text{exact}=\frac{(d+2)!}{2d^{\frac{d+2}{2}}}$ is the sphere comparison constant before rounding. Using $E_n$ and the tabulated flat-space bounds $H_n^\text{max}$ as inputs, one obtains
\begin{equation}
	F_d\!\left(\frac{E_n}{512}\right)
	\left[\nu_{d+2}^\text{exact}(H_n^\text{max})^9+10^{-4}\right]
	<\nu_{d+2}(H_n^\text{max})^9\,.
\end{equation}
Hence, the internal control parameter satisfies
\begin{equation}
	h\leq\left[\frac{d\nu_{d+2}}{2n(n-1)}\right]^{\frac{1}{9}}H_n^\text{max}\,.
\end{equation}
For $1\leq H_n^\text{hyp}<2$, the ratio bound gives
\begin{equation}
	h<2\max\!\left(1,x_*\sqrt{\frac{100}{99}}\right)
\end{equation}
instead. For $H_n^\text{hyp}<1$, at least one required physical length is sub-Planckian and $H_n^\text{hyp}$ itself is below the allowed ceiling. Since $H_n^\text{hyp}\leq h$, the bound on $H_n^\text{hyp}$ therefore holds at every scalar critical point, while the bound on $h$ requires $H_n^\text{hyp}\geq1$.} \\

\noindent To sum up, we have proven rigorous scalar-field analytic bounds and provided estimates for universal numerical bounds for the control parameters of M-theory compactifications on hyperbolic manifolds with Casimir energy. The results forbid parametric control and severely limit numerical control.

\section{Numerical and parametric bounds for flat manifolds}\label{sec:flat_manifolds}

Having discussed hyperbolic manifolds in detail, in this section we include some considerations on flat manifolds, namely Ricci-flat in \cref{sec:ricci-flat} and Riemann-flat in \cref{sec:RFMs}. This setting is very interesting phenomenologically, because it seems to be able to avoid the usual geometric arguments against (parametric) scale separation \cite{DeLuca:2021ojx, Coudarchet:2023mfs, Cribiori:2024jwq, Bedroya:2025fie, Bedroya:2025ltj, Cribiori:2026btb}. Still, other arguments based on holography and Swampland constraints \cite{Collins:2022nux, Montero:2022ghl, Apers:2022zjx, Lust:2022lfc, Perlmutter:2024noo, Bena:2024are, Bedroya:2025fie, Bedroya:2025ltj, Bobev:2025yxp, Revello:2026yla, Revello:2026opc, Revello:2026eqp, Perlmutter:2026kbf} seem to provide stronger restrictions. (See, however, \cite{Andriot:2025cyi,Apers:2026lgi, Apers:2026ahu, Aggarwal:2026jan} for some counterpoints.) In the context of this work, one could hope that this setting could circumvent the no-go results we found for hyperbolic manifolds in \cref{sec:casimir_riemann_surfaces,sec:hyperbolic_manifolds}. As we shall see, we will not succeed---to the contrary, we will extend our no-go theorem to the Riemann-flat case and show that the expectation based on the Swampland ideas are valid also for this class.  
There are only a finite number of Riemann-flat manifolds in every dimension. This makes scale separation with arbitrarily small $V>0$ impossible to realize in this class, unless there is a family of critical points, which we will exclude. Nevertheless, one may consider a looser sense of scale separation which is not parametric but would potentially lead to a small enough $V \approx 10^{-122}$ (in Planck units). In this section, by scale separation we only mean this latter possibility.

\subsection{Ricci-flat manifolds}\label{sec:ricci-flat}

\begin{figure}[!ht]
    \centering
    \includegraphics[width=0.65\linewidth]{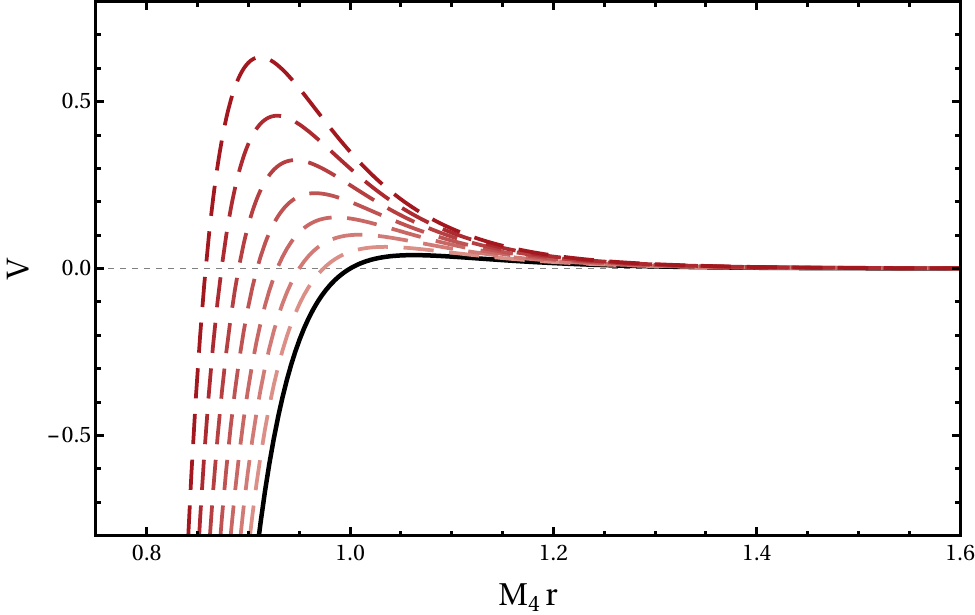} \\
    \includegraphics[width=0.65\linewidth]{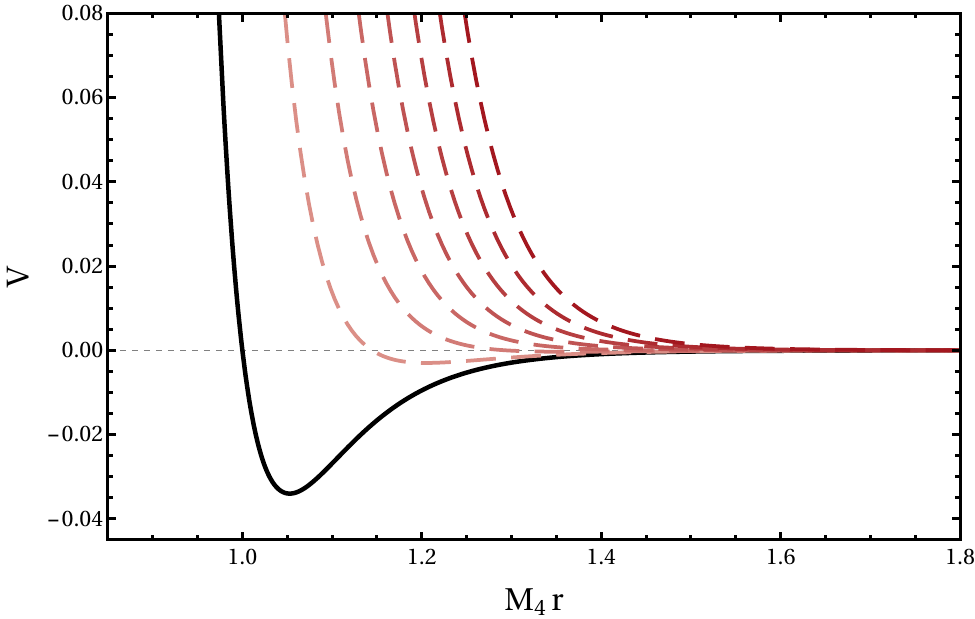}
    \caption{A depiction of the Einstein-frame effective potential for the volume radius $r$ (fixing the other moduli, if any) that compares $d=4, n=7$ Casimir compactifications on Ricci-flat manifolds with flux. Top plot: four-form flux, leading to \ac{dS}; increasing the flux (dashed lines) \emph{decreases} the critical radius. Bottom plot: seven-form flux, leading to \ac{AdS}; increasing the flux (dashed lines) \emph{increases} the critical radius while also increasing scale separation \cite{Luca:2022inb, Aparici:2025kjj, Aparici:2026qme}. The coefficients were chosen for visual clarity, but the general shape does not change when using the correct coefficients.}
    \label{fig:pot_plot}
\end{figure}

For general Ricci-flat manifolds (which are also believed to be finite in number), one might hope to avoid the problems with this weak sense of scale separation. Indeed, when the internal manifold is curved, a straightforward scaling argument \cite{Bedroya:2025ltj} shows that the total potential at a critical point will scale like the internal curvature term, thereby preventing scale separation (at least parametrically). This competition is depicted in \cref{fig:pat_plat}. In this sense, Ricci-flat manifolds provide a natural candidate to avoid this argument, and are indeed used, e.g., in the DGKT proposal \cite{DeWolfe:2005uu}; to get a critical point, the effect of curvature can be replaced by fluxes, providing a different two-term structure \cite{Bento:2025wuu,ValeixoBento:2025qih} that we depict in \cref{fig:pot_plot}. In this way, one also sees that such \ac{dS} critical points are pushed to smaller radii at larger flux (at fixed averaged Casimir energy density). This is the opposite situation as for the \ac{AdS} solutions of \cite{Luca:2022inb, Aparici:2025kjj, Aparici:2026qme}, where increasing the flux makes the solution (parametrically) scale-separated. \\

\noindent However, as we shall see in more detail in \cref{sec:RFMs}, for \ac{dS} solutions having control of the \ac{EFT} is necessary for scale separation in the first place: the ratio of internal-to-external length scales is controlled by a combination of flux and Casimir terms which can only be small when the radius is large in Planck units. While a detailed understanding of the Casimir energy for general Ricci-flat manifolds is out of reach at present, one might hope to obtain better control by seeking a sequence of manifolds for which the order of the first non-vanishing curvature invariant in the M-theory effective action increases. For example, we might start from a Ricci-flat manifold with non-vanishing $R^4$ term, then take a Ricci-flat manifold with vanishing $R^4$ term but non-vanishing $D^4R^4$ term, and so on. This would effectively go in a similar direction to increasing the Casimir energy. However, upon inspection, this method cannot work to produce large numerical or parametric hierarchies within this two-term balance mechanism: the Casimir energy density scales with the volume radius as $r^{-11}$, and a \ac{dS} critical point in the $r$ direction can only exist if the curvature term overcomes it in the large-volume limit. This forbids, e.g., manifolds with vanishing $R^4$-terms. Within the well-understood toolkit of \ac{EFT}, the only other available ingredient is flux, whose complications lead one to simplify the kind of Ricci-flat manifold one chooses to consider. This is the main motivation behind the study of compactifications on \acp{RFM} supported by fluxes and Casimir energy \cite{Luca:2022inb,Bento:2025wuu,DallAgata:2025jii,Aparici:2025kjj,ValeixoBento:2025qih,Aparici:2026wdm, Aparici:2026qme}, to which we shall now turn.

\subsection{Riemann-flat manifolds}\label{sec:RFMs}

A simple setting to look for \ac{dS} solutions is provided by \acp{RFM}. Solutions of supergravity with Casimir energy on \acp{RFM} have been the subject of significant recent interest \cite{Luca:2022inb,DallAgata:2025jii,Bento:2025wuu,ValeixoBento:2025qih,Aparici:2025kjj,Aparici:2026wdm,Aparici:2026qme}, but the only parametrically controlled ones are of the \ac{AdS} type. In this section we explore the possibility of a controlled \ac{dS} solution by seeking uniform upper bounds on the Casimir energy (and the resulting \ac{EFT} control parameters). This setup requires fluxes; we thus first restrict to rigid \acp{RFM} to retain some mathematical control on this class of (possible) solutions. Then we will extend the analysis to also exclude that viable extrema of the Casimir potential be realized by a sequence of non-rigid critical points, as we found for Riemann surfaces in \cref{sec:beyond_rigid_riemann_surfaces}. Due to the presence of fluxes, we will first cover settings in which the moduli extremize each term individually, and then generalize to critical points of the combined effective potential. \\

\noindent Since there is no meaningful curvature radius for an \ac{RFM} $M_n$, the counterpart to the curvature radius $r$ discussed in \cref{sec:setup} is here replaced by the ``volume radius'' $r^{n} \equiv \text{Vol}(M_n)$. Before proceeding, as we already noted parametric control cannot be achieved in any case: there are finitely many \acp{RFM} of any given dimension (up to isomorphism), and the flux numbers allowed by a \ac{dS} critical point work against parametric control \cite{Bento:2025wuu}. Nevertheless, in the spirit of the numerical bounds obtained for hyperbolic manifolds, we shall seek uniform numerical bounds for \ac{EFT} control parameters in the case of rigid \acp{RFM}. With some refinements, the same strategy also excludes the possibility that parametric control be achieved by the existence of infinite sequences of critical points for a given \ac{RFM}. We will first address points that extremize the Casimir and flux potentials, and then extend the treatment to all critical points. \\

\noindent \textbf{\ac{dS} solutions on \acp{RFM}.} We now briefly review the solutions of \cite{Bento:2025wuu} in our notation. Slightly modifying the flux normalization of \cite{Bento:2025wuu} by taking $[G_4] \in H^4(M_n, \mathbb{Z})$ to be integrally quantized\footnote{More precisely, for our purposes only the non-torsional part of the flux matters. Thus, we can neglect the subtleties related to shifted Dirac quantization \cite{Freed:2019sco}.}, the lower-dimensional effective Lagrangian reads\footnote{Recall that our eleven-dimensional Planck mass is rescaled by a factor of $2^{-\frac{1}{9}}$ relative to the normalization of \cite{Bento:2025wuu}. This modifies the ``gauge coupling'' coefficient in the flux term.}
\begin{eqaed}\label{eq:11d_RFM_effective_action}
    \int_{M_n} \!\! d^n y \left(M_{11}^9 R_{d} - \left(\frac{\pi^2}{2} \right)^{\!\frac{1}{3}} M_{11}^3 \norm{G_4}^2 + \frac{A}{r^{11}}\right) ,
\end{eqaed}
Letting $g = r^2 \widehat{g}$, so that the volume form induced by $\widehat{g}$ is normalized to unit volume, and $G_4 = N^a \, \omega_a$ in terms of a unit-period basis of cohomology representatives, the average flux term over the internal manifold is $\langle\norm{G_4}^2\rangle_{M_n} = r^{-8} \mathcal{Q}(N)$, in terms of the quadratic form
\begin{eqaed}\label{eq:quadratic_form_flux}
    \mathcal{Q}(N) & \equiv N^a N^b \int_{M_n} \omega_a \wedge \star_{\widehat{g}} \, \omega_b \\
    & = \frac{N^{a}N^{b}}{\abs{\Gamma}} \int_{T^n} p^*(\omega_a) \wedge \star_{\widehat{g}} \, p^*(\omega_b)
\end{eqaed}
which in general depends on flux numbers and shape moduli. In \cref{eq:quadratic_form_flux} we pulled back the four-forms via the projection $p : T^n \to M_n$ as in \cite{Bento:2025wuu}. Hence, the dimensionally reduced field equations for the $\text{dS}_d$ curvature radius $L$ and the internal volume radius $r$ follow by extremizing the reduced action
{
\begin{eqaed}\label{eq:reduced_system_RFM}
	S=\omega_d L^d r^n\left(M_{11}^9\,\frac{d(d-1)}{L^2}-\left(\frac{\pi^2}{2}\right)^{\!\frac{1}{3}}\frac{M_{11}^3}{r^8}\,\mathcal Q(N)+\frac{A(x)}{r^{11}}\right)-b\log(\mu r)\,.
\end{eqaed}
Here we have set $v=1$ for the unit-volume metric. For later use, define $q_\text{flux}=(\pi^2/2)^{1/3}\mathcal Q(N)$, with the flux normalization of \cref{eq:11d_RFM_effective_action}. The scaling formulae below first refer to the flat-background approximation, in which this global zero-mode term has vanishing density and $A$ is independent of $x$.
}
Up to easily computed numerical prefactors\footnote{The prefactor for the volume radius is bounded between $0.89$ and $0.92$ for $4 \leq n \leq 7$.}, the resulting solution takes the form
\begin{eqaed}\label{eq:RFM_solution}
    M_{11}r & \propto \left( \frac{A}{\mathcal{Q}(N)} \right)^{\!\frac{1}{3}} \\
    M_{11}L & \propto \frac{A^{\frac{4}{3}}}{\mathcal{Q}(N)^{\frac{11}{6}}}\,,
\end{eqaed}
which once again shows that $A$ must be large to have control on the \ac{EFT}. In contrast with \cref{sec:setup}, however, the ratio
\begin{eqaed}\label{eq:scale_sep_RFM}
    x_*^2 \equiv \frac{r^2}{L^2} \propto \frac{\mathcal{Q}(N)^3}{A^2}
\end{eqaed}
shows that control can also lead to scale separation. In the context of a no-go theorem, in such cases we need not include the \ac{dS} curvature in the Casimir-energy computation; analogously to the flat-space approximations in the preceding sections, we will begin with this more interesting case in the following. For curved external space the heat-kernel comparison bounds $\mathcal A(x)=A(x)-\frac{x}{d}A'(x)$, evaluated at the actual stationary ratio. The common-scale term also enters the two radius equations. Writing $R=M_{11}r$ and $T=M_{11}L$, we have
\begin{equation}
\begin{aligned}
	d(d-1)R^9x^2&=\frac{d}{2(d+1)}\left[\mathcal A(x)+\frac{b x^d}{3\omega_d}\right],\\
	q_\text{flux}R^3&=\frac{3d}{2(d+1)}\mathcal A(x)+\frac{(d-2)b x^d}{6(d+1)\omega_d}.
\end{aligned}
\end{equation}
Write $Y=d(d-1)R^9x^2$, $Z=q_\text{flux}R^3$ and $B=\frac{b x^d}{\omega_d}$. For one scalar $b=1$. More generally, when flux-dependent terms in the kinetic operators are neglected, the estimate $|b|\lesssim x^n+s^{-n}$, proved in \cref{app:bounding_sugra_casimir}, makes $|B/Y|\lesssim T^{-9}+(Rs)^{-n}T^{-(d-2)}$. Hence, the scaling relations in \cref{eq:RFM_solution} survive in a controlled limit.
\\

\noindent Before proceeding with the analysis, let us remark that the present scalar-field case does not contain couplings with the flux in the Laplacian. These couplings can arise in the full supergravity case; we address this subtlety in the parametric analysis of \cref{app:bounding_sugra_casimir}. \\

\noindent \textbf{The kinematics of rigid \acp{RFM}.} A compact \ac{RFM} of dimension $n$ is determined by a faithful quotient $M_n = T^n/\Gamma = \mathbb{R}^n/B$, with $T^n = \mathbb{R}^n/\Lambda$ defined by a lattice $\Lambda$. These groups fit into an exact sequence of the form
\begin{eqaed}\label{eq:ses}
    0 \longrightarrow \Lambda \longrightarrow B \longrightarrow \Gamma \longrightarrow 0
\end{eqaed}
with the ``Bieberbach group'' $B$ torsion-free\footnote{The torsion-free condition on $B$ ensures that the \ac{RFM} be smooth. In string/M-theory one often has control over singularities of various types, which could modify the conclusions of our arguments.} and finite crystallographic holonomy group $\Gamma$. We have $\Lambda = B \cap \mathbb{R}^n$ and $\Gamma = B/\Lambda$. If a metric $g$ on $\mathbb{R}^n$ descends to a rigid metric on $M_n$, the enhanced symmetry group $\text{Aut}_B(g)$ of (the linear part of) isometries preserving $B$ must freeze all flat-metric deformations except for the volume. Hence, its defining representation on $\mathbb{R}^n$ must be irreducible, for otherwise reciprocal rescalings of two orthogonal invariant subspaces would provide an invariant modulus. Furthermore, since $\text{Aut}_B(g)$ preserves lengths, the linear span of the set of shortest lattice vectors in $\Lambda$ must be the full representation space $\mathbb{R}^n$, and thus there are $n$ linearly independent such vectors $v_1 , \dots , v_n \in \Lambda$ with shortest length $\norm{v_i} \equiv \lambda$. Over the integers, these vectors generate a sublattice of index $j \geq 1$ in $\Lambda$. As a result, we have
\begin{eqaed}\label{eq:vol_RFM_bound}
    \text{Vol}(M_n) & = \frac{1}{\abs{\Gamma}} \, \text{Vol}(\mathbb{R}^n/\Lambda) = \frac{1}{j \abs{\Gamma}} \, \abs{\det(v_1, \dots, v_n)} \leq \frac{\lambda^n}{\abs{\Gamma}}
\end{eqaed}
by Hadamard's inequality on matrix determinants applied to the unit cell. Given an element $g = (a,Q) \in B \backslash \{\text{id} \}$, the rotation holonomy $Q$ has a finite order $q=\text{ord}(Q)$ and at least an eigenvalue\footnote{If $Q$ had no eigenvector with eigenvalue $+1$, $Q-I_n$ would be invertible, and thus $g$ would have a fixed point.} $+1$. Hence, the projector $\Pi_{\ker(Q-I_n)}$ is non-zero, and $g^q = (q\Pi_{\ker(Q-I_n)}a,I_n)$ defines a lattice vector in $\Lambda$ of length $q\norm{\Pi_{\ker(Q-I_n)}a} \geq \lambda$, which is at most\footnote{We have $\Pi_{\ker(Q-I_n)}a = \frac{1}{q}\sum_{k=0}^{q-1}Q^k a$, whose norm is bounded by $\norm{a}$.} $q$ times the translation length of $g$. If the latter is taken to be the systole length $\text{sys}(M_n) \equiv r\,s$, we have
\begin{eqaed}\label{eq:RFM_systole}
    s & \geq \frac{\lambda}{qr} \geq \frac{\abs{\Gamma}^\frac{1}{n}}{q} \geq q^{\frac{1}{n}-1}
\end{eqaed}
for $q$ the order of the corresponding holonomy. We can bound the order $q$ of rotation holonomies $Q \in \Gamma \subset \text{GL}(n,\mathbb{Z})$ as follows. Since its eigenvalues are roots of unity, and at least one of them is unity, the characteristic polynomial $P_Q$ must take the form
\begin{eqaed}\label{eq:Q_char_poly}
    P_Q(x) = (x-1) \prod_{m|q} \Phi_m(x)^{p_m}
\end{eqaed}
in terms of the cyclotomic polynomials $\Phi_m$ of divisors of $q$. Since $\deg \Phi_m = \varphi(m)$ is the Euler totient function, we have
\begin{eqaed}\label{eq:degree_constraint}
    \sum_{m|q} p_m \, \varphi(m) & = n - 1\,, \\
    \sum_{m|q : m>1} p_m \, \varphi(m) & \leq n - 1\,.
\end{eqaed}
From \cref{eq:degree_constraint} we can determine a bound $q \leq q_n^\text{max}$, which is given by the least common multiple of the $m$ appearing in a product of the form of \cref{eq:Q_char_poly}, for any given $n$. This translates into a lower bound $s \geq (q_n^\text{max})^{\frac{1}{n}-1}$ on the systole length. In turn, \cref{eq:vol_RFM_bound}, relates the systole length to the volume via \cref{eq:RFM_systole}. In our setting $4 \leq n \leq 7$, since the solution must be supported by M-theory flux. Hence, the maximum value of $m$ that can appear in the product of \cref{eq:Q_char_poly} is the maximum $m$ such that $\varphi(m) \leq 6$, which translates into $m \leq 18$. \\

\noindent In this range, the bounds $m\leq m_n^\text{max}$ on the divisor in \cref{eq:Q_char_poly}, $q \leq q_n^\text{max}$ on the order of $Q \in \Gamma$ and $s \geq s_n^\text{min}$ on the systole length are reported in \cref{tab:order_bound}. Finally, using \cref{eq:RFM_systole} we can also bound the order $\abs{\Gamma} \leq (s q_n^\text{max})^n$ in terms of the (dimensionless) systole length.
\begin{table}[ht!]
    \centering
    \begin{tabular}{c|c|c|c}
        $n$ & $m_n^\text{max}$ & $q_n^\text{max}$ & $s_n^\text{min} = (q_n^\text{max})^{\frac{1}{n}-1}$ \\
        \hline
        4 & $6$ & $6$ & $0.26$ \\
        5 & $12$ & $12$ & $0.13$ \\
        6 & $12$ & $12$ & $0.12$ \\
        7 & $18$ & $30$ & $0.05$ \\
    \end{tabular}
    \caption{The bound $\text{ord}(Q)$ derived from \cref{eq:degree_constraint} on the order of holonomy elements $Q \in \Gamma$ and dimensionless systole length $s$ (rounded down) for rigid \acp{RFM} of dimensions $4 \leq n \leq 7$.}
    \label{tab:order_bound}
\end{table}
Analogous results appeared in \cite{Bento:2025wuu} when discussing a class of seven-dimensional \acp{RFM} with supersymmetry-breaking spin structure. However, we need not discuss this subtlety in our case, since we want to obtain a uniform upper bound on \ac{EFT} control parameters. As such, even if these bounds were saturated by \acp{RFM} that do not break supersymmetry, it would simply mean that the corresponding \ac{dS} solution does not exist. Of course, implementing this condition would presumably strengthen the resulting bounds. \\

\noindent \textbf{Constraints on fluxes from rigidity.} As we have remarked, introducing fluxes is necessary for a solution to exist in our setting. In particular, there must be $G_4$-flux for the \ac{dS} critical point to exist; for $n=7$ one can also include $G_7$-flux, but its contribution is removed by the axion runaway discussed in \cite{Bento:2025wuu,ValeixoBento:2025qih}. The flux contribution to the potential can break the enhanced symmetry group $\text{Aut}_B(g)$ to a subgroup, but the Casimir contribution will still be stationary at $g$ (except for the overall volume). Hence, the shape moduli $u$ must extremize the flux contribution separately. In order for this to happen, the variation of $\norm{G_4}^2$ must vanish. From this we learn that the internal part of the energy-momentum tensor induced by $G_4$ must be isotropic, since its contraction with traceless (i.e., volume-preserving) deformations must vanish. Thus, in particular, its contraction with its own (internal) traceless part must vanish. This isotropy condition constrains the possibilities in an interesting way:

\begin{itemize}
    \item For $n=4$ there is only one choice for $G_4$ (in the orientable case).
    \item For $n=5$, a class $[G_4] \in H^4(M,\mathbb{Z})$ (up to a normalization \cite{Bento:2025wuu,ValeixoBento:2025qih}) has a Hodge-dual representative vector $v$, so that the traceless part of the internal energy-momentum is proportional to $v \otimes v - \frac{\norm{v}^2}{5} g$. But this tensor only vanishes for $v=0$, which rules out a critical point in this class.
    \item For $n=6$, analogous considerations lead to a dual two-form $\omega$ whose components satisfy $6 \, \omega_{ij}\omega^{jk} = - \, \omega^2 \, \delta_i^k$. Raising an index with $g$, the resulting endomorphism squares to a negative multiple of the identity, hence defining an orthogonal almost-complex structure on $M_6$ (which is trivially integrable due to flatness). Therefore, such a $G_4$-flux defines a K\"{a}hler structure on $M_6$. Together with flux quantization, the overall scale of $G_4$ is then fixed by picking the flux quanta corresponding to the largest stabilized radius \cite{Bento:2025wuu,ValeixoBento:2025qih}.
    \item For $n=7$, the holonomy preserves a positive $G_2$-structure compatible with the metric, as shown below. As in the preceding case, the overall scale of $G_4$ is determined by maximizing the radius over the flux-quantized possibilities. \\
\end{itemize}

\noindent \textbf{Control bounds for rigid \acp{RFM}.} We now proceed to derive an upper bound on the Casimir parameter $A$. Once again, we consider the contribution of a single scalar, up to the renormalization of the \ac{UV}-divergent part (which is equivalent to the presence of higher-dimensional supersymmetry); for an \ac{RFM} defined by $M_n = \mathbb{R}^n/B$, as in \cref{eq:ses}, in the flat-space approximation one has \cite{Bento:2025wuu,DallAgata:2025jii}
\begin{eqaed}\label{eq:casimir_RFM_formula}
    A = \frac{945}{64\pi^5} \int_{M_n} \text{vol}_{\widehat{g}}(y) \sum_{b \in B \backslash \{ \text{id} \}} \frac{1}{\text{dist}_{(\mathbb{R}^n,\widehat{g})}(y, b \cdot y)^{11}}\,,
\end{eqaed}
since the identity term is completely removed by higher-dimensional supersymmetry due to flatness. Here we are employing the rescaled unit-volume metric $\widehat{g}$ and the corresponding volume form $\text{vol}_{\widehat{g}}$, which only depends on the shape moduli (which are assumed to be at a critical point). To be pedantic, \cref{eq:casimir_RFM_formula} contains an abuse of notation: we are really integrating over a fundamental domain for $M_n$ inside the universal cover $\mathbb{R}^n$. Analogously to the hyperbolic case, the orbit of a given point $y$ is $s$-separated, and thus the number $N_y(L)$ of group elements $b$ moving $y$ by a $\widehat{g}$-distance less than $L$ can be bounded by packing Euclidean balls of radius $s/2$ into the ball of radius $L+s/2$ centered at $y$. However, we are able to get much stronger bounds thanks to the exact expressions for the scalar-field Casimir energy that are available in this case. As anticipated, using \cref{eq:heat_kernel_ratio_bound} one can apply these expressions to get a bound on the modified parameter $\mathcal{A}$ of \cref{eq:A_combination_curved}. By inserting the upper bound in \cref{eq:K_Sd_global_bound} in \cref{eq:heat_kernel_ratio} and maximizing over the argument, one finds improved constants $\nu_{d+2}^\text{imp} \approx (6.42,3.93,2.63,1.95)$ for $4 \leq n \leq 7$ with respect to \cref{eq:heat_kernel_ratio_bound}. This accounts for the contribution from the \ac{dS} curvature to the Casimir potential in the winding part; the identity part contains the \ac{UV}-subtracted sphere heat kernel and the analytically continued terms (which in the case are scaleless and thus vanish), but it is universal and bounded numerically by $10^{-4}x^{11}$ in all cases. \\

\noindent We now turn to the flux term $\mathcal{Q}(N)$, for which we seek a lower bound on account of \cref{eq:RFM_solution}. We recall that, in the rigid case, there is no solution for $n=5$; for $n=4$ we simply have $\mathcal{Q}(N) = N_\text{cover}^2/\abs{\Gamma}^2 \geq 1/\abs{\Gamma}^2$ in terms of a quantized flux on the covering torus\footnote{This is a weaker bound than $\mathcal{Q}(N) = N^2_\text{RFM} \geq 1$ in terms of flux quanta on the quotient, but it is useful to write a single expression for all cases, as we do below.}. For $n=6$ (resp. $n=7$), as we have seen, interesting K\"{a}hler (resp. $G_2$) geometry appears, allowing us to obtain ``arithmetic'' bounds on $\mathcal{Q}(N)$. Anticipating this strategy, we use Dirac quantization to evaluate $\mathcal{Q}(N)$, by writing $p^*(G_4) = \frac{1}{\abs{\Gamma}}\iota_N \text{vol}$ as the contraction of the volume form on the covering torus with a flux multivector $N \in \Lambda^{n-4}(\Lambda) \simeq H^4(T^n,\mathbb{Z})$ on the covering torus. As a result, using \cref{eq:quadratic_form_flux} one finds $\mathcal{Q}(N) = \norm{N}^2/\abs{\Gamma}^2$. \\

\noindent We now address the remaining cases:

\begin{itemize}
    \item For $n=6$, on the one hand the bivector $N \in \Lambda^2(\Lambda)$ gives $1/6 N\wedge N\wedge N = k \, V_\Lambda$, a positive-integer multiple of the unit-cell multivector $V_\Lambda = \abs{\Gamma} \, \text{vol}_{\widehat{g}}^\sharp$ of $\Lambda$. On the other hand, the K\"{a}hler structure discussed above gives $N = \mathfrak{n} J^\sharp$ in terms of the complex structure, and $\frac{1}{6} \, J^\sharp \wedge J^\sharp \wedge J^\sharp = \text{vol}_{\widehat{g}}^\sharp$. Since $\norm{J}^2 = \frac{1}{2} \tr(J^T J) = 3$, we obtain $\mathfrak{n}^3 = k \abs{\Gamma} \geq \abs{\Gamma}$ and $\mathcal{Q}(N) \geq 3 \, \abs{\Gamma}^{-\frac{4}{3}}$.

    \item 
    For $n=7$, write $N = \mathfrak{n} \, \varphi^\sharp$, with $\norm{\varphi}^2=7$. Isotropy implies that $\langle\iota_u\varphi,\iota_v\varphi\rangle=3\widehat{g}(u,v)$, so in a suitable $\widehat{g}$-orthonormal frame, $\varphi$ takes the standard definite or split form\footnote{Along each volume-preserving deformation generated by a symmetric traceless matrix, the squared norm is convex. Since isotropy forces its first derivative to vanish, the original form minimizes the norm along that deformation. This also holds after complexification, given that imaginary variations have vanishing real pairing with the real form. Its non-zero complex orbit is therefore closed and distinguished from zero by a homogeneous invariant \cite{BohmLafuente2017}. Such invariants are proportional to powers of the degree-seven polynomial characterizing stable forms, which therefore cannot vanish \cite{BenedettiEtAl2017}. The standard forms also minimize the norm, and such minima within each real orbit differ only by an orthogonal transformation \cite{BohmLafuente2017,CortesEtAl2009}.}. Its cubic bilinear form satisfies
    \begin{eqaed}\label{eq:G2-associator}
    	\frac{1}{6} \, \iota_u \varphi \wedge \iota_v \varphi \wedge \varphi = h(u,v) \, \text{vol}_{\widehat{g}}\,,
    \end{eqaed}
\end{itemize}
where $h(u,v)=\widehat{g}(Su,v)$ and $S$ is self-adjoint with respect to $\widehat{g}$, with $S^2=I$. Choosing the overall sign of $\varphi$ and $\mathfrak{n}$ leaves $N$ unchanged and makes $h$ either positive definite or of signature $(4,3)$. Since $N$ is integral, for any two integral one-forms $\alpha, \beta \in \Lambda^*$ we have
\begin{eqaed}\label{eq:N_cubic_invariant}
	\iota_\alpha N \wedge \iota_\beta N \wedge N & = k(\alpha,\beta) \, \abs{\Gamma} \, \text{vol}_{\widehat{g}}^\sharp \\
	& = 6\mathfrak{n}^3 \, h^{-1}(\alpha,\beta) \, \text{vol}_{\widehat{g}}^\sharp
\end{eqaed}
for some non-degenerate integral symmetric bilinear form $k$. Given that $S^2=I$, raising both indices of $h$ with $\widehat{g}$ yields $h^{-1}$. In a lattice cobasis $\{e^i\}_{i=1}^7$, the matrix $h^{-1}(e^i,e^j)$ has determinant with absolute value $\abs{\Gamma}^{-2}$. Using $\abs{\det k}\geq1$ leads to\footnote{Stronger arithmetic constraints on $\det k$ would result in tighter bounds. However, at this level of generality, this does not significantly improve the resulting control parameters.}
\begin{eqaed}\label{eq:G2-determinant}
	(6\abs{\mathfrak{n}}^3)^7 \, \abs{\Gamma}^{-2} = \abs{\Gamma}^7 \, \abs{\det k} \geq \abs{\Gamma}^7\,.
\end{eqaed}
As a result, since $\mathcal{Q}(N) = 7\mathfrak{n}^2/\abs{\Gamma}^2$, we finally arrive at
\begin{eqaed}\label{eq:G2_Q_bound}
    \mathcal{Q}(N) \geq \frac{7}{6^{\frac{2}{3}}\abs{\Gamma}^{\frac{8}{7}}}\,.
\end{eqaed}

\noindent In the definite case, $\varphi$ defines itself a positive $G_2$-structure compatible with $\widehat{g}$. In the split case, let $E_-$ be the three-dimensional negative eigenspace of $S$. For the standard split form, reversing the sign of the component along $E_-$ gives a positive three-form compatible with $\widehat{g}$. We therefore set $\Phi=\varphi-2\varphi|_{\Lambda^3E_-}$, where the restricted term is evaluated by projecting each of its three arguments orthogonally onto $E_-$. Holonomy preserves $\varphi$ and $\widehat{g}$ and, as a consequence, also $S$, $E_-$ and this projection. It therefore also preserves $\Phi$. The holonomy arguments below use this positive $G_2$-structure. \\

\noindent To sum up, we obtained the following bounds on the flux quadratic form on rigid \acp{RFM}:%
\begin{eqaed}\label{eq:Q_final_bound}
    \mathcal{Q}(N) \geq \begin{cases}
    \abs{\Gamma}^{-2}\,, \qquad n = 4\,, \\
    3 \, \abs{\Gamma}^{-\frac{4}{3}}\,, \qquad n = 6\,, \\
    7 \, 6^{-\frac{2}{3}} \, \abs{\Gamma}^{-\frac{8}{7}}\,, \qquad n = 7\,,
\end{cases}
\end{eqaed}
which can be summarized as $\mathcal{Q}(N) \geq Q_n \, \abs{\Gamma}^{-\frac{8}{n}}$. Combining \cref{eq:Q_final_bound,eq:RFM_systole}, we can obtain bounds on the \ac{EFT} control parameters of \cref{eq:RFM_solution}, the most relevant of which we denote $H_n^\text{RFM} = \min(M_{11}r, M_{11}L, M_{11}r s)$. In practice, $H_n^\text{RFM} \leq M_{11}r s$ is dominated by the physical systole length, which resonates with the findings of \cite{Bento:2025wuu,Aparici:2026wdm}. \\

\noindent When bounding fluxes, it is useful to exploit the fact that a rigid \ac{RFM} is either a torus ($\Gamma$ is trivial) or has vanishing first Betti number\footnote{To see this, observe that the $\Gamma$-invariant subspace $(\mathbb{R}^n)^{\Gamma}$ is also invariant under the (possibly enhanced) symmetry group $\text{Aut}_B(g)$, which normalizes $\Gamma$. By irreducibility, this subspace must vanish unless $\Gamma$ is trivial. But then the first Betti number $b_1(M_n) = \dim (\mathbb{R}^n)^\Gamma = 0$.}. Later on, when extending the no-go theorem to a parametric result for all critical points of the Casimir potential, we will not make use of this fact.

\begin{itemize}
    \item For $n=4$, the flux is a top-rank form. Therefore, $N^2/\abs{\Gamma}^2 \geq 1$ is a stricter lower bound, stemming from the fundamental class of the \ac{RFM}. This rescales the control parameter by $(q_4^\text{max})^{-\frac{2}{3}}$, resulting in a $\approx 70\%$ removal. This result also follows from the following stronger reasoning: since $M_4$ must be orientable for a solution to exist, if $\Gamma$ is non-trivial we have $b_1(M_4)=0$, and thus the Gauss-Bonnet theorem implies that $0=\chi(M_4)=2+b_2(M_4)$, which is impossible. Therefore, $\Gamma$ is trivial and $M_4 = T^4$ is a torus. As a result, setting $\abs{\Gamma}=1$ in \cref{eq:vol_RFM_bound,eq:RFM_systole} gives $\lambda \geq 1$ and $s \geq 1/6$. But for a torus $s = \lambda \geq 1$, which is a stronger bound.

    \item For $n=6$, if $M_6$ is not a torus, the flat K\"{a}hler structure has a dramatic interplay with $b_1(M_6)=0$: flat K\"{a}hler manifolds are classified, and the possible holonomies with $b_1(M_6)=0$ are $\Gamma = \mathbb{Z}_2 \times \mathbb{Z}_2$ or the order-eight dihedral group $\Gamma = D_8$ \cite{dekimpe}. Including the case $M_6 = T^6$, we learn that $q_6^\text{max} \leq 4$ rather than $12$\footnote{More specifically, the rotation generator $r$ and its inverse have order $q=4$, while the other non-trivial elements have order $q=2$.}.

    \item For $n=7$, a substantial improvement stems from the $G_2$-structure: the cyclotomic constraints of \cref{eq:Q_char_poly,eq:degree_constraint} must comply with the $G_2$-structure, which decomposes the fundamental representation $\mathbf{7} = \mathbf{1} \oplus \mathbf{3} \oplus \overline{\mathbf{3}}$ with respect to the $\text{SU}(3)$ stabilizer of the invariant axis. Thus, the eigenvalues of $Q \in \Gamma$ take the form $(1,u_1,u_1^{-1},u_2,u_2^{-1},u_3,u_3^{-1})$ with $u_1 u_2 u_3 = 1$. Including the integrality constraint lowers the bound on the order to $q_7^\text{max} = 12$ by listing the possibilities. In fact, we can derive a much stronger statement: $q_7^\text{max} = 2$. Indeed, by the main theorem of \cite{Lutowski2021Flat}, over the rationals the holonomy representation on $\Lambda \otimes \mathbb{Q}$ has one homogeneous\footnote{A representation is \emph{homogeneous} if its irreducible components are isomorphic. A general representation decomposes as the direct sum of isotypic components, each of which is homogeneous.} component only if $M_7 = T^7$. Since $\text{Aut}_B(g)$ preserves $\Gamma$, its action must permute the distinct isotypic components
\begin{eqaed}\label{eq:isotypic_decomposition}
        \Lambda \otimes \mathbb{Q} = \bigoplus_{i=1}^I V_i\,.
    \end{eqaed}
    In particular, all isotypic components in one orbit have the same dimension. Since the whole real $\text{Aut}_B(g)$-representation $\mathbb{R}^7$ is irreducible by rigidity, there is only one orbit, and therefore
    \begin{eqaed}\label{eq:equal_dims_7}
        I \times \dim (V_i) = 7\,.
    \end{eqaed}
    The theorem of \cite{Lutowski2021Flat} tells us that $I=1$ if and only if $M_7=T^7$, for which the Casimir coefficient has no enhancement from holonomy sectors. For a non-trivial \ac{RFM}, we thus learn that $I=7$ and thus the faithful rational holonomy representation is a sum of (non-trivial\footnote{If one character were trivial, the $\Gamma$-invariant subspace would contain a real line, contradicting $b_1(M_7) = 0$.}) distinct characters! This tells us that $\Gamma = \mathbb{Z}_2^r$ for some $r \leq 7$. On the one hand, there are $2^r-1 \geq 7$ non-trivial characters, and thus $r \geq 3$; on the other hand, the maximal rank of abelian 2-subgroups of $G_2$ is three, hence $r \leq 3$. We conclude\footnote{We can also prove that $\Gamma = \mathbb{Z}_2^3$ more directly as follows. Any two distinct character lines are orthogonal, so their cross product in $\mathbb{R}^7$ is orthogonal to both and carries the product of their characters by $G_2$-equivariance: $Q(u \times v)= (Qu)\times (Qv) = \chi_u(Q)\chi_v(Q)(u \times v)$. Since the holonomy representation is faithful, the seven characters (plus the identity) must therefore generate the whole character group $\Gamma^\vee = \mathbb{Z}_2^r$, from which $2^r=8$.} that $\Gamma = \mathbb{Z}_2^3$, and all $Q \in \Gamma$ have order $q=2$.
\end{itemize}

\noindent \textbf{Further improvements from Casimir branes.} Finally, we can get even more mileage by using the exact expression for the Casimir coefficient derived in \cite{Bento:2025wuu}. For $n=6,7$ we can make further use of the detailed arithmetic geometry of \acp{RFM} with K\"{a}hler and $G_2$-structure. The exact scalar-field Casimir energy on $\mathbb{R}^{d-1,1} \times M_n$ can be decomposed into a sum over holonomies $Q \in \Gamma$ (we denote $q = \text{ord}(Q)$) by choosing a lift $b_Q = (u_Q, Q) \in B_Q$ in the fiber $B_Q$ over the projection $B \to \Gamma=B/\Lambda$, and taking the projection $v_Q=\Pi_{\ker(Q-I_n)}u_Q$ \cite{Bento:2025wuu}. Each contribution is controlled by the invariant torus $T^Q \equiv (\mathbb{R}^n)^{\langle Q \rangle}/\Pi_{\ker(Q-I_n)}\Lambda$, whose dimension we denote by $k_Q$, according to \cite{Bento:2025wuu}
\begin{eqaed}\label{eq:A_casimir_branes}
    A = \frac{945}{64\pi^5} \sum_{v \in \Lambda \backslash \{0\} } \frac{1}{\norm{v}^{11}} + \sum_{Q \in \Gamma \backslash \{I_n\}} \frac{\Gamma(\frac{d+k_Q}{2})}{2\pi^\frac{d+k_Q}{2}} \, \frac{\text{Vol}(T^Q)}{\abs{\Gamma}} \sum_{v \in \Pi_{\ker(Q-I_n)}\Lambda} \frac{1}{\norm{v_Q + v}^{d+k_Q}}\,,
\end{eqaed}
where the overall identity contribution was isolated for convenience. This expression is heavily constrained by the K\"{a}hler/$G_2$-structure. In somewhat more physical terms, we are exploiting the properties of ``Casimir branes'' \cite{Bento:2025wuu} (where the density of Casimir energy is large in the internal manifold) that preserve the K\"{a}hler/$G_2$-structure:

\begin{itemize}
    \item For $n=6$, in the case $\Gamma = \mathbb{Z}_2 \times \mathbb{Z}_2$ the holonomy representation $\mathbb{R}^6 \simeq \mathbb{C} \oplus \mathbb{C} \oplus \mathbb{C}$ splits into a direct sum of non-trivial complex characters, and each $Q \in \Gamma \backslash \{I_6\}$ must have a $+1$ eigenvalue. This forces the characters to be the three distinct characters, defined by the charges $(+,-), (-,+), (-,-)$. As a result, any $Q$ has eigenvalues $(1,1,-1,-1,-1,-1)$ and therefore $k_Q = \dim \ker (Q - I_6) = 2$. For $\Gamma = D_8$, the holonomy representation cannot split into complex lines, for otherwise the square $r^2$ of the rotation generator would act non-faithfully (as the identity). Put differently, (complex) one-dimensional characters factor through the abelianization $(D_8)_\text{ab} = D_8/\langle r^2 \rangle \simeq \mathbb{Z}_2 \times \mathbb{Z}_2$. Therefore, the holonomy representation of $D_8$ splits into a complex line and the two-dimensional irreducible complex representation. Out of the non-trivial complex-line characters, the only one for which all $Q \in D_8$ fix an axis assigns $\chi(r) =+1$ to the rotation generator\footnote{This is because the rotation generator $r$ has no $+1$ eigenvalue in the irreducible complex two-dimensional representation.} and therefore $\chi(s)=-1$ on the reflection generator. By inspection, once again $k_Q = 2$. Therefore, we learn that $k_Q = 2$ in all cases.

    \item For $n=7$, due to the decomposition $\mathbf{7} = \mathbf{1} \oplus \mathbf{3} \oplus \overline{\mathbf{3}}$ of the fundamental representation of $G_2$ under the stabilizer $\text{SU}(3)$ of the invariant axis of $Q$, only $k_Q=1$ and $k_Q=3$ are possible for $Q \neq I_7$. The case $k_Q=1$ can then be excluded by the fact that the holonomy representation of $\Gamma = \mathbb{Z}_2^3$ consists of the seven distinct non-trivial characters: any non-trivial element $Q \in \Gamma$ evaluates to $+1$ on half of the eight characters, and to $-1$ on the other\footnote{To see this, let $Q=(Q_1,Q_2,Q_3) \in \mathbb{Z}_2^3=\{0,1\}^3$ with $Q_1 = 1$ (without loss of generality), and let $\chi(Q) = (-1)^{x \cdot Q}=(-1)^{x_1}(-1)^{x_2Q_2+x_3Q_3}$ with $x \in \mathbb{Z}_2^3$. Precisely half of the characters have $x_1=0$.}. Hence, the eigenvalues of $Q$ in the holonomy representation are $(+,+,+,-,-,-,-)$, that is $k_Q=3$.
\end{itemize}

\noindent We can thus obtain a much stronger bound on \cref{eq:A_casimir_branes} by leveraging these considerations. Firstly, no vector $v_Q+v$ in the sum can vanish: a lift $b_Q = (u_Q+u,Q) \in B_Q$ with $\Pi_{\ker(Q-I_n)}u = v$ satisfies $b_Q^q = (q(v_Q+v),I_n) \in B$, which cannot be trivial by torsion-freeness. Recalling that the minimal length of lattice vectors is $\lambda$, the minimal length of non-zero vectors $v=\Pi_{\ker(Q-I_n)}u \in \Pi_{\ker(Q-I_n)}\Lambda$ is bounded below by $\lambda/q$, since $qv = \sum_{j=0}^{q-1} Q^j u \in \Lambda$. We also have $\norm{v_Q+v} \geq s$, so we can write $\norm{v_Q+v} \geq \max(s,\lambda/q)$. Similarly, by the same token of \cref{eq:vol_RFM_bound}, the volume $\text{Vol}(T^Q) \leq \lambda^{k_Q}$ since the projection onto the invariant subspace of $Q$ does not increase norms and there is a basis of $k_Q$ vectors of length at most $\lambda$ coming from the projection of a lattice basis. \\

\noindent The identity contribution can be estimated by packing in $n$ dimensions, but using the minimal lattice length $\lambda$ rather $s$. This gives
\begin{eqaed}\label{eq:A_id_lattice_packing}
    \sum_{v \in \Lambda \setminus \{ 0 \} } \frac{1}{\norm{v}^{11}} & \leq 11 \int_\lambda^\infty \ell^{-12} \left(1 + \frac{2\ell}{\lambda}\right)^{\!n} d\ell \\
    & = 11 \, \lambda^{-11} \int_1^\infty x^{-12} \left(1+2x\right)^{\!n} dx = K_n \, \lambda^{-11}\,.
\end{eqaed}
The sum over the projected lattice can also be estimated by packing, now balls of radius $\lambda/(2q)$ in $k_Q$ dimensions. This yields
\begin{eqaed}\label{eq:A_w_lower-dim_packing}
    \sum_{v \in \Pi_{\ker(Q-I_n)}\Lambda} \frac{1}{\norm{v_Q + v}^{d+k_Q}} & \leq (d+k_Q) \int_{\max(s,\lambda/q)}^\infty \ell^{-d-1-k_Q} \left(1+\frac{2q\ell}{\lambda}\right)^{\!k_Q} d\ell \\
    & = (d+k_Q) \left(\frac{q}{\lambda}\right)^{\!d+k_Q} \int_{\max(1,qs/\lambda)}^\infty x^{-d-1-k_Q} \left(1+2x\right)^{\!k_Q} dx \\
    & \equiv (d+k_Q) \left(\frac{q}{\lambda}\right)^{\!d+k_Q} \mathcal{I}_{k_Q}(\max(1,qs/\lambda))\,.
\end{eqaed}
These integrals are easily evaluated using the Euler beta function. \\

\noindent We now have all the ingredients at our disposal. Putting them together with \cref{eq:vol_RFM_bound,eq:RFM_systole}, we obtain functions of $1/q_n^\text{max} \leq s/\lambda \leq 1$ that we can maximize to obtain improved bounds; for $n=6$, the maximum lies at $s/\lambda = 1$ for $\Gamma=\mathbb{Z}_2 \times \mathbb{Z}_2$ and at $s/\lambda = 1/4$ for $\Gamma = D_8$, while for $n=7$ the maximum lies at $s/\lambda = 1$. The results are collected in \cref{tab:RFM_control_bounds_improved_again}. \\

{\noindent To check the common-scale correction to these finite bounds, recall that $H_n^\text{RFM}=\min(R,Rs,T)$. For a flat internal space, the subtracted identity integral converges at both endpoints, and no additional finite terms from analytic continuation are required. Scaling $t$ by $x^{-2}$ gives $\mathcal A_\text{id}(x)\propto x^{11}$. Using $x_0=\sqrt{\frac{n(n-1)}{d(d-1)}}$ as a computational reference, the constants $C_d$ and $b_k$ defined in \cref{eq:C_d_constants,eq:local_expansion_higher-dim} give
\begin{equation}
	\frac{|\mathcal A_\text{id}(x)|}{x^{11}}
	\leq x_0^{-11}\left(\frac{C_d}{(4\pi)^{n/2}}+
	\frac{1}{(4\pi)^{11/2}}\left[\frac{\widehat\nu_{d+2}}{11}+
	\sum_{k=0}^{5}\frac{\abs{b_k} x_0^{2k}}{11-2k}\right]\right)<10^{-3}
\end{equation}
for $n=4,6,7$, with sharper sphere bounds $\widehat\nu_{d+2}=(6.41476,2.62713,1.94380)$ in that order. The two terms come from integrating over $t\leq1$ and $t\geq1$, respectively. For $N_\text{sc}$ identical scalars, both the identity term and $b$ are multiplied by $N_\text{sc}$. On $H_n^\text{RFM}\geq2$, their contributions relative to $Y$ are bounded by
\begin{equation}
\begin{aligned}
	\delta_{N_\text{sc}}&=\frac{N_\text{sc}}{512}\left[
	\frac{10^{-3}}{2(d^2-1)}+\frac{1}{6(d^2-1)\omega_d}\right],\\
	q_\text{flux}(M_{11}r)^3&\leq\frac{3d}{2(d+1)}
	\frac{N_\text{sc}\widehat\nu_{d+2}A_\text{flat}}{1-\delta_{N_\text{sc}}} \, .
\end{aligned}
\end{equation}
Indeed, the identity and logarithmic terms in the first radius equation contribute at most $\frac{N_\text{sc}10^{-3}}{2(d^2-1)T^9}$ and $\frac{N_\text{sc}}{6(d^2-1)\omega_dR^nT^{d-2}}$, while $Z\leq3Y$. Direct substitution gives $\widehat\nu_{d+2}/(1-\delta_{N_\text{sc}})<\nu_{d+2}^\text{imp}$ for both $N_\text{sc}=1$ and $256$. One can then check thet the curved scalar bounds remain unchanged. \\
}

\begin{table}[ht!]
    \centering
    \begin{tabular}{c|c|c}
        $n$ & $H_n^\text{RFM}$ & $H_n^\text{phys}$ \\
        \hline
        4 & $1.6$ & $18.8$ \\
        5 & -- & -- \\
        6 & $3.6$ & $31$ \\
        7 & $4.3$ & $34$ \\
    \end{tabular}
    \caption{Numerical bounds (rounded up) for $H_n^\text{RFM}=\min(M_{11}r,M_{11}rs,M_{11}L)$ for rigid \acp{RFM}. The middle column uses the flat-background approximation while the rightmost column uses the heuristic estimate $256$-scalar comparison, as well as the numerical prefactor $\nu_{d+2}^\text{imp}$. As remarked in the main text, this provides a heuristic expectation on the bounds stemming from the full supergravity Casimir energy, which we analyze in \cref{app:bounding_sugra_casimir}.}
    \label{tab:RFM_control_bounds_improved_again}
\end{table} 

\noindent \textbf{Beyond rigid \acp{RFM}.} As discussed above, the fact that the dimensionless systole length $s$ disappears from the bounds on the \ac{EFT} control parameter suggests a generalization of the bounds beyond rigid \acp{RFM}, encompassing all critical points of the Casimir potential. Once more, as in \cref{sec:beyond_rigid_riemann_surfaces}, we use the scalar-field Casimir potential as an approximation; while the upper bounds on the potential itself are not expected to be violated by the partial boson-fermion cancellation in supergravity, the bounds on the gradient might be affected in the opposite direction. Indeed, the asymptotic scalings we will derive do change in the supergravity case; nevertheless, as we show in \cref{app:bounding_sugra_gradient}, the argument can be modified accordingly. At any rate, to the extent that the preceding bounds require rigidity for their derivation, we will not use them in what follows. \\

\noindent Due to the arithmetic complications of the Riemann-flat landscape beyond rigid settings, we aim for a parametric no-go theorem. In addition, the contribution from the \ac{dS} curvature further complicates matters, mostly due to the zero-mode problem already encountered in \cref{sec:beyond_rigid_riemann_surfaces}. Therefore, in the following we shall begin by restricting to the flat-space approximation, which is physically justified in the presence of scale separation. As hinted at by the examples we have studied, we do not expect the conclusion to be significantly affected. Indeed, the zero-mode with $l = 0$ on the Wick-rotated \ac{dS} gives a mild modification due to the exponents in the Schwinger integral, while the higher angular momenta $l > 0$ are heavily (exponentially in $l^2$) suppressed. At any rate, we will extend our proof by including the \ac{dS} curvature and addressing the zero-mode problem as we have done in \cref{sec:beyond_rigid_riemann_surfaces} for Riemann surfaces. We shall begin by turning the numerical bounds of \cref{tab:RFM_control_bounds_improved_again} into parametric bounds for critical points that extremize the Casimir and flux potentials individually; the extension to general critical points will follow. \\

\noindent We proceed in analogy with \cref{sec:beyond_rigid_riemann_surfaces}. In order for $s$ to disappear from the \ac{EFT} control parameter $H_n^\text{RFM}$, the bound $\abs{\Gamma} \leq (s q_n^\text{max})^n$ is crucial. This bound can be traced back to \cref{eq:vol_RFM_bound}, which in turn stems from the identification of $n$ linearly independent shortest-length vectors in the lattice $\Lambda$. This step requires rigidity\footnote{Indeed, kinematically one can engineer a family of arbitrarily squished tori with $s \to 0$ but trivial $\Gamma$.}; as we now show, extremizing the Casimir potential leads to a similar condition on the order $\abs{\Gamma}$ of the holonomy group. In more detail, for a given \ac{RFM}, letting $\lambda_1 \leq \dots \leq \lambda_n$ be the successive minimum lengths such that there are $k$ linearly independent lattice vectors with length $\leq \lambda_k$, we only need prove $\lambda_n/\lambda_1 \leq O(1)$ uniformly for critical points of the Casimir potential; in other words, critical \acp{RFM} may be said to be ``rounded''. This excludes the possibility that infinite sequences of such points in moduli space violate the no-go by resembling a highly anisotropic ``sausage''. Then, since there are finitely many isomorphism classes of \acp{RFM} in any given dimension $n$, the uniform parametric bound extends to the whole landscape with $4\leq n \leq 7$. \\

\noindent When the defining representation $\mathbb{R}^n$ of the (possibly enhanced) symmetry group $\text{Aut}_B(g)$ is irreducible, it follows that $\lambda_1 = \lambda_n$. Therefore, without loss of generality we can focus on cases in which the defining representation is reducible. Assume (by way of contradiction) that there is a sequence of Casimir critical points with $\lambda_n/\lambda_1 \to \infty$. In this case, there is a $k < n$ such that $\lambda_{k+1}/\lambda_k \to \infty$ along a sequence, since $\lambda_n/\lambda_1$ can be written as a telescopic product of successive ratios. In other words, we consider the first parametric gap in length within the sequence of successive minima. An observation that will prove useful later is that, since $\text{Vol}_{\widehat{g}}(M_n) = 1$, whenever $\lambda_n/\lambda_1 \to \infty$ we also have $\lambda_1 \to 0$, because the product of the $\lambda_i$ is bounded by the volume of the covering torus by Minkowski's second theorem on lattices. \\

\noindent Once again, since lattice lengths are preserved by $\Gamma$ (and, more generally, by a possibly enhanced symmetry $\text{Aut}_B(g)$), there is an orthogonal decomposition $\mathbb{R}^n = W_k \oplus W_k^\perp$ into invariant subspaces, where $W_k$ is the linear span of $k$ linearly independent lattice vectors $\{v_1, \dots, v_k \}$ of length up to $\lambda_k$ and $\dim W_k = k$. Then, the shortest length $\lambda_k^\perp$ of vectors in the projection $\Pi_{W_k^\perp}\Lambda$ also diverges relative to $\lambda_k$. To see this, observe that any lattice vector $u$ with a non-zero projection to $W_k^\perp$ must have length at least $\lambda_{k+1}$, for otherwise the linearly independent set $\{v_1, \dots, v_k, u \}$ would span $W_k$. Then, for such $u$ and $v \in \Lambda \cap W_k$, the difference $u-v$ is also such a vector, and we can choose $v$ such that
\begin{eqaed}\label{eq:shortest_lattice_projection}
    \norm{\Pi_{W_k^\perp}u}^2 & = \norm{\Pi_{W_k^\perp}(u-v)}^2 \\
    & = \norm{u-v}^2 - \norm{\Pi_{W_k}u-v}^2 \\
    & \geq \lambda_{k+1}^2 - \frac{k^2}{4} \, \lambda_k^2
\end{eqaed}
by choosing the components of $v$ in the basis $\{v_1 , \dots , v_k \}$ to be within $\pm 1/2$ of the components of $\Pi_{W_k}u$. As a result, since \cref{eq:shortest_lattice_projection} holds for any lattice vector $u$ with a non-zero projection to $W_k^\perp$, we have
\begin{eqaed}\label{eq:shortest_Wperp}
    \lambda_k^\perp \geq \sqrt{\lambda_{k+1}^2 - \frac{k^2}{4} \, \lambda_k^2} \gg \lambda_k
\end{eqaed}
along the sequence of critical points. \\

\noindent In this situation, we are led to consider the volume-preserving flat-metric deformation of $\widehat{g}$ defined by
\begin{eqaed}\label{eq:RFM_metric_def}
    \widehat{g}(\sigma) \equiv e^{2\sigma} \, \widehat{g}\Big|_{W_k} + e^{-\frac{2k}{n-k}\sigma} \, \widehat{g}\Big|_{W_k^\perp}\,.
\end{eqaed}
Using \cref{eq:casimir_RFM_formula}, and removing the numerical prefactor defining $A \equiv \frac{945}{64\pi^5} \, a(\sigma)$,
\begin{eqaed}\label{eq:A_RFM_deformation}
    a(\sigma) & = \int_{M_n} \text{vol}_{\widehat{g}}(y) \sum_{b \in B \backslash \{ \text{id} \} } \frac{1}{\left( e^{2\sigma} \norm{\Pi_{W_k}(y - b \cdot y)}^2 + e^{-\frac{2k}{n-k}\sigma} \norm{\Pi_{W_k^\perp}(y - b \cdot y)}^2  \right)^{\!\frac{11}{2}}} \\
    & \equiv \int_{M_n} \text{vol}_{\widehat{g}}(y) \sum_{b \in B \backslash \{ \text{id} \} } \frac{1}{\left( e^{2\sigma} \, \delta^2_{W_k}(b,y) + e^{-\frac{2k}{n-k}\sigma} \, \delta^2_{W_k^\perp}(b,y) \right)^{\!\frac{11}{2}}}\,.
\end{eqaed}
Thus, the directional derivative of the Casimir potential along the deformation of \cref{eq:RFM_metric_def}, evaluated at a critical point in the (putative) sequence, is proportional to
\begin{eqaed}\label{eq:dA_RFM_deformation}
    a'(0) & = - \, 11 \int_{M_n} \text{vol}_{\widehat{g}}(y) \sum_{b \in B \backslash \{ \text{id} \} } \frac{\delta^2_{W_k}(b,y) - \frac{k}{n-k} \, \delta^2_{W_k^\perp}(b,y)}{\left( \delta^2_{W_k}(b,y) + \delta^2_{W_k^\perp}(b,y) \right)^{\!\frac{13}{2}}} \\
    & \equiv - 11 \left(a_{W_k} - \frac{k}{n-k} \, a_{W_k^\perp} \right) ,
\end{eqaed}
where $a_{W_k}, a_{W_k^\perp} \geq 0$. In fact, the contribution to $a_{W_k}$ from the Bieberbach elements $b = (\pm v_1,I_n)$ gives
\begin{eqaed}\label{eq:dA_Wk_term}
    a_{W_k} \geq 2 \, \lambda_1^{-11}\,.
\end{eqaed}
To bound $a_{W_k^\perp}$, we split the sum according to whether the distance $\text{dist}_{\mathbb{R}^n,\widehat{g}}(y, b \cdot y) \leq \alpha \lambda_1$ for some $\alpha > 1$. The contribution $a_{W_k^\perp}^\text{long}$ with distances $> \alpha \lambda_1$ is bounded by
\begin{eqaed}\label{eq:dA_Wkperp_long_bound}
    a_{W_k^\perp}^\text{long} & \leq \int_{M_n} \text{vol}_{\widehat{g}}(y) \sum_{b \in B \backslash \{ \text{id} \}}^\text{long} \frac{1}{\text{dist}_{\mathbb{R}^n,\widehat{g}}(y, b \cdot y)^{11}} \\
    & = \int_{M_n} \text{vol}_{\widehat{g}}(y) \int_{\alpha \lambda_1}^\infty \ell^{-11} \, dN_y(\ell) \\
    & \leq 11 \int_{\alpha \lambda_1}^\infty \ell^{-12} \left(1 + \frac{2q_n^\text{max} \ell}{\lambda_1} \right)^{\!n} d\ell \\
    & = 11 \, \lambda_1^{-11} \int_{\alpha}^\infty u^{-12} \left(1 + 2q_n^\text{max} u \right)^{\!n} du \\
    & \leq 11 \, \lambda_1^{-11} \int_{\alpha}^\infty u^{n-12} \left(1 + 2q_n^\text{max}\right)^{\!n} du \\
    & = \frac{11}{11-n} \, \frac{(1+2q_n^\text{max})^n \alpha^n}{(\alpha \lambda_1)^{11}} \\
    & \equiv K_n^\text{long} \, \alpha^{n-11} \lambda_1^{-11}\,,
\end{eqaed}
where we used Euclidean packing with the lower bound $s \geq \lambda_1/q_n^\text{max}$ on the dimensionless systole\footnote{This bound is the generalization of \cref{eq:RFM_systole} to the non-rigid case, where lengths are measured with respect to the metric $\widehat{g}$ (i.e., in units of the volume radius $r$).}. Thus, we can (and will) choose $\alpha = O(1)$ large enough so that $\frac{k}{n-k} \, a_{W_k^\perp}^\text{long} < \lambda_1^{-11}$. \\

\noindent Finally, the remaining contribution $a_{W_k^\perp}^\text{short}$ summing over terms with distances $\leq \alpha \lambda_1$ can be bounded as follows. Any individual summand is bounded by $s^{-11} \leq (q_n^\text{max})^{11}/\lambda_1^{11}$. Each term in the sum is such that $\delta_{W_k^\perp}(b,y) \leq \alpha \lambda_1$. The sum over such Bieberbach group elements can be organized into holonomy sectors $Q \in \Gamma = B/\Lambda$ and lifts $B_Q$ thereof. Choosing a lift $b_Q = (v_Q, Q) \in B_Q$ for each $Q \in \Gamma$, any other lift $b=(v_Q+v,Q)$ is parametrized by a lattice vector $v \in \Lambda$. This decomposition is useful because
\begin{eqaed}\label{eq:Wkperp_displacement_Wperp-torus}
    \delta_{W_k^\perp}((v_Q+v,Q),y) & = \norm{\Pi_{W_k^\perp}(b \cdot y - y)} \\
    & = \norm{\Pi_{W_k^\perp}(b_Q \cdot y - y) + \Pi_{W_k^\perp}v}
\end{eqaed}
can only be bounded by $\alpha \lambda_1$ for at most one $\Pi_{W_k^\perp}v \in \Pi_{W_k^\perp}\Lambda$, because $\norm{\Pi_{W_k^\perp}v} \geq \lambda_k^\perp \gg \lambda_1$ along the sequence due to \cref{eq:shortest_Wperp}. Hence, the number of lifts $b \in B$ of $Q \in \Gamma$ that contribute to the sum is bounded by the number of lattice vectors $v \in W_k \cap \Lambda$, since two lifts $b_1, b_2$ differ by such a vector. We have $\norm{\Pi_{W_k}(b \cdot y - y)} \leq \alpha \lambda_1$ and ${b_1 \cdot y - b_2 \cdot y} \geq \lambda_1$. Therefore, the number of lifts of $Q$ contributing to the sum is bounded by packing balls of radius $\lambda_1/2$ in $W_k$ into a ball of radius $\alpha \lambda_1$ in $W_k$. In the sum over $Q$ we can exclude the cases where $Q|_{W^\perp_k}$ acts trivially, for otherwise the transverse displacement vanishes. This sum is invariant under $\Gamma$ (whose action conjugates $Q$), so its integral over $M_n$ is well defined and can be lifted to the covering torus, where the contributing region for each $Q$ comprises points admitting short displacements. We can maximize this bound over $Q$, which gives
\begin{eqaed}\label{eq:dA_Wkperp_short_bound}
    a_{W_k^\perp}^\text{short} \leq \frac{(q_n^\text{max})^{11}}{\lambda_1^{11}} \left(1+2\alpha\right)^{\!k} \max_{Q \in \Gamma : Q|_{W^\perp_k} \neq \text{id}} \text{Vol}_{\widehat{g}}(M_n^\text{short}(Q))\,.
\end{eqaed}
The volume of the region $M_n^\text{short}(Q)$, defined by
\begin{eqaed}\label{eq:short_Mn_region}
    M_n^\text{short}(Q) \equiv \{x \in T^n : \exists b \in B_Q, \norm{b \cdot x - x} \leq \alpha \lambda_1 \}\,,
\end{eqaed}
is the volume of points in the covering torus such that there is a Bieberbach lift $b \in B_Q$ realizing a displacement $\leq \alpha \lambda_1$, maximized over $Q \in \Gamma$. \\

\noindent In order to bound the volume factor in \cref{eq:dA_Wkperp_short_bound}, we use the fact that, after projecting the displacement vectors to $W_k^\perp$, eventually along the sequence of critical points there is at most one lift $b \in B_Q$ with a ``short'' displacement. This leads us to consider the torus $T_\perp \equiv W_k^\perp/\Pi_{W_k^\perp} \Lambda$, whose systole is $\lambda_k^\perp \gg \lambda_1$ due to \cref{eq:shortest_Wperp}. Any lift descends to an affine isometry $f_Q : T_\perp \to T_\perp$ that only depends on $Q$, and we are interested in the volume of the region displaced less than $\alpha \lambda_1 \ll \lambda_k^\perp$ by $f_Q$. Because some power of $b$ is a lattice translation, $f_Q$ has finite order $\text{ord}(f_Q)$; if it had no fixed point, lifting to a translation $\widetilde{f}_Q^{\text{ord}(f_Q)}$ on $W_k^\perp$ would define a geodesic on $T_\perp$ whose length would parametrically exceed $\alpha \lambda_1$. (This follows from the same argument leading to \cref{eq:RFM_systole}.) Therefore, $f_Q$ must have a fixed point; equivalently, it acts as the projection of $Q$ around a fixed point. We thus learn that the volume of $M_n^\text{short}(Q)$ is bounded in terms of the volume of the region $T_\perp^\text{short}(Q)$ in $T_\perp$ displaced by $f_Q$ less than $\alpha \lambda_1$ (maximized over $Q \in \Gamma$). This region is defined on the cover $W_k^\perp$ by the vectors $w$ satisfying
\begin{eqaed}\label{eq:Wkperp_cover_region_short}
    \norm{\Pi_{W_k^\perp}((Q-I_n)w)} \leq \alpha \lambda_1\,.
\end{eqaed}
More precisely, since $\text{vol}_{\widehat{g}}(M_n) = 1$, the volume $\text{vol}_{\widehat{g}}(M_n^\text{short}(Q))$ is bounded by the \emph{relative} volume,
\begin{eqaed}\label{eq:relative_volume_bound}
    \frac{\text{vol}_{\widehat{g}}(M_n^\text{short}(Q))}{\abs{\Gamma}} \leq \frac{\text{Vol}_{\widehat{g}}(T^\text{short}_\perp(Q))}{\text{Vol}_{\widehat{g}}(T_\perp)}\,,
\end{eqaed}
since the fraction (relative measure) of points in $M_n$ satisfying the short-displacement condition maps to a fraction of points in $T_\perp$ satisfying the analogous condition\footnote{Mathematically, this property is encoded in the pushforward of the normalized Haar measure induced by the projection on the covering torus $T^n \to T_\perp$.}, which for $\lambda_k^\perp \gg \lambda_1$ can be stated as \cref{eq:Wkperp_cover_region_short}. Since the systole of $T_\perp$ is $\lambda_k^\perp \gg \lambda_1$, the volume transverse to the fixed locus of $f_Q$, whose codimension $p \geq 1$, is bounded above by $O(\alpha \lambda_1)^p$ in $T_\perp^\text{short}(Q)$, and below by $(\lambda_k^\perp)^p$ in $T_\perp$ (up to a constant\footnote{Valid constants can be determined by pushing the normalized Haar measure via the projection onto the orthogonal complement of the fixed-point locus $\ker(\Pi_{W_k^\perp}Q-I_{n-k})$ on the cover. This leads to the volume of a ball divided by the volume of the transverse torus, which can be in turn bounded below by a systole-sized ball, whose radius is bounded below by a fraction of $\lambda_k^\perp$.}). As a result, since the orders of elements $Q \in \Gamma$ are uniformly bounded as in \cref{tab:order_bound}, we obtain
\begin{eqaed}\label{eq:final_RFM_vol_parametric_bound}
    \text{vol}_{\widehat{g}}(M_n^\text{short}(Q)) \leq O\!\left(\left(\frac{\lambda_1}{\lambda_k^\perp}\right)^{\!p}\right) ,
\end{eqaed}
which, combined with \cref{eq:dA_Wk_term,eq:dA_Wkperp_long_bound,eq:dA_Wkperp_short_bound}, implies that \cref{eq:dA_RFM_deformation} is eventually negative along a sequence with $\lambda_n \gg \lambda_1$. In fact, a slightly stronger conclusion from these asymptotic scalings\footnote{This can be seen from \cref{eq:dA_Wkperp_long_bound}, which is valid for any split $\alpha$, and taking $\alpha \to \infty$ in the logarithmic derivative of \cref{eq:asymptotics_dA_sigma}.} is that along the sequence
\begin{eqaed}\label{eq:asymptotics_dA_sigma}
    \partial_\sigma \log A\Big|_{\sigma=0} = -11 + o(1)\,.
\end{eqaed}
Although the uniform constant can in principle be computed by bounding \cref{eq:relative_volume_bound} using Euclidean balls, the resulting expressions would need to be maximized over $\abs{\Gamma}$ and the codimension $p$ of the fixed locus in $T_\perp$, leading to very large numbers. \\

\noindent To sum up, we showed that the smallest and largest successive minima of the translation lattice cannot form a parametric hierarchy at critical points of the Casimir potential. This tells us that $\lambda_n \leq C_n \, \lambda_1$ for some uniform constant $C_n$, and thus the bound on $\abs{\Gamma}$ as a function of the systole that we employed in the rigid setting applies up to (a power of) $C_n$. The resulting parametric bounds on the \ac{EFT} control parameter---the physical systole length in eleven-dimensional Planck units---do not depend on $s$, and therefore apply to any critical point of the Casimir potential, at least insofar as the flat-space approximation is valid. \\

\noindent \textbf{Including \ac{dS} curvature.} Finally, we extend the parametric no-go result to settings without scale separation. In this case the ratio $x = r/L$ between the internal and \ac{dS} radii extremizes to an $O(1)$ value which depends, \emph{a priori}, on the hierarchy between the Casimir and flux term. In order to obtain a parametric bound on $\lambda_1/\lambda_n$ at a critical point of the Casimir potential, in order to extend the preceding parametric no-go theorem to this case, we follow the strategy employed in \cref{sec:beyond_rigid_riemann_surfaces} to include the \ac{dS} curvature in the gradient of the Casimir potential. The main lesson of \cref{sec:beyond_rigid_riemann_surfaces} in this respect is that the zero-mode contribution from the \ac{dS} heat kernel $K_{S^d}(t) = 1 + \dots$ ought to be treated separately. \\

\noindent To begin with, we discuss the non-zero-modes. Applying the analog of \cref{eq:K_S9_upper_bounds,eq:K_Sd_global_bound} to any dimension, along with the Li-Yau inequality for positive curvature \cite{LiYau1986} we learn that $K_{S^d}(t)-1$ is bounded from above by the flat-space heat kernel (up to a dimension-dependent constant), while $K_{S^d}(t)$ is bounded by the flat-space heat kernel from below (again, up to a dimension-dependent constant). In the following, we will employ the notation $\lesssim$ to denote parametric inequalities, i.e., up to such positive, parametrically $O(1)$ factors. This suggests that we can apply these parametric lower bounds to the whole Casimir gradient, without isolating the zero-mode contribution; for an upper bound, we will treat it separately. While the upper bound on $K_{S^d}(t)-1$ does not apply to $K_{S^d}(t)$ globally, due to the zero-mode, it applies for $t$ below some $O(1)$ scale. This suggests that the problem with the zero-mode will arise in the \ac{IR} region $t \geq t_*$ for some transition scale, which is indeed the case; as we shall see, it will be convenient to pick $t_* = \lambda_1^2$. \\

\noindent Hence, up to a separate discussion for the zero-mode contribution, we can recycle the decomposition in \cref{eq:dA_RFM_deformation} by multiplying each term by a suitable constant, and taking the cutoff $\alpha$ of \cref{eq:dA_Wkperp_long_bound,eq:dA_Wkperp_short_bound} to be sufficiently large so that the (contribution from non-zero-modes to the) total directional derivative $a'(0)$ at the would-be critical point remains negative for $\lambda_n \gg \lambda_1$. Then, all that is left is to bound the zero-mode term. For the full kernel negative estimate, we retain only the two translations $b=(\pm v_1,I_n)$ in \cref{eq:dA_Wk_term}. The zero-mode contribution of the retained pair must therefore be subtracted from the total zero-mode term before applying an absolute bound. \\

\noindent Tracking the relative numerical factors from the modification of the Schwinger integral, the net zero-mode contribution to the directional derivative along the metric deformation of \cref{eq:RFM_metric_def} then contains a (regularized) integral
\begin{eqaed}\label{eq:dA_RFM_zm_term}
    a'_\text{z.m.}(0) = - \, K_0 \, x^{d} \int_{M_n} \text{vol}_{\widehat{g}}(y) \sum_{b \in B \backslash \{ \text{id} \} } \frac{\delta^2_{W_k}(b,y) - \frac{k}{n-k} \, \delta^2_{W_k^\perp}(b,y)}{\left( \delta^2_{W_k}(b,y) + \delta^2_{W_k^\perp}(b,y) \right)^{\!\frac{n+2}{2}}} \Bigg|_{\text{reg}}
\end{eqaed}
where the relative constant is $K_0 = \frac{16n}{945 \pi} \Gamma(6-\frac{n}{2})\Gamma(\frac{n}{2})$. This expression manifests the fact that the scale-separation limit $x \ll 1$ reduces to the preceding results. The zero-mode contribution of this pair is $-2K_0x^d\lambda_1^{-n}$, so the remaining correction is bounded above by $\abs{a^{\prime}_{\text{z.m.}}(0)}+2K_0x^d\lambda_1^{-n}$. As in the case of Riemann surfaces studied in \cref{sec:beyond_rigid_riemann_surfaces}, the packing estimate diverges. However, it is more practical to recast \cref{eq:dA_RFM_zm_term} into a heat-kernel expression. Up to numerical prefactors,
\begin{eqaed}\label{eq:dA_RFM_zm_heat-kernel}
    a'_\text{z.m.}(0) & \propto \int_0^\infty \frac{dt}{t} \, \partial_{\sigma} K_{M_n(\sigma)}(t)\bigg|_{\sigma=0} \\
    & \propto \int_{M_n} \text{vol}_{\widehat{g}}(y) \int_0^\infty \frac{dt}{t^{1+\frac{n}{2}}} \sum_{b \in B \backslash \{ \text{id} \} } \partial_\sigma e^{-\frac{\text{dist}_{\mathbb{R}^n,\widehat{g}(\sigma)}(y,b \cdot y)^2}{4t}}\Bigg|_{\sigma = 0}\,,
\end{eqaed}
where the derivative is taken with respect to the deformation parameter $\sigma$ introduced in \cref{eq:RFM_metric_def}. Notice that, due to the nature of the metric deformation, the derivative at $\sigma=0$ can be bounded (up to a constant) by replacing it with multiplication with the exponent itself. Schematically, in terms of the orthogonal decomposition $\mathbb{R}^n = W_k \oplus W_k^\perp$ identified in the preceding analysis,
\begin{eqaed}\label{eq:schematic_distance_derivative}
    \abs{\partial_\sigma \text{dist}^2}\big|_{\sigma = 0} & = \abs{2 \, \text{dist}^2_{W_k} - \frac{2k}{n-k} \, \text{dist}^2_{W_k^\perp}} \\
    & \leq \text{const.} \times \text{dist}^2\,.
\end{eqaed}
An analogous orthogonal decomposition of the Laplacian operator shows that the quadratic form $\abs{\langle f,\partial_\sigma \Delta_{\widehat{g}(\sigma)}\big|_{\sigma=0}f \rangle} \lesssim \langle f,\Delta_{\widehat{g}}f\rangle$, where the symbol $\lesssim$ denotes inequality up to positive, parametrically $O(1)$ constants. Therefore, the heat-kernel trace is bounded according to
\begin{eqaed}\label{eq:heat_kernel_spectral_trace_RFM_bound}
    \abs{\partial_{\sigma} K_{M_n(\sigma)}(t)}\big|_{\sigma=0} & \lesssim t \tr(\Delta_{\widehat{g}} \, e^{-t \Delta_{\widehat{g}}}) \\
    & = - \, t \partial_t K_{M_n}(t)\,.
\end{eqaed}
Furthermore, we recall that, by \cref{eq:dA_Wk_term,eq:dA_Wkperp_long_bound,eq:dA_Wkperp_short_bound}, the contribution from the non-zero-modes (more precisely, with the zero mode included for the selected pair) to the directional derivative $a'(0)$ is negative and $O(\lambda_1^{-11})$. \\

\noindent These observations reveal a simple strategy. Splitting the Schwinger integral at the natural scale\footnote{In practice, this choice allows us to bound the relevant Schwinger integrals more easily.} $t=\lambda_1^2$, the \ac{UV} part of \cref{eq:dA_RFM_zm_heat-kernel} can be bounded using $s \gtrsim \lambda_1$ and \cref{eq:schematic_distance_derivative} via the same packing method used in the flat-space case\footnote{Strictly speaking, when integrating by parts in \cref{eq:dA_RFM_zm_UV_bound} to apply the packing bound on $N_y(\ell)$, the negative $\ell$-derivative of the other integrand factor---the combination multiplying $N_y(\ell)$---is negative in the range (if any) between $s$ and $2\sqrt{t}$. Since that contribution is negative, we can remove it from the upper bound. Then, we apply the packing estimate and re-extend the lower integration extremum down to $s$.},
\begin{eqaed}\label{eq:dA_RFM_zm_UV_bound}
    \abs{a'_\text{z.m.}(0)}\big|_\text{UV} & \lesssim \int_0^{\lambda_1^2} \frac{dt}{t^{1+\frac{n}{2}}} \int_{M_n} \text{vol}_{\widehat{g}}(y) \int_s^\infty \frac{\ell^2}{t} \, e^{-\frac{\ell^2}{4t}} \, dN_y(\ell) \\
    & \lesssim \int_0^{\lambda_1^2} \frac{dt}{t^{1+\frac{n}{2}}} \int_s^\infty \frac{\ell}{t} \, \frac{\ell^2}{4t} \, \left( 1 + \frac{2\ell}{s} \right)^{\!n} e^{-\frac{\ell^2}{4t}} \, d\ell \\
    & \lesssim \int_0^{\lambda_1^2} \frac{dt}{t^{1+\frac{n}{2}}} \int_{s}^\infty \frac{\ell^{n+3}}{t^2s^n} \, e^{-\frac{\ell^2}{4t}} \, d\ell \\
    & \lesssim \lambda_1^{-n} \int_0^{\lambda_1^2} \frac{dt}{t} \int_s^\infty \frac{d\ell}{\ell} \left(\frac{\ell}{\sqrt{t}}\right)^{n+4} e^{-\frac{\ell^2}{4t}} \\
    & \lesssim \lambda_1^{-n} \ll \lambda_1^{-11}\,.
\end{eqaed}
On the other hand, the \ac{IR} part of \cref{eq:dA_RFM_zm_heat-kernel} with $t \geq t_*$ contains the problematic divergence when bounded in this fashion, due to the large-$\ell$ behavior. However, it can be bounded via \cref{eq:heat_kernel_spectral_trace_RFM_bound}, so that the otherwise divergent behavior is avoided by the packing estimate. In detail\footnote{Strictly speaking, the scalar-field heat kernel itself contains the identity contribution, which is cancelled by higher-dimensional supersymmetry in the full supergravity expression. Still, including it only adds $O(\lambda_1^{-n})$ and thus does not affect the conclusion of \cref{eq:dA_RFM_zm_IR_bound}.},
\begin{eqaed}\label{eq:dA_RFM_zm_IR_bound}
    \abs{a'_\text{z.m.}(0)}\big|_\text{IR} & \lesssim - \int_{\lambda_1^2}^\infty dt \, \partial_t K_{M_n}(t) \\
    & = K_{M_n}(\lambda_1^2) - 1 \\
    & \lesssim \int_{M_n} \text{vol}_{\widehat{g}}(y) \int_s^\infty \frac{e^{-\frac{\ell^2}{4\lambda_1^2}}}{\lambda_1^n} \, dN_y(\ell) \\
    & \leq \int_s^\infty \frac{\ell}{2\lambda_1^2} \, \frac{e^{-\frac{\ell^2}{4\lambda_1^2}}}{\lambda_1^n} \left( 1 + \frac{2\ell}{s} \right)^{\!n} d\ell \\
    & \lesssim \int_s^\infty \frac{\ell^{n+1}}{\lambda_1^{n+2}s^n} \, e^{-\frac{\ell^2}{4\lambda_1^2}} \, d\ell \\
    & \lesssim \lambda_1^{-n} \ll \lambda_1^{-11}\,.
\end{eqaed}
Putting \cref{eq:dA_RFM_zm_UV_bound,eq:dA_RFM_zm_IR_bound} together, we established that $\lambda_n \lesssim \lambda_1$ for critical points even accounting for the \ac{dS} curvature. Therefore, the parametric version of the numerical bounds in \cref{tab:RFM_control_bounds_improved_again} derived for rigid \acp{RFM} actually extends to all critical points of the Casimir potential. Moreover, the methods we employed can be extended to all critical points of the full effective potential, as we shall see shortly. As we have remarked, in principle it is possible to compute numerical bounds, but the resulting numbers will likely be extremely large due to the arithmetic nature of the \ac{RFM} landscape. Given the results obtained in \cite{Bento:2025wuu,Aparici:2026wdm}, it appears unlikely that such huge numbers are saturated in actual examples. As also mentioned in \cref{sec:beyond_rigid_riemann_surfaces}, we reiterate that the partial Bose-Fermi cancellation in supergravity could work against a numerical bound on the gradient, but similar asymptotic scalings lead to the same parametric statement in \cref{eq:asymptotics_dA_sigma} as we discuss in \cref{app:bounding_sugra_gradient}. When the ratio $x=r/L=O(1)$ of radii is away from the scale-separated regime, the full gradient $\nabla A$ is in general modified, since the \ac{dS} kernel depends on $x$---we will address this shortly. \\

\noindent \textbf{General critical points.} The methods we used to establish that a critical point of the Casimir potential must be ``rounded'' can also be employed to extend our parametric no-go to general critical points, i.e., extrema of the full effective potential. This is because any controlled solution requires that $A \gg 1$ irrespective of the flux contribution, which leads us to the ``sausage'' regime driven by the deformation of \cref{eq:RFM_metric_def}. In this regime, one can orthogonally decompose
\begin{eqaed}\label{eq:G4_sausage_decomposition}
    G_4 = \sum_{j=0}^4 G_4^{(j)}\,, \qquad G_4^{(j)} \in \Lambda^j(W_k^*) \otimes \Lambda^{4-j}((W_k^\perp)^*)
\end{eqaed}
into terms with definite exponential scalings in $\sigma$, namely
\begin{eqaed}\label{eq:G4_sausage_norms}
    \norm{G_4^{(j)}}^2_{\widehat{g}(\sigma)} = e^{\beta_j \sigma} \norm{G_4^{(j)}}^2_{\widehat{g}(0)}\,, \qquad \beta_j \equiv -2j + \frac{2k}{n-k}(4-j)\,. 
\end{eqaed}
As a result, since $\mathcal{Q}(N) = \sum_{j=0}^4 \norm{G_4^{(j)}}^2_{\widehat{g}(\sigma)}$, we have
\begin{eqaed}\label{eq:dlogQ_sigma}
    \partial_\sigma \log \mathcal{Q}(N)\Big|_{\sigma=0} = \frac{\sum_{j=0}^4 \beta_j \norm{G_4^{(j)}}^2_{\widehat{g}(0)}}{\sum_{j=0}^4 \norm{G_4^{(j)}}^2_{\widehat{g}(0)}} \geq -8\,.
\end{eqaed}
In terms of the dimensionless radion $\rho$ and the shape moduli $m$, the dimensionally reduced effective potential, including the contribution from the $G_4$-flux, reads \cite{Bento:2025wuu,Aparici:2025kjj}
\begin{eqaed}\label{eq:dd_eff_potential_einstein_frame}
    M_d^{-d} V = e^{-\frac{dn}{d-2} \rho} \left(\mathcal{Q}_N(m) \, e^{(n-8)\rho} - \mathcal{C}(m) \, e^{-d \rho} \right)
\end{eqaed}
where $\mathcal{C} \propto A$ and $\mathcal{Q}_N \propto \mathcal{Q}(N)$ include all numerical prefactors from the dimensional reduction for convenience. Anticipating the approach of \cref{sec:potential_slopes}, we recast \cref{eq:dd_eff_potential_einstein_frame} as
\begin{eqaed}\label{eq:dd_eff_potential_recast}
    M_d^{-d}V & = \mathcal{Q}_N \, e^{-\alpha \rho} \left(1 - \, \frac{\mathcal{C}}{\mathcal{Q}_N} \, e^{-3\rho} \right) \\
    & \equiv \frac{\mathcal{Q}_N \, e^{-\alpha \rho}}{1+\chi}\,, \qquad \alpha \equiv \frac{2n}{d-2} + 8\,.
\end{eqaed}
Then, the differential $d_{(\rho,\sigma)} V$ in the $(\rho,\sigma)$ plane simplifies to
\begin{eqaed}\label{eq:dlogV_RFM}
    d_{(\rho,\sigma)}\log V = (3\chi - \alpha) \, d\rho + (1+\chi) \, d_{(\rho,\sigma)}\log \mathcal{Q}_N - \chi \, d_{(\rho,\sigma)}\log \mathcal{C}\,,
\end{eqaed}
where we accounted for the $\rho$-dependence of the Casimir potential in the presence of the external \ac{dS} factor. However, the contribution from the radial direction is negligible: the \ac{dS} heat kernel depends on the combination $x^2t = r^2 t /L^2$, which in the \ac{UV} regime where $t \leq \lambda_1^2$ is negligible for $x=O(1)$ and $\lambda_1 = o(1)$. The \ac{IR} regime\footnote{More precisely, one can split the Schwinger integral into three regions divided at $t = \lambda_1^2$ and $t=1$, and bound the sphere heat kernel in terms of its flat-space counterpart accordingly (up to the zero-mode contribution). Since the \ac{UV} region $t \leq \lambda_1^2$ already contributes the $\lambda_1^{-11}$ scaling, and the upper bound does not affect it.} splits into exponentially suppressed $l > 0$ modes, and the zero-mode. The latter is $O(\lambda_1^{-n}) \ll \lambda_1^{-11}$, consistently with the derivative scalings of \cref{eq:dA_RFM_zm_UV_bound,eq:dA_RFM_zm_IR_bound}; since it is homogeneous\footnote{The only $x$-dependence in the zero-mode contribution is in the inverse volume factor $1/\text{Vol}(S^d) \propto x^{d}$.} in $x$, its derivative at $x=O(1)$ does not modify the scalings of \cref{eq:asymptotics_dA_sigma}. \\

\noindent A critical point then fixes $\chi = \alpha/3 + o(1)$, and thus also
\begin{eqaed}\label{eq:dm_critical_point_RFM}
    d_\sigma\log \mathcal{Q}_N = \frac{\alpha}{\alpha+3} \, d_\sigma\log \mathcal{C} + o(1)\,.
\end{eqaed}
On account of \cref{eq:asymptotics_dA_sigma,eq:dlogQ_sigma}, in the controlled regime \cref{eq:dm_critical_point_RFM} requires that
\begin{eqaed}\label{eq:RFM_critical_point_controlled_condition}
    \partial_\sigma\log \mathcal{Q}(N)\bigg|_{\sigma=0} = - \frac{11\alpha}{\alpha+3} + o(1)
\end{eqaed}
along a sequence. However, we have
\begin{eqaed}\label{eq:RFM_critical_point_contradiction}
    \frac{11\alpha}{\alpha+3} = \frac{22}{3} \, \frac{12-n}{11-n} > 8\,,
\end{eqaed}
which leads to a contradiction. This argument also shows that large fluxes do not help to achieve a controlled regime. \\

\noindent Altogether, our parametric no-go result extends to all critical points of \ac{RFM} compactifications of M-theory.

\section{Potential slopes in the controlled region}\label{sec:potential_slopes}

In the preceding sections we have studied in detail the possibility of \ac{dS} critical points. However, it is worth examining the alternative that a realistic model of our universe be realized away from a critical point. The most immediate obstruction to this is the tendency of potentials in string/M-theory compactifications to be too steep in the controlled region of field space \cite{Bedroya:2019snp, Bedroya:2022tbh, Bedroya:2025fie, Bedroya:2025ltj, Etheredge:2026cha}. This feature is related to the difficulties in realizing a cosmologically viable accelerated expansion \cite{Andriot:2023wvg,Shiu:2023fhb,Shiu:2023nph,Andriot:2024jsh,Casas:2024xqy,Andriot:2026lac}; however, the dark dimension scenario \cite{Montero:2022prj} suggests a natural coupling between dark energy and dark matter \cite{Andriot:2025los}, which provides a tantalizing---if partial---agreement with observational data \cite{Agrawal:2019dlm, Bedroya:2025fwh}. \\

\noindent While critical points, stable or otherwise, are a surefire way to obtain small slopes for a viable cosmological model, the state of affairs that we have described thus far prompts us to look beyond critical points. In principle, the presence of the Casimir energy could allow for different slopes $\norm{\nabla V}/V$ of the effective potential, depending on where in field space control is achieved. For instance, it is straightforward to see that the slope can never go below the asymptotic value
\begin{eqaed}\label{eq:asymptotic_slope_value}
    c_d \equiv \frac{2}{\sqrt{d-2}}
\end{eqaed}
if the overall volume direction dominates---which would not be tenable for cosmological application in our universe. Among the settings we were able to investigate in detail in this paper, the two options are hyperbolic Riemann surfaces and (non-rigid) \acp{RFM}, since they admit deformations that become light in the controlled region that we will now turn to away from the asymptotic regime. As in the preceding sections, we focus on the Casimir potential produced by scalar fields for simplicity. The parametric arguments can be extended to the supergravity Casimir potential via the methods developed in \cref{app:bounding_sugra_gradient}, while the numerical bounds do not apply. Nevertheless, one might expect them to provide a ballpark estimate within some orders of magnitude.

\subsection{Slopes for Riemann surfaces}\label{sec:slope_riemann_surfaces}

We begin with the case of Riemann surfaces. For our purposes in this section, we need the dimensionally reduced effective action including kinetic terms for the radion and complex-structure fields. From \cref{eq:eff_action}, the Einstein-frame result in terms of the nine-dimensional Planck scale $M_9$ reads
\begin{eqaed}\label{eq:9d_reduction_riemann_surface_action}
    S_{9\text{d}} = \int d^9x \sqrt{-g} \left( \frac{M_9^7}{2} \left(R - \, \frac{18}{7} \, (\partial \rho)^2 - G_{ij}(m) \, \partial m^i \cdot \partial m^j \right) - V \right) ,
\end{eqaed}
where $r=M_9^{-1}e^{\rho}$ is the dynamical curvature radius, $m$ denotes complex-structure moduli, and the effective potential\footnote{The dependence on the \ac{dS} curvature affects the solution, but appears in the lower-dimensional effective action as curvature terms. Nonetheless, as we have seen, its multiplicative contributions are bounded by $O(1)$ even without scale separation---to the extent that the dimensional reduction makes sense.}
\begin{eqaed}\label{eq:9d_eff_potential_einstein_frame}
    M_9^{-9} V = e^{-\frac{18}{7} \rho} - \mathcal{C}(m) \, e^{-\frac{81}{7} \rho} 
\end{eqaed}
contains the Casimir energy $\mathcal{C}(m)$ in units of $r$. The kinetic metric $G_{ij}(m)$ is proportional to the Weil-Petersson metric; see, e.g., \cite{Schlenker:2020ptx} for the relative normalization. \\

\noindent Since the radion direction is orthogonal to the complex-structure moduli, the squared slope in nine-dimensional Planck units can be written as
\begin{eqaed}\label{eq:dV_squared_riemann_surfaces}
    \lambda_V^2 \equiv M_9^7 \frac{\norm{\nabla V}^2}{V^2} = \frac{63}{2} \left(\chi - \frac{2}{7}\right)^2 + \chi^2 \norm{\nabla \log \mathcal{C}}^2_G\,,
\end{eqaed}
where we defined the convenient combination $\chi \equiv \frac{\mathcal{C}e^{-9\rho}}{1-\mathcal{C}e^{-9\rho}}$. Regardless of the complex-structure moduli, $\chi$ (as a function of the radion) minimizes the slope at $\chi_* = 9(\frac{63}{2}+\norm{\nabla \log\mathcal{C}}^2_G)^{-1}$, at which
\begin{eqaed}\label{eq:min_slope_chi_riemann_surfaces}
    \lambda_V^2 \geq \lambda_\text{min}^2 \equiv \frac{18}{7} \, \frac{\norm{\nabla \log \mathcal{C}}^2_G}{\frac{63}{2} + \norm{\nabla \log \mathcal{C}}^2_G}\,.
\end{eqaed}
This result reduces to the existence of a critical point when $\mathcal{C}$ is extremized, whence $\lambda_\text{min} = 0$. Moreover, \cref{eq:min_slope_chi_riemann_surfaces} also shows that $\norm{\nabla \log \mathcal{C}}_G$ cannot be too large: otherwise $\lambda_\text{min} \to \sqrt{18/7}$ would exceed the asymptotic value $c_9 = 2/\sqrt{7}$. Specifically, it must be
\begin{eqaed}\label{eq:slope_condition_dlogC}
    \delta \equiv \norm{\nabla \log \mathcal{C}}^2_G < 9\,.
\end{eqaed}
In order to achieve this in the controlled region of field space, we know that we cannot take a pure large-volume limit, since the slope would be dominated by the curvature term and exceed the asymptotic value of \cref{eq:asymptotic_slope_value}. The \ac{EFT} control parameters are $M_{11}r$ and $M_{11}r s$, where $s$ is the dimensionless systole length of the Riemann surface. Away from critical points, the general expressions can be recast as
\begin{eqaed}\label{eq:EFT_control_parameters_off-shell_riemann_surfaces}
    M_{11}r & = \left(\frac{\mathcal{C}}{2v} \, \frac{1+\chi}{\chi}\right)^{\!\frac{1}{9}}\,, \\
    M_{11}r s & = s(m) \left(\frac{\mathcal{C}}{2v} \, \frac{1+\chi}{\chi}\right)^{\!\frac{1}{9}}\,,
\end{eqaed}
consistently with our findings in \cref{eq:solution} for critical points where they depend on the ratio $\mathcal{C}/v$. \\

\noindent From \cref{eq:dV_squared_riemann_surfaces}, we observe that the positive quantity $\chi$ can be neither large nor small, for otherwise $\lambda_V$ would exceed the asymptotic value irrespective of the combination $\delta$ of \cref{eq:slope_condition_dlogC}. Therefore, $\chi$ must remain at $O(1)$ values. This implies that $\mathcal{C}(m)/v$ must become large, which is the same control condition as for critical points. The bounds on $\mathcal{C}/v$ derived in \cref{sec:casimir_riemann_surfaces} only use the existence of a uniform lower bound on $s$. Hence, $\mathcal{C}/v$ can only be large in a ``pinching'' degeneration of the Riemann surface, for which $s \to 0$. \\

\noindent \textbf{Potential slope for pinching degenerations.} A quick asymptotic estimate can be performed as follows. From \cref{sec:beyond_rigid_riemann_surfaces}, in particular using \cref{eq:riera,eq:dA_bound_2,eq:dA_curved_bound} and restoring the factors of $v$, we see that $\abs{\nabla\mathcal{C}\cdot \nabla s} \gtrsim s^{-10}$. From \cref{eq:zeta_riemann_s,eq:winding_A_curved}, we further have $\mathcal{C} \lesssim v \, s^{-10}$. Then, using \cref{eq:riera} in the pinching limit \cite{Riera2005,wolpert2008behaviorgeodesiclengthfunctionsteichmueller,Wolpert2012Geodesic}, by Cauchy-Schwarz we have
\begin{eqaed}\label{eq:delta_estimate}
    \delta \gtrsim \frac{1}{vs} \to \infty
\end{eqaed}
due to the normalization of the (Weil-Petersson) kinetic metric\footnote{The relative normalization of $1/v$ arises by writing the Weil-Petersson metric as an inner product of metric deformations \cite{Schlenker:2020ptx}, so that the overall factor of $v$ absorbed into the nine-dimensional Planck scale appears in the denominator. Then \cref{eq:slope_condition_dlogC} contains a contraction with the inverse kinetic metric, leading to \cref{eq:delta_estimate}.}, $G \propto \frac{2}{v} \, G_\text{WP}$. This tells us that for any given genus $g$, where $v = 4\pi (g-1)$, the quantity $\delta$ of \cref{eq:slope_condition_dlogC} must diverge in a pinching degeneration, and thus the potential slope $\lambda_V$ in \cref{sec:slope_riemann_surfaces} will exceed the asymptotic value $c_9$ of \cref{eq:asymptotic_slope_value}. \\

\noindent The question is thus whether one can compensate for this effect by simultaneously taking a family of Riemann surfaces with $g \to \infty$. However, we argue that this is not the case: in a pinching degeneration, short simple geodesics become disjoint by the collar lemma, and their number $n_\text{pinch}$ is bounded by $3g-3$. Therefore, even in the most favorable case in which all disjoint necks shrink at the same rate, $\mathcal{C} \sim n_\text{pinch} \, s^{-10}$ and $\norm{\nabla \mathcal{C}}_\text{WP}^2 \sim n_\text{pinch} \, s^{-21}$ are modified\footnote{The squared norm $\norm{\nabla \mathcal{C}}^2_G$ scales linearly because of the Riera formula \cite{Riera2005,wolpert2008behaviorgeodesiclengthfunctionsteichmueller,Wolpert2012Geodesic}. Intuitively, since pinching geodesics are disjoint, the cross-terms do not contribute.} in such a way that
\begin{eqaed}\label{eq:n_pinch_delta_riemann_surfaces}
    \delta \sim \frac{v}{n_\text{pinch} s} \gtrsim s^{-1} \to \infty\,.
\end{eqaed}
An argument for why the subleading cross-terms cannot accumulate to change this conclusion is that the collars around the relevant short geodesics are disjoint, and infinitesimal moduli deformations are supported within each of them. Their overlap in Teichm\"{u}ller space is computed from harmonic representatives, but the harmonic projection on Beltrami differentials cannot increase norms. We will sharpen this argument shortly, using the results of \cref{app:geodesic_length_variations}; for the time being, we conclude that in the parametrically controlled regime the potential slope $\lambda_V$ of \cref{eq:dV_squared_riemann_surfaces} exceeds the asymptotic value $c_9$ in \cref{eq:asymptotic_slope_value}. \\

\noindent \textbf{Numerical bounds below the asymptotic slope.} Having argued for a parametric condition on the slope, the next natural question is how much numerical control can be achieved while remaining within the region $\lambda_V < c_9$. Fortunately, in the restricted setting of Riemann surfaces, this problem can be tackled rather explicitly. As shown in \cref{eq:slope_condition_dlogC}, this requires that $\delta < 9$. From \cref{eq:dV_squared_riemann_surfaces}, we also see that
\begin{eqaed}\label{eq:lambdaV_bound_9-10_riemann_surfaces}
    \lambda_V^2 \geq \frac{63}{2}\left(\chi - \frac{2}{7}\right)^{\!2}\,,
\end{eqaed}
so that $\lambda_V < c_9$ implies
\begin{eqaed}\label{eq:chi_bounds_slope_riemann_surfaces}
    \frac{6-2\sqrt{2}}{21} < \chi < \frac{6+2\sqrt{2}}{21}\,.
\end{eqaed}
In turn, on account of \cref{eq:EFT_control_parameters_off-shell_riemann_surfaces}, the control parameter is thus bounded by
\begin{eqaed}\label{eq:EFT_control_parameter_off-shell_riemann_surfaces_chi-bound}
    H(s) \leq \min(1,s) \left(\frac{11+3\sqrt{2}}{2} \, \frac{\mathcal{C}}{2v}\right)^{\!\frac{1}{9}}\,,
\end{eqaed}
which means that we must make the interplay between the Casimir coefficient and its gradient quantitative. \\

\noindent To this end, we derive a numerical version of \cref{eq:delta_estimate,eq:n_pinch_delta_riemann_surfaces} via the results of \cref{app:geodesic_length_variations}. The numerical bounds of \cref{eq:dA_bound,eq:dA_bound_2} on the directional derivatives $\langle \nabla \mathcal{C},\nabla \ell_\alpha\rangle$ along systolic variations for $\ell_\alpha = s < 1$ can be combined with the norm bound of \cref{eq:WP_norm_dC_bound} to show that the cross-terms between disjoint geodesics indeed do not contribute to the bound. Including the numerical prefactors of $\frac{1}{4096\pi^5} \times \frac{2}{\pi} = \frac{1}{2048\pi^6}$ from the rescaling of the Casimir term and gradient bounds in \cref{eq:dA_bound_2}, we find
\begin{eqaed}\label{eq:dC_systolic_num_bound}
    - \, \langle \nabla \mathcal{C}, \nabla \ell_\alpha\rangle & \geq \frac{1}{2048\pi^6} \left( \frac{469578}{s^{10}} - \frac{3642}{s^2} \right) \geq \frac{0.2366}{s^{10}}\,,
\end{eqaed}
from which, using \cref{eq:WP_norm_dC_bound} and the normalization $G = \frac{2}{v} \, G_\text{WP}$ on the kinetic metric of \cref{eq:9d_reduction_riemann_surface_action}, we obtain
\begin{eqaed}\label{eq:norm_dC_sys_bound}
    \norm{\nabla \mathcal{C}}_G^2 \geq \frac{0.0287}{s^{21}} \, v \, N_\text{sys}\,, 
\end{eqaed}
in terms of the number $N_\text{sys}$ of systoles. However, this simple result is not sharp enough to obtain a good bound for $\delta$. In order to do that, we need to extend the bounds of \cref{sec:casimir_riemann_surfaces,sec:beyond_rigid_riemann_surfaces} to account for the distribution of short geodesics: otherwise, there could be a few systoles and many geodesics just barely longer. To remedy this, we observe that \cref{eq:dA_bound} can be derived for any $\ell_\alpha < 1$, and thus \cref{eq:dA_bound_2} applies replacing $s$ with $\ell_\alpha$ on the right-hand side. As a result,
\begin{eqaed}\label{eq:dC_general_bound}
    - \, \langle \nabla \mathcal{C}, \nabla \ell_\alpha \rangle \geq 0.2366 \, \ell_\alpha^{-10}\,,
\end{eqaed}
so that
\begin{eqaed}\label{eq:norm_dC_general_l_bound}
    \norm{\nabla \mathcal{C}}^2_G \geq \frac{0.0287}{2} \, v \sum_{\alpha \in \mathcal{P} : \ell_\alpha < 1} \ell_\alpha^{-21}\,. 
\end{eqaed}

\noindent The next step is to bound the Casimir parameter $\mathcal{C}$ itself. Since the identity contribution to $\mathcal{C}/v$ is tiny, we implicitly consider the winding term in what follows. Anticipating a generalization of \cref{eq:F_kernel} to contain the sum of $\ell_\alpha^{-10}$, we apply Cauchy-Schwarz to the vectors with components $\ell_\alpha^{-10}$ and $1$ for each $\alpha$, which leads to the useful estimate
\begin{eqaed}\label{eq:sum_lengths_powers_cauchy-schwarz}
    \left(\sum_{\alpha \in \mathcal{P} : \ell_\alpha < 1} \ell_\alpha^{-10}\right)^{\!2} & \leq N(1) \sum_{\alpha \in \mathcal{P} : \ell_\alpha < 1} \ell_\alpha^{-20} \\
    & \leq N(1) \sum_{\alpha \in \mathcal{P} : \ell_\alpha < 1} \ell_\alpha^{-21} \\
    & \leq 2 \cdot (3g-3) \sum_{\alpha \in \mathcal{P} : \ell_\alpha < 1} \ell_\alpha^{-21} \\
    & = \frac{3v}{2\pi} \sum_{\alpha \in \mathcal{P} : \ell_\alpha < 1} \ell_\alpha^{-21}
\end{eqaed}
in terms of the number of geodesics of length below $1$, which are simple and pairwise disjoint\footnote{At first glance, the homological bound $N(1)/2 \leq 3g-3$ for simple pairwise disjoint unoriented geodesics might seem strange, since naively one could, e.g., move a geodesic going through a handle in $\Sigma_g$ into a nearby disjoint one. However, with the hyperbolic metric on $\Sigma_g$, these two curves cannot both be geodesics. If they were, since together they bound an annulus $A \subset \Sigma_g$,  the Gauss-Bonnet theorem for sectional curvature $K=-1$ would tell us that $0 \neq -\text{Area}(A)=2\pi \chi(A)-\int_{\partial A}k_g ds=0$.}. In order to obtain this sort of upper bound on $\mathcal{C}$, we refine the packing estimate of \cref{eq:F_kernel} by splitting the sum over geodesics at the cutoff length $\ell_\alpha=1$. Since $\lim_{z \to 0^+}z^5K_5(z)=384$ and $z^5K_5(z)$ is decreasing, we have
\begin{eqaed}\label{eq:refined_bound}
    \sum_{\alpha \in \mathcal{P} : \ell_\alpha < 1} \sum_{k \geq 1} \frac{K_5(k\ell_\alpha/2)}{k^5\ell_\alpha^4 \sinh(k\ell_\alpha/2)} & \leq \sum_{\alpha \in \mathcal{P} : \ell_\alpha < 1} \sum_{k \geq 1} \frac{24576}{k^{11}\ell_\alpha^{10}} = 24576 \, \zeta(11) \sum_{\alpha \in \mathcal{P} : \ell_\alpha < 1} \ell_\alpha^{-10}\,.
\end{eqaed}
This takes care of the ``short'' geodesics. For the ``long'' ones, i.e., those with $\ell_\alpha \geq 1$, we can recycle the bound in \cref{eq:A_bound} (restricted to this contribution\footnote{This is because, even if the systole length is smaller than what we used for \cref{eq:A_bound}, geodesics of length $\ell_\alpha \geq 1$ intersect the thick part of the surface by the collar lemma.}), obtaining
\begin{eqaed}\label{eq:packing_above_u}
    \mathcal{C} \leq 26 \, v + \frac{24576 \, \zeta(11)}{2048\pi^5} \, \sum_{\alpha \in \mathcal{P} : \ell_\alpha < 1} \ell_\alpha^{-10}\,.
\end{eqaed}
Combining this result with \cref{eq:norm_dC_general_l_bound,eq:sum_lengths_powers_cauchy-schwarz} we arrive at
\begin{eqaed}\label{eq:delta_general_l_bound}
    \delta \geq \frac{0.03 \, S_{10}^2}{(26 + 0.04 \, S_{10})^2}\,, \qquad S_{10} \equiv \frac{1}{v} \sum_{\alpha \in \mathcal{P} : \ell_\alpha < 1} \ell_\alpha^{-10}\,.
\end{eqaed}
Then, the \ac{TCC} asymptotic bound can be violated as in \cref{eq:slope_condition_dlogC} only for $S_{10} < 1466.1$ (rounding the previous decimals favorably toward improving control). This translates into $\mathcal{C}/v \leq 85$, which bounds \cref{eq:EFT_control_parameter_off-shell_riemann_surfaces_chi-bound} according to
\begin{eqaed}\label{eq:final_EFT_control_bound_off-shell_riemann_surfaces}
    H(s) \leq 2\,.
\end{eqaed}
At fixed shape, \cref{eq:chi_bounds_slope_riemann_surfaces} confines the canonically normalized radion to an $O(1)$ interval in lower-dimensional Planck units. Indeed, the critical radion $\rho \sim \frac{1}{9}\log g$ for large genus $g \gg 1$, but an $O(1)$ variation is enough to vary the slope $\lambda_V$ by $O(1)$. The bound in \cref{eq:final_EFT_control_bound_off-shell_riemann_surfaces} on the \ac{EFT} control parameter excludes a parametrically controlled limit in the region $\lambda_V<c_9$. This also aligns with the considerations of \cite{vandeHeisteeg:2023uxj} and supersymmetric examples \cite{Klawer:2021ltm}.

\subsection{Slopes for Riemann-flat manifolds}\label{sec:slope_RFMs}

We now turn to \acp{RFM}. In this case, the computation is complicated by the arithmetic structure of the torus moduli space and fluxes. Including the relevant kinetic terms, in the notation of \cref{sec:RFMs}, the counterparts of \cref{eq:9d_reduction_riemann_surface_action,eq:9d_eff_potential_einstein_frame} are given by
\begin{eqaed}\label{eq:dd_reduction_RFM_action}
    S_\text{red} = \int d^dx \sqrt{-g} \left( \frac{M_d^{d-2}}{2} \left(R - \, \frac{n(D-2)}{d-2} \, (\partial \rho)^2 - \frac{1}{4} \, \tr((\widehat{g}^{-1}(m) \partial \widehat{g}(m))^2) \right) - V \right) ,
\end{eqaed}
where $D=d+n=11$, and now $r=M_d^{-1}e^{\rho}$ is the dynamical volume radius, whereas $m$ denotes the volume-preserving shape moduli of the \ac{RFM} $M_n = \mathbb{R}^n/B = T^n/\Gamma$. The resulting effective potential is given by \cref{eq:dd_eff_potential_einstein_frame,eq:dd_eff_potential_recast}. The resulting differential
\begin{eqaed}\label{eq:dlogV_RFM_2}
    d\log V = (3\chi - \alpha) \, d\rho + (1+\chi) \, d\log \mathcal{Q}_N - \chi \, d\log \mathcal{C}
\end{eqaed}
yields the slope
\begin{eqaed}\label{eq:slope_RFM}
    \lambda_V^2 & = \frac{d-2}{n} \left(\chi - \frac{2(d+1)}{d-2}\right)^2 + \norm{(1+\chi) \, \nabla_m\log \mathcal{Q}_N - \chi \, \nabla_m\log \mathcal{C}}^2_\text{shape}\,.
\end{eqaed}
As in \cref{sec:slope_riemann_surfaces}, we conclude that $\chi$ must remain at positive\footnote{For all $d+n \leq 11$, this follows from the same analysis leading to \cref{eq:chi_bounds_slope_riemann_surfaces}.} $O(1)$ values in order for $\lambda_V$ to be below the asymptotic value $c_d$ of \cref{eq:asymptotic_slope_value}. Moreover, again as in \cref{eq:EFT_control_parameters_off-shell_riemann_surfaces}, for $\chi = O(1)$ we have 
\begin{eqaed}\label{eq:EFT_control_parameters_off-shell_RFM}
    M_{11}r & \sim \left(\frac{\mathcal{C}}{\mathcal{Q}_N}\right)^{\!\frac{1}{3}}\,,
\end{eqaed}
which forces $\mathcal{C} \gg \mathcal{Q}_N$. In particular, the Casimir coefficient must be large and positive\footnote{The positivity condition follows from the fact that $\chi > 0$ and that we consider regions in which the potential is positive.}; in light of the analysis in \cref{sec:RFMs}, this requires that the \ac{RFM} be highly anisotropic, with a hierarchy among the standard sequence of successive minima of lattice lengths. As shown in \cref{sec:RFMs}, this generates volume-preserving deformations that rescale orthogonal invariant subspaces. We are thus led to consider this class of limits in the following. \\

\noindent \textbf{Potential slope for the relevant class of \ac{RFM} limits.} To go beyond these qualitative considerations, we can rule out a class of limits in which the metric is parametrized by orthogonal radii
\begin{eqaed}\label{eq:special_RFM_metric}
    ds^2_{M_n}(y) = \sum_{i=1}^n e^{2\rho_i} \, dy_i^2\,.
\end{eqaed}
Within this family\footnote{If the Bieberbach group of the \ac{RFM} does not allow all these deformations, we simply set the corresponding moduli to constants.}, it is simpler to control whether all cycles become large, and thus whether the \ac{EFT} is under complete control. Moreover, the slope of the potential is (bounded below by) an explicit expression as follows. \\

\noindent The field-space line element in these coordinates takes the form \cite{Bento:2025wuu}
\begin{eqaed}\label{eq:RFM_field-space_metric_special}
    \frac{1}{d-2} \left(d \log \sqrt{\det(g)}\right)^2 + \frac{1}{4} \tr((g^{-1} dg)^2) = \frac{1}{d-2}\left(\sum_{i=1}^n d\rho_i\right)^{\!2} + \sum_{i=1}^n d\rho_i^2\,.
\end{eqaed}
Rather than detailing the complete proof, we only sketch the necessary arguments here; we relegate a more detailed exposition of the techniques to \cref{app:bounding_sugra_gradient}.  Working in the flat-space approximation and neglecting flux-dependent terms in the supergravity kinetic operators, proceed by contradiction. Assume a sequence of positive-potential configurations whose shortest lattice length tends to infinity in eleven-dimensional Planck units, while the potential slope remains below $\gamma_d=c_d\sqrt{\frac{8}{10-d}}$ by a fixed amount. The radial estimate of \cref{eq:slope_RFM} confines $\chi$ to a compact positive interval, and therefore bounds the flux-to-Casimir ratio. Choose a lattice basis adapted to the different length scales and use Poisson summation to separate the dominant contributions into groups of comparable scales. Comparing the degrees with which they depend on the inverse of the length, eight for the flux density and eleven for the Casimir density, together with flux quantization, forces every non-zero flux component to involve directions longer than the dominant Casimir shifts. Letting $p \leq n-1 = 10 -d$ be the number of such directions, consider varying the metric by scaling them homogeneously, similarly as in \cref{eq:RFM_metric_def}. Along these variations, the flux logarithmic decay rate is at least the Casimir rate, so their cancellation cannot reduce the limiting directional slope. Combining the estimate of the flux decay rate of $\frac{2p}{d-2}+2$ with the physical field-space norm $\sqrt{\frac{p^2}{d-2}+p}$ for these metric variations yields $\liminf\lambda_V\geq\gamma_d$ for $4\leq d\leq7$, contradicting the assumption. Therefore, $\lambda_V \geq c_d$ along the limits encoded in \cref{eq:special_RFM_metric}. It would be interesting to investigate whether this conclusion extends beyond this simple class of limits uniformly across \acp{RFM}.

\section{Conclusions}\label{sec:conclusions}

In this paper we have studied the possibility of controlled \ac{dS} solutions in M-theory compactifications supported by Casimir energy. In the spirit of \cite{Bento:2025wuu,Aparici:2026wdm}, the ingredients involved in these constructions are rather simple, and are accessible within the low-energy \ac{EFT} description of supergravity. However, as may have been anticipated from the strong form of the Swampland \ac{dS} conjectures, as we have seen, there is a basic tension that seemingly prevents the resulting solutions from achieving good control, meaning that the curvature radius and the shortest cycles cannot be large in eleven-dimensional Planck units. This tension occurs because the Casimir energy is kinematically controlled by the size of internal cycles, and in turn its value determines them dynamically. This translates into a trade-off, whereby short cycles enhance the Casimir energy, which leads to long cycles, which in turn decreases the Casimir energy. The natural expectation is that this trade-off be balanced at $O(1)$ values in eleven-dimensional Planck units, since there are no other scales in the problem. \\

\noindent A possible loophole to this mechanism could have been global topology: if the internal manifold can be tuned along a discrete family in such a way that the extremized Casimir energy (or rather, its averaged density) be large, there might be a way to concoct controlled \ac{dS} saddles in M-theory. This could happen parametrically, if the improvement from global topology were arbitrarily good, or it could happen merely numerically. In the latter case, one is naturally led to ask what the largest obtainable numbers can be. \\

\noindent In this paper, assuming that the supergraviton Casimir energy behave similarly as that of scalar fields, we answered this question for a number of types of compactifications, with the following results:

\begin{itemize}
    \item For Riemann surfaces in \cref{sec:casimir_riemann_surfaces} and higher-dimensional (thus rigid) hyperbolic manifolds in \cref{sec:hyperbolic_manifolds}, there is a numerical upper bound on the control parameters of the \ac{EFT}.

    \item For \acp{RFM} that are isolated by enhanced symmetries, we also obtained a numerical upper bound in \cref{sec:RFMs}. For general \acp{RFM}, we derived a parametric bound of how much control can be achieved at a critical point of the two-derivative \ac{EFT}.
\end{itemize}

\noindent Combining the critical-point bounds of \cref{sec:RFMs} with finiteness of the affine types, there is a numerical upper bound; our loose estimates give very large numbers, but it would be interesting to obtain (and perhaps saturate) stronger numerical bounds in this context. The same goes for the full supergravity counterpart of these results: in \cref{app:bounding_sugra_casimir} we extend them from the scalar-field case to the supergraviton, but only at the parametric level. As mentioned in \cref{sec:setup}, we derived these bounds in the absence of warping. If such effects are present and significant, we cannot exclude critical points; however, in this case it is not clear that an \ac{EFT} analysis can be trusted in the first place \cite{DeLuca:2021pej}. \\

\noindent In \cref{sec:potential_slopes} we allowed the possibility of a rolling solution away from a critical point. Here the \ac{TCC} provides a bound for the asymptotic slope of the potential in field space, but a realistic accelerating cosmology should lie below this threshold---in other words, the potential should decay slowly enough. For Riemann surfaces, we derived a numerical bound on how well the \ac{EFT} can be controlled within the region of field space in which the \ac{TCC} bound is violated. For \acp{RFM}, we obtained a parametric bound for metric variations in which the Casimir potential becomes large; these directions in moduli space are available in a controlled regime for the \ac{EFT}. \\

\noindent While the quantitative bounds were derived in these specific settings, we highlighted a general lesson for Casimir compactifications that seems to suggest that \ac{dS} constructions with small vacuum energy require more exotic ingredients, such as singularities or strong quantum-gravitational effects. Indeed, the only known examples of (probably maximum) \ac{dS} which can be argued to exist are in strongly quantum regimes \cite{Chen:2025rkb, Baykara:2026jzs}. The exploration of these genuinely \ac{UV} ingredients, and in particular whether they can lead to parametrically small $V$, is essential to a broader understanding of the quantum-gravity landscape and, more importantly in the purview of this work, for getting closer to finding a realistic model of our universe within it. \\

\noindent The lessons we have learned in this paper, while reaffirming the difficulties in realizing quasi-dS in string theory also point to some possible directions for further exploration. To begin with, we emphasize once more that our results can exclude the existence of \emph{controlled} solutions within the scope of the \ac{EFT} setup we discussed. This leaves open the possibility of establishing the \emph{existence} of quasi-\ac{dS} solutions supported by the ingredients we studied, along the lines of \cite{Mohseni:2025tig, Chen:2025rkb, Baykara:2026jzs}. If this can be done, the notion of parametric control itself might be incompatible with a realistic model of our universe. This is not necessarily problematic---after all, physical quantities have specific numerical values, large or small as they may be. In this context, a prospect toward a scenario akin to the dark dimension \cite{Montero:2022prj} is to realize a five-dimensional supersymmetric bulk where supersymmetry is broken at the boundaries by non-perturbatively suppressed effects. For example, one could imagine heterotic M-theory\footnote{In the context of the dark dimension scenario, this is a natural setup which has already been discussed in \cite{Schwarz:2024tet, Reig:2025gch, Blumenhagen:2026rgu} (see also \cite{Deffayet:2024hug}).} compactified on a Calabi-Yau threefold at a point of enhanced symmetry in moduli space, leading to critical points of the sort studied in \cite{Chen:2025rkb, Baykara:2026jzs}. \\

\noindent A concrete possibility is M-theory on a Calabi–Yau threefold times \(S^1/\mathbb Z_2\), with an anti-periodic circle, starting at an isolated symmetry point in the CY moduli space, whose existence will be established in a forthcoming work\footnote{Based on a work to appear by A. Castellano, S. Chen and C. Vafa.}. Antiperiodic fermions on the covering circle generate a negative Casimir energy, while boundary gaugino condensates with a non-zero relative source mismatch can provide a competing positive contribution. At fixed CY moduli and to leading order in weak warping, the four-dimensional Einstein-frame potential for the interval length \(\ell\) takes the form 
\begin{eqaed}\label{eq:HW_potential}
    V(\ell) \sim \frac{B}{\ell^3} - \, \frac{C}{\ell^6} \, ,
\end{eqaed}
where \(B,C>0\) correspond respectively to non-perturbative gaugino condensate at the boundaries of the interval, and the Casimir effect in the bulk. This potential has a positive maximum (i.e. unstable \ac{dS}) at \(\ell_*^3=2C/B\); an exponentially small condensate contribution \(B=B_0e^{-2S_{\rm gc}}\) therefore produces an exponentially large interval (even with Casimir coefficient $C = O(1)$). Defining \(m_\text{KK}\) as the first massive-graviton mass in the same Einstein frame, the critical-point energy satisfies (after converting $m_\text{KK}$ to Einstein frame dividing by $\ell_*^{1/2})$
\begin{eqaed}\label{eq:Lambda_dd}
    \Lambda\equiv V_*\sim m_\text{KK}^4 \, ,
\end{eqaed}
realizing the dark-dimension\footnote{Relations of the type $m_\text{KK}\sim \Lambda^{1/d}$ hold in any such scenario where the small term $B$ has a lower-power denominator than the Casimir term, and are insensitive to its precise power.} relation \(m_{\text{ KK}}\sim\Lambda^{1/4}\). Consequently, \(H/m_{\text{KK}}\sim m_{\text{KK}}/M_4\ll1\) for a sub-Planckian KK scale, allowing substantial scale separation without requiring a metastable minimum. Completing this construction requires compatible boundary gauge bundles and condensate phases, as well as satisfying the remaining moduli equations.  Ideally, even if technically the five-dimensional critical points would lie in the middle of moduli space, attaining values of, say, 20 units for canonically normalized moduli would translate due to non-perturbative nature of the corrections into extreme numerical suppressions for $B$, as needed in this construction.  For example, there is a rigid Calabi-Yau (a "banana nano-manifold" \cite{bryan2025enumerativegeometryarithmeticbanana} involving an elliptic surface with an $E_8$ lattice) with Hodge numbers $(h^{11},h^{21})=(4,0)$ where the ratio of the B/A periods of the isolated rigid complex structure is $\tau\approx 16.0\ i $, leading to $|q|^2\sim 5\times 10^{-88}$ (this example was suggested and computed by GPT). \\

\noindent If five-dimensional bulk supersymmetry plays a role in the dark-energy scale, it is natural to consider \(m_{3/2}\sim\Lambda^{1/4}\).  Intriguingly, as observed in the context of the dark dimension scenario
in \cite{Anchordoqui:2023oqm} (see also \cite{Banks:2000fe,Banks:2002wj,Castellano:2021yye,Cribiori:2021gbf}), in the four-dimensional supergravity description, the smallness of \(V=F_\text{tot}^{\,2}-3m_{3/2}^{\,2}M_4^2\) with respect to the F-term requires \(F_\text{tot}\sim m_{3/2}M_4\). Assuming at most order-one couplings of the standard-model brane to the longitudinal gravitino, the visible-sector splittings satisfy (setting $M_4=1$)
\[
m_\text{soft}\lesssim\sqrt{F_{\rm tot}}
\overset{V\sim 0}{\sim} \sqrt{m_{3/2}}\sim {\Lambda}^{1/8}
\sim \text{TeV} \, ,
\]
suggesting supersymmetry near the TeV scale, given that the supersymmetry-breaking mass splitting cannot be any smaller than the TeV scale experimentally. This would involve supersymmetry breaking originating on the standard-model brane and transmitted to the bulk, differently from the global mechanism discussed above. The two pictures might be connected if the two boundaries coincide, as in the recent proposals for non-supersymmetric heterotic strings in \cite{Baykara:2026vdc} (see also \cite{Faraggi_2008}). \\

\noindent This paradigm is similar to several attempts to find realistic models in string phenomenology, such as the Large Volume Scenario (which could have a natural link to dark dimension scenario with one mesoscopic dimension) or various uplift mechanisms \cite{Kachru:2003aw, Burgess:2003ic, Balasubramanian:2005zx, Westphal:2006tn, Cicoli:2012fh}. On the other hand, Swampland considerations on the nature of quasi-\ac{dS} \cite{Obied:2018sgi, Bedroya:2019snp, Bedroya:2022tbh, Montero:2022prj}, together with recent observational evidence \cite{DESI:2025zgx, Bedroya:2025fwh}, might ease the burden of these constructions: neither metastable minima nor true parametric control might be needed or even possible in this context \cite{Lust:2022lfc}. However, as hinted at above, what might be needed is some genuinely quantum-gravitational input that relates \ac{UV} and \ac{IR} physics in a way that \ac{EFT} cannot fathom. Moving towards this goal may lead to a model of our universe within the string landscape and a better understanding of its deep theoretical underpinnings.

\section*{Acknowledgements}

We would like to thank M. Aparici, B. Bento, A. Castellano and M. Montero for valuable discussions.\\

\noindent I.B. is grateful to Harvard University and the Harvard Swampland Initiative for their hospitality during the development of this work.
This research is supported in part by a grant from the Simons Foundation (602883) and
the DellaPietra Foundation.
\appendix

\section{Packing arguments for geodesic counts}\label{app:packing}

Here we provide a geometric argument to bound the number of prime geodesics in hyperbolic manifolds.

\subsection{Riemann surfaces}\label{app:riemann_surfaces}

The number $N(L)$ of prime geodesics with length below $L$ can be estimated as follows. Cover $\Sigma_g$ by balls of radius $\ell_0 \leq \text{sys}(\Sigma_g)$. The number of center points (a maximal $\ell_0$-separated set) can be estimated by observing that the (open) balls of radius $\ell_0/2$ around these points are disjoint. These balls embed in the (unit-curvature-radius) hyperbolic plane, and thus their area is given by $\text{Area}(\rho) = 4\pi\sinh^2(\rho/2)$ with $\rho = \ell_0$ or $\ell_0/2$. As a result, we can use $M=(g-1)/\sinh^2(\ell_0/4)$ points. Since the balls of radius $\ell_0$ cover $\Sigma_g$, each prime geodesic $\gamma \in \mathcal{P}$ with $\ell_\gamma < L$ passes within distance $\ell_0$ of one of the center points $p$, and thus induces a loop based at $p$ of length bounded by $L+2\ell_0$. This loop lifts to a curve in the hyperbolic plane, whose other endpoint is also the center of a ball of radius $\ell_0/2$. Thus, the number of choices for the endpoint is bounded by
\begin{eqaed}\label{eq:lifts_number}
    \frac{\text{Area}(L+2\ell_0+\ell_0/2)}{\text{Area}(\ell_0/2)} = \frac{\sinh^2(L/2+5\ell_0/4)}{\sinh^2(\ell_0/4)}\,.
\end{eqaed}
Multiplying by the number of basepoints $M$ available from the covering, choosing $\ell_0 = 0.9$ we obtain\footnote{In this range the function is decreasing, so this is an upper bound relative to the smallest possible systole length.}
\begin{eqaed}\label{eq:constant_C}
    N(L) & \leq \frac{g-1}{\sinh^2(\ell_0/4)} \, \frac{\sinh^2(L/2+5\ell_0/4)}{\sinh^2(\ell_0/4)} \\
    & \leq (g-1) \, \frac{e^{5\ell_0/2}}{4\sinh^4(\ell_0/4)} \, e^L \\
    & \leq 4\pi (g-1) \times 72 \, e^L\,.
\end{eqaed} \\

\noindent \textbf{Packing argument for crossing geodesics.} Another useful packing estimate for Riemann surfaces bounds the number of prime geodesics $\gamma$ crossing a fixed systole $\alpha$, with $\ell_\alpha = \text{sys}(\Sigma_g)=s$. Pick a point $p \in \alpha$. Then, choosing an intersection point\footnote{The multiplicity of intersection for a given geodesic has been already estimated below \cref{eq:crossings_angles}.} for a geodesic $\gamma$ crossing $\alpha$ induces a loop based at $p$ of length bounded by $s/2 + \ell_\gamma + s/2 = \ell_\gamma + s$, by traversing the shortest arc on $\alpha$ between an intersection point and $p$. Lifting these loops to the hyperbolic-plane cover, the separation between the endpoints remains bounded by $\ell_\gamma + s$, and open balls of radius $s/2$ around possible endpoints are disjoint (since they are connected by an isometry element whose displacement is at least $s$) and fit in a ball of radius $\ell_\gamma + s + s/2 = \ell_\gamma + 3s/2$ around the (lift of the) initial point. Therefore, the number $N_\text{cross}(L)$ of prime geodesics crossing a fixed systole $\alpha$ of length $\ell_\gamma < L$ is bounded according to
\begin{eqaed}\label{eq:crossing_packing_bound}
    N_\text{cross}(L) & \leq \frac{\text{Area}(L+3s/2)}{\text{Area}(s/2)} \\
    & = \frac{\sinh^2(L/2+3s/4)}{\sinh^2(s/4)} \\
    & \leq \frac{4e^{\frac{3s}{2}}}{s^2} \, e^L\,.
\end{eqaed}
Furthermore, assuming $s \leq 1$, we have
\begin{eqaed}\label{eq:crossing_packing_bound_18}
    N_\text{cross}(L) \overset{s \leq 1}{\leq} \frac{18}{s^2} \, e^L\,.
\end{eqaed}
In fact, this bound can be applied for any $\ell_\alpha < 1$ since it simply follows from the collar lemma. This generalization is used in \cref{sec:slope_riemann_surfaces}.

\subsection{Higher dimensions}\label{app:higher_dim}

For the higher-dimensional case the logic is analogous (except that we take the basepoint for packing in the thick part of the manifold), but the formula for the volumes of balls generalizes to $\text{Vol}(\ell) = \frac{2\pi^{\frac{n}{2}}}{\Gamma(n/2)}\int_0^\ell \sinh^{n-1}(u) du$. We have\footnote{In fact, the prime geodesic theorem \cite{prime_geodesic_theorem} improves the bound by a factor of $1/L$ asymptotically for large $L$. However, numerically, it turns out that the most important contribution to the upper bound comes from $L=O(1)$. At any rate the theorem applies to a fixed manifold rather than providing a uniform estimate.}
\begin{eqaed}\label{eq:N_bound_ndim}
    N(L) - N(\ell_0) \leq \frac{v}{\text{Vol}(\ell_0/2)} \, \frac{\text{Vol}(L + 5\ell_0/2)}{\text{Vol}(\ell_0/2)}\,.
\end{eqaed}
Using $\int_0^\ell \sinh^{n-1}(u) du \leq \int_0^\ell e^{(n-1)u} du \leq \frac{e^{(n-1)\ell}}{n-1}$, we obtain
\begin{eqaed}\label{eq:vol_bound}
    N(L) - N(\ell_0) & \leq \frac{\Gamma(n/2)v}{2\pi^{\frac{n}{2}}(n-1)} \, \frac{e^{(n-1)\frac{5\ell_0}{2}}}{\left( \int_0^{\ell_0/2}\sinh^{n-1}(u)du \right)^2} \, e^{(n-1)L} \\
    & \equiv k_n^\text{packing} \, v \, e^{(n-1)L}\,,
\end{eqaed}
where $\ell_0$ is taken to be the maximum between the (lower bound for the) Margulis constant $\mu_n$ and the systole length $s$, at least when $s<1$ where this subtlety works to our advantage. All in all, we shall choose $\ell_0=\max(\mu_n,\min(s,1))$. (More precisely, the case $n=4$ is an exception, as we will discuss shortly.) This is because for $s < \mu_n$ we can use the ``thick-thin'' decomposition of the manifold $M_n$ and the Margulis-tube decomposition of the ``thin'' part to perform the packing estimate: geodesics longer than $\ell_0$ intersect the ``thick'' part which can be covered as before and admits disjoint embedded balls of radius $\ell_0/2$. Then, $N(s)=0$ while $N(\mu_n)$ is twice the number of Margulis-tube components in $M_n$. For $3 \leq n \leq 5$, this can be bounded as follows. There are constants $l_n$ such that prime geodesics of length below $l_n$ are the cores of solid tubes $T_n$ with length-dependent radii $r_n(\ell_\gamma)$, which are proven to be pairwise disjoint for $n<6$ \cite{Meyerhoff_1987,Cao1998,Kellerhals2001,Kellerhals_2003,Kellerhals2004}. We have
\begin{eqaed}\label{eq:length_tubes}
    l_3 & = \frac{\sqrt{3}}{4\pi} \log^2( \sqrt{2} + 1) \approx 0.107\,, \\
    l_4 & = \frac{\sqrt{3}}{4\pi} \, \log^2\!\left(\frac{17}{9}\right) \approx 0.056\,, \\
    l_5 & = \frac{\sqrt{3}}{9\pi} \approx 0.061\,.
\end{eqaed}
For $n=3$ we can thus choose $\mu_3 \geq 0.104$ as the length cutoff, leading to $r_3 \approx 0.2268$ \cite{Meyerhoff_1987}. For $n=4$ we have $\mu_4 \approx 0.0612 > l_4$, so we have to pick $l_4$ as a cutoff, for which $r_4 \approx 0.590$. For $n=5$, the length cutoff $l_5$ for pairwise disjoint tubes coincides with the bound for $\mu_5$ \cite{Kellerhals_2003}, and thus we can choose this cutoff, leading to $r_5 = \sinh^{-1}(\frac{1}{2}) \approx 0.481$. \\

\noindent The constants $r_n$ are the values of $r_n(\ell_\gamma)$ at the upper length cutoffs. For these length-dependent radii, the tube volumes decrease with core length. Evaluating $\text{Vol}(T_n)$ at the upper length cutoffs yields $N(\mu_n)\leq 2v/\text{Vol}(T_n)$ for $n=3,5$, leading to
\begin{eqaed}\label{eq:Nmu_bound_3-5}
    N(\mu_3) & \leq 2 \cdot 59 \, v\,, \\
    N(\mu_5) & \leq 2 \cdot 53 \, v\,,
\end{eqaed}
where the factor of two stems from summing over oriented prime geodesics. As for $n=4$, pairwise disjointness of the $T_4$ tubes is only proven up to $l_4 < \mu_4$, this method bounds $N(l_4) \leq 2v/\text{Vol}(T_4) \approx 2 \cdot 17.55 \, v < 36 \, v$. To bound $N(\mu_4) = N(l_4) + (N(\mu_4) - N(l_4))$, using \cref{eq:vol_bound} applied to $\ell_0 = l_4$ results in large bounds for $N(\mu_4)/v$. However, there is no need to include this contribution, since $l_4$ is a perfectly good Margulis scale to obtain pairwise disjoint tubes and estimate $N(L)$, so for $n=4$ we pick $\ell_0 = l_4$. All in all, we denote this additional contribution as
\begin{eqaed}\label{eq:vol_bound_2}
    \frac{N(L)}{v} & \leq k_n^\text{packing} \, e^{(n-1)L} + k_n^\text{Margulis}
    & \equiv k_n \, e^{(n-1)L}\,,
\end{eqaed}
where the overall parameter $k_n$ depends on the systole length and $L$ according to the above specifications. For $n=6,7$ we could not determine an explicit numerical bound for $k_n^\text{Margulis}$ in this way. To remedy this, we now provide an improved packing argument for short geodesics.

\subsection{Improved packing argument for short geodesics}\label{app:packing_short}

The source of large numbers in the packing estimates is the denominator of \cref{eq:N_bound_ndim,eq:vol_bound}, which contains a squared volume of the separating balls. Furthermore, the left-hand side shows the dependence on the unknown constants $N(\mu_n)$. Performing a packing estimate at the systole scale removes the latter unknown, but potentially worsens the denominator dramatically, and thus the bound in \cref{eq:harmonic_control} for $s \ll \mu_n$. To remedy this, we employ a different method. \\

\noindent Since balls of radius $s/2$ around points $p, g \cdot p$ in the covering $\mathbb{H}^n$ connected by a non-trivial element $g \in \Gamma$ are disjoint\footnote{In fact, the whole orbit $\Gamma \cdot p$ is $s$-separated, by invariance of the distance.}, we can bound the number $\mathcal{N}_p(L_0)$ of elements of $\Gamma$ (excluding the identity) that displace a lift of $p \in M_n = \mathbb{H}^n/\Gamma$ by less than some distance $L_0$ according to
\begin{eqaed}\label{eq:moving_number}
    \mathcal{N}_p(L_0) \leq \frac{\text{Vol}(L_0+s/2)}{\text{Vol}(s/2)} - 1
\end{eqaed}
by packing balls of radius $s/2$ inside the ball of radius $L_0+s/2$ centered at (some lift of) $p$. On the other hand, $\mathcal{N}_p(L_0)$ is the sum over geodesics of the indicator function of a solid ellipsoid (times its axis) $E_g(L_0) \subset \mathbb{H}^n/\langle g \rangle$ analogous to that defined in \cref{eq:ellipsoid}, except that the Margulis constant is replaced by $L_0$. Hence, integrating over $M_n$ with its canonical measure $m$, unfolding to $\mathbb{H}^n/\langle g_\gamma \rangle$ for each conjugacy class and using \cref{eq:volume_ellipsoid}, we find
\begin{eqaed}\label{eq:integrating_N_over_M}
    \int_{M_n} \mathcal{N}_p(L_0) \, dm(p) & = \sum_{k \geq 1} \sum_{\gamma \in \mathcal{P}} \text{Vol}(E_{g_{\gamma}^k}(L_0)) \\
    & \geq \frac{(2\pi)^\frac{n-1}{2}}{\Gamma(\frac{n+1}{2})}\sum_{k \geq 1}\sum_{\gamma \in \mathcal{P}:k\ell_\gamma < L_0} \frac{\ell_\gamma}{D(\gamma^k)} \left(\cosh(L_0) - \cosh(k\ell_\gamma)\right)^{\frac{n-1}{2}} \\
    & \geq \frac{(2\pi)^\frac{n-1}{2}}{\Gamma(\frac{n+1}{2})}\sum_{k \geq 1}\sum_{\gamma \in \mathcal{P}:k\ell_\gamma < L_0} \ell_\gamma \, \frac{\left(\cosh(L_0) - \cosh(k\ell_\gamma)\right)^{\frac{n-1}{2}}}{(2 \cosh(k\ell_\gamma/2))^{n-1}} \\
    & \equiv \frac{(2\pi)^\frac{n-1}{2}}{\Gamma(\frac{n+1}{2})}\sum_{\gamma \in \mathcal{P}} \sum_{1 \leq k < L_0/\ell_\gamma} \ell_\gamma \, f(k\ell_\gamma)\,.
\end{eqaed}
The combination $f(k\ell_\gamma)$ in the fraction is a decreasing function of its argument $k\ell_\gamma$. Hence, we can bound
\begin{eqaed}\label{eq:integral_bound}
    \sum_{1 \leq k < L_0/\ell} \ell \,  f(k\ell) & \geq \sum_{1 \leq k < L_0/\ell} \int_{k\ell}^{\min((k+1)\ell,L_0)} f(z) \, dz \\
    & \geq \int_{\ell}^{L_0} f(z) \, dz\,.
\end{eqaed}
Inserting \cref{eq:integral_bound} into \cref{eq:integrating_N_over_M}, we can further restrict to prime geodesics of length bounded by $L$ and lower bound the integral accordingly; the resulting sum over prime geodesics factors out into $N(L)$. Combining \cref{eq:integrating_N_over_M,eq:moving_number}, we obtain
\begin{eqaed}\label{improved_packing_bound}
    \frac{(2\pi)^\frac{n-1}{2}}{\Gamma(\frac{n+1}{2})} \, \frac{N(L)}{v} \int_L^{L_0} f(z) \, dz \leq \frac{\text{Vol}(L_0+s/2)}{\text{Vol}(s/2)} - 1\,.
\end{eqaed}
In summary, we have successfully removed a factor of volume from the denominator with respect to \cref{eq:N_bound_ndim}. Optimizing over $L_0 > L$,
\begin{eqaed}\label{eq:N_bound_improved}
    \frac{N(L)}{v} \leq \frac{\Gamma(\frac{n+1}{2})}{(2\pi)^\frac{n-1}{2}} \inf_{L_0 > L} \frac{\frac{\text{Vol}(L_0+s/2)}{\text{Vol}(s/2)} - 1}{\int_L^{L_0} \left(\frac{\cosh(L_0) - \cosh(z)}{(2\cosh(z/2))^2}\right)^{\!\frac{n-1}{2}} \, dz}\,.
\end{eqaed}

\section{Teichm\"{u}ller variations for Riemann surfaces}\label{app:laplacian_bounds_geodesic_variations}

In this appendix we collect some more specialized results on the geometry of Riemann surfaces and their Teichm\"{u}ller moduli space. In \cref{app:laplacian_bounds} we derive bounds on the variation of the Laplacian determinant along geodesic-length deformations, which we use in \cref{sec:beyond_rigid_riemann_surfaces} to show that critical points of the Casimir potential have a uniform lower bound on the systole length. In \cref{app:geodesic_length_variations} we derive analogous bounds for the logarithmic slope of the Casimir potential, used in \cref{sec:slope_riemann_surfaces} to bound \ac{EFT} control parameters in the region where the \ac{TCC} bound of \cref{eq:asymptotic_slope_value} is violated.

\subsection{Bounds on the Laplacian determinant on Riemann surfaces}\label{app:laplacian_bounds}

When showing in \cref{sec:beyond_rigid_riemann_surfaces} that the critical points of the Casimir energy on a Riemann surface have a uniform lower bound on the systole length, the inclusion of the external \ac{dS} curvature in the heat kernel causes some trouble, due to the zero-mode on the corresponding Euclidean sphere. The contribution of this mode to the directional derivative of the Casimir potential in \cref{eq:dA_contraction} is given by \cref{eq:dA_zeta} in terms of the derivative of the Laplacian determinant, or equivalently the Selberg zeta function \cite{DHoker1986}. In this appendix we provide bounds for this quantity. \\

\noindent To begin with, the results of \cite{Zograf1987ALI} applied to the scalar Laplacian give
\begin{eqaed}\label{eq:logZ_variation}
    X (\log Z'_{\Sigma_g}(1)) & = - 2\operatorname{Re} \int_{\Sigma_g} R^{(0)} \, \mu_X\,,
\end{eqaed}
for the variation induced by a real vector field $X$ on Teichm\"{u}ller space. In the notation of \cite{Zograf1987ALI, takhtajan1991local}, $R^{(0)} = \partial_z \partial_{z'}(G_1(z,z')-Q_1(z,z')) dz dz'|_{z'=z} \in \Omega^{2,0}(\Sigma_g)$ is a quadratic differential obtained subtracting the Green's functions $G_1$ (resp. $Q_1$) on $\Sigma_g = \mathbb{H}^2/\Gamma$ (resp. $\mathbb{H}^2$) for the scalar Laplacian, while $\mu_X \in \Omega^{-1,1}(\Sigma_g)$ is the Beltrami differential associated to $X$ via the Kodaira-Spencer map. The strategy is then to bound $R^{(0)}$ pointwise (i.e., in the tensor $L^\infty$-norm induced by the hyperbolic metric on $\Sigma_g$),
\begin{eqaed}\label{eq:holder_bound}
    \abs{X(\log Z'_{\Sigma_g}(1))} \leq 2 \norm{R^{(0)}}_{L^\infty_\text{hyp}(\Sigma_g)} \, \norm{\mu_X}_{L^1_\text{hyp}(\Sigma_g)}\,.
\end{eqaed}
This approach is also convenient for the following reason. The Beltrami differential $\mu_{X_\alpha}$ associated with the gradient of a geodesic length $X_\alpha = \nabla \ell_\alpha$, with $\ell_\alpha = \text{sys}(\Sigma_g)$, is written as \cite{wolpert2008behaviorgeodesiclengthfunctionsteichmueller,Wolpert2012Geodesic}
\begin{eqaed}\label{eq:beltrami_geodesic}
    \mu_{X_\alpha} = \frac{2}{\pi} \, \overline{\Theta}_\alpha (ds^2)^{-1}\,,
\end{eqaed}
where $ds^2$ is the hyperbolic line element, and $\Theta_\alpha$ is the Poincar\'{e} series
\begin{eqaed}\label{eq:poincare_series}
    \Theta_\alpha = \sum_{g \in \langle g_\alpha \rangle \backslash \Gamma} g^*\!\left(\frac{dz}{z}\right)^{\!2}\,.
\end{eqaed}
This allows us to bound the hyperbolic $L^1$-norm by unfolding,
\begin{eqaed}\label{eq:unfolding}
    \norm{\mu_{X_\alpha}}_{L^1_\text{hyp}(\Sigma_g)} & \leq \frac{2}{\pi} \sum_{g \in \langle g_\alpha \rangle \backslash \Gamma} \int_{\Sigma_g} g^*\abs{\frac{dz}{z}}^2 \\
    & = \frac{2}{\pi} \int_{\mathbb{H}^2/\langle g_\alpha \rangle} \abs{\frac{dz}{z}}^2 \\
    & = 2 \, \text{sys}(\Sigma_g)\,.
\end{eqaed}
where in the last equality we chose, e.g., the fundamental domain $1 \leq \abs{z} \leq e^{\text{sys}(\Sigma_g)}$ to evaluate the integral. \\

\noindent We now turn to the $L^\infty$-norm term in \cref{eq:holder_bound}. In the notation of \cite{Zograf1987ALI, takhtajan1991local}, $R^{(0)} = - \frac{R_S - R_F
}{6\pi}$ is the difference between the projective connections associated with the Schiffer kernel $S$ and the Fuchsian uniformization of $\Sigma_g$ \cite{Fay1977, tyurin}. A direct computation of this expression leads us back to \cref{eq:dA_zm}, but we can exploit the geometric nature of this quantity in a different fashion. Choosing local Fuchsian coordinates $z$ in an embedded disk $D \subset \Sigma_g$ around some point at $z=0$, we have $R_F = 0$ and $R_S(z) = 6 \left(S(z,z') - \frac{1}{(z-z')^2}\right)dz dz'\bigg|_{z'=z}$ is the regular part $6H_S(z,z)dz^2$ of the Schiffer kernel $L_{\Sigma_g} = -\frac{1}{2\pi i}S(z,z') \, dz dz'$ \cite{Zograf1987ALI}. The subtracted singular part is the hyperbolic kernel $L_D$ for the disk. The goal is to bound $H_S(0,0)$. To this end, we consider the Schiffer comparison operator \cite{schippers2018plemeljsokhotskiisomorphismquasicirclesriemann}
\begin{eqaed}\label{eq:schiffer_operator}
    & T(D,D) : \overline{\mathcal{H}_1(D)} \to \mathcal{H}_1(D)\,, \\
    & [T(D,D) \, \overline{\omega}](z) = \int_{D} \left(L_{\Sigma_g}|_D - L_D\right) \wedge  \overline{\omega} 
\end{eqaed}
acting on the (conjugated) Bergman space $\overline{\mathcal{H}_1(D)}$ square-integrable anti-holomorphic one-forms, where the integral is understood to act on the second argument of the kernels. Remarkably, $T(D,D)$ satisfies a completeness/unitarity relation of the form (see \cite{schippers2018plemeljsokhotskiisomorphismquasicirclesriemann}, Theorem 3.13)
\begin{eqaed}\label{eq:T_unitarity}
    T(D,D)^\dagger T(D,D) + T(D,\Sigma_g \backslash D)^\dagger T(D,\Sigma_g \backslash D) + S(D)^\dagger S(D) = 1\,,
\end{eqaed}
where $S(D)$ is another comparison operator. In particular, the operator norm
\begin{eqaed}\label{eq:operator_norm}
    \norm{T(D,D)}_\text{op} \leq 1\,.
\end{eqaed}
With this key result at hand, we proceed as follows. We have
\begin{eqaed}\label{eq:T_dz}
    [T(D,D) \, \overline{dz}] & = \int_{D} \left(L_{\Sigma_g}(z,w) - L_D(z,w)\right) \wedge \overline{dw} \\
    & = - \frac{1}{2\pi i} \int_D H_S(z,w) \, dz \wedge dw \wedge \overline{dw} \\
    & = R^2 \, H_S(z,0) \, dz\,,
\end{eqaed}
where $R$ is the coordinate radius of the disk, i.e., the radius measured in the flat area form $d\Re w \, d\Im w = \frac{i}{2} \, dw \wedge \overline{dw}$ around the center point. With the conventions of \cite{schippers2018plemeljsokhotskiisomorphismquasicirclesriemann}, the tensor $L^2$-norm is $\norm{dz}^2_{L^2_\text{hyp}(D)} = \frac{i}{2}\int_D dz \wedge \overline{dz} = \pi R^2$. Thus, writing $z = x + iy$,
\begin{eqaed}\label{eq:T_ratio_norm}
    \frac{\norm{T(D,D) \, \overline{dz}}^2_{L^2_\text{hyp}(D)}}{\norm{dz}^2_{L^2_\text{hyp}(D)}} & = \frac{iR^2}{2\pi} \int_D \abs{H_S(z,0)}^2 \, dz \wedge \overline{dz} \\
    & = \frac{R^2}{\pi} \int_D \abs{H_S(z,0)}^2 \, dx \wedge dy \\
    & \geq R^4 \, \abs{H_S(0,0)}^2\,,
\end{eqaed}
where we applied the mean-value inequality on $D$ to $H_S(z,0)$. Together with \cref{eq:T_unitarity}, \cref{eq:T_ratio_norm} leads to
\begin{eqaed}\label{eq:H_S_bound}
    \abs{H_S(0,0)} \leq \frac{1}{R^2}\,,
\end{eqaed}
so that $R^{(0)} = - \frac{1}{\pi} \, H_S \, dz^2$
satisfies
\begin{eqaed}\label{eq:R0_bound}
    \norm{R^{(0)}}_{L^\infty_\text{hyp}(\Sigma_g)} \leq \frac{1}{4\pi} \, \coth^2\!\left(\frac{\text{sys}(\Sigma_g)}{4}\right) ,
\end{eqaed}
since \cref{eq:H_S_bound} holds for any embedding radius below the injectivity radius at the central point of $D$, and thus in particular for the minimal injectivity radius given by $\text{sys}(\Sigma_g)/2$. Indeed, the coordinate radius $R = \tanh \frac{r_\text{geod}}{2}$ in terms of the geodesic radius $r_\text{geod}$ of the embedded disk $D$. Hence, optimizing over $R$ one obtains \cref{eq:R0_bound}. The additional factor of four in the denominator arises by writing the tensor norm in local Fuchsian coordinates on $D$, with line element $ds^2 = \frac{4\abs{dz}^2}{(1-\abs{z}^2)^2}$. \\

\noindent \textbf{Combining the norm bounds.} Combining \cref{eq:unfolding,eq:R0_bound}, we finally obtain that the directional derivative of the (logarithm of the) Selberg zeta function along the direction of a systole-length gradient is bounded according to 
\begin{eqaed}\label{eq:final_selberg-zeta_bound}
    \abs{X_{\alpha}(\log Z'_{\Sigma_g}(1))} \leq \frac{\text{sys}(\Sigma_g)}{\pi} \, \coth^2\!\left(\frac{\text{sys}(\Sigma_g)}{4}\right) .
\end{eqaed}
This result is used in \cref{eq:dA_zeta_final_bound} to extend our main no-go result to all critical points in the complex-structure moduli space of hyperbolic Riemann surfaces, going beyond the rigid case.

\subsection{Bounds on geodesic-length variations}\label{app:geodesic_length_variations}

Following the approach of \cite{wolpert2008behaviorgeodesiclengthfunctionsteichmueller,Wolpert2012Geodesic}, we assume $s < 1$, so that geodesics $\alpha$ of length $\ell_\alpha < 1$ are the cores of pairwise disjoint collars of half-width $r_\text{collar} \geq \sinh^{-1}(1/\sinh(1/2)) \approx 1.4$. Lifting $\alpha$ to the cyclic cover $\langle g_\alpha \rangle \backslash \mathbb{H}^2$, the embedded collar has local cylindrical coordinate $z=e^{u+i\theta} = e^w$ with $0 \leq u < \ell_\alpha$ and $\theta$ bounded in a range $\theta_{\ell_\alpha} < \theta < \pi - \theta_{\ell_\alpha}$ around the core $\theta = \pi/2$. The deck transformation $g_\alpha : w \mapsto w + \ell_\alpha$. Using the line element $ds^2 = \frac{du^2+d\theta^2}{\sin^2\theta}$, we have $\theta_{\ell_\alpha} < \ell_\alpha$, so we can use the smaller collar $C_\alpha$, defined by $\ell_\alpha < \theta < \pi - \ell_\alpha$, for later convenience \cite{Wolpert2012Geodesic}. Then, a local harmonic Beltrami differential (representing a local metric deformation) takes the form $\mu = \overline{\varphi} \, (ds^2)^{-1}$ where $\varphi(w) \, dw^2$ is a holomorphic quadratic differential. Thus, $\varphi$ is periodic on the cylinder and we can write a Fourier expansion of the form
\begin{eqaed}\label{eq:fourier_beltrami}
    \varphi(w) = \sum_{n \in \mathbb{Z}} \varphi_n \, e^{2\pi i n w/\ell_\alpha}\,.
\end{eqaed}
The corresponding geodesic-length variation is given by $d\ell_\alpha(\mu) = \frac{2}{\pi}\Re(\mu, \Theta_\alpha)$ \cite{wolpert2008behaviorgeodesiclengthfunctionsteichmueller} in terms of the Poincaré series $\Theta_\alpha$ in \cref{eq:poincare_series}. By unfolding to the cyclic cover $\langle g_\alpha\rangle\backslash \mathbb{H}^2$ the pairing of \cite{wolpert2008behaviorgeodesiclengthfunctionsteichmueller}, one sees that only the zero-mode of \cref{eq:fourier_beltrami} enters, since
\begin{eqaed}\label{eq:length_variation_unfolding}
    d\ell_\alpha(\mu) & = \frac{2}{\pi} \Re \frac{i}{2} \int_{\langle g_\alpha\rangle \backslash \mathbb{H}^2} \overline{\varphi(w)} \sin^2\theta \, dw \wedge \overline{dw} \\
    & = \frac{2}{\pi} \, \Re \int_{0}^{\pi} d\theta \int_0^{\ell_\alpha} du \,  \overline{\varphi(u+i\theta}) \sin^2\theta \\
    & = \frac{2}{\pi} \, \ell_\alpha \, \Re \varphi_0 \int_{0}^{\pi} \sin^2 \theta \, d\theta \\
    & = \ell_\alpha \, \Re \varphi_0\,.
\end{eqaed}
With respect to the local hyperbolic area element $dA_\text{hyp} = du d\theta/\sin^2\theta$, the tensor $L^2$-norm of a harmonic Beltrami differential restricted to the subcollar $C_\alpha$ can be estimated according to
\begin{eqaed}\label{eq:subcollar_beltrami_norm}
    \norm{\mu}^2_{L^2_\text{hyp}(C_\alpha)} & = \int_{C_\alpha} \abs{\varphi(u+i\theta)}^2 \sin^2\theta \, du \, d\theta \\
    & = \ell_\alpha \sum_{n \in \mathbb{Z}} \abs{\varphi_n}^2 \int_{\ell_\alpha}^{\pi - \ell_\alpha} e^{-4\pi n \frac{\theta}{\ell_\alpha}} \, \sin^2\theta \, d\theta \\
    & \geq \ell_\alpha \, \abs{\varphi_0}^2 \int_{\ell_\alpha}^{\pi - \ell_\alpha} \sin^2\theta \, d\theta \\
    & = \ell_\alpha \, \abs{\varphi_0}^2 \frac{\pi - 2\ell_\alpha + \sin(2\ell_\alpha)}{2} \\
    & \overset{\ell_\alpha < 1}{\geq}  \frac{\pi - 2 + \sin(2)}{2} \,  \ell_\alpha \, \abs{\varphi_0}^2 \\
    & \equiv K_\text{collar} \, \ell_\alpha \, \abs{\varphi_0}^2\,.
\end{eqaed}
Combining \cref{eq:length_variation_unfolding,eq:subcollar_beltrami_norm} we obtain
\begin{eqaed}\label{eq:length_variation_vs_norm_bound}
    \abs{d\ell_\alpha(\mu)}^2 & \leq \ell_\alpha^2 \, \abs{\varphi_0}^2 \leq K_\text{collar}^{-1} \, \ell_\alpha \norm{\mu}^2_{L^2_\text{hyp}(C_\alpha)}\,.
\end{eqaed}
Now, we can use the fact that the collars $C_\alpha$ are disjoint. Summing \cref{eq:length_variation_vs_norm_bound} over geodesics $\alpha$ of length $\ell_\alpha < 1$ yields (recall that the sum over $\alpha \in \mathcal{P}$ counts \emph{oriented} prime geodesics)
\begin{eqaed}\label{eq:beltrami_norm_bound_sum_geodesics}
    \norm{\mu}^2_{L^2_\text{hyp}(\Sigma_g)} & \geq \norm{\mu}^2_{L^2_\text{hyp}(\mathop{\bigcup}_{\alpha \in \mathcal{P} : \ell_\alpha < 1} C_\alpha)} \\
    & = \frac12 \sum_{\alpha \in \mathcal{P} : \ell_\alpha < 1} \norm{\mu}^2_{L^2_\text{hyp}(C_\alpha)} \\
    & \geq \frac{K_\text{collar}}{2} \sum_{\alpha \in \mathcal{P} : \ell_\alpha < 1} \frac{\abs{d\ell_\alpha(\mu)}^2}{\ell_\alpha}\,.
\end{eqaed}
For the purposes of \cref{sec:slope_riemann_surfaces}, we are led to consider the harmonic Beltrami differential $\mu_{\nabla\mathcal{C}}$ associated with the gradient of the Casimir coefficient $\mathcal{C}$ as a function on Teichm\"{u}ller space. Then, \cref{eq:beltrami_norm_bound_sum_geodesics} bounds its Weil-Petersson norm according to
\begin{eqaed}\label{eq:WP_norm_dC_bound}
    \norm{\nabla\mathcal C}^2_\text{WP} \geq \frac{K_\text{collar}}{2} \sum_{\alpha \in \mathcal{P} : \ell_\alpha < 1} \frac{\abs{\langle \nabla \mathcal{C}, \nabla \ell_\alpha \rangle}^2}{\ell_\alpha}\,.
\end{eqaed}
This result can be combined with the explicit bounds on the directional derivatives of $\mathcal{C}$ along geodesic-length variations, obtained in \cref{sec:beyond_rigid_riemann_surfaces}, to bound the potential slope in \cref{sec:slope_riemann_surfaces}.

\section{Spectral analysis of supergraviton fluctuations}\label{app:spectral_analysis_supergraviton_fluctuations}

In this appendix we develop a spectral analysis of the one-loop determinants for the eleven-dimensional supergraviton on $S^d \times M_n$, where $M_n$ is a compact Einstein manifold of non-positive Ricci curvature. Eventually we will restrict to hyperbolic manifolds and \acp{RFM}, in order to extend the bounds in the main text to parametric bounds for the actual supergravity effective potential. In principle, these methods can adapt to any direct product of Einstein manifolds, and can be used for perturbative stability analyses as well. \\

\noindent \textbf{Off-shell supergraviton fluctuations.} The BRST complex of the eleven-dimensional supergraviton decomposes as follows. Letting $X = S^d \times M_n$, the spectrum comprises the following fields.

\begin{itemize}
    \item The graviton $h$, a symmetric covariant rank-two tensor; that is, $\mathcal{V}_h = \text{Sym}^2(T^*X)$.

    \item The Faddeev-Popov ghost for small diffeomorphisms $c$; that is, $\mathcal{V}_{c} = T^*X$.

    \item The (fluctuations of the\footnote{The full three-form potential $C_3$ behaves as a (higher) connection, but its fluctuations are tensors.}) three-form potential $C_3$, along with its full set of ghosts; that is, $\mathcal{V}_{C_3} = \oplus_{p=0}^3 \Lambda^p(T^*X)$.

    \item The gravitino $\psi$; that is, $\mathcal{V}_\psi = \mathcal{S} \otimes T^*X$ where $\mathcal{S} \to X$ is the spinor bundle of spacetime.

    \item The supersymmetry ghost $\gamma$; that is, $\mathcal{V}_{\gamma} = S$.
\end{itemize}

\noindent The fluctuation operators for fermions $D_\text{fermion} = \mathcal{D}^\dagger \mathcal{D}$ arise by squaring a Dirac-type operator $\mathcal{D}$, leading to a non-negative operator in each BRST sector. Similarly, upon gauge-fixing the three-form (and its ghosts) by the transversality condition $d^\dagger C_3=0$, the gauge-fixed fluctuation (Hodge) operator $d^\dagger d + d d^\dagger$ is a semi-positive definite Laplace-like operator. As for the graviton sector, it is convenient to choose the Lorenz gauge $Kh \equiv -2\nabla \cdot h = 0$ rather than the more usual De Donder gauge. (Notice that $K^\dagger$ acts on vectors with the usual normalization for infinitesimal diffeomorphisms, and its kernel defines Killing vectors.) This way, the induced fluctuation operator $KK^\dagger$ on the ghost is automatically non-negative. The graviton itself then decomposes into vector (moduli Killing vectors, due to the gauge condition) and (horizontal) tensor modes\footnote{To see this, notice that writing $h = H + K^\dagger \xi$, imposing the gauge condition on the horizontal part $K H = 0$ implies that $KK^\dagger \xi = K h$. The operator $KK^\dagger$ is self-adjoint and semi-positive definite (and elliptic), and its kernel is $\ker K K^\dagger = \ker K^\dagger$ are the Killing vectors, since $\langle \xi, KK^\dagger \xi \rangle = \norm{K^\dagger \xi}^2$. Hence, $KK^\dagger$ is invertible on the orthogonal complement of Killing vectors, to which $Kh$ belongs; as a result, the equation has a unique solution modulo Killing vectors, and $h$ admits the claimed orthogonal decomposition.} $\xi, H$ according to $h = H + K^\dagger \xi$. As a result, as in the usual transverse-longitudinal decomposition of gauge fields in the Lorenz gauge, the gauge-fixing term $\frac{1}{\alpha}\norm{Kh}^2=\frac{1}{\alpha}\norm{KK^\dagger\xi}^2$ only affects the fluctuation operator for vector modes $\xi$, and makes it positive for sufficiently small gauge-fixing parameter $\alpha$. In other words, in the Landau gauge the trace-log formula contains the positive semi-definite operator $KK^\dagger$ and the fluctuation operator $D_H$ acting on horizontal tensor modes $H$ with $KH=0$. The latter does not give the full one-loop Casimir energy by itself, but it is the only contribution that does not come from a manifestly non-negative operator. This is relevant for bounding the Casimir energy (and therefore the \ac{EFT} control parameters) by analogy with the scalar-field case.

\subsection{The anatomy of graviton fluctuations}\label{app:anatomy_graviton_fluctuations}

Having chosen a gauge in which the only potentially problematic fluctuations are gravitons, we now approach the computation of the corresponding eigenvalues on $S^d \times M_n$ in steps. First, we write the complete fluctuation operator for gravitons in the chosen gauge. Then, we decompose modes according to the product-manifold structure. Then, on each Einstein manifold factor we further perform a York decomposition of modes. The spectral analysis can then be carried out by tensor, vector and scalar harmonics, whose eigenvalues will be related to those of the full fluctuation operator. \\

\noindent \textbf{The fluctuation operator for horizontal graviton modes.} Imposing the gauge condition $KH=0$, we obtain a complicated Lichnerowicz-like fluctuation operator, whose eigenvalues must be bounded away from zero for the bounds to be extended to the full supergravity spectrum. Notice that the zero-modes are killed by the modified Casimir coefficient $\mathcal{A}$, while negative modes contribute to the determinant magnitude through their eigenvalues in absolute value after steepest-descent contour rotation (compatibly with the relative homology class to the boundary of field space). Thus, the task is to bound the first few eigenvalues of this operator, since near-zero combinations of positive sphere eigenvalues and negative internal eigenvalues can dominate $A_\text{IR}$ as we shall we later. To proceed, we write
\begin{eqaed}\label{eq:symmetric_bundle_split}
    \text{Sym}^2(T^*\!(S^d \times M_n)) = [\text{Sym}^2(T^*\!S^d) \otimes \mathcal{O}_{M_n}] \!\oplus\! [T^*\!S^d \otimes T^*\!M_n] \!\oplus\! [\mathcal{O}_{S^d} \otimes \text{Sym}^2(T^*\!M_n)]\,,
\end{eqaed}
where $\mathcal{O}$ denotes scalars. The quadratic fluctuation operator $D_H$ for horizontal gravitons is given by the quadratic Lagrangian $\mathcal{L}^{(2)} = \frac{1}{2} \langle h , D_H h\rangle$, with \cite{Dahl_2022, Bastianelli:2022pqq, Basile:2024oms}
\begin{eqaed}\label{eq:horizontal_graviton_action}
    \mathcal{L}^{(2)} & = M_D^{D-2} \, \bigg( -\frac{1}{4} h^{\mu \nu}(\nabla^2-R)h_{\mu \nu} + \frac{1}{8} \overline{h}(\nabla^2-R)\overline{h} - \frac{1}{8} \abs{\nabla \overline{h}}^2 \\
    & - \frac{1}{2} R_{\mu \nu} \left(h^{\mu \rho} {h_\rho}^\nu - \overline{h} h^{\mu \nu} \right) - \frac{1}{2} R_{\mu \nu \rho \sigma} h^{\mu \rho} h^{\nu \sigma} \bigg)
\end{eqaed}
in our normalization, where $\overline{h} \equiv g^{\mu \nu} h_{\mu \nu}$. After canonically normalizing $h$, in the intrinsic notation of \cite{Dahl_2022} we find\footnote{More precisely, the right-hand side of \cref{eq:horizontal_graviton_operator} should be post-composed with the projector onto the horizontal subspace defined by the gauge condition $Kh=0$. We will take this into account later on.}
\begin{eqaed}\label{eq:horizontal_graviton_operator}
    D_H h = \Delta_\text{L}h - 2\left(\text{Ric} \circ h + h \circ \text{Ric}\right) + R \, h + \langle \text{Ric}, h \rangle \, g - \left(\Delta + \frac{R}{2}\right) \overline{h} \, g + \text{Ric} \, \overline{h}\,,
\end{eqaed}
where $\Delta_\text{L}$ is the Lichnerowicz operator. We can now decompose the horizontal fluctuation $h$ according to \cref{eq:symmetric_bundle_split}. Since the split is parallel, all the terms in \cref{eq:horizontal_graviton_operator} that do not involve the trace $\overline{h} = \tr h$ are block-diagonal with respect to it. In particular, the Lichnerowicz operator $\Delta_\text{L}$ on $S^d \times M_n$ acts as $\Delta_\text{L} \otimes \text{id} + \text{id} \otimes \Delta$ on external tensors and $\Delta \otimes \text{id} + \text{id} \otimes \Delta_\text{L}$ on internal tensors, since the $\nabla^2$ term is simply the sum of the two ordinary Laplacians. As for vector-vector fluctuations, the Riemann term in $\Delta_\text{L}$ acts as zero (there are no non-zero components with mixed indices), and the two Ricci endomorphisms arrange into covector Hodge operators, resulting in $\Delta_\text{L}|_\text{vector-vector} = \Delta_1 \otimes \text{id} + \text{id} \otimes \Delta_1$. Since both manifold factors are Einstein, the other non-trace terms simplify in terms of the blocks $h_{S^d}, h_{M_n}$ and the corresponding block traces $\overline{h} = \overline{h}_{S^d} + \overline{h}_{M_n}$. (The mixed components contract to zero with the background metric $g = g_{S^d} \oplus g_{M_n}$.) Similarly, the last term in \cref{eq:horizontal_graviton_operator} simplifies to constants multiplying $g_{S^d} \, \overline{h}$ and $g_{M_n} \, \overline{h}$. All in all, it is convenient to collect the first three terms of \cref{eq:horizontal_graviton_operator} into a block-diagonal operator that does not involve block traces, while the remaining three terms involve them. This means that these terms do not contribute for fluctuations whose block traces both vanish. \\

\noindent \textbf{Decomposing into sectors.} Under the orthogonal decomposition of \cref{eq:symmetric_bundle_split}, the operator $D_H$ behaves in a simple fashion because the metric is block-diagonal and each factor is Einstein. We can decompose fluctuations on each factor according to the spectrum of convenient canonical operators, namely $\Delta_\text{L}$ for tensors, the ordinary Laplacian $\Delta$ for scalars, and the Hodge operator $\Delta_1 = d^\dagger d + d d^\dagger = -\nabla^2 + \text{Ric} \, \circ$ for vectors (more precisely covectors). We choose eigenmodes for each of these representations for $S^d$ and $M_n$ separately, with the following notation. Tensors with $\Delta_\text{L}$ eigenvalues $\lambda$ (resp. $\mu$) on $S^d$ (resp. $M_n$) are denoted $T_\lambda$ for $S^d$ and $\widetilde{T}_\mu$ for $M_n$; similarly, vectors are denoted $V_\lambda , \widetilde{V}_\mu$ and scalars $S, \widetilde{S}_\mu$. Altogether, we have

\begin{itemize}
    \item External tensors: $h(x,y) = T_\lambda(x) \widetilde{S}_{\mu}(y)$.

    \item Vector-vector fluctuation: $h(x,y) = V_\lambda(x) \widetilde{V}_{\mu}(y)$.

    \item Internal tensors: $h(x,y) = S_\lambda(x) \widetilde{T}_{\mu}(y)$.
\end{itemize}

\noindent Each term is a field on an Einstein manifold, which can be expanded according to a York decomposition. In particular, covectors can be decomposed as\footnote{Notice that in this notation $v$ is a scalar.} $V = V^\perp + dv$, while tensors on a $D$-dimensional space can be decomposed as
\begin{eqaed}\label{eq:tensor_york}
    T_{ab} & = T_{ab}^\text{TT} + \nabla_a T_b^\perp + \nabla_b T_a^\perp + \left(\nabla_a \nabla_b - \frac{1}{D} \, g_{ab} \, \nabla^2\right) \mathcal{T} + \frac{1}{D} \, g_{ab} \, \tau \\
    & \equiv T_{ab}^\text{TT} + (K^\dagger T^\perp)_{ab} + (Q \mathcal{T})_{ab} + \frac{1}{D} \, g_{ab} \, \tau\,,
\end{eqaed}
where we defined the traceless Hessian operator $Q\mathcal{T} \equiv \frac{1}{2}K^\dagger d\mathcal{T} + \frac{1}{D} \, g \, \Delta \mathcal{T}$. The above index-free notation is useful to simplify the ensuing computations, thanks the intertwining operator identities\footnote{The Einstein condition is used to prove the last three identities in \cref{eq:operator_identities}. In particular, the last identity follows from the definition of $Q$ by applying the preceding ones a few times, while the fourth and fifth are adjoints of each other and follow by applying Bianchi identities to the Riemann tensor that arises from commuting covariant derivatives.}
\begin{eqaed}\label{eq:operator_identities}
    \Delta_1(dv) & = d \Delta v\,, \\
    \Delta_\text{L} (g \, \tau) & = g \, \Delta \tau \\
    \tr \Delta_\text{L}T & = \Delta \tr T\,, \\
    K\Delta_\text{L}T & = \Delta_1 K T\,, \\
    \Delta_\text{L} K^\dagger T^\perp & = K^\dagger \Delta_1 T^\perp\,, \\
    \Delta_\text{L} Q \mathcal{T} & = Q \Delta \mathcal{T}\,.
\end{eqaed}
Based on the preceding discussion, modes whose block traces both vanish do not mix under the decomposition of \cref{eq:symmetric_bundle_split}. Hence, in terms of the further York decomposition over each manifold factor, the only mixings occur for scalar-scalar modes, of which \emph{a priori} there are five: two from the external-tensor sector, two from the internal-tensor sector, and one from the vector-vector sector. The gauge condition produces two equations---one per manifold factor---generically reducing this to three independent scalar modes. As for the remaining York sectors, we need to study how the gauge condition acts. Transverse-traceless tensors, as well as transverse vector-vector modes, are automatically horizontal. Thus, the only non-trivial conditions left to account for arise in sectors of the form\footnote{The symbol $\odot \equiv \frac{1}{2}(\otimes + \otimes^\text{op})$ denotes the symmetric tensor product.} $K^\dagger T^\perp(x) \widetilde{S}(y) + V^\perp(x) \odot d\widetilde{v}(y)$ and its counterpart under exchange of the two manifold factors. Diagonalizing the canonical differential operators $\Delta_\text{L}, \Delta_1, \Delta$ over both manifold factors by choosing harmonic modes for each, for each choice of eigenvalues the gauge condition imposes a linear constraint on the two coefficients of the above linear combination, leaving one. Solving this constraint and projecting the scalar-scalar sector onto the horizontal subspace finally leaves us with several unmixed sectors and a $3 \times 3$ mixed scalar-scalar sector. \\

\noindent \textbf{Diagonalizing the fluctuation operator.} The above preparatory work simplifies the ensuing analysis significantly. As anticipated above, we combine the split of \cref{eq:symmetric_bundle_split} with the York decomposition of (co)vector and tensor modes, and take each field in the ansatz to be a harmonic of a canonical differential operator on either $S^d$ or $M_n$, namely the Lichnerowicz operator $\Delta_\text{L}$ for tensors, the Hodge operator $\Delta_1$ for covectors, and the regular Laplacian $\Delta$ for scalars. We shall denote the eigenvalues on $S^d$ (resp. $M_n$) by $\lambda$ (resp. $\mu$), and distinguish tensors, covectors and scalars by subscripts, e.g., $\lambda_\text{t},\lambda_\text{v},\lambda_\text{s}$. Moreover, superscripts will denote the corresponding component in the York decomposition: covectors have a transverse and longitudinal component, whose eigenvalues of the corresponding operators will be denoted $\lambda_\text{v}^\perp, \lambda_\text{v}^0$ (and the same with $\mu$ replacing $\lambda$) respectively; tensors have transverse-traceless, transverse-vector, traceless Hessian and trace components, whose eigenvalues of the corresponding operators will be denoted $\lambda_\text{t}^\text{TT}, \lambda_\text{t}^{\perp}, \lambda_\text{t}^\text{H}, \lambda_\text{t}^\text{tr}$ (and the same with $\mu$ replacing $\lambda$) respectively. In total, before imposing the gauge condition, we thus have $4\cdot 1 + 1 \cdot 4+ 2 \cdot 2=12$ harmonics\footnote{Notice that the eigenvalue labels we defined highlight where the corresponding harmonics come from in the York decomposition of each sector of \cref{eq:symmetric_bundle_split}, but the actual harmonics can be the same. For instance, $\lambda_\text{s}$ and $\lambda_\text{t}^\text{tr}$ both refer to scalar harmonics, albeit of different origin as graviton fluctuation modes.}. Our strategy will thus be to exploit \cref{eq:operator_identities} to reduce the action of \cref{eq:horizontal_graviton_operator} to these eigenvalues, and diagonalize each orthogonal sector.

\subsection{Spectral analysis of graviton fluctuations}\label{app:spectral_analysis}

Using the notation of the preceding section, with the combined split-York decomposition we have six sectors, labelled T-S, S-T, V-V, V-S, S-V, S-S. The corresponding harmonics, and the pairs of external (for $S^d$) and internal (for $M_n$) eigenvalues of $\Delta_\text{L}, \Delta_1, \Delta_0$, are listed in \cref{tab:split-york_sectors}.

\begin{table}[ht!]
    \centering
    \begin{tabular}{c|c|c|c}
        Label & Harmonics & $S^d$ eigenvalues & $M_n$ eigenvalues \\
        \hline
        T-S & $T^\text{TT}\widetilde{S}$ & $\lambda_\text{t}^\text{TT}$ & $\mu_\text{s}$ \\
        S-T & $S \, \widetilde{T}^\text{TT}$ & $\lambda_\text{s}$ & $\mu_\text{t}^\text{TT}$ \\
        V-V & $V^\perp \, \widetilde{V}^\perp$ & $\lambda_\text{v}^\perp$ & $\mu_\text{v}^\perp$ \\
        V-S & $(K^\dagger T^\perp) \widetilde{S}, V^\perp \, d\widetilde{v}$ & $\lambda_\text{t}^\perp, \lambda_\text{v}^\perp$ & $\mu_\text{s}, \mu_\text{v}^0$ \\
        S-V & $S (K^\dagger \widetilde{T}^\perp), dv \, \widetilde{V}^\perp$ & $\lambda_\text{s}, \lambda_\text{v}^0$ & $\mu_\text{t}^\perp, \mu_\text{v}^\perp$ \\
        S-S & $(Q\mathcal{T})\widetilde{S}, g_{S^d}\tau \, \widetilde{S}, dv \, d\widetilde{v}, S (Q\widetilde{\mathcal{T}}), S \, g_{M_n} \widetilde{\tau}$ & $\lambda_\text{t}^\text{H}, \lambda_\text{t}^\text{tr}, \lambda_\text{v}^0, \lambda_\text{s}, \lambda_\text{s}$ & $\mu_\text{s}, \mu_\text{s}, \mu_\text{v}^0, \mu_\text{t}^\text{H}, \mu_\text{t}^\text{tr}$ \\
    \end{tabular}
    \caption{The split-York decomposition of graviton fluctuations on $S^d \times M_n$, with the corresponding eigenvalue labels of the canonical operators $\Delta_\text{L}, \Delta_1, \Delta$ in the notation of \cref{app:anatomy_graviton_fluctuations}. The arguments $x,y$, the eigenvalue subscripts and the (symmetric) tensor product of vectors in the harmonics are suppressed to lighten the notation, and the pairs of external ($S^d$) and internal ($M_n$) eigenvalues are listed in corresponding orders. Notice that eigenvalues coming from different sectors are labelled differently, but their values may coincide---for instance, $\lambda_\text{s}$ and $\lambda_\text{t}^\text{tr}$ both label eigenvalues of scalar harmonics, albeit arising from different sectors with respect to the split in \cref{eq:symmetric_bundle_split}.}
    \label{tab:split-york_sectors}
\end{table}

\noindent From the decomposition of \cref{tab:split-york_sectors}, we see that the modes decompose according to $12=1+1+1+2+2+5$. \\

\noindent \textbf{Imposing the gauge condition.} We now implement the divergenceless condition $Kh=0$ on each sector. For the T-S, S-T and V-V sectors, this gauge condition is automatically satisfied. For the V-S and S-V sectors, we compute the action of $K$ on a linear combination of harmonics; for instance, for the V-S sector, one finds
\begin{eqaed}\label{V-S_gauge_operator}
    K \left( \alpha (K^\dagger T^\perp)\widetilde{S}+ \beta \, V^\perp \odot d\widetilde{v} \right) = \alpha (KK^\dagger T^\perp)\widetilde{S} + \frac{\beta}{2} \, V^\perp \, K(d\widetilde{v})\,.
\end{eqaed}
Writing out the operators in components, one sees that $K(d\widetilde{v})=2\Delta \widetilde{v} = 2 \mu_\text{v}^0 \, \widetilde{v}$, and
\begin{eqaed}\label{eq:V-S_simplification}
    (KK^\dagger T^\perp)_i & = -2 \, \nabla^j(\nabla_i T^\perp_j + \nabla_j T^\perp_i) \\
    & = 2 \left( \Delta T^\perp - \text{Ric} \circ T^\perp \right)_i \\
    & = 2 \left(\Delta_1 T^\perp - 2 \, \text{Ric} \circ T^\perp \right)_i \\
    & = 2 \left(\lambda_\text{t}^\perp - \, 2(d-1)x^2 \right) T^\perp_i\,.
\end{eqaed}
Here we used the fact that we are working in units of the internal radius $r$, so that the external Einstein constant is proportional to $r^2/L^2 = x^2$. Putting these relations together, the gauge condition for this sector reads
\begin{eqaed}\label{eq:V-S_gauge_condition}
    2 \, \alpha \left(\lambda_\text{t}^\perp - \, 2(d-1)x^2 \right) T^\perp \, \widetilde{S} + \beta \, \mu_\text{v}^0 \, V^\perp \, \widetilde{v} = 0\,.
\end{eqaed}
Analogously, for the S-V sector, a generic linear combination $\alpha \, S(K^\dagger \widetilde{T}^\perp) + \beta \, dv \odot \widetilde{V}^\perp$ ought to satisfy
\begin{eqaed}\label{eq:S-V_gauge_condition}
    2 \, \alpha \left(\mu_\text{t}^\perp - \, 2 \kappa_{M_n} \right) S \, \widetilde{T}^\perp + \beta \, \lambda_\text{v}^0 \, v \, \widetilde{V}^\perp = 0\,,
\end{eqaed}
where the Einstein constant $\kappa_{M_n}$ equals $-(n-1)$ for $M_n$ a hyperbolic manifold of curvature radius $r$, and vanishes for \acp{RFM}. Let us first consider the case in which \cref{eq:V-S_gauge_condition,eq:S-V_gauge_condition} are solved by each of the two terms vanishing individually. In this case, either the ansatz coefficients $\alpha, \beta$ vanish or the factors they multiply vanish. When the second term has vanishing eigenvalue, i.e., $\mu_\text{v}^0 = 0$ in \cref{eq:V-S_gauge_condition} or $\lambda_\text{v}^0 = 0$ in \cref{eq:S-V_gauge_condition}, the corresponding scalars $\widetilde{v}, v$ are harmonic functions on a compact manifold, thus constants. Hence, their gradients---and therefore the corresponding graviton fluctuations---vanish. In this case, the remaining first term must vanish by itself. If $\alpha = 0$, the whole graviton fluctuation trivially vanishes; otherwise, we have $KK^\dagger T=0$ (resp. $KK^\dagger \widetilde{T}=0$) in \cref{eq:V-S_gauge_condition} (resp. \cref{eq:S-V_gauge_condition}). Contracting with $T^\perp$ (resp. $\widetilde{T}^\perp$), we learn that $T^\perp$ (resp. $\widetilde{T}^\perp$) is Killing, and thus does not contribute to the graviton fluctuation due to the York decomposition of \cref{eq:tensor_york}. Conversely, if the vector multiplying $\alpha$ is Killing (or if $\alpha=0$) the scalar multiplying $\beta$ is constant. \\

\noindent As a result, the only non-trivial graviton fluctuations in the V-S and S-V sectors satisfying the gauge condition must solve \cref{eq:V-S_gauge_condition,eq:S-V_gauge_condition} by compensating two non-zero terms. Since the combined V-S or S-V harmonics appearing in \cref{eq:V-S_gauge_condition,eq:S-V_gauge_condition} are orthogonal unless the corresponding eigenvalues are the same, this can only happen when, in \cref{eq:V-S_gauge_condition} (resp. \cref{eq:S-V_gauge_condition}), we have $\lambda_\text{t}^\perp = \lambda_\text{v}^\perp$ and $\mu_\text{s} = \mu_\text{v}^0$ (resp. $\lambda_\text{s} = \lambda_\text{v}^0$ and $\mu_\text{t}^\perp = \mu_\text{v}^\perp$). This amounts to a single independent mode, both in the V-S and S-V sectors, after gauge fixing; its eigenvalue can be determined by orthogonal projection as follows. \\

\noindent \textbf{Horizontal projections of vector-scalar modes.} For concreteness we stick with the V-S sector; the results for the S-V sector can be obtain \emph{mutatis mutandis}. From the above considerations, to have non-trivial horizontal fluctuations satisfying \cref{eq:V-S_gauge_condition} the two harmonics must be the same, both for the vector and scalar mode. We choose to denote them by $V^\perp$ for the vector on $S^d$ and $\widetilde{v}$ for the scalar on $M_n$, and we normalize them to unity according to $\norm{V^\perp}_{L^2(S^d)} = 1, \norm{\widetilde{v}}_{L^2(M_n)} = 1$. With these ingredients we build the graviton fluctuation
\begin{eqaed}\label{eq:V-S_pre-horizontal_fluctuation}
    h_\text{v-s} = (K^\dagger V^\perp)\widetilde{v} + \beta \, V^\perp \odot d\widetilde{v}
\end{eqaed}
and impose \cref{eq:V-S_gauge_condition}. This fixes $\beta = -\frac{2}{\mu_\text{v}^0}(\lambda_\text{v}^\perp - 2(d-1)x^2)$. Since the two terms in \cref{eq:V-S_pre-horizontal_fluctuation} are orthogonal, the squared norm gives
\begin{eqaed}\label{eq:V-S_horizontal_norm}
    \norm{h_\text{v-s}}^2_{L^2(S^d \times M_n)} & = \norm{K^\dagger V^\perp}^2_{L^2(S^d)} + \frac{\beta^2}{2} \, \norm{d\widetilde{v}}^2_{L^2(M_n)} \\
    & = \langle V^\perp, KK^\dagger V^\perp\rangle_{L^2(S^d)} + \frac{\beta^2}{2} \langle \widetilde{v}, \Delta \widetilde{v}\rangle_{L^2(M_n)} \\
    & = 2\left(\lambda_\text{v}^\perp - 2(d-1)x^2\right) + \frac{\beta^2}{2} \, \mu_\text{v}^0\,.
\end{eqaed}
Since the horizontal subspace of the V-S sector is one-dimensional, to obtain the eigenvalue $\Lambda_\text{v-s}$ of the projection of the fluctuation operator $D_H$ we need only compute
\begin{eqaed}\label{eq:V-S_horizontal_eigenvalue}
    \Lambda_\text{v-s} = \frac{\langle h_\text{v-s}, D_H h_\text{v-s} \rangle_{L^2(S^d \times M_n)}}{\norm{h_\text{v-s}}^2_{L^2(S^d \times M_n)}}\,.
\end{eqaed}
Using \cref{eq:horizontal_graviton_operator} and the intertwining identities in \cref{eq:operator_identities}, we obtain
\begin{eqaed}\label{eq:lichnerowicz_action_v-s}
    \Delta_\text{L}(K^\dagger V^\perp \widetilde{v}) & = \Delta_\text{L}^{S^d}(K^\dagger V^\perp) \widetilde{v} + (K^\dagger V^\perp)\Delta^{M_n} \widetilde{v} \\
    & = K^\dagger (\Delta_1^{S^d}V^\perp) \widetilde{v} + (K^\dagger V^\perp) \Delta^{M_n}\widetilde{v} \\
    & = (\lambda_\text{v}^\perp + \mu_\text{v}^0)(K^\dagger V^\perp) \widetilde{v}\,, \\
    \Delta_\text{L}(V^\perp \odot d\widetilde{v}) & = (\Delta_1^{S^d} V^\perp) \odot d\widetilde{v} + V^\perp \odot \Delta_1^{M_n}(d\widetilde{v}) \\
    & = (\Delta_1^{S^d} V^\perp) \odot d\widetilde{v} + V^\perp \odot d(\Delta^{M_n} \widetilde{v}) \\
    & = (\lambda_\text{v}^\perp + \mu_\text{v}^0) (V^\perp \odot d\widetilde{v})\,.
\end{eqaed}
All in all, $(\Delta_\text{L} + R) h_\text{v-s} = (\lambda_\text{v}^\perp + \mu_\text{v}^0+R) h_\text{v-s}$ gives the action of the first two terms in \cref{eq:horizontal_graviton_operator}, where the Ricci scalar evaluates to $R = d(d-1)x^2 + n \kappa_{M_n}$. The remaining non-zero terms act via Ricci endomorphisms, which give
\begin{eqaed}\label{eq:V-S_D_H_Ricci_terms}
    \text{Ric} \circ h_\text{v-s} + h_\text{v-s} \circ \text{Ric} = 2(d-1)x^2 \, (K^\dagger V^\perp) \widetilde{v} + \left((d-1)x^2 + \kappa_{M_n} \right) \beta \, V^\perp \odot d\widetilde{v}\,.
\end{eqaed}
Putting these results together, we obtain
\begin{eqaed}\label{eq:V-S_D_H_total}
    D_H h_\text{v-s} & = \left(\lambda_\text{v}^\perp + \mu_\text{v}^0+R -4(d-1)x^2\right) (K^\dagger V^\perp) \widetilde{v} \\
    & + \left(\lambda_\text{v}^\perp + \mu_\text{v}^0+R -2(d-1)x^2 -2\kappa_{M_n} \right) \beta \, V^\perp \odot d\widetilde{v} \\
    & = \left(\lambda_\text{v}^\perp + \mu_\text{v}^0+ (d-4)(d-1)x^2 + n \kappa_{M_n}\right) (K^\dagger V^\perp) \widetilde{v} \\
    & + \left(\lambda_\text{v}^\perp + \mu_\text{v}^0+(d-2)(d-1)x^2 + (n-2)\kappa_{M_n} \right) \beta \, V^\perp \odot d\widetilde{v}\,.
\end{eqaed}
Finally, \cref{eq:V-S_horizontal_eigenvalue} evaluates to
\begin{eqaed}\label{eq:V-S_horizontal_eigenvalue_computation}
    \Lambda_\text{v-s} = \lambda_\text{v}^\perp + \mu_\text{v}^0 + (d-2)(d-1)x^2 + (n-2)\kappa_{M_n} - \, \frac{2\mu_\text{v}^0 ((d-1)x^2-\kappa_{M_n})}{\lambda_\text{v}^\perp + \mu_\text{v}^0 - 2(d-1)x^2}\,.
\end{eqaed}
The analogous computation for the S-V sector then yields
\begin{eqaed}\label{eq:S-V_horizontal_eigenvalue}
    \Lambda_\text{s-v} = \mu_\text{v}^\perp + \lambda_\text{v}^0 + (d-2)(d-1)x^2 + (n-2)\kappa_{M_n} - \, \frac{2\lambda_\text{v}^0 (\kappa_{M_n}-(d-1)x^2)}{\mu_\text{v}^\perp + \lambda_\text{v}^0 - 2\kappa_{M_n}}\,.
\end{eqaed}

\subsection{Positivity in non-scalar sectors}\label{app:positivity_non-scalar_sectors}

We can now show the positivity of the spectrum of these local operators, before moving on to the more involved problem of scalar-scalar fluctuations. \\

\noindent \textbf{Positivity in the tensor-scalar and vector-vector sectors.} As discussed above, the Lichnerowicz operator has a simple action on the T-S, S-T, and V-V sectors, all of which are automatically horizontal and thus require no gauge fixing. Using the formalism we have developed in the preceding section, we can compute the action of \cref{eq:horizontal_graviton_operator} on each sector as follows.

\begin{itemize}
    \item \textbf{T-S sector:} tensor spherical harmonics have eigenvalues $\lambda_\text{t}^\text{TT} = (l(l+d-1)+2(d-1))x^2$ with $l \geq 2$ \cite{rubin-ordonez} under the Lichnerowicz operator on $S^d$ \cite{Dahl_2022}. Hence, the total Lichnerowicz operator $\Delta_\text{L}$ on $S^d \times M$ acts as $\Delta_\text{L} h_\text{t-s} = (\lambda_\text{t}^\text{TT}+\mu_\text{s})h_\text{t-s}$. Adding the other contributions in \cref{eq:horizontal_graviton_operator}, we obtain $D_H h_\text{t-s} = (\lambda_\text{t}^\text{TT}+\mu_\text{s}+R-4(d-1)x^2)h_\text{t-s}$. The tensor eigenvalue is larger than the last term, and $R \geq 0$ both for \acp{RFM} and hyperbolic manifolds where $x_*^2 = \frac{n(n-1)}{d(d-1)}$. Therefore, the overall eigenvalue is positive.

    \item \textbf{S-T sector:} analogously as before, we have $D_H h_\text{s-t} = (\lambda_\text{s} + \mu_\text{t}^\text{TT}+R-4\kappa_{M_n})h_\text{s-t}$. Since $M_n$ is Einstein, the Lichnerowicz operator $\Delta_\text{L}$ is related to the Einstein operator $\Delta_\text{E}$ (see, e.g., \cite{Dahl_2022}) by $\Delta_\text{L} = \Delta_\text{E} + 2\kappa_{M_n}$; furthermore, for the cases of interest (\acp{RFM} and hyperbolic manifolds) the smallest eigenvalue of $\Delta_\text{E}$ is bounded below by $- \, \frac{n-2}{n-1}\kappa_{M_n}$ \cite{Kr_ncke_2014}, and thus $\mu_\text{t}^\text{TT} \geq \frac{n}{n-1} \kappa_{M_n}$. As a result, the overall eigenvalue is again positive.

    \item \textbf{V-V sector:} for these modes, $\Delta_\text{L} = \Delta^{S^d}_1 \otimes \text{id} + \text{id} \otimes \Delta^{M_n}_1$ as discussed above. Hence, $D_H h_\text{v-v} = (\lambda_\text{v}^\perp + \mu_\text{v}^\perp + R -2(d-1)x^2-2\kappa_{M_n})h_\text{v-v}$. By non-negativity of $KK^\dagger$ and \cref{eq:V-S_simplification}, we have $\lambda_\text{v}^\perp\geq2(d-1)x^2$. Since $\mu_\text{v}^\perp\geq0$, the overall eigenvalue is at least $R-2\kappa_{M_n}>0$ in both cases.
\end{itemize}

\noindent \textbf{Positivity in the vector-scalar sector.} Again in light of \cref{eq:V-S_horizontal_norm}, and its counterpart for the S-V sector, non-trivial modes have $\lambda_\text{v}^\perp > 2(d-1)x^2$ in the V-S sector; similarly $\mu_\text{v}^\perp > 2\kappa_{M_n}$ in the S-V sector. Then, for \acp{RFM} we have $\kappa_{M_n}=0$ and these eigenvalues are non-negative\footnote{For the S-V sector, non-negativity is manifest with $\kappa_{M_n}=0$. For the V-S sector, the last fraction is bounded according to $\frac{2\mu_\text{v}^0(d-1)x^2}{\lambda_\text{v}^\perp + \mu_\text{v}^0 - 2(d-1)x^2} \leq 2(d-1)x^2$, which leads to $\Lambda_\text{v-s}^\text{RFM} \geq \lambda_\text{v}^\perp - 2(d-1)x^2 + \mu_\text{v}^0 + (d-2)(d-1)x^2 \geq 0$.} for any $x$. \\

\noindent For hyperbolic manifolds, we first work at the reference ratio $x=x_*\equiv\sqrt{\frac{n(n-1)}{d(d-1)}}$, with $\kappa_{M_n}=-(n-1)$. The corrected field equations approach this ratio in a controlled limit. The extension of the following spectral bounds to a fixed neighborhood of $x_*$ is given at the end of this appendix. It is convenient to define the positive combinations $\widehat{\lambda}_\text{v}^\perp \equiv \lambda_\text{v}^\perp - 2(d-1)x_*^2$ and $\widehat{\mu}_\text{v}^0 \equiv \mu_\text{v}^0 + \lambda_\text{v}^\perp - 2(d-1)x^2$. Then, \cref{eq:V-S_horizontal_eigenvalue_computation} simplifies to
\begin{eqaed}\label{eq:V-S_eigenvalue_simplification}
    \Lambda_\text{v-s} & = \widehat{\mu}_\text{v}^0 - 2(d-1)x_*^2 + \frac{2\widehat{\lambda}_\text{v}^\perp (d+n)(n-1)}{d \widehat{\mu}_\text{v}^0} \\
    & = \frac{d (\widehat{\mu}_\text{v}^0-(d-1)x_*^2)^2 + 2(d+n)(n-1)\widehat{\lambda}_\text{v}^\perp - d(d-1)^2x_*^4}{d \widehat{\mu}_\text{v}^0}\,.
\end{eqaed}
Using the fact that $\lambda_\text{v}^\perp = (l(l+d-1)+d-2)x_*^2$ with $l \geq 1$ (see, e.g., \cite{CHODOS1984412, Bonifacio:2020xoc}), and that the Killing vectors with $l=1$ are excluded, we learn that $\lambda_\text{v}^\perp \geq 3d x_*^2$, and thus the term proportional to $\widehat{\lambda}_\text{v}^\perp$ is larger in absolute value than the last term in \cref{eq:V-S_eigenvalue_simplification}; this shows that $\Lambda_\text{v-s} > 0$. As for the S-V sector, the last fraction in \cref{eq:S-V_horizontal_eigenvalue} is non-negative. Hence, since $\mu_\text{v}^\perp \geq 0$ and $\lambda_\text{v}^0 = l(l+d-1)x_*^2 \geq dx_*^2$ by excluding the zero-mode scalar due to the above considerations, we have
\begin{eqaed}\label{eq:S-V_eigenvalue_bound}
    \Lambda_\text{s-v} \geq dx_*^2 +(d-2)(d-1)x_*^2 - (n-2)(n-1) > 0\,.
\end{eqaed}
All that is left is to analyze the scalar-scalar sector, which comprises five independent modes before imposing the gauge condition.

\subsection{Spectral bounds in the scalar-scalar sector}\label{app:spectral_bounds_scalar-scalar_sector}

In the S-S sector the analysis is more involved, since there are five modes (as displayed in \cref{tab:split-york_sectors}) and the gauge condition requires a projection onto a (generically) three-dimensional subspace. Thanks to the intertwining identities of \cref{eq:operator_identities}, the operator $D_H$ of \cref{eq:horizontal_graviton_operator} is block-diagonal in the S-S harmonic basis; as in the preceding section, the upshot is that we can pick a single pair $S, \widetilde{S}$ with $\Delta^{S^d} S=\lambda_\text{s} S$ and $\Delta^{M_n} \widetilde{S} = \mu_\text{s} \widetilde{S}$. The ansatz for the graviton fluctuation then reads
\begin{eqaed}\label{eq:S-S_graviton_fluctuation}
    h_\text{s-s} = \alpha_{Q} \, (QS)\widetilde{S} + \alpha_\text{tr} \, g_{S^d} \, S \, \widetilde{S} + \alpha_{dd} \, dS \odot d\widetilde{S} + \alpha_{\widetilde{Q}} \, S(Q\widetilde{S}) + \alpha_{\widetilde{\text{tr}}} \, g_{M_n} \, S \, \widetilde{S}\,.
\end{eqaed}
Using the same methods as in the preceding section, we compute
\begin{eqaed}\label{eq:S-S_useful_identities}
    K((QS)\widetilde{S}) & = 2 \left( \frac{d-1}{d} \, \lambda_\text{s} - (d-1)x_*^2 \right) dS \, \widetilde{S}\,, \\
    K(g_{S^d} \, S \, \widetilde{S}) & = -2 \, dS \, \widetilde{S}\,, \\
    K(dS \odot d\widetilde{S}) & = \lambda_\text{s} \, S \, d\widetilde{S} + \mu_\text{s} \, dS \, \widetilde{S}\,, \\
    K(S (Q\widetilde{S})) & = 2 \left( \frac{n-1}{n} \, \mu_\text{s} - \kappa_{M_n} \right) S \, d\widetilde{S}\,, \\
    K(g_{M_n} \, S \, \widetilde{S}) & = -2 \, S \, d\widetilde{S}\,.
\end{eqaed}
The gauge condition $K h_\text{s-s}=0$ then becomes
\begin{eqaed}\label{eq:S-S_gauge_condition}
    & \left(2 \left( \frac{d-1}{d} \, \lambda_\text{s} - (d-1)x_*^2 \right) \alpha_Q - 2 \alpha_\text{tr} + \mu_\text{s} \, \alpha_{dd} \right) dS \, \widetilde{S} = 0\,, \\
    & \left( 2 \left( \frac{n-1}{n} \, \mu_\text{s} - \kappa_{M_n} \right) \alpha_{\widetilde{Q}} -2 \alpha_{\widetilde{\text{tr}}} + \lambda_\text{s} \, \alpha_{dd} \right) S \, d\widetilde{S} = 0\,,
\end{eqaed}
or equivalently\footnote{When the scalar eigenvalues $\lambda_\text{s}$ (resp. $\mu_\text{s}$) vanishes, $S$ (resp. $\widetilde{S}$) is a harmonic function on a compact manifold, hence a constant, thus its gradient vanishes.}
\begin{eqaed}\label{eq:S-S_gauge_condition_simplified}
    & C_\text{s} \, \lambda_\text{s} \equiv \left(2 \left( \frac{d-1}{d} \, \lambda_\text{s} - (d-1)x_*^2 \right) \alpha_Q - 2 \alpha_\text{tr} + \mu_\text{s} \, \alpha_{dd} \right) \lambda_\text{s} = 0\,, \\
    & \widetilde{C}_\text{s} \, \mu_\text{s} \equiv \left( 2 \left( \frac{n-1}{n} \, \mu_\text{s} - \kappa_{M_n} \right) \alpha_{\widetilde{Q}} -2 \alpha_{\widetilde{\text{tr}}} + \lambda_\text{s} \, \alpha_{dd} \right) \mu_\text{s} = 0\,.
\end{eqaed}
Our goal is now to compute the action of the operator $D_H$ on \cref{eq:S-S_graviton_fluctuation} and project it onto the horizontal subspace defined by \cref{eq:S-S_gauge_condition_simplified}. Denoting by $\mathcal{K}_\text{s-s}$ the $2 \times 5$ coefficient matrix stemming from this linear condition in the five-component array $\alpha \equiv (\alpha_Q, \alpha_\text{tr}, \alpha_{dd}, \alpha_{\widetilde{Q}}, \alpha_{\widetilde{\text{tr}}})$, \cref{eq:S-S_gauge_condition_simplified} can be recast as the matrix equation $\mathcal{K}_\text{s-s} \alpha = 0$. \\

\noindent \textbf{The Gram matrix of scalar-scalar fluctuations.} To begin with, we compute the Gram matrix of fluctuations, since they are not normalized. Taking normalized scalar harmonics, with $\norm{S}_{L^2(S^d)} = \norm{\widetilde{S}}_{L^2(M_n)} = 1$, we need only compute the squared norms of the five modes in \cref{eq:S-S_graviton_fluctuation}, since they are mutually orthogonal. As a result, the Gram matrix $G_\text{s-s}$ is the diagonal matrix $\text{diag}(N_Q, N_\text{tr}, N_{dd}, N_{\widetilde{Q}}, N_{\widetilde{tr}})$, where\footnote{In more detail for $N_Q$, and analogously for $N_{\widetilde{Q}}$, using the definition of $Q$ and the fact that $QS$ is traceless we have $N_Q = \langle QS, QS\rangle_{L^2(S^d)} = \frac{1}{2} \langle KQS, dS \rangle_{L^2(S^d)} = \left(\frac{d-1}{d} \, \lambda_\text{s} - (d-1)x_*^2\right) \lambda_\text{s}$. The last step follows from \cref{eq:S-S_useful_identities}.}
\begin{eqaed}\label{eq:S-S_squared_norms}
    N_Q & = \left(\frac{d-1}{d} \, \lambda_\text{s} - (d-1)x_*^2\right) \lambda_\text{s}\,, \\
    N_\text{tr} & = d\,, \\
    N_{dd} & = \frac{1}{2} \, \lambda_\text{s} \mu_\text{s}\,, \\
    N_{\widetilde{Q}} & = \left(\frac{n-1}{n} \, \mu_\text{s} - \kappa_{M_n}\right) \mu_\text{s}\,, \\
    N_{\widetilde{tr}} & = n\,.
\end{eqaed}
Finally, proceeding as in the preceding section for vector-scalar modes, we compute the matrix elements of $D_H$ in the basis defined by \cref{eq:S-S_graviton_fluctuation}. This will allow us to compute the orthogonal projection onto the horizontal subspace and work out the gauge-fixed spectrum. \\

\noindent \textbf{The matrix elements of the graviton fluctuation quadratic form in the scalar-scalar sector.} To begin with, analogously as for the V-S and S-V sectors, we observe that the first three terms of \cref{eq:horizontal_graviton_operator} yield a diagonal $5 \times 5$ matrix. In particular, as before, using \cref{eq:operator_identities}, we obtain $(\Delta_\text{L} + R) h_\text{s-s} = \left( \lambda_\text{s} + \mu_\text{s} + R \right) h_\text{s-s}$. Analogously as in \cref{eq:V-S_D_H_Ricci_terms}, the Ricci endomorphism terms then act according to
\begin{eqaed}\label{eq:ricci_endomorphism_s-s_action}
    \text{Ric} \circ \left((QS)\widetilde{S}\right) + \left((QS)\widetilde{S}\right) \circ \text{Ric} & = 2(d-1)x_*^2 \, (QS)\widetilde{S}\,, \\
    \text{Ric} \circ \left(g_{S^d} \, S \, \widetilde{S}\right) + \left(g_{S^d} \, S \, \widetilde{S}\right) \circ \text{Ric} & = 2(d-1)x_*^2 \, g_{S^d} \, S \, \widetilde{S}\,, \\
    \text{Ric} \circ \left(dS \odot d\widetilde{S}\right) + \left(dS \odot d\widetilde{S}\right) \circ \text{Ric} & = \left((d-1)x_*^2 + \kappa_{M_n}\right) dS \odot d\widetilde{S}\,, \\
    \text{Ric} \circ \left(S(Q\widetilde{S})\right) + \left(S(Q\widetilde{S})\right) \circ \text{Ric} & = 2\kappa_{M_n} \, S(Q\widetilde{S})\,, \\
    \text{Ric} \circ \left(g_{M_n} \, S \, \widetilde{S}\right) + \left(g_{M_n} \, S \, \widetilde{S}\right) \circ \text{Ric} & = 2\kappa_{M_n} \, g_{M_n} \, S \, \widetilde{S}\,.
\end{eqaed}
As for the remaining three terms in \cref{eq:horizontal_graviton_operator}, we have
\begin{eqaed}\label{eq:S-S_D_H_mixing_trace_terms}
    \langle \text{Ric}, h_\text{s-s}\rangle \, g - \left(\Delta + \frac{R}{2}\right) \overline{h}_\text{s-s} \, g & + \text{Ric} \, \overline{h}_\text{s-s} = -d\Big(\lambda_\text{s} + \mu_\text{s} + \frac{R}{2}-2(d-1)x_*^2\Big) \alpha_\text{tr} \, g_{S^d} \, S \widetilde{S} \\
    & -n\Big(\lambda_\text{s} + \mu_\text{s} + \frac{R}{2} - 2\kappa_{M_n}\Big) \alpha_{\widetilde{\text{tr}}} \, g_{M_n} \, S \widetilde{S} \\
    & -n\Big(\lambda_\text{s} + \mu_\text{s} + \frac{R}{2} - (d-1)x_*^2 - \kappa_{M_n}\Big) \alpha_{\widetilde{\text{tr}}} \, g_{S^d} \, S \widetilde{S} \\
    & -d\Big(\lambda_\text{s} + \mu_\text{s} + \frac{R}{2} - \kappa_{M_n} - (d-1)x_*^2\Big) \alpha_\text{tr} \, g_{M_n} \, S \widetilde{S}\,,
\end{eqaed}
displaying the mixing due to the trace modes. We now have all the necessary ingredients: together, \cref{eq:S-S_squared_norms,eq:ricci_endomorphism_s-s_action,eq:S-S_D_H_mixing_trace_terms} define the $5 \times 5$ Gram matrix $G_\text{s-s}$ and the $5 \times 5$ fluctuation-operator matrix $\mathcal{H}_\text{s-s}$ whose entries are the coefficients of $\langle h_\text{s-s}, D_H h_\text{s-s}\rangle_{L^2(S^d \times M_n)}$ in the harmonic basis of \cref{eq:S-S_graviton_fluctuation}. Furthermore, the gauge condition of \cref{eq:S-S_gauge_condition_simplified} defines a $2 \times 5$ matrix $\mathcal{K}_\text{s-s}$ as discussed above. \\

\noindent \textbf{The horizontal spectral problem in the scalar-scalar sector.} Altogether, the matrices $\mathcal{H}_\text{s-s}, G_\text{s-s}$ and $\mathcal{K}_\text{s-s}$ define the spectral problem for the horizontal projection of the S-S sector entirely within the vector space $\mathbb{R}^5$ parametrized by the array $\alpha$. That is, we seek non-trivial fluctuations (i.e., those with $\langle \alpha, G_\text{s-s}\alpha\rangle_{\mathbb{R}^5} > 0$) such that $\mathcal{K}_\text{s-s}\alpha = 0$. Letting $P$ be a $5 \times 3$ matrix whose columns span $\ker \mathcal{K}_\text{s-s}$, the physical spectrum is given by the solutions $\Lambda_\text{s-s}$ to the (horizontally projected) secular equation
\begin{eqaed}\label{eq:S-S_secular_equation}
    \det(P^\text{t}(\mathcal{H}_\text{s-s} - \Lambda_\text{s-s}G_\text{s-s})P) = 0\,,
\end{eqaed}
as can be seen, e.g.,\footnote{Another derivation is as follows. In the generic case in which $\lambda_\text{s} > dx_*^2,\mu_\text{s} > 0$ (namely, we exclude zero-modes and the first scalar spherical harmonic, which is annihilated by the traceless Hessian operator $Q$), the Gram matrix $G_\text{s-s}$ is invertible. Hence, we can write the orthogonal projector on the horizontal subspace as $\Pi_{\ker \mathcal{K}_\text{s-s}} = I_5 - G_\text{s-s}^{-1} \mathcal{K}_\text{s-s}^\text{t} (\mathcal{K}_\text{s-s} G_\text{s-s}^{-1} \mathcal{K}_\text{s-s}^\text{t})^{-1} \mathcal{K}_\text{s-s}$ by solving the defining equations $\Pi_{\ker \mathcal{K}_\text{s-s}}^2 = \Pi_{\ker \mathcal{K}_\text{s-s}}, \mathcal{K}_\text{s-s} \Pi_{\ker \mathcal{K}_\text{s-s}} = 0$ and requiring self-adjointness with respect to the inner product induced by $G_\text{s-s}$. In terms of $P$, we have $\Pi_{\ker \mathcal{K}_\text{s-s}} = P(P^\text{t} G_\text{s-s}P)^{-1}P^\text{t} G_\text{s-s}$. Then, the projected eigenvalue equation $\Pi_{\ker \mathcal{K}_\text{s-s}} G_\text{s-s}^{-1}\mathcal{H}_\text{s-s} P \alpha = \Lambda_\text{s-s}P \alpha$ leads to the secular equation in \cref{eq:S-S_secular_equation}.} by the constrained extremization over $\widehat{\alpha} \in \mathbb{R}^3$ defined by the Lagrangian $\langle P \widehat{\alpha}, \mathcal{H
}_\text{s-s} P \widehat{\alpha} \rangle_{\mathbb{R}^5} - \Lambda_\text{s-s}(\langle P\widehat{\alpha}, G_\text{s-s} P\widehat{\alpha} \rangle_{\mathbb{R}^5} -1)$. The generic situation is that the horizontal subspace is three-dimensional; this happens when $\lambda_\text{s} > dx_*^2,\mu_\text{s} > 0$---that is, excluding zero-modes and the first scalar spherical harmonic. In this case we do not need specify such a matrix $P$; indeed, the requirements that $\mathcal{K}_\text{s-s} \alpha = 0$ and that $(\mathcal{H}_\text{s-s}-\Lambda_\text{s-s}G_\text{s-s})\alpha \in (\ker \mathcal{K}_\text{s-s})^\perp = \Im \mathcal{K}_\text{s-s}^\text{t}$ allow us to recast \cref{eq:S-S_secular_equation} according to
\begin{eqaed}\label{eq:S-S_secular_equation_simplified}
    \det\mqty(\mathcal{H}_\text{s-s} - \Lambda_\text{s-s} G_\text{s-s} & \mathcal{K}_\text{s-s}^\text{t} \\ \mathcal{K}_\text{s-s} & 0) = 0\,.
\end{eqaed}
However, even in the generic case, solving the secular equation for any scalar-scalar harmonic is rather involved, especially given that we merely want to uniformly bound the eigenvalues away from zero. \\

\noindent Fortunately, since the Gram matrix is diagonal, we can proceed in a simpler way than by direct computation. Any exceptional mode that makes $G_\text{s-s}$ singular can simply be removed from the ansatz of \cref{eq:S-S_graviton_fluctuation} by deleting the corresponding rows and columns\footnote{In more intrinsic terms, the horizontal subspace of physical fluctuations is really $\ker \mathcal{K}_\text{s-s}/(\ker \mathcal{K}_\text{s-s} \cap \ker G_\text{s-s})$.}. This reduces the $5 \times 5$ to, say, a $q \times q$ system, and the analog of the $5 \times 3$ matrix $P$ is now a $q \times p$ matrix with $p \leq \min(q,3)$. To avoid cluttering the ensuing expressions, we denote the reduced matrices with the same symbols by abuse of notation. As a result, the counterpart of the secular polynomial of \cref{eq:S-S_secular_equation} is the determinant of a $p \times p$ matrix. In this case $P^\text{t} G_\text{s-s} P$ is positive definite, and the quantities $\Lambda_\text{s-s}$ are the eigenvalues of $(P^\text{t} G_\text{s-s} P)^{-1}P^\text{t} \mathcal{H}_\text{s-s} P$. With this proviso, we proceed as follows. \\

\noindent \textbf{Bounding the spectrum of the scalar-scalar fluctuations.} Our goal is to uniformly bound the eigenvalues $\Lambda_\text{s-s}$ of fluctuations in the horizontal subspace away from zero. Reiterating the general strategy for the reader's convenience, this is because small eigenvalues are the main obstacle to extending our bounds on the scalar-field Casimir energy to the eleven-dimensional supergraviton: the steepest-descent evaluation of the semiclassical supergravity path integral takes care of negative modes; hence, we need to control the \ac{IR} behavior of the Schwinger integral, which is dominated by small eigenvalues. As discussed below \cref{eq:sugra_heat_kernel}, for the sectors in the full BRST complex in which the fluctuation operator (in some gauge) is non-negative we can recycle the techniques we developed for the scalar-field case. The upshot of the preceding analysis is that the scalar-scalar sector of graviton fluctuations can \emph{a priori} fail this requirement. Still, if we can bound the eigenvalues away from zero we can isolate the potentially problematic contribution to the Casimir energy and bound it separately. Given the spectral problem formulated above, our strategy is to lower-bound the determinant of $(P^\text{t} G_\text{s-s} P)^{-1}P^\text{t} \mathcal{H}_\text{s-s} P$ (which gives the product of the eigenvalues) and upper-bound the individual eigenvalues (in absolute value). This will lead to a uniform bound of each eigenvalue away from zero. \\

\noindent We begin by upper-bounding the individual eigenvalues. From the discussion around \cref{eq:ricci_endomorphism_s-s_action}, we observe that the action of $D_H$ contains terms involving $\lambda_\text{s}, \mu_\text{s}$ and terms involving curvatures, which are uniformly bounded. As such, we can write
\begin{eqaed}\label{eq:S-S_D_H_upper_bound}
    \abs{\langle h_\text{s-s}, D_H h_\text{s-s} \rangle_{L^2(S^d \times M_n)}} \lesssim \left(1+\lambda_\text{s} + \mu_\text{s}\right) \norm{h_\text{s-s}}^2_{L^2(S^d \times M_n)}\,,
\end{eqaed}
where $\lesssim$ indicates that there exists a (uniform) positive prefactor realizing the inequality. Since the Gram matrix induces the $L^2$-norm in the harmonic basis of \cref{eq:S-S_graviton_fluctuation}, we obtain the upper bound
\begin{eqaed}\label{eq:S-S_eigenvalue_upper_bound}
    \abs{\Lambda_\text{s-s}} \lesssim 1 + \lambda_\text{s} + \mu_\text{s}\,.
\end{eqaed}
Next, we seek a lower bound on $\det\left((P^\text{t} G_\text{s-s} P)^{-1}P^\text{t} \mathcal{H}_\text{s-s} P\right) = \det(P^\text{t} \mathcal{H}_\text{s-s} P)/\det(P^\text{t} G_\text{s-s} P)$. We first address the non-generic cases where $G_\text{s-s}$ is singular, which comprise scalar spherical harmonics with $l=0,1$ (recall that $\lambda_\text{s} = l(l+d-1)x_*^2$ on $S^d$) and internal modes with $\mu_\text{s} = 0$. Some of these cases are finitely many, and it suffices to check that the determinant is non-zero. For infinite families, we want a lower bound in terms of the scalar eigenvalues mirroring that of \cref{eq:S-S_eigenvalue_upper_bound}. We proceed in order of increasing complexity.

\begin{itemize}
    \item The case $\lambda_\text{s} = \mu_\text{s} = 0$ corresponds to a two-dimensional graviton fluctuation subspace spanned by $(\alpha_\text{tr}, \alpha_{\widetilde{\text{tr}}})$ and $S, \widetilde{S}$ constant. This subspace is horizontal, so that we can choose $P=I_2$, and the reduced Gram matrix is $G_\text{s-s} = \text{diag}(d,n)$. In both of the relevant cases for $\kappa_{M_n}$, the reduced matrix $\mathcal{H}_\text{s-s}$ is non-singular.

    \item The case $\lambda_\text{s} = dx_*^2, \mu_\text{s}=0$ corresponds to the first scalar spherical harmonic ($l=1$). Since the norm $N_Q$ vanishes, on account of \cref{eq:S-S_squared_norms}, we have $QS=0$, and once again only the two block traces parametrized by $(\alpha_\text{tr}, \alpha_{\widetilde{\text{tr}}})$ survive. However, in this case $S$ is not constant; the gauge condition of \cref{eq:S-S_gauge_condition_simplified} implies $\alpha_\text{tr} = 0$. The relevant determinant $(\mathcal{H}_\text{s-s})_{55}/(G_\text{s-s})_{55}$ is non-zero.

    \item The case in which $\mu_\text{s}=0$ is accompanied by scalar spherical harmonics with $l \geq 2$ corresponds to the subspace spanned by $(\alpha_Q, \alpha_\text{tr}, \alpha_{\widetilde{\text{tr}}})$. The gauge condition of \cref{eq:S-S_gauge_condition_simplified} fixes $\alpha_\text{tr} = \left(\frac{d-1}{d}\lambda_\text{s} - (d-1)x_*^2\right)  \alpha_Q$, and thus we can choose
    \begin{eqaed}\label{eq:P_matrix_mu=0_l2}
        P = \mqty(1 & 0 \\ \frac{d-1}{d}\lambda_\text{s} - (d-1)x_*^2 & 0 \\ 0 & 1)\,.
    \end{eqaed}
    The reduced Gram determinant $\det(P^\text{t} G_\text{s-s}P) = n(d-1)(\lambda_\text{s}-(d-1)x_*^2)(\lambda_\text{s}-dx_*^2)$ is non-zero and quadratic in $\lambda_\text{s}$, while the reduced matrix $\mathcal{H}_\text{s-s}$ is block-diagonal. Combining this with \cref{eq:P_matrix_mu=0_l2}, we find
    \begin{eqaed}\label{eq:H_matrix_mu=0_l2_det}
        \det(P^\text{t} \mathcal{H}_\text{s-s} P) & = \left(H_{11} + \left(\frac{d-1}{d} \, \lambda_\text{s} - (d-1)x_*^2\right)^{\!2} H_{22} \right) H_{55} \\
        & - \left(\frac{d-1}{d} \, \lambda_\text{s} - (d-1)x_*^2\right)^{\!2} \! H_{25}^2\,.
    \end{eqaed}
    This expression grows as $\lambda_\text{s}^4$ for large $\lambda_\text{s}$, and thus its ratio to the reduced Gram determinant is lower-bounded in absolute value by $\lambda_\text{s}^2$ up to a constant. Moreover, for $2\leq n\leq7$ and $d=11-n$, expanding \cref{eq:H_matrix_mu=0_l2_det} in powers of $l-2$ gives strictly negative coefficients, including the constant term, in both cases $\kappa_{M_n}=0,-(n-1)$. Hence it is non-zero for every $l\geq2$. This exhausts the exceptional cases with $\mu_\text{s}=0$.

    \item The case $l=0, \mu_\text{s} > 0$ corresponds to a subspace spanned by $(\alpha_\text{tr}, \alpha_{\widetilde{Q}}, \alpha_{\widetilde{\text{tr}}})$. The gauge condition of \cref{eq:S-S_gauge_condition_simplified} fixes $\alpha_{\widetilde{\text{tr}}} = \left(\frac{n-1}{n} \mu_\text{s} - \kappa_{M_n}\right) \alpha_{\widetilde{Q}}$. Along the lines of the preceding cases, we thus choose
    \begin{eqaed}\label{eq:P_matrix_mu_l=0}
        P = \mqty(1 & 0 \\ 0 & 1 \\ 0 & \frac{n-1}{n} \mu_\text{s} - \kappa_{M_n})\,.
    \end{eqaed}
    The reduced Gram determinant $\det(P^\text{t} G_\text{s-s} P) = d(\mu_\text{s}-\kappa_{M_n})((n-1)\mu_\text{s}-n\kappa_{M_n})$ is non-zero and grows as $\mu_\text{s}^2$ for large $\mu_\text{s}$. Once again, the reduced matrix $\mathcal{H}_\text{s-s}$ is block-diagonal and
    \begin{eqaed}\label{eq:H_matrix_mu_l=0_det}
        \det(P^\text{t} \mathcal{H}_\text{s-s} P) & = \left(H_{44} + \left(\frac{n-1}{n} \, \mu_\text{s} - \kappa_{M_n}\right)^{\!2} H_{55} \right) H_{22} \\
        & - \left(\frac{n-1}{n} \, \mu_\text{s} - \kappa_{M_n}\right)^{\!2} \! H_{25}^2\,.
    \end{eqaed}
    grows as $\mu_\text{s}^4$ for large $\mu_\text{s}$. Furthermore, it is non-zero in any dimension for any $\mu_\text{s} > 0$ and in both cases $\kappa_{M_n} = 0, -(n-1)$. More precisely, the numerator determinant approaches zero for small $\mu_\text{s}$ when $\kappa_{M_n}=0$, but the ratio of determinants is uniformly bounded away from zero. The upshot is that the determinant ratio is lower-bounded in absolute value by $\mu_\text{s}^2$ up to a constant.

    \item The case $l=1, \mu_\text{s} > 0$ corresponds to a subspace spanned by $(\alpha_\text{tr}, \alpha_{dd}, \alpha_{\widetilde{Q}}, \alpha_{\widetilde{\text{tr}}})$, since once again $QS = 0$. The gauge condition of \cref{eq:S-S_gauge_condition_simplified} fixes $\alpha_\text{tr} = \frac{\mu_\text{s}}{2} \, \alpha_{dd}$, and
    \begin{eqaed}\label{eq:alpha_trtilde_l=1_mu}
        \alpha_{\widetilde{\text{tr}}} = \frac{\lambda_\text{s}}{2} \, \alpha_{dd} + \left(\frac{n-1}{n} \mu_\text{s} - \kappa_{M_n}\right) \alpha_{\widetilde{Q}}\,.
    \end{eqaed}
    The resulting $4 \times 2$ matrix $P$ reads
    \begin{eqaed}\label{eq:P_matrix_mu_l=1}
        P = \mqty(\frac{\mu_\text{s}}{2} & 0 \\ 1 & 0 \\ 0 & 1 \\ \frac{\lambda_\text{s}}{2} & \frac{n-1}{n} \mu_\text{s} - \kappa_{M_n})\,.
    \end{eqaed}
    Once again the Gram determinant is non-zero, but this time we have $\det(P^\text{t} G_\text{s-s} P) = \frac{\mu_\text{s}}{4}((n-1)\mu_\text{s} - n\kappa_{M_n})\left((2\lambda_\text{s} + d\mu_\text{s})(\mu_\text{s}-\kappa_{M_n})+\lambda_\text{s}^2\right)$ which grows as $\mu_\text{s}^4$. The ratio of determinants is also uniformly bounded away from zero, and the numerator grows as $\mu_\text{s}^6$, thus once more giving a lower bound of $\mu_\text{s}^2$ up to a constant on the (absolute value of the) ratio.
\end{itemize}

\noindent All in all, we learn that the ratio of determinants is lower-bounded in absolute value by $\text{const.} \cdot (1+\lambda_\text{s}+ \mu_\text{s})^p$ in all the exceptional cases, with $p=1,2$ the dimension of the horizontal subspace of physical modes. We are left with the generic case $l \geq 2, \mu_\text{s} > 0$, in which $p=3$. In this case we have the full set $\alpha$ coefficients, pertaining to the ansatz in \cref{eq:S-S_graviton_fluctuation}, and the gauge condition of \cref{eq:S-S_gauge_condition_simplified} can be used to fix
\begin{eqaed}\label{eq:S-S_generic_gauge_alpha}
    \alpha_\text{tr} & = \left(\frac{d-1}{d} \lambda_\text{s} - (d-1)x_*^2\right) \alpha_{Q} + \frac{\mu_\text{s}}{2} \, \alpha_{dd}\,, \\
    \alpha_{\widetilde{\text{tr}}} & = \frac{\lambda_\text{s}}{2} \, \alpha_{dd} + \left(\frac{n-1}{n} \mu_\text{s} - \kappa_{M_n}\right) \alpha_{\widetilde{Q}}\,.
\end{eqaed}
The corresponding matrix $P$ takes the form
\begin{eqaed}\label{eq:S-S_generic_P-matrix}
    P = \mqty(1 & 0 & 0 \\ \frac{d-1}{d} \lambda_\text{s} - (d-1)x_*^2 & \frac{\mu_\text{s}}{2} & 0 \\ 0 & 1 & 0 \\ 0 & 0 & 1 \\ 0 & \frac{\lambda_\text{s}}{2} & \frac{n-1}{n} \mu_\text{s} - \kappa_{M_n})\,.
\end{eqaed}
The Gram determinant evaluates to
\begin{eqaed}\label{eq:S-S_generic_Gram_det}
    \det(P^\text{t} G_\text{s-s} P) & = \frac{d-1}{4} \, \lambda_\text{s} \mu_\text{s} \left(\lambda_\text{s} - dx_*^2\right)\left((n-1)\mu_\text{s} - n \kappa_{M_n}\right) \\
    & \times \left((\lambda_\text{s}-(d-1)x_*^2)(\lambda_\text{s}+2\mu_\text{s}-2\kappa_{M_n})+\mu_\text{s}(\mu_\text{s}-\kappa_{M_n})\right)\,,
\end{eqaed}
which is manifestly positive in the cases of interest. Moreover, the last bracket grows as $(\lambda_\text{s}+\mu_\text{s})^2$, so without loss of generality we can write $\det(P^\text{t} G_\text{s-s}P) \sim \lambda_\text{s}^2 \mu_\text{s} (\mu_\text{s}-\kappa_{M_n}) (\lambda_\text{s} + \mu_\text{s})^2$ to take into account limits in which $\mu_\text{s} \to 0$ while $\lambda_\text{s} \to \infty$. The numerator determinant $\det(P^\text{t} \mathcal{H}_\text{s-s} P)$ is more involved, but from the analysis of the exceptional cases we can guess---and verify explicitly---that it vanishes when $\lambda_\text{s}=0,dx_*^2$ and when $\mu_\text{s}=0,n \kappa_{M_n}/(n-1)$. Thus we can write
\begin{eqaed}\label{eq:S-S_Hss_det_factor}
    \det(P^\text{t} \mathcal{H}_\text{s-s} P) = \lambda_\text{s} (\lambda_\text{s}-dx_*^2) \mu_\text{s} ((n-1)\mu_\text{s}-n\kappa_{M_n}) \, \text{D}
\end{eqaed}
for some polynomial $\text{D}$. For all $2 \leq n \leq 7$ and in both cases $\kappa_{M_n} = 0, -(n-1)$, by writing $\lambda_\text{s} = l(l+d-1)x_*^2$ with $l=2+u$ one finds polynomials in $u$ and $\mu_\text{s}$ with all negative coefficients. The limit $\mu_\text{s} \to 0$ yields a degree-ten polynomial in $u$ whose constant term is non-zero as well. In addition, $\text{D} \sim \mu_\text{s}^5$ for large $\mu_\text{s}$ at fixed $u$. As a result, $\text{D}$ is uniformly bounded away from zero, and we have $\abs{D} \geq \text{const.}(1+u^{10}+\mu_\text{s}^5) \geq \text{const.}(1+\lambda_\text{s} + \mu_\text{s})^5$. Combined with \cref{eq:S-S_generic_Gram_det}, we learn that
\begin{eqaed}\label{eq:S-S_det_ratio_bound}
    \abs{\frac{\det(P^\text{t} \mathcal{H}_\text{s-s} P)}{\det(P^\text{t} G_\text{s-s} P)}} \gtrsim (1+\lambda_\text{s}+\mu_\text{s})^p
\end{eqaed}
with $p=3$ in the general case, as befits the pattern hinted at by the exceptional cases. \\

\noindent Finally, combining \cref{eq:S-S_eigenvalue_upper_bound,eq:S-S_det_ratio_bound} we obtain the sought-after result: letting $\{\Lambda_a\}_{a \in \mathcal{I}}$ denote the set of scalar-scalar eigenvalues we labelled by $\Lambda_\text{s-s}$ thus far, we have
\begin{eqaed}\label{eq:S-S_eigenvalue_final_bound}
    \abs{\Lambda_a} & = \frac{\prod_{b \in \mathcal{I}} \abs{\Lambda_b}}{\prod_{b \in \mathcal{I} : b \neq a} \abs{\Lambda_b}} \\
    & = \frac{1}{\prod_{b \in \mathcal{I} : b \neq a} \abs{\Lambda_b}} \abs{\frac{\det(P^\text{t} \mathcal{H}_\text{s-s} P)}{\det(P^\text{t} G_\text{s-s} P)}} \\
    & \gtrsim \frac{(1+\lambda_\text{s}+\mu_\text{s})^p}{(1+\lambda_\text{s}+\mu_\text{s})^{p-1}} \\
    & = 1+\lambda_\text{s}+\mu_\text{s}\,.
\end{eqaed}
To sum up, \cref{eq:S-S_eigenvalue_upper_bound,eq:S-S_eigenvalue_final_bound} provide the two-sided bound
\begin{eqaed}\label{eq:S-S_eigenvalue_two-sided_bound}
    \text{const.}^{(1)} \cdot \left(1 + \lambda_\text{s} + \mu_\text{s} \right) \leq \abs{\Lambda_\text{s-s}} \leq \text{const.}^{(2)} \cdot \left(1 + \lambda_\text{s} + \mu_\text{s} \right) 
\end{eqaed}
on the scalar-scalar eigenvalues of horizontal graviton fluctuations. This bound can be employed to control the \ac{IR} behavior of the supergravity heat kernel. In this form, \cref{eq:S-S_eigenvalue_two-sided_bound} is not uniform as $x \to 0$, which could happen in \acp{RFM}. However, when $\kappa_{M_n} = 0$ the eigenvalue problem becomes homogeneous in $\lambda_\text{s}, \mu_\text{s}$; hence, factoring out $x^2$ and applying \cref{eq:S-S_eigenvalue_two-sided_bound} to the reduced problem (where effectively $x=1$) yields the analogous bound with $x^2 + \lambda_\text{s} + \mu_\text{s}$ in place of $1 + \lambda_\text{s} + \mu_\text{s}$, and uniform constants. \\

\noindent\textbf{Extension away from the reference ratio.} For hyperbolic $M_n$, the preceding bounds remain uniform in a fixed neighborhood of $x_*$. Indeed, after Gram normalization, the scalar-gradient block of $G=KK^\dagger$ is bounded below by a positive multiple of $1+\lambda_\text{s}+\mu_\text{s}$. The determinant of the generic two-dimensional block has positive coefficients as a polynomial in $\mu_\text{s}$, and its polynomial value at $\mu_\text{s}=0$ equals $8(\lambda_\text{s}-(d-1)x^2)(\lambda_\text{s}+2(n-1))$. The active $l=0$ block is $4(\mu_\text{s}+n-1)$. The Gram-normalized $K$ and $x\partial_xK$ have norm $\lesssim\sqrt{1+\lambda_\text{s}+\mu_\text{s}}$. Differentiating the orthogonal projection $\Pi=I-K^\dagger G'^{-1}K$, where the prime omits the Killing kernel, therefore gives
\begin{equation}
	\norm{x\partial_x\Pi}\lesssim1 \, ,\qquad
	\norm{x\partial_x(\Pi D_H\Pi)}\lesssim1+\lambda_\text{s}+\mu_\text{s} \, ,
\end{equation}
with the horizontal spaces identified by unitary transport. Together with \cref{eq:S-S_eigenvalue_two-sided_bound}, this preserves its lower bound, up to a fixed factor. Vanishing basis vectors are deleted as above. In the V-S sector the corresponding projection depends only on the direction $(\sqrt{\widehat\lambda},\sqrt\mu)$, with $\widehat\lambda \equiv \lambda_\text{v}^\perp-2(d-1)x^2$. Its logarithmic derivative has norm $\sqrt{\widehat\lambda\mu}/(\widehat\lambda+\mu)\leq\frac12$; the Killing case gives only a pure-gauge mixed mode. Thus no inverse small internal eigenvalue occurs. The other non-scalar bounds follow directly from their displayed eigenvalues. These estimates justify using the reference spectrum for the nearby stationary ratios, rather than assuming exact equality to $x_*$.

\section{Bounding the supergravity Casimir potential}\label{app:bounding_sugra_casimir}

In this appendix we extend the bounds obtained from the scalar-field Casimir energy to the actual spectrum of eleven-dimensional supergravity\footnote{It is worth mentioning that the Casimir energy of minimal supergravity on $\mathbb{R}^3 \times S^1$ was studied in \cite{Tong:2014era}.}. The one-loop Casimir energy functional for the full eleven-dimensional supergraviton on $\text{dS}_d \times M_n$ is a gauge-dependent quantity, since the geometry does not solve the classical field equations. However, its value at a critical point---that is, a solution of the quantum effective action---is gauge-independent, up to higher-loop corrections. Notice that in this setting, unless scale-separation is achieved, the flat-space approximation is not well-defined, since there is no critical point on $\mathbb{R}^{d-1,1} \times M_n$ even for the quantum effective action (approximated by the Einstein-Hilbert, flux, and Casimir terms). \\

\noindent In our case, $M_n$ is hyperbolic or flat (i.e., a \ac{RFM}). Most of the ensuing discussion applies to any compact Einstein manifold of non-positive curvature, but we will stick with this terminology for some mathematical steps and for uniformity with the rest of the paper. For the \ac{RFM} case, there is also $G_4$-flux in the background, which enters the kinetic operators for fluctuations; however, in a parametrically controlled regime, we can neglect this contribution. To see why, we observe that the classical energy density $\rho_\text{flux} \propto \norm{G_4}^2 = \mathcal{Q}(N)$ of the $G_4$-flux defines a mass scale $\mu_\text{flux}$ via $\rho_\text{flux} \equiv M_{11}^9 \mu_\text{flux}^2$. This mass scale is what can replace derivatives in the kinetic operators; as a result, the one-loop Casimir energy expressed in terms of $\mu_\text{flux}$ contains no powers of the Planck scale $M_{11}$. \\

\noindent By parity invariance in the external spacetime, the first correction to the (averaged) Casimir energy density $A = \mathcal{C}/v$ due to background fluxes entering the kinetic operators scales as $\mu_\text{flux}^2/\ell^9$, where $\ell$ is some internal length scale which satisfies $M_{11} \ell \gg 1$ by assumption. Hence, this correction is subleading with respect to the classical flux term, which is also strictly positive due to Dirac quantization. The flux-induced change in the contribution of a fixed set of non-zero light modes, whose physical eigenvalues remain of order $L^{-2}$ with and without flux, is $O((v r^n L^d)^{-1})$. Relative to the classical flux term, this is suppressed by $O((v(M_{11}r)^n(M_{11}L)^{d-2})^{-1})$. The same suppression holds for first logarithmic radius and regular shape derivatives away from additional zero eigenvalues. Since the curvature radius $r$ of the internal manifold extremizes to $M_{11} r \propto (A/\mathcal{Q}(N))^{1/3}$, parametric control requires $A \gg \mathcal{Q}(N)$, which means that there is no cancellation that would be sensitive to generically subleading corrections, except possibly for flat directions. In that case, however, not only would the contributions we are discussing be relevant, but also the two-loop terms will be relevant---in fact, they are of the same order, since the relative suppression $(M_{11}\ell)^{-9}$ is the same. Our analysis is restricted to one-loop level. \\

\noindent With these premises, the (renormalized) dimensionless Casimir coefficient $A$ is given, in terms of the heat kernel traces for the two factors, by the Schwinger integral
\begin{eqaed}\label{eq:sugra_renormalized_A}
    A & = \frac{x^d}{2\omega_d v} \int_0^{t_*} \frac{dt}{t} \left(K^\text{SUGRA}_{S^d \times M_n}(t) - [K^\text{SUGRA}_{S^d \times M_n}(t)]_\text{UV} \right) + \frac{x^d}{2\omega_d v} \int_{t_*}^\infty \frac{dt}{t} \, K^\text{SUGRA}_{S^d \times M_n}(t) \\
    & {+\frac{x^d}{2\omega_d v}\sum_{k=0}^{5}\frac{a_k\,t_*^{k-\frac{11}{2}}}{k-\frac{11}{2}}-\frac{b x^d}{2\omega_d v}(\log t_*+\gamma_\text{E})} \\
    & \equiv A_\text{id} + A_\text{UV} + A_\text{IR}-\frac{b x^d}{2\omega_d v}(\log t_*+\gamma_\text{E})\,,
\end{eqaed}
where $\omega_d$ is the volume of the unit $d$-sphere and $t_*$ is an arbitrary splitting time in the fixed-reference zeta prescription of \cref{eq:scaling_action} \cite{Vassilevich:2003xt}. With zero-modes omitted, the subtraction is
\begin{equation}
    [K^\text{SUGRA}_{S^d\times M_n}(t)]_\text{UV}
    =\sum_{k=0}^{5}a_k\,t^{k-\frac{11}{2}}-b\,,
\end{equation}
where $a_k$ are the local heat-kernel coefficients. As in \cref{eq:schwinger_split}, the finite terms in \cref{eq:sugra_renormalized_A} restore the analytically continued integrals of these subtractions, making the result independent of $t_*$. Here $\gamma_\text{E}$ is Euler's constant. The local sum is included in $A_\text{id}$. At fixed kernel dimensions, the finite $b$-term is constant in the action and drops out of $\mathcal A$. More precisely, \cref{eq:sugra_renormalized_A} should exclude zero-modes and rotate negative modes according to the steepest-descent procedure that defines the semiclassical path integral \cite{Witten:2010cx}. For a fixed number of zero-modes, their subtraction from the heat trace is independent of $x$ and therefore does not contribute to $\mathcal{A}=A-\frac{x}{d}\partial_x A$. This statement concerns variation of $x$ at fixed $r$. The full functional also contains $-b\log(\mu r)$, as in \cref{eq:scaling_action}; its common-scale derivative must be retained. The \ac{UV} part $A_\text{UV}$ is defined by separating the identity term $A_\text{id}$, which for the manifolds of interest is universal---it only depends on the universal cover $\widetilde{M_n}$. This term is also the one containing the subtraction of local terms (and thus of \ac{UV}-divergences). In local coordinates $x$ on $S^d$ and $y$ on $M_n$, the supergravity heat kernel takes the schematic form
\begin{eqaed}\label{eq:sugra_heat_kernel}
    K^\text{SUGRA}_{S^d \times M_n}(t;(x,y),(x',y')) = \!\!\!\! \sum_{(f_\text{ext},f_\text{int}) \in \text{fields}(S^d \times M_n)} \!\!\!\! K^{f_\text{ext}}_{S^d}(t;x,x') \, K^{f_\text{int}}_{M_n}(t;y,y')\,, 
\end{eqaed}
where the set of tensor/spinor representations $\text{fields}(S^d \times M_n)$ associated with the spacetime frame bundle is labeled by pairs $(f_\text{ext}, f_\text{int})$. More precisely, as we shall explore in detail later, there can be non-trivial mixings between the two manifold factors in the total heat kernel; we keep \cref{eq:sugra_heat_kernel} in this form for ease of presentation. For each Laplace-type sector with label $f$ representing an associated bundle $\mathcal{V}_f \to S^d$ or $\mathcal{V}_f \to M_n$, there is a corresponding fluctuation operator $D_f = (\nabla + A_f)^\dagger (\nabla + A_f) + \mathcal{E}_f$, where $A_f$ is a connection on $\mathcal{V}_f$ and $\mathcal{E}_f$ an endomorphism on $\mathcal{V}_f$. When deriving spectral bounds, it is convenient to choose a gauge and work with the BRST complex of the supergraviton. As a result of gauge fixing, the horizontal fluctuations will not decouple, and the kinetic operator have to be diagonalized. We will discuss this in detail in the following. \\

\noindent Having developed the necessary spectral analysis in \cref{app:spectral_analysis_supergraviton_fluctuations}, we now turn to bounding the Casimir energy for eleven-dimensional supergravity on $\text{dS}_d \times M_n$. More precisely, in the Lorenz gauge introduced in \cref{app:spectral_analysis_supergraviton_fluctuations}, we will obtain parametric bounds on the \ac{EFT} control parameter $\mathcal{A}^\text{SUGRA}$ defined by\footnote{Recall that the combination $A^\text{SUGRA}(x) - \frac{x}{d} \partial_x A^\text{SUGRA}(x)$ contains the $x$-derivative of the combination $K^\text{SUGRA}/\text{Vol}(S^d) \propto \frac{x^d}{\omega_d} \, K^\text{SUGRA}$. Simplifying, there remains only a derivative acting on the heat kernel.}
\begin{eqaed}\label{eq:A_sugra_control_parameter}
    \mathcal{A}^\text{SUGRA} & = -\frac{x^{d+1}}{2d\omega_d v} \int_0^{t_*} \frac{dt}{t} \, \partial_x\left(K^\text{SUGRA}_{S^d \times M_n}(t) - [K^\text{SUGRA}_{S^d \times M_n}(t)]_\text{UV}\right) \\
    & - \frac{x^{d+1}}{2d\omega_d v} \int_{t_*}^\infty \frac{dt}{t} \, \partial_x K^\text{SUGRA}_{S^d \times M_n}(t) \\
    & {-\frac{x^{d+1}}{2d\omega_d v}\sum_{k=0}^{5}\frac{(\partial_x a_k)\,t_*^{k-\frac{11}{2}}}{k-\frac{11}{2}}} \\
    & \equiv \mathcal{A}^\text{SUGRA}_\text{id} + \mathcal{A}^\text{SUGRA}_\text{UV} + \mathcal{A}^\text{SUGRA}_\text{IR}\,,
\end{eqaed}
where we defined the analog decomposition of \cref{eq:sugra_renormalized_A} with respect to a fixed $t_*$. {The finite local sum is included in $\mathcal A^\text{SUGRA}_\text{id}$, making the full expression independent of $t_*$. In this prescription the completely flat identity contribution is scaleless and vanishes, as in \cref{eq:casimir_RFM_formula}. For an \ac{RFM}, the remaining identity density is proportional to $L^{-11}$, since $L$ is the only geometric length in the identity heat-kernel density. Its logarithmic radius derivatives have the same scaling, and it is independent of the internal shape.} Then, the identity term is local and universal, hence uniformly bounded. Where possible, the strategy is to reduce to the case of a scalar field, which we already study in detail. We discuss each contribution separately, starting with the sectors whose local fluctuation operator is non-negative. \\

{\noindent\textbf{The common-scale contribution.} With flux-dependent terms in the kinetic operators neglected, the coefficient $b$ is controlled by zero-mode counts, even when it does not vanish. A local Laplace-type kinetic operator on the closed eleven-dimensional product has $\zeta_D(0)=-\dim\ker D$. For the horizontal graviton, put $G=KK^\dagger$, $\Pi=I-K^\dagger G'^{-1}K$ and $\mathcal H=\Pi D_H\Pi|_{\text{Im}\,\Pi}$. The canonical trace for odd-class\footnote{Here ``odd-class'' refers to the parity $\sigma_j(x,-\xi)=(-1)^j\sigma_j(x,\xi)$ of the homogeneous symbol terms; differential operators, their elliptic inverses and the projection above have this property.} pseudodifferential operators is cyclic in odd dimension and vanishes on the identity \cite{KontsevichVishik1994}. Consequently
\begin{equation}
	\tr\Pi=\dim\ker G,\qquad
	\zeta_{|\mathcal H|}(0)=\dim\ker G-\dim\ker\mathcal H.
\end{equation}
Applying the trace theorem to $\mathcal H^2$, extended elliptically to the complementary subspace, proves the formula also with infinitely many conformal negative modes. The BRST weights thus express $b$ as a fixed linear combination of finite kernel dimensions, including residual gauge kernels. \\

\noindent These counts have a uniform geometric bound. To see this also for the projected graviton, set $q=\lambda_\text{s}+\mu_\text{s}$, $k=\max((d-1)x^2,|\kappa_{M_n}|)$ and $h_0=gS\widetilde S$, with normalized scalar harmonics. Then \cref{eq:horizontal_graviton_operator} gives, on an S-S block,
\begin{equation}
	\mathcal H=q\left(I-\Pi h_0(\Pi h_0)^\text{t}\right)+\Pi E\Pi,
	\qquad\norm{E}\leq\left(4+3D+\frac{D^2}{2}\right)k<100k.
\end{equation}
For $q\geq4k$, the gradient ghost matrix obeys $G\geq qI$. Since $Kh_0=-2d(S\widetilde S)$, this implies $\norm{\Pi h_0}^2\geq D-4=7$. The principal bracket has absolute eigenvalues at least one, so no kernel occurs for $q>100k$. The traceless blocks obey the analogous curvature threshold directly. It is therefore enough to count low eigenvalues on complete tensor bundles, where scalar heat-kernel domination applies \cite{Hess:1977gb,LiYau1986}. \\

\noindent For hyperbolic $M_n$, with $s \equiv \text{sys}(M_n)$, choose $u=\min(1,1/x)$ and heat time $u^2$. The sphere trace is $\lesssim(xu)^{-d}$, while an embedded ball of radius $\min(u,s)/2$ gives the internal bound $\lesssim vu^{-n}\max(1,(u/s)^n)$. Thus
\begin{equation}
	\abs{b}\lesssim vx^{-d}u^{-11}\max\left(1,\left(\frac us\right)^n\right).
\end{equation}
Substitution in \cref{eq:x_extremum}, with $H_n^\text{hyp}=\min(M_{11}r,M_{11}L,M_{11}rs)$, yields
\begin{equation}
	\frac{\abs{d(d-1)x^2-n(n-1)}}{1+x^2}\lesssim (H_n^\text{hyp})^{-9}.
\end{equation}
Because $n<9$, this bound is uniform even when the volume grows and the systole shrinks simultaneously. Its left-hand side has positive limits at $x=0$ and $x=\infty$, so every controlled family satisfies $x^2=x_*^2+O((H_n^\text{hyp})^{-9})$. The spectral extension established at the end of \cref{app:spectral_analysis_supergraviton_fluctuations} then applies. The geometric spin and tangent-twisted Dirac operators also retain a positive external sphere gap \cite{HommaTomihisa2020}; their squared product operators add the non-negative internal square. \\

\noindent For an \ac{RFM}, all curvature endomorphisms instead scale as $x^2$. Taking heat time $x^{-2}$ gives $|b|\lesssim v(x^n+s^{-n})$. The additional terms in the flux balance are therefore relatively $O(K_{\text{RFM}}^{-9})$, where $K_{\text{RFM}}=\min(M_{11}L,M_{11}rs)$. No cancellation of $b$ is required. These conclusions apply when flux-dependent terms in the kinetic operators are neglected. Flux-dependent operators can have an additional scale derivative, so the preceding restriction away from extra zero eigenvalues remains necessary.
}

\subsection{Bounding non-negative sectors}\label{app:bounding_non-negative_sectors}

For the three-form, the gravitino and their ghosts, the local fluctuation operator $D$ is non-negative in the Lorenz gauge. $D=(\nabla+A)^\dagger (\nabla+A)+\mathcal{E}$ is a Laplace-like operator whose endomorphism is bounded below by a multiple $-E$ of the identity, with $E \geq 0$, since it is composed of curvature tensors. {The diffeomorphism ghost is also non-negative, but requires the separate reduction given below.} For the \ac{UV} winding term, we can bound \cite{Hess:1977gb, OuhabazOuhabaz+2005, güneysu2012generalizedschrodingersemigroups}
\begin{eqaed}\label{eq:kato_HSU_bound}
    \abs{K^\text{SUGRA}_{S^d \times M_n}(t;(x,y),(x',y')} \lesssim e^{E \, t} \, K_{S^d \times M_n}(t;(x,y),(x',y'))
\end{eqaed}
in terms of the heat kernel of the scalar Laplacian. Importantly, this bound holds pointwise. These operators split\footnote{After gamma-trace gauge-fixing, the gravitino operator is a Dirac operator \cite{Esposito:1996se} whose induced Laplace-like operator splits on $S^d \times M_n$.} as $D = x^2 D_{S^d} \otimes \text{id} + \text{id} \otimes D_{M_n}$, and thus acting with $-x\partial_x$ on the heat kernel brings a factor of $2tx^2D_{S^d}$. Using the fact that $z e^{-z} \leq e^{-z/2}$ for $z \geq 0$, and that\footnote{This follows from the fact that the ratio between these functions of $z$ is continuous and has finite limits at small and large $z$, and is thus bounded by a constant. The reverse inequality also holds, since $K_{S^d}$ is decreasing in its argument.} $K_{S^d}(z/2) \lesssim K_{S^d}(z)$, we bound the Casimir parameter for a given sector $\text{f}$ as
\begin{eqaed}\label{eq:x-derivative_HSU}
    \abs{\mathcal{A}_\text{UV}^f} & \lesssim \frac{x^d}{d\omega_d v} \int_{M_n} \text{vol}_{M_n}(y) \int_0^{t_*} \frac{dt}{t} \, tx^2 D_{S^d} \, K_{S^d}^f(t) \sum_{g \in \pi_1(M_n) : g \neq \text{id}} \abs{K_{\widetilde{M_n}^f}(t; y, g \cdot y)} \\
    & \lesssim \frac{x^d}{d\omega_d v} \int_{M_n} \text{vol}_{M_n}(y) \int_0^{t_*} \frac{dt}{t} \, e^{Et} \, K_{S^d}(t) \sum_{g \in \pi_1(M_n) : g \neq \text{id}} \abs{K_{\widetilde{M_n}}(t; y, g \cdot y)} \\
    & \lesssim e^{E t_*} \, \mathcal{A}_\text{UV}^\text{scalar}\,.
\end{eqaed}
\noindent At fixed $t_*$ and for the bounded values of $x$ considered here, the last inequality follows because $K_{S^d}(t)/[-x\partial_xK_{S^d}(t)]$ is continuous for $x^2t>0$ and tends to $1/d$ as $x^2t \to 0$. Hence, it is bounded on $0 < t\leq t_*$. The implicit constants may depend on dimension, sector and the fixed upper bounds for $t_*$ and $x^2t_*$, but not on the internal volume or systole. As for the \ac{IR} contribution to \cref{eq:A_sugra_control_parameter}, notice that the $x$-derivative removes a factor of $t$ from the integrand, and there is no split between identity and winding terms in the trace. Expressing it in terms of a sum over eigenvalues $\Lambda = x^2 \lambda + \mu$ of the fluctuation operator $D_f$, we have
\begin{eqaed}\label{eq:A_IR_sector}
    \mathcal{A}_\text{IR}^f & \propto \frac{x^d}{d\omega_d v} \int_{t_*}^\infty dt \sum_{\Lambda \in \sigma(D_f)} x^2 \lambda \, e^{-\Lambda \, t} \\
    & = \frac{x^d}{d\omega_d v} \sum_{\Lambda \in \sigma(D_f)} \frac{x^2 \lambda}{x^2 \lambda + \mu} \, e^{-\Lambda t_*}\,.
\end{eqaed}
For non-negative external and internal operators, with zero-modes omitted as following the prescription above, the fraction lies between zero and one. Applying \cref{eq:kato_HSU_bound} at $t=t_*$ yields
\begin{eqaed}
	\abs{\mathcal{A}_\text{IR}^f}
	&\lesssim e^{Et_*}\frac{x^d}{d\omega_d v}K_{S^d}(t_*)K_{M_n}(t_*) \\
	&\lesssim e^{Et_*}\mathcal{A}_\text{IR}^\text{scalar}\,.
\end{eqaed}
The second inequality is obtained by restricting the positive scalar integral in \cref{eq:A_sugra_control_parameter} to $[t_*,2t_*]$. On this interval, $-x\partial_xK_{S^d}(t)\gtrsim K_{S^d}(t_*)$ for bounded $x^2t_*$, including its zero limit. Furthermore, monotonicity gives $K_{M_n}(t)\geq K_{M_n}(2t_*)$. The fixed lower bound on the internal Ricci curvature means that the diagonal scalar heat kernel at $t_*$ is at most a fixed multiple of its value at $2t_*$ \cite[Theorem~2.3]{LiYau1986}. Integrating this pointwise comparison over $M_n$ gives $K_{M_n}(t_*)\lesssim K_{M_n}(2t_*)$, with a constant depending only on dimension, $t_*$ and the Ricci lower bound. \\

\noindent \textbf{The diffeomorphism ghost.} The Lorenz operator defined in \cref{app:spectral_analysis_supergraviton_fluctuations} is given by
\begin{equation}
    G\equiv KK^\dagger=2(\Delta_1-2\operatorname{Ric}+dd^\dagger) \, .
\end{equation}
The additional second-order term $dd^\dagger$ mixes external and internal gradients. For scalar harmonics with $\lambda_\text{s}=l(l+d-1)x^2>0$ and $\mu_\text{s}>0$, its action on the normalized modes $\lambda_\text{s}^{-1/2}dS\,\widetilde S$ and $\mu_\text{s}^{-1/2}S\,d\widetilde S$ is
\begin{equation}
	\frac{G}{2}=
	\begin{pmatrix}
		2\lambda_\text{s}+\mu_\text{s}-2(d-1)x^2 & \sqrt{\lambda_\text{s}\mu_\text{s}}\\
		\sqrt{\lambda_\text{s}\mu_\text{s}} & \lambda_\text{s}+2\mu_\text{s}-2\kappa_{M_n}
	\end{pmatrix} .
\end{equation}
Introducing $\widehat\lambda_\text{s}\equiv\lambda_\text{s}-(d-1)x^2\geq\lambda_\text{s}/d$, direct Laplace inversion gives
\begin{equation}
	\begin{aligned}
		R_l(\mu_\text{s})&\equiv x\partial_x\log\det G
		=\int_0^\infty w_l(t)e^{-\mu_\text{s}t}\,dt \, ,\\
		|w_l(t)|&\leq4\widehat\lambda_\text{s}[1+(d-1)x^2t]e^{-\widehat\lambda_\text{s}t}
		\lesssim\lambda_\text{s}e^{-\lambda_\text{s}t/(4d)} \, .
	\end{aligned}
\end{equation}
The internal transverse modes have eigenvalues $2(\lambda_\text{s}+\mu_\text{v}^\perp-2\kappa_{M_n})$. Hodge decomposition recombines these, both on the quotient and the cover, with the mixed sector into full scalar and one-form traces. The constant-mode correction cancels because $R_l(0)=2+2\lambda_\text{s}/(\lambda_\text{s}-2\kappa_{M_n})$, where $2=x\partial_x\log(4\widehat\lambda_\text{s})$ comes from the genuine $\mu_\text{s}=0$ mode. The $l=0$ and spherical Killing modes have vanishing $x$-derivative, while the other external transverse modes have eigenvalues and multiplicities comparable to the scalar ones up to fixed multiplicative constants. \\

\noindent Applying \cref{eq:kato_HSU_bound} to $\Delta_1-2\kappa_{M_n}=\nabla^\dagger\nabla-\kappa_{M_n}$ bounds the winding contribution to $\mathcal A^\text{ghost}$, integrated over all proper times, in absolute value by scalar winding integrals with a fixed rescaling of the external heat time. The comparison following \cref{eq:x-derivative_HSU} controls the part with $t\leq t_*$ by $\mathcal A_\text{UV}^\text{scalar}$. For $t\geq t_*$, the scalar winding trace satisfies $0\leq W(t)\leq K_{M_n}(t_*)$. By integrating the external factor and using the inequality following \cref{eq:A_IR_sector} we obtain the scalar \ac{IR} bound. \\

\noindent Finally, the mixed matrix satisfies $G\gtrsim(\lambda_\text{s}+\mu_\text{s})\operatorname{id}$ and $\|x\partial_xG\|\lesssim\lambda_\text{s}+\mu_\text{s}$, meaning that $|x\partial_x\Lambda|/\Lambda\lesssim1$. The remaining modes with $x\partial_x\Lambda\neq0$ obey analogous estimates. The original proper time \ac{IR} tails, with weights $(x\partial_x\Lambda/\Lambda)e^{-\Lambda t_*}$, are hence bounded by scalar heat traces evaluated at fixed multiples of $t_*$ on the quotient and cover. The scalar cover identity trace is at most the quotient trace, so the inequality following \cref{eq:A_IR_sector} applies. Subtracting the winding tail from this full winding contribution then yields
\begin{equation}
	\begin{aligned}
		|\mathcal A_\text{UV}^\text{ghost}|&\lesssim
		\mathcal A_\text{UV}^\text{scalar}+\mathcal A_\text{IR}^\text{scalar},\\
		|\mathcal A_\text{IR}^\text{ghost}|&\lesssim\mathcal A_\text{IR}^\text{scalar}.
	\end{aligned}
\end{equation}
For fixed $t_*$ and bounded $x$, the constants are independent of volume and systole, including $x\to0$ for \acp{RFM}. \\

\noindent Non-negativity of the fluctuation operator also applies to graviton fluctuations, except for the scalar-scalar sector. Indeed, as discussed in \cref{app:spectral_analysis_supergraviton_fluctuations,app:positivity_non-scalar_sectors} the only problematic sector is the scalar-scalar one, since it contains mixings as well as negative eigenvalues. As a result, the corresponding heat kernel contains a non-local operator supported on the negative-eigenvalue subspace, after performing the standard steepest-descent procedure \cite{Witten:2010cx}. At first glance, gauge fixing and the York decomposition obscures the locality of the construction, which prevents a straightforward application of the above reasoning. Instead, the inequality must be established after recombining the projected sectors into complete bundle traces by the Hodge and York decompositions. For V-S and V-V, the scalar-zero-mode corrections cancel. Indeed, their eigenvalues both approach $\lambda_\text{v}^\perp + R-2(d-1)x^2 - 2\kappa_{M_n}$ as the internal eigenvalue tends to zero. The rational V-S formula of \cref{eq:V-S_horizontal_eigenvalue_computation} admits the uniform UV expansion used below in \cref{eq:S-S_S-V_UV_expansion_det}, so its remaining scalar trace is handled by the same determinant argument. The S-T, S-V and S-S sectors are recombined below around \cref{eq:BRST_cancellation_det}, keeping their exceptional mode corrections. The remaining finite-dimensional Killing mode corrections and proper time tails are controlled by \cref{eq:A_IR_sector}.

\subsection{Bounding the scalar-scalar sector}\label{app:bounding_scalar-scalar_sector}

As apparent from the analysis of \cref{app:spectral_bounds_scalar-scalar_sector}, the scalar-scalar sector is more involved, and we cannot straightforwardly recycle the above techniques. To proceed, it is simpler to start with the \ac{IR} part: \cref{eq:A_IR_sector} applies \emph{mutatis mutandis} with the eigenvalues taken in absolute value. Then, despite the presence of negative modes, \cref{eq:S-S_eigenvalue_two-sided_bound} implies that
\begin{eqaed}\label{eq:S-S_x-derivative_bound}
    {\frac{\abs{x\partial_x \abs{\Lambda_\text{s-s}}}}{\abs{\Lambda_\text{s-s}}} \lesssim 1}\,,
\end{eqaed}
since the $x$-derivative in the numerator does not affect the upper bound of \cref{eq:S-S_eigenvalue_two-sided_bound}\footnote{The condition $Kh=0$ depends on $x$, meaning that the allowed fluctuation subspace changes when $x$ varies. After normalizing the basis with the Gram matrix, let $\Pi$ denote the orthogonal projector onto this subspace. The derivative of the restricted quadratic form acts both on the operator and on the projections of both arguments. The gauge condition of \cref{eq:S-S_gauge_condition_simplified} and the norms in \cref{eq:S-S_squared_norms} yield $\norm{x\partial_x\Pi}\lesssim1$, such that the projection terms obey the same upper bound $1+\lambda_\text{s}+\mu_\text{s}$ as the operator term. For \acp{RFM}, the homogeneity discussed below \cref{eq:S-S_eigenvalue_two-sided_bound} gives the analogous estimate with $x^2$ in place of $1$ and uniformly as $x\to0$.}, while the corresponding lower bound for the denominator removes the scaling in \cref{eq:S-S_x-derivative_bound}. \\

\noindent Finally, in order to bound the \ac{UV} winding part, we have to express the sum over windings in a convenient fashion. The scalar-scalar harmonics are indexed by a pair of scalar eigenvalues {$\lambda \equiv \lambda_\text{s} = l(l+d-1)x^2$} and $\mu \equiv \mu_\text{s}$. We denote the total eigenvalues by $\Lambda_a$, where $a \in \mathcal{I}_{l}(\mu)$ indexes up to three eigenvalues (which is the generic case). It is useful to express the Schwinger integral according to
\begin{eqaed}\label{eq:S-S_UV_schwinger}
	\mathcal{A}_\text{UV}^\text{s-s} & \propto \int_0^{t_*} dt \int_{[0,\infty)} \sum_{l \geq 0, a \in \mathcal{I}_{l}(\mu)} D_l^{(d)} \, x\partial_x \abs{\Lambda_a} \, e^{-\abs{\Lambda_a}t} \, d\Pi_\mu^\text{winding} \\
	& \equiv \int_{[0,\infty)} f(\mu) \, d\Pi_\mu^\text{winding}\,,
\end{eqaed}
where $D_l^{(d)} = \frac{(2l+d-1)\Gamma(l+d-1)}{\Gamma(d) \Gamma(l+1)}$ are the degeneracies of scalar spherical harmonics. Denote the scalar spectral counting measure of $M_n$, including multiplicities, by $d\Pi_\mu^\text{total}$ and the scalar spectral projection kernel on the universal cover by $d\Pi_\mu(y,y')$. Define $d\Pi_\mu^\text{identity}=\int_{M_n}\text{vol}_{M_n}(y)\,d\Pi_\mu(y,y)$ and $d\Pi_\mu^\text{winding}=d\Pi_\mu^\text{total}-d\Pi_\mu^\text{identity}$ on $[0,\infty)$. For scalar heat kernels, this difference gives the winding trace by the method of images. The spectral pairing involving $f(\mu)$ is understood through the renormalized Schwinger prescription of \cref{eq:sugra_renormalized_A}, with the spherical sum and internal spectral integral performed before the proper-time integration. The most straightforward way to bound \cref{eq:S-S_UV_schwinger} would be to use \cref{eq:S-S_eigenvalue_two-sided_bound}, reducing it to a scalar heat kernel; this unfortunately does not work, because the winding measure $d\Pi_\mu^\text{winding}$ is not guaranteed to be positive. It would be upon adding the identity contribution, but we would have to treat the \ac{UV}-divergent terms explicitly. Alternatively, in the main text, in principle this expression is amenable to an exact trace formula\footnote{More precisely, because of the exceptional cases analyzed in \cref{app:spectral_bounds_scalar-scalar_sector}, $f_t(\mu)$ is discontinuous at $\mu=0$ and the formula requires a correction of the form $f_t^\text{exceptional}(0) - \lim_{\mu \to 0^+}f_t(\mu)$.}, but bounding the Fourier transform seems more difficult. It is more effective to make direct use of the explicit expressions for the fluctuation determinants, without solving complicated secular equation to obtain the eigenvalues. \\

\noindent To this end, notice that the Schwinger integration in $f(\mu)$, for a single pair $(l,\mu)$, leads to
\begin{eqaed}\label{eq:S-S_UV_Schwinger_integration}
    \int_0^{t_*} dt \sum_{a \in \mathcal{I}_l(\mu)} x\partial_x \abs{\Lambda_a} \, e^{-\abs{\Lambda_a}t} = x\partial_x \log \prod_{a \in \mathcal{I}_l(\mu)} \abs{\Lambda_a} - \sum_{a \in \mathcal{I}_l(\mu)} \frac{x\partial_x \abs{\Lambda_a}}{\abs{\Lambda_a}} \, e^{-\abs{\Lambda_a}t_*}\,.
\end{eqaed}
The first term of \cref{eq:S-S_UV_Schwinger_integration} contains the determinant ratio
\begin{eqaed}\label{eq:D_det_ratio}
    \mathcal{D}_{l}(\mu) \equiv \frac{\det(P^\text{t}\mathcal{H}_\text{s-s}P)}{\det(P^\text{t} G_\text{s-s} P)} = \prod_{a \in \mathcal{I}_l(\mu)} \Lambda_a
\end{eqaed}
for each sector labelled by $l,\mu$. The normalization removes the spurious zeros corresponding to unphysical modes removed by gauge-fixing. The second term of \cref{eq:S-S_UV_Schwinger_integration} has a similar structure to the \ac{IR} Casimir parameter: to bound it, one can add and subtract the identity contribution $d\Pi_\mu^\text{identity}$ to the measure of \cref{eq:S-S_UV_schwinger}. The integral against the total (positive) spectral measure of the internal Laplacian is bounded by a scalar heat kernel via \cref{eq:S-S_eigenvalue_two-sided_bound} for the same reason leading to \cref{eq:A_IR_sector,eq:S-S_x-derivative_bound}. The remaining identity term also has a positive measure, and thus \cref{eq:S-S_eigenvalue_two-sided_bound} is bounded universally by the heat kernel (density) of the covering space $\widetilde{M_n}$. Hence, we only need bound the first term of \cref{eq:S-S_UV_Schwinger_integration}, which is determined by the determinant ratios $\mathcal{D}_{l}(\mu)$ we computed in \cref{app:spectral_bounds_scalar-scalar_sector}. \\

\noindent The strategy is then to isolate the terms that dominate the \ac{UV} regime, i.e., those that cannot be bounded by replacing the spectral measure $d\Pi_\mu^\text{winding}$ by the total (positive) spectral measure because they would lead to \ac{UV} divergences. Instead, we seek to express these terms directly in terms of scalar-field heat kernels, and bound the remainder separately. To this end, adding the identity contribution in \cref{eq:A_sugra_control_parameter} we can recast $\mathcal{A}^\text{s-s}_\text{id}+\mathcal{A}^\text{s-s}_\text{UV}$ in a convenient fashion: schematically, letting $f(\mu) = f_\text{UV}(\mu) + f_\text{rest}(\mu)$ denote the aforementioned decomposition, we have
\begin{eqaed}\label{eq:A_sugra_UV+id}
    \mathcal{A}^\text{s-s}_\text{id}+\mathcal{A}^\text{s-s}_\text{UV} & = \left[\int_{\sigma(\Delta^{\widetilde{M_n}})} f(\mu) \, d\Pi_\mu^\text{identity}\right]_\text{renormalized} + \int_{{[0,\infty)}} f(\mu) \, d\Pi_\mu^\text{winding} \\
    & = \left[\int_{\sigma(\Delta^{\widetilde{M_n}})} f_\text{UV}(\mu) \, d\Pi_\mu^\text{identity}\right]_\text{renormalized} + \int_{{[0,\infty)}} f_\text{UV}(\mu) \, d\Pi_\mu^\text{winding} \\
    & + \int_{{[0,\infty)}} f_\text{rest}(\mu) \, d\Pi_\mu^\text{total}\,.
\end{eqaed}
In other words, the simpler structure of $f_\text{UV}(\mu)$ will allow us to express it in terms of (the winding parts of) scalar-field heat kernels---which we have already bounded---without affecting the spectral measure. In light of \cref{eq:S-S_UV_Schwinger_integration}, we can analyze the determinant ratio of \cref{eq:D_det_ratio} directly. \\

\noindent \textbf{Decoding a divergence.} Before moving on, there is a subtlety about \cref{eq:A_sugra_UV+id} that is worth pointing out: the exceptional tower of modes with $\mu=0$ and $l \geq 1$ discussed in \cref{app:spectral_bounds_scalar-scalar_sector} brings additional \ac{UV} divergences that are not cancelled within the S-S sector, but cancel when the S-S and S-V traces are recombined. This can occur because the horizontal projection does not express \cref{eq:D_det_ratio} in terms of local differential operators. To wit, for $l \geq 1$, from \cref{eq:S-V_horizontal_eigenvalue} we have\footnote{For $l=1$, one must compute $\mathcal{D}_1(\mu)$ according to the exceptional modes and their eigenvalues.}
\begin{eqaed}\label{eq:BRST_cancellation_det}
    \lim_{\mu \to 0} \mathcal{D}_{l}(\mu) = \mathcal{D}_{l}(0) \, \Lambda_\text{s-v}\,,
\end{eqaed}
where the right-hand side is evaluated at $\mu_\text{v}^\perp = 0$ and $\lambda_\text{v}^0 = \lambda = l(l+d-1)x^2$. When the transverse one-form trace is written as the full one-form trace minus the scalar trace, the constant scalar must be added back because its gradient vanishes. The resulting S-V term cancels the exceptional S-S correction in \cref{eq:BRST_cancellation_det}, with the same spherical multiplicity and graviton determinant weight. The ensuing treatment therefore applies to the recombined traces. \\

\noindent \textbf{The \ac{UV} part of the determinant ratio.} After Schwinger integration, the \ac{UV} regime corresponding to $t \to 0$ is $\omega \equiv 1 + \lambda + \mu \gg 1$. This also applies to the exceptional sectors discussed in \cref{app:spectral_bounds_scalar-scalar_sector}. Indeed, the derivative part of the fluctuation operator $D_H$ acts as $\omega$ on S-S modes, while the curvature terms remain bounded. As a result, due to \cref{eq:S-S_eigenvalue_two-sided_bound}, we can expand \cref{eq:D_det_ratio} around the leading contribution $\mathcal{D}_{l}(\mu) \propto \prod_{a \in \mathcal{I}_l(\mu)} \omega$. To this end, we define $\lambda \equiv \theta \, \omega$ and $\mu \equiv (1-\theta)\omega - 1$ and expand $x\partial_x \log \abs{\mathcal{D}_l(\mu)}$ for $\omega \gg 1$ uniformly in $0 \leq \theta \leq 1$. The resulting expansion takes the schematic form
\begin{eqaed}\label{eq:S-S_S-V_UV_expansion_det}
    x\partial_x \log \abs{\mathcal{D}_l(\mu)} = \sum_{k=0}^5 \frac{\text{poly}_k(\theta)}{\omega^k} + \mathcal{R}_k(\theta,\omega)\,,
\end{eqaed}
where $\text{poly}$ denotes a polynomial and $\abs{\mathcal{R}_k(\theta,\omega)} \lesssim \omega^{-6}$. The choice of truncating \cref{eq:S-S_S-V_UV_expansion_det} at this order has the same rationale as the renormalization of the heat kernel: in eleven dimensions, the remainder $\mathcal{R}_k(\theta,\omega)$ can be integrated against the full spectral measure (including the identity part) without \ac{UV}-divergences. \cref{eq:S-S_S-V_UV_expansion_det} follows from the fact that $\mathcal{D}_l(\mu)$ is a ratio of polynomials, and factoring out the leading factors of $\omega$ one can expand the remaining fraction according to $1 + \text{poly}(\theta)/\omega + \dots$ as above. Then one can apply $x\partial_x \theta = 2\theta(1-\theta)$, and analogously for $\omega$. \\

\noindent Each polynomial then comprises terms of the form $\theta^m/\omega^k = \lambda^m/\omega^{m+k}$. Applying Schwinger's trick $1/x^k = \frac{1}{\Gamma(k)}\int_0^\infty dt \, t^{k-1} \, e^{-x t}$ results in scalar-field heat kernels, after summing over spherical harmonics and integrating against the spectral measure $d\Pi_\mu^\text{winding}$ into $K_{M_n}^\text{winding}(t)$. With this, one can then bound the spherical part by iteratively using again that $z \, e^{-z} \leq \frac{2}{e}e^{-z/2}$ for $z\geq0$ and rescaling the argument $t$ (up to introducing numerical prefactors) as done previously\footnote{In the case of \acp{RFM}, the horizontal projection removes $b_1(M)$ additional modes, which are Killing (co)vectors coming from S-V harmonics with $\mu_\text{v}^\perp = 0$. For $l\geq1$, this subtraction is cancelled by the $b_1(M)$ correction in the York decomposition of the S-T trace. Indeed, $\Lambda_\text{s-v}|_{\mu_\text{v}^\perp=0} = \Lambda_\text{s-t}|_{\mu_\text{t}^\text{TT}=0}$. The remaining $l=0$ correction is finite-dimensional and is bounded by \cref{eq:A_IR_sector}.}. \\

\noindent \textbf{The remainder part of the determinant ratio.} The only thing left to obtain a parametric bound for the supergravity Casimir parameter $\mathcal{A}^\text{SUGRA}$ is to bound the remainders $\mathcal{R}_k(\theta, \omega)$ of \cref{eq:S-S_S-V_UV_expansion_det}. Because the combination $\theta$ is bounded, we have $\abs{\mathcal{R}_k(\theta,\omega)} \lesssim \omega^{-6}$. Therefore, by including the identity term as in \cref{eq:A_sugra_UV+id}, we can bound the integral against the full spectral measure according to
\begin{eqaed}\label{eq:remainder_total_spectral_integral}
    \sum_{l \geq 0} D_l^{(d)}\int_{{[0,\infty)}} \abs{\mathcal{R}_k(\theta, \omega)} \, d\Pi_\mu^\text{total} & \lesssim \Tr_{S^d \times M_n} (1+\Delta^{S^d} + {\Delta^{M_n}})^{-6} \\
    & \propto \int_0^\infty dt \, t^5 \, e^{-t} \, K_{S^d}(t) \, K_{M_n}(t)\,.
\end{eqaed}
Putting everything together, we obtain a parametric bound for the full supergravity Casimir parameter $\mathcal{A}^\text{SUGRA}$ in terms of the bounds derived in the scalar-field case

\subsection{Bounding gradients of the supergravity potential}\label{app:bounding_sugra_gradient}

In the preceding sections we established that the scalar-field bounds on the \ac{EFT} control parameters extend---at least parametrically---to the full supergravity spectrum. This implies that parametrically controlled critical points of the Casimir potential must approach a degeneration limit $s \equiv \text{sys}(M_n) \to 0$. For rigid manifolds, this is enough; here we provide a brief argument to include more general critical points for Riemann surfaces and \acp{RFM}. \\

\noindent \textbf{Casimir gradients on Riemann surfaces.} In the case of Riemann surfaces, the leading short-geodesic contribution arises from the \ac{UV} region of the heat kernel. In this limit, the curvature endomorphisms are subleading, and the leading contribution is proportional to the holonomies of short geodesics. Antiperiodic holonomies contribute a total of $128+128=256$, while periodic ones contribute $128-128=0$ to leading order. To treat periodic, antiperiodic and mixed degeneration limits uniformly, fix a sufficiently small length cutoff $\ell_0$. Each primitive closed geodesic with $\ell_\gamma<\ell_0$ is the central geodesic, or core, of a disjoint collar. Let $\sigma$ parameterize a deformation through hyperbolic metrics that lengthens all these cores simultaneously. The infinitesimal change of the metric, measured relative to the metric itself, can be bounded independently of genus and of the core lengths. The deformation can be chosen so that $\frac{1}{2}\leq\partial_\sigma\log\ell_\gamma\leq\frac{3}{2}$ for every short core\footnote{Multiply the angular metric variation (along the closed curves around the collar) by a smooth function equal to one near the core and zero near the boundary. This confines the curvature error $f$ to the transition region. Linearized uniformization solves $(\Delta+2)u=f$, and the maximum principle bounds $u$ uniformly. In the part of the collar between the core and the transition region, where $f=0$ and $|q|\leq w$, comparison with a constant multiple of $\frac{\cosh q}{\cosh w}$ gives an $O(e^{-w})$ bound at the core, since $(\Delta+2)\cosh q=\frac{1}{\cosh q}>0$. Here $q$ is signed distance from the core and $w\sim\log\frac{1}{\ell_\gamma}$, so this correction tends to zero for short cores.}. We separate the leading antiperiodic contribution from the remainder by writing
\begin{equation}
	A=A_{\text{ap}}+A_{\text{rem}},\qquad
	A_{\text{ap}}=\frac{c}{v}\sum_{\substack{\gamma\ \text{antiperiodic}\\\ell_\gamma<\ell_0}}\ell_\gamma^{-10},\qquad c>0,
\end{equation}
where each geometric core is counted once. The positive, $x$-independent coefficient $c$ includes a factor of two for the two directions around each core and the sum $\sum_{j=1,3,5,\ldots}\frac{1}{j^{11}}$ over repeated traversals. Only odd $j$ contribute at leading order: after $j$ traversals of an antiperiodic core, the fermion sign is $(-1)^j$, giving the supergravity weight $128[1-(-1)^j]$. The curvature and lower-order operator corrections to the leading flat-space contribution enter the short-time heat-kernel expansion through combinations such as $tR$ and $t\mathcal E$, since proper time has dimensions of length squared. Since the winding factor is $e^{-\frac{(j\ell_\gamma)^2}{4t}}$, each extra power of $t$ gives two powers of $\ell_\gamma$ after Schwinger integration. The first correction is therefore $O(\ell_\gamma^{-8})$. The geodesic prefactor also has a relative correction of order $(j\ell_\gamma)^2$, which gives the same power. A logarithmic length derivative preserves these powers. Applying the finite subtraction of \cref{eq:S-S_S-V_UV_expansion_det} to the normalized log determinants leaves $\mathcal R=O(\omega^{-6})$ and $\partial_\mu\mathcal R=O(\omega^{-7})$, where $\omega=1+\lambda+\mu$. The uniform bound on the relative metric variation ensures that the shape derivative of the remainder is still trace-class. Packing estimates then give an $O(vs^{-2})$ bound for its trace and for the contribution away from short cores. Small scalar eigenvalues satisfy $|\partial_\sigma\mu|\lesssim\mu$, so their logarithmic derivatives remain bounded. Their shape-dependent $x^9\log\mu$ terms cancel from $\mathcal A$, while the fermionic product operators retain the positive sphere gap. {Consequently, uniformly in a fixed neighborhood of $x_*=\frac{1}{6}$, which contains every sufficiently controlled stationary ratio,}
\begin{equation}
	|\partial_\sigma A_{\text{rem}}|
	+\left|A_{\text{rem}}-\frac{x}{9}\partial_x A_{\text{rem}}\right|
	\lesssim 1+s^{-2}+\frac{1}{v}\sum_{\ell_\gamma<\ell_0}\ell_\gamma^{-8}
	\lesssim 1+s^{-8},
\end{equation}
where the last inequality uses the collar bound of $3g-3$ short geometric cores and $v=4\pi(g-1)$. At a shape critical point, $\partial_\sigma A=0$. Since $\partial_\sigma A_{\text{ap}}\leq-5A_{\text{ap}}$, it follows that $5A_{\text{ap}}\leq|\partial_\sigma A_{\text{rem}}|$. Using also $\partial_x A_{\text{ap}}=0$, the bound above gives $\mathcal A\lesssim1+s^{-8}$. Consequently, $M_{11}rs\lesssim s(1+s^{-8})^{\frac{1}{9}}\to0$, even when the core lengths approach zero at different rates. \\

\noindent \textbf{Potential gradients on Riemann-flat manifolds.} Suppose that a parametrically controlled sequence of critical points exists. Select a subsequence and group the successive lattice minima into blocks of dimensions $m_a$, with comparable lengths within each block and diverging ratios between adjacent blocks. Holonomy maps lattice vectors to lattice vectors of the same length. It therefore preserves all linear combinations of the vectors shorter than a chosen length, as well as the perpendicular directions. Choosing this length between successive blocks shows that each block can be rescaled independently without breaking the holonomy symmetry. Denote their lengths in eleven-dimensional Planck units by $u_a$, with $u_{\min}=\min_a u_a$ differing from $M_{11}rs$ by bounded factors of positive sign. Consider
\begin{equation}
	Q=(M_{11}r)^{-8}\mathcal Q_N,\qquad
	C=(M_{11}r)^{-11}\mathcal C,\qquad
	c=\frac{\alpha}{\alpha+3},\qquad \delta=\frac{2}{3d},
\end{equation}
where $C$ is the flat space Casimir density, with the sign convention of \cref{eq:dd_eff_potential_recast}. First, assume that $C>0$. Using a lattice basis adapted to these length scales, write the leading flux density and the positive Casimir contributions as
\begin{equation}
	Q_0=\sum_I q_I\prod_a u_a^{-b_{Ia}},\qquad
	C_0=\sum_J c_J\prod_a u_a^{-a_{Ja}},\qquad
	\sum_a b_{Ia}=8,\qquad \sum_a a_{Ja}=11.
\end{equation}
Here $a_{Ja}\geq0$, every non-zero $q_I\gtrsim1$, and $c_J>0$ with $\sum_Jc_J\lesssim1$. Indeed, the  mixing between blocks is suppressed by the ratios between the shorter to longer lengths, leaving positive quadratic forms in integral flux periods on bounded, non-degenerate block metrics. Each non-zero coefficient below is therefore bounded below due to quantization, even when the flux integers vary. The Casimir sum is obtained by averaging over the rotated directions and then performing a Poisson resummation along the shorter fixed directions\footnote{{The integrated image kernel is of degree eleven in the inverse length. The bounded order of the holonomy group bounds the denominators of the fractional lattice shifts and hence the resummed coefficients.}}. The errors $Q-Q_0$ and $C-C_0$, and their first logarithmic block derivatives, become negligible relative to $Q_0$ and $C_0$, respectively. \\

\noindent Denote the exponent averages weighted by their contributions to $Q_0$ and $C_0$ by $\bar b_a$ and $\bar a_a$. Radial stationarity and the shape equations of \cref{eq:dm_critical_point_RFM} imply, respectively,
\begin{equation}
	\frac{Q}{C}=\frac{1}{c}+\varepsilon_r,
	\qquad \bar b_a=c\bar a_a-\delta m_a+\varepsilon_a,
\end{equation}
where $\varepsilon_r$ and $\varepsilon_a$ tend to zero along the sequence. Applying the log-sum inequality and the coefficient bounds to these positive sums gives
\begin{equation}
	\log\frac{Q_0}{C_0}
	\geq\sum_a(\bar a_a-\bar b_a)\log u_a-B
	\geq3\log u_{\min}-B.
\end{equation}
Here $B$ is a fixed constant along the sequence. The second inequality follows because, sufficiently far along the sequence,
\begin{equation}
\begin{aligned}
	\bar a_a-\bar b_a&=(1-c)\bar a_a+\delta m_a-\varepsilon_a>0,\\
	\sum_a(\bar a_a-\bar b_a)&=11-8=3.
\end{aligned}
\end{equation} Hence $Q/C\gtrsim u_{\min}^3$, contradicting radial stationarity as $u_{\min}\to\infty$. \\

\noindent To include external curvature, set $S=M_{11}rs$ and $T=M_{11}L$. Continuing to neglect flux-dependent terms in the kinetic operators, denote any correction to the radius or logarithmic block-shape equations, in the density normalization used above, by $\mathcal R$. Then
\begin{equation}
	\abs{\mathcal R}\leq C_{\text{rem}}\left(T^{-2}S^{-9}+T^{-11}+T^{-d}S^{-n}\right),
\end{equation}
where $C_{\text{rem}}$ is independent of the growing length scales.
The first term is obtained from the extra proper-time factor accompanying the curvature contribution and the packing estimate. The last two terms follow from heat-kernel bounds on the number of modes whose absolute eigenvalues are comparable to $T^{-2}$ or smaller. The same estimate controls the contribution of the common-scale logarithm to these equations. For $T\leq S$, the full one-loop radius response is bounded by a constant times $T^{-11}$, incapable of balancing the classical curvature term of order $T^{-2}$ in a controlled limit. Hence $T>S$. Since $d+n=11$,
\begin{equation}
	\frac{\abs{\mathcal R}}{T^{-2}}\leq3C_{\text{rem}}S^{-9}\longrightarrow0 \, .
\end{equation}
Thus the corrections become negligible relative to the classical curvature term. If all flat-space contributions vanish, radius balance already excludes control. Otherwise, it bounds $Q$ and $C$ above and below by positive multiples of $T^{-2}$. The relative errors in the flat-space stationarity relations are, consequently, vanishing, and the preceding contradiction applies.

\begin{acronym}

\tooltipacro{QFT}{quantum field theory}
\acrodefplural{QFT}{quantum field theories}

\tooltipacro{EFT}{effective field theory}
\acrodefplural{EFT}{effective field theories}

\tooltipacro{CFT}{conformal field theory}
\acrodefplural{CFT}{conformal field theories}

\tooltipacro{UV}{ultraviolet}
\tooltipacro{IR}{infrared}

\tooltipacro{dS}{de Sitter}
\tooltipacro{AdS}{anti-de Sitter}

\tooltipacro{RFM}{Riemann-flat manifold}
\acrodefplural{RFM}{Riemann-flat manifolds}

\tooltipacro{KK}{Kaluza-Klein}
\tooltipacro{ESC}{emergent string conjecture}
\tooltipacro{TCC}{trans-Planckian censorship conjecture}

\end{acronym}

\bibliographystyle{ytphys}
\baselineskip=.95\baselineskip
\bibliography{ref}

\end{document}